\documentclass[12pt,twoside,a4paper,final]{report}

\usepackage{sidecap}
\usepackage{amssymb,amsmath,amsfonts} 
\usepackage{graphicx,color} 
\usepackage{booktabs} 
\usepackage{rotating} 
\usepackage{textcomp} 
\usepackage{epstopdf}
\usepackage{lscape}
\usepackage{setspace}
\usepackage{chapterbib}
\usepackage{changepage}
\usepackage{titlesec}
\usepackage{color}

\usepackage{fancyhdr}
\usepackage[left=20mm, top=20mm, bottom=20mm, right=20mm]{geometry}
\usepackage[font=onehalfspacing]{caption}

\usepackage{type1cm}
\usepackage{lettrine}

\usepackage{changepage}

\titleformat{\chapter}[display] {\bfseries\normalfont\Large\filcenter} {\titlerule[1pt]
\vspace{1pt}
\titlerule
\vspace{1pc}
 \LARGE\MakeUppercase{\chaptertitlename} \thechapter}
{1pc} {\titlerule
\vspace{1pc}
  \Huge}

\titlespacing{\chapter}{0pt}{*0}{*8}

\onehalfspace
\newcommand {\scr}[1] {\mathcal{#1}}
\newcommand {\tbf}[1] {\textbf{#1}}
\newcommand {\tit}[1] {\textit{#1}}

\newcommand {\trm}[1] {\textrm{#1}}

\newcommand {\impm} {\rightarrow}   

\newcommand {\del} {\partial}
\newcommand{\eqn}[1]{\begin{eqnarray} #1 \end{eqnarray}}

\newcommand {\bra}[1] {\langle #1 |}
\newcommand {\ket}[1] {| #1 \rangle}
\newcommand {\braket}[2] {\langle #1 | #2 \rangle} 
\newcommand {\bk}[1] {\langle #1 \rangle} 
 
\newcommand {\trc}[1] {\text{Tr}\{#1\} }
\newcommand {\ptrc}[2] {\text{Tr}_{#1}\{#2\} }
\newcommand {\brac}[1] {\left(#1\right)}
\newcommand {\sqbrac}[1] {\left[#1\right]}
\newcommand {\densop}[2] {| #2 \rangle \langle #1 | } 
\newcommand {\invard}[1] {\mathrm{d}\tilde{\textbf{#1}} }
\newcommand {\iinvard}[1] {\mathrm{d}\tilde{\textrm{#1}} }

\begin{document}
\vspace{30mm}
\begin{titlepage}
\begin{center}
\includegraphics[width=0.7\textwidth]{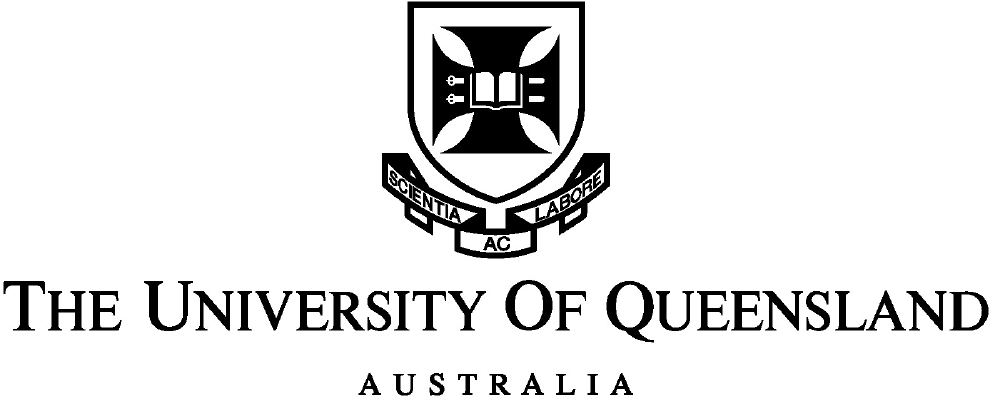}\\[1cm]    
\vspace{20mm}

{ \Huge \bf Causality Violation\\[0.4cm] and Nonlinear Quantum Mechanics}\\[0.4cm]
\vspace{-10mm}
\vspace{20mm}
{\large
\begin{center}
Jacques~L.~Pienaar\\
B.Sc. (Hons I) The University of Melbourne\\
\vspace{20mm}
\end{center}
}
\vspace{20mm}

\emph{A thesis submitted for the degree of Doctor of Philosophy at \\
The University of Queensland in 2013}\\ 
School of Mathematics and Physics
\vspace{5mm}

\end{center}
\end{titlepage}

\newpage

\chapter*{Abstract}
\addcontentsline{toc}{section}{Abstract}
\pagenumbering{roman}
\noindent It is currently unknown whether the laws of physics permit time travel into the past. While general relativity indicates the theoretical possibility of causality violation, it is now widely accepted that a theory of quantum gravity must play an essential role in such cases. As a striking example, the logical paradoxes usually associated with causality violation can be resolved by quantum effects. We ask whether the explicit construction of a theory that allows causality violation might in turn teach us something about quantum gravity. Taking the toy model of Deutsch\footnote{Deutsch, D. Quantum mechanics near closed timelike lines. Phys. Rev. D 44, 3197Ð3217 (1991).} as a starting point, in Part~\ref{part:LEN} we argue that, despite being a nonlinear modification of quantum mechanics, the model does not imply superluminal signalling and its predictions can be operationally verified by experimenters within an appropriate ontological setting. In Part~\ref{part:HEN} we extend the model to relativistic quantum fields. The detailed outline is described below.

\section*{Thesis outline}
The thesis is divided into two parts, according to the physical setting. The first part is based on the formalism of standard quantum mechanics, which might be referred to as ``nonrelativistic" because all observers and (massive) systems of interest are assumed to be approximately at rest in the laboratory frame. However, the formalism does contain at least one `relativistic' element: light is assumed to travel at a finite speed between systems that are separated by large distances in space. Within this regime, therefore, it is legitimate to consider the problem of superluminal signalling between distant systems, leading us to refer to this regime simply as ``low energy". The second part of the thesis is based on the formalism of relativistic quantum field theory, particularly quantum optics. This formalism extends to situations where the systems of interest are in relativistic motion with respect to each other and the laboratory, in which case it becomes desirable to use a formalism whose physical predictions are invariant under Lorentz transformations. Accordingly, we refer to this as the ``high energy" regime. In each part, we first review the standard formalism and then introduce modifications of the formalism in order to accommodate possible violations of causality, in a manner consistent with Deutsch's model. The first part is primarily concerned with assessing the consistency of Deutsch's original model, as indicated by the extent to which it respects operational notions of locality and verifiability. The second part is concerned with generalising Deutsch's model to include the spatial and temporal properties of quantum fields interacting with the causality-violating region.

\subsection*{Part~\ref{part:LEN}: The low energy regime} 
Chapter~\ref{ChapQM} gives an overview of standard quantum mechanics in the nonrelativistic regime, including the quantum circuit formalism and the operational formalism. In Chapter~\ref{ChapDeutsch}, we review Deutsch's toy model of causality violation and the surrounding literature on the topic. In order to better understand the physical interpretation of the model, we place it in the context of the broader literature on nonlinear modifications to quantum mechanics. In Chapter~\ref{ChapNonlinBox} we identify Deutsch's model as an example of a \tit{nonlinear box}. We show that nonlinear boxes are amenable to an operational description, which allows us to clarify whether they allow superluminal signalling and whether their predictions can be verified in principle. We conclude that Deutsch's model is non-signalling and verifiable, the latter condition being contingent on an appropriate \tit{ontological model}, an example of which is briefly discussed.

\subsection*{Part~\ref{part:HEN}: The high energy regime}
In Chapter~\ref{ChapFields}, we briefly review relativistic quantum field theory and quantum optics. Using this formalism, we describe relativistic effects in general quantum circuits. This provides a basis for future work on quantum information and quantum computation in relativistic settings, while also providing us with the formal tools needed for the following chapters. In Chapter~\ref{ChapCTC1} we apply these tools to quantum states of light interacting with a closed timelike curve. We find that even in trivial cases where the CTC does not produce any interactions between the past and future parts of the field, the CTC can still be exploited to violate Heisenberg's uncertainty relation for quantum optics, allowing the perfect cloning of coherent states. Finally, in Chapter~\ref{ChapCTC2} we propose a generalisation of Deutsch's model that extends its regime of applicability to fields whose temporal uncertainty exceeds the size of the temporal jump induced by the CTC. This allows us to consider scenarios in which the CTC becomes too small to have any experimentally observable effects, in which limit we smoothly recover standard quantum optics. We conjecture that our generalised formalism might also provide an alternative model for describing the effect of gravitational time-dilation on entangled quantum systems. Based on this conjecture, we draw a connection to the theory of ``event operators" proposed earlier in the literature\footnote{Ralph, T. C., Milburn, G. J. \& Downes, T. Quantum connectivity of space-time and gravitationally induced decorrelation of entanglement. Phys. Rev. A 79, 022121 (2009).}, and discuss the related possibility of an experimental test in Earth's gravitational field.


\newpage

\chapter*{Declaration by author}
\addcontentsline{toc}{section}{Declaration by author}

\noindent This thesis is composed of my original work, and contains no material previously published or written by another person except where due reference has been made in the text. I have clearly stated the contribution by others to jointly-authored works that I have included in my thesis.\\

\noindent I have clearly stated the contribution of others to my thesis as a whole, including statistical assistance, survey design, data analysis, significant technical procedures, professional editorial advice, and any other original research work used or reported in my thesis. The content of my thesis is the result of work I have carried out since the commencement of my research higher degree candidature and does not include a substantial part of work that has been submitted to qualify for the award of any other degree or diploma in any university or other tertiary institution. I have clearly stated which parts of my thesis, if any, have been submitted to qualify for another award.\\

\noindent I acknowledge that an electronic copy of my thesis must be lodged with the University Library and, subject to the General Award Rules of The University of Queensland, immediately made available for research and study in accordance with the \textit{Copyright Act 1968}.\\

\noindent I acknowledge that copyright of all material contained in my thesis resides with the copyright holder(s) of that material. Where appropriate I have obtained copyright permission from the copyright holder to reproduce material in this thesis.

\newpage
\section*{Publications during candidature}
\addcontentsline{toc}{section}{List of Publications}
\subsubsection*{Peer-reviewed publications}
{\small
\begin{enumerate}
\item{\textsl{Space-time qubits}, \\  
Pienaar, J. L., Myers, C. R., Ralph, T. C. 
\textsl{Physical Review A} {\bf 84}, 022315 (2011). (\textsl{arXiv:1101.4250v2})}
\item{\textsl{Quantum fields on closed timelike curves}, \\  
Pienaar, J. L., Myers, C. R., Ralph, T. C. 
\textsl{Physical Review A} {\bf 84}, 062316 (2011). (\textsl{arXiv:1110.3582v2})}
\item{\textsl{Open timelike curves violate HeisenbergÕs uncertainty principle}, \\  
Pienaar, J. L., Myers, C. R., Ralph, T. C. 
\textsl{Physical Review Letters} {\bf 110}, 060501 (2013). (\textsl{arXiv:1206.5485v2})}
\end{enumerate}
}
\subsubsection*{Online e-print publications}
{\small
\begin{enumerate}
\setcounter{enumi}{3}
\item{\textsl{The preparation problem in nonlinear extensions of quantum theory}, e-print \textsl{arXiv:1206.2725v1}, (2012).} \\
Cavalcanti, E. G., Menicucci, N. C., Pienaar, J. L.
\end{enumerate}
}
\vspace{10mm}

\section*{Publications included in this thesis}
No publications included.

\vspace{10mm}

\section*{Contributions by others to this thesis}

The results described in Chapter~\ref{ChapNonlinBox} are based on joint work undertaken with Dr. Nicolas Menicucci and Dr. Eric Cavalcanti of the University of Sydney (see Ref. [4] above). All three authors contributed equally to this work. For the remaining work in this thesis, credit is to be shared equally between the author and his advisory team, Dr. Casey Myers and Prof. Timothy Ralph.

\vspace{10mm}

\section*{Statement of parts of the thesis submitted to qualify for the award of another degree}
None.

\newpage

\newpage
\chapter*{Acknowledgements}
\addcontentsline{toc}{section}{Acknowledgements}
\vspace{-10mm}
{ \footnotesize
Let me begin by acknowledging the support of the Australian Government for awarding me an Australian Postgraduate Award (APA), without which this work would not have been possible. In addition, I owe much to the advice and support of my parents; thanks Mum and Dad. Thanks to Professors Lloyd Hollenberg, Andrew Greentree and Jeffrey McCallum at the University of Melbourne for giving me excellent advice about pursuing a PhD. Thanks to uncle Bart and aunt Pat for looking after me until I found my feet in Brisbane. Special thanks to Aggie Branczyk and Charles Meaney for stimulating discussions, Rob Pfeifer for providing the expertise in our physics reading group and Matt Broome, Leif Humbert and the rest of the motley crew from Teach at the Beach. Thanks to Guifre Vidal for his hospitality and humour, thanks to Andrew Doherty for running the QFT reading group, thanks to Gerard Milburn for being an excellent host and for hanging out with me one time at the James Squire brewhouse. I owe a debt of gratitude to Professor Ping Koy Lam for being kind enough to host me at the Australian National University during my stay in Canberra; I cannot express how valuable this experience was to the completion of my PhD. Thanks also to the quantum theory group at ANU for making me feel welcome, particularly Craig Savage for enlightening discussions, Helen Chrzanowski for lending me her laptop and to Michael Hush for introducing me to the subtleties of the many-worlds interpretation. Thanks to Nick Menicucci and Eric Cavalcanti for hosting me at the University of Sydney and for being good mentors and friends, not to mention excellent collaborators. Thanks also to the rest of the quantum theory group at the University of Sydney for making my time in Sydney extremely lively. Thanks to all of my old friends, Aviral, Brett, Daniel, Eric, Grant and Sunil for keeping me grounded in the real world. Thanks to the members of LEG for showing up week after week, especially Andrew Birrell and Nathan Walk for sounding out even the craziest ideas. Thanks to the rest of the QOQI group at UQ, particularly Saleh Rahimi-Keshari for teaching me how to mix shisha tobacco. Thanks to Vincent Lam for introducing me to the philosophy of physics, which has influenced this thesis in innumerable subtle ways. Thanks to Ruth Forrest, Danielle Faccer and Kaerin Gardner whose efficiency and organisational abilities have contributed to the progress of academia more than they might realise. Thanks to everyone in the physics department at UQ with whom I have had only a passing correspondence, but who contributed to the rich academic environment that continues to thrive within the walls of Parnell and the Annexe. Thanks to Joel Corney and Murray Kane for their support and patience. Thanks to my associate advisor Casey Myers, for his sober and methodic approach to physics that has saved me from many a blunder and taught me the value of patience and persistence. These attributes, balanced by less sober anecdotes and wry depictions of academia after a few beers, made working with Casey an educational and entertaining experience. When I was asking around for advice, I was told that every PhD student reaches a point at which they feel their work is irrelevant, that they are sick of looking at it, and it is the job of a good supervisor to bring the student through this difficult stage. My own PhD was not without its complications, but never once did I feel that what I was doing was irrelevant or unimportant, never once did I think that my problems were insurmountable and not for a moment did I stop enjoying myself. I relished every second of my PhD, for which I have my advisor Tim Ralph to thank -- for his tactfulness and wisdom in matters both personal and academic, and his enduring enthusiasm for our work. Thanks Tim!
}

\newpage

\subsection*{Keywords}
quantum optics, general relativity, quantum computation, quantum information

\vspace{10mm}

\section*{Australian and New Zealand Standard Research Classifications (ANZSRC)}
ANZSRC code: 020603, Quantum Information, Computation and Communication, 40\% \\
ANZSRC code: 020604, Quantum Optics, 40\% \\
ANZSRC code: 020699, Quantum Physics not elsewhere classified, 20\% 

\vspace{10mm}

\section*{Fields of Research (FoR) Classification}
FoR code: 0206, Quantum Physics, 80\% \\
FoR code: 0299 Other Physical Sciences 20\% 

\newpage
\thispagestyle{plain}
\mbox{}

\clearpage

\tableofcontents

\listoffigures

\newpage
\chapter*{List of Abbreviations used in the thesis}
\addcontentsline{toc}{section}{List of Symbols}
\vspace{-10mm}
The following is a list of notational conventions and symbols used in this thesis. It is not comprehensive, but may serve as a useful guide to those who become lost in the forest of indices, hats and greek letters. Whenever exceptions apply, the different usage of the symbol will be made clear from the context.\\

{\small
\begin{itemize}
\item{Script letters $\mathcal{H},\mathcal{M},\mathcal{N}$ are reserved for abstract spaces and maps on those spaces. The details will be made clear from the context.}
\item{The greek letter $\rho$ is used almost exclusively to denote a density matrix representing a quantum state.}
\item{Generally $i$ stands for the imaginary number $\sqrt{-1}$, but it is frequently also used as an index, as in $\delta_{i,j}$, in cases where confusion is unlikely to occur.}
\item{We use some common mathematical notation for sets and logic, including curly brackets $\{ ... \}$ to denote a set of elements, and the symbols $\in, \subset, \forall, ...$ whose meaning is expected to be familiar to the reader, as are standard terms like ``union" and ``intersection" pertaining to sets.}
\item{All standard mathematical notation corresponding to linear algebra is assumed knowledge, particularly the trace operation $\trc{...}$ and tensor product $\otimes$.}
\item{The Levi-Civita symbol $\epsilon_{i j k ...}$ is equal to:\\
\eqn{
\epsilon_{i j k ...} := \begin{cases} 1 & \trm{if (i,j,k,...) is an even permutation of (1,2,3,...)} \, , \\
 -1 & \trm{if (i,j,k,...) is an odd permutation of (1,2,3,...)} \, , \\
 0 & \trm{otherwise}
\end{cases} \, \\
}
}
\item{The symbol $:=$ means ``is defined to be".}
\end{itemize}
}

Abbreviations:
{\small
\begin{itemize}
\item{\tbf{iff}: if and only if}
\item{\tbf{GR}: general relativity}
\item{\tbf{QM}: quantum mechanics}
\item{\tbf{CTC}: closed time-like curve}
\item{\tbf{OTC}: `open' time-like curve}
\item{\tbf{QFT}: quantum field theory}
\item{\tbf{SPDC}: spontaneous parametric down converter/conversion}
\end{itemize}
}

\newpage

\begin{quote}
\em
[W]hy have physicists found it so hard to create a quantum theory of gravity? [...] The real heart of the matter is that general relativity is a theory of spacetime itself, and so a quantum theory of gravity is going to have to be talking about superpositions over spacetime and fluctuations of spacetime. One of the things you'd expect such a theory to answer is whether closed timelike curves can exist. So quantum gravity seems `CTC-hard', in the sense that it's at least as hard as determining if CTCs are possible! [...] Of course, this is just one instantiation of a general problem: that no one really has a clear idea of what it means to treat spacetime itself quantum mechanically.
\end{quote}
Scott Aaronson\\

\begin{quote}
\em
\raggedright
What is space and what is time? This is what the problem of quantum gravity is about.
\end{quote}
Lee Smolin

\pagenumbering{arabic}

\pagestyle{fancy}
\renewcommand{\sectionmark}[1]{\markboth{\thesection.\ #1}{}}

\chapter*{Introduction}
\addcontentsline{toc}{part}{Introduction}
\label{ChapINTRO}

\section*{Why study causality violation?}
\addcontentsline{toc}{section}{Why study causality violation?} 

The Stanford Encyclopedia of Philosophy has an excellent article on the topic of time travel in modern physics, focussing primarily on the situation in classical physics\cite{SEP}. After exhaustively demonstrating that there exist self-consistent solutions for a wide range of scenarios involving time travel, the authors conclude with the following cautionary statement:\\
\tit{``If time travel entailed contradictions then the issue would be settled.[...] But if the only requirement demanded is logical coherence, then it seems all too easy. A clever author can devise a coherent time-travel scenario in which everything happens just once and in a consistent way. This is just too cheap: logical coherence is a very weak condition, and many things we take to be metaphysically impossible are logically coherent."}
The authors give the example of Aristotle, who conjectured that water was infinitely divisible. While not logically inconsistent, Aristotle's view is incorrect according to modern chemistry -- but this was determined by empirical facts and not by any logical or conceptual analysis. On one hand, there is nothing unscientific about Aristotle's conjecture; science progresses by making hypotheses and doing experiments to eliminate the wrong ones. On the other hand, a theory must be motivated by more than just the requirement that it be self-consistent.

There are at least two reasons to try and construct an explicit model of causality violation. One simple reason is that if such a construction fails, it might strengthen the case for ruling out time travel as unphysical, or even lead to a no-go theorem along these lines. However, the results of the present thesis indicate that, if one is willing to accept some transgressions of the accepted laws of quantum mechanics (QM), there may be no obvious theoretical reason to rule out causality violation. But where does that leave us? The answer lies in the second reason for looking at causality violation: we do not yet understand how causal structures enter quantum mechanics at the most fundamental level.

To date, the accepted formulations of quantum mechanics rely on the \tit{classical} concept of a background spacetime, which in turn dictates the causal structure that is to be obeyed by the theory. If, however, we wish to consider the back-action on the metric of massive quantum systems or formulate quantum mechanics at the Planck scale, it no longer makes sense to invoke a classical background. But what must take its place? One answer might be the ``spinfoam" of loop quantum gravity (LQG), but this presupposes that violations of causality are forbidden. Is this supposition justified in the absence of a given classical background? One way to find out would be to investigate similar structures that allow more general causal structures, based upon formulations of quantum mechanics in causality-violating metrics. Results in the literature already suggest that, in the absence of any externally imposed causal structure, quantum mechanics might allow ``indefinite" causal structures that are not consistent with any familiar classical background metric\cite{ORE12}. Hence a study of causality violation might improve our understanding of what kinds of causal structures might appear in quantum gravity.

Finally, we note that the formalism developed in this thesis may ultimately lead to testable predictions that could confirm or deny the possibility of such effects in nature. While it is not inconceivable that causality violations occurring at the Planck scale might leave a detectable signature, there is also the question of whether one might see deviations from the predictions of standard quantum field theory in gravitational fields due to a theory based upon Deutsch's model. This possibility will be discussed briefly at the end of the thesis.

\section*{Locality, causality and free will}
\addcontentsline{toc}{section}{Locality, causality and free will} 
\lettrine[lines=3]{I}{n} physics, there is a vague principle called \tit{locality} that has many different incarnations. Broadly speaking, locality is the requirement that the state of affairs at one location should be independent of the state of affairs at some spatially remote location. The specifics naturally depend on what one means by the ``state of affairs" and by ``independent". In the present thesis, we consider locality to mean that, roughly speaking, two events in spacetime can be causally connected only if it is possible for a beam of light to propagate from one to the other through the background spacetime (see Sec \ref{SecSignal}). This requirement is often called \tit{signal locality} in the literature.

The notion of locality is closely connected to two other mysterious concepts in physics: the concepts of free will and the concept of causality. To see the connection, note that locality requires that the state of affairs at one location should not change in response to experimental interventions on systems at remote locations. Implicit in the concept of an ``experimental intervention" is the idea that the experimenter has the ``free will" to choose their actions. Without this assumption, it is not clear how to define causal influences resulting from experimental intervention, which is a key part of any physical model. By formulating our models in an operational way (see Sec \ref{SecOperational}) we implicitly assume throughout this thesis that experimenters have free will. We deliberately and explicitly avoid regimes in which this assumption may not apply, such as when the experimenters themselves are sent back in time. Furthermore, since signal locality is known to be upheld empirically in both the regimes of classical general relativity (GR) and of quantum mechanics (QM), we will take the position that any future theory of quantum gravity should also have this feature. We will use the term \tit{superluminal signalling}, or sometimes just `signalling', to refer to a violation of signal locality. 

Interestingly, there are two other arguments commonly levelled against superluminal signalling in physical theories. One argument is that it conflicts with relativity. This is somewhat misleading; while special relativity prevents any system of positive or zero mass from ever exceeding the speed of light, it does not prevent the existence of hypothetical particles such as tachyons, whose rest mass is imaginary, from propagating permanently at superluminal speed. While this might violate signal locality, it does not contradict the postulates of special relativity, nor does it appear to contradict general relativity. Usually what is meant by ``conflict with relativity" is actually ``conflict with signal locality", which is more of an empirical fact about relativistic physics than a part of its foundations. 

The second argument usually levelled against superluminal signalling is that it can be used to send signals back in time, thereby violating causality and leading to unphysical effects. For example, an effect from the future can prevent its own cause in the past, leading to a logical paradox. Surprisingly, simple mechanical situations exhibiting such paradoxes are rare in the literature and tend to be rather contrived; far more common is the opposite problem of \tit{under}-specification. In such cases, there typically exists an infinite set of consistent solutions for a given set of initial conditions and no physical rule for assigning probabilities to the different outputs. Contrary to popular conception, it is this ``information paradox" that plagues time travel in the physics literature far more than the problem of logical paradoxes\cite{SEP}. 

Given these unphysical effects, it seems reasonable to forbid causality violation as a matter of principle and, by implication, superluminal signalling. However, if it is causality violation rather than superluminal signalling that lies at the root of the problem, then there is a problem with general relativity, for it is well known that there exist solutions to Einstein's field equations that permit closed time-like curves (CTCs) in which the geometry of spacetime itself contains closed causal loops. To avoid the troubling effects of time travel, therefore, we must find a physical reason to forbid such metrics from arising. Most attempts along these lines argue that the matter distributions that create such metrics are unphysical, but these arguments must inevitably account for the quantum properties of matter, and we are led into the mysterious regime of quantum gravity.

Remarkably, once we confine ourselves to quantum systems, the above argument against superluminal signalling no longer applies. Toy models of quantum mechanics in the presence of causality violation, such as that due to Deutsch\cite{DEU91}, indicate that logical paradoxes do not occur for quantum systems, in the sense that the dynamical laws plus the initial conditions always have a unique consistent solution\footnote{Strictly speaking the solution is non-unique for a set of measure zero in the space of possible initial conditions, but even in those cases there is a natural way to select a unique solution, see Sec \ref{SecInfParadox} and Sec \ref{SecEquivCirc} for details.}. Somehow, the laws of quantum mechanics open a side door to the possibility of a consistent treatment of causality violation. If there are no longer any paradoxes associated with causality violation, then we cannot use this as an argument against superluminal signalling. Indeed, some toy models of causality violation for quantum systems do violate signal locality\cite{RAL12,LLO11}. 

In the present thesis, we aim to develop a toy model of causality violation that respects signal locality. Our reason for retaining signal locality is not because superluminal signalling is in conflict with relativity (it isn't) and not because it leads to causality violation, but because it would represent a degree of nonlocality not seen in either the quantum or relativistic regimes so far. Simplicity then demands that such effects should also be absent from any model which lies in the intersection of these regimes. While this is sufficient for the present thesis, we note that there may be a further, deeper reason to enforce signal locality even in the presence of causality violation. For any theory that takes place against the background of a given spacetime metric, the possible causal relationships between events are fixed \tit{a priori} by the prescribed metric. Even a metric that contains a CTC does not permit a signal between \tit{any} pair of events in spacetime, but only between those events that are connected by a timelike path through the metric. Since a violation of signal locality constitutes a violation of this constraint, it represents a disagreement between the causal structure implied by the theory and the causal structure prescribed by the background metric. One might therefore make a case in favour of signal locality from the point of view of logical consistency, while still allowing for violations of causality, although we will not pursue this line of argument further in the present work.

\clearpage
\pagestyle{plain}
\pagebreak
\renewcommand\bibname{{\LARGE{References}}} 
\bibliographystyle{refs/naturemagmat2012}
\addcontentsline{toc}{section}{References} 

\newpage
\thispagestyle{plain}
\mbox{}
\part{Low energy}
\label{part:LEN}

\pagestyle{fancy}
\renewcommand{\sectionmark}[1]{\markboth{\thesection.\ #1}{}}

\chapter{Standard Quantum Mechanics}
\label{ChapQM}

\begin{quote}
\em
Our best theories are not only truer than common sense, they make more sense than common sense.
\end{quote}
David Deutsch\\

\begin{quote}
\em
Quantum mechanics makes absolutely no sense.
\end{quote}
Roger Penrose\\

\newpage

\section*{Abstract}
We assume that the reader is familiar with basic quantum mechanics, particularly Dirac notation and linear algebra. The present chapter serves as a reminder about certain elements of quantum mechanics that will be used frequently throughout the thesis, while drawing attention to those conceptual aspects of quantum mechanics that will be called into question when the theory is modified to include causality violation and nonlinear dynamics. Sections \ref{SecSDM} and \ref{SecCSM} cover the basic postulates and the treatment of kinematics and dynamics using the density matrix approach that is ubiquitous in quantum information theory. Sec \ref{SecQCirc} reviews the circuit representation of the quantum formalism and Sec \ref{SecOperational} reviews the operational approach to quantum mechanics, which is important for the considerations of Chapter \ref{ChapNonlinBox}. We refer the reader to Ref. \cite{NIE} for more details on quantum mechanics and quantum circuits, and Ref. \cite{SPE05} for details on the operational formalism. 

\section{States, dynamics and measurements\label{SecSDM}}

Quantum mechanics can be broken down into three main elements: states, dynamics and measurements. The \tit{state} describes the properties of a physical system prepared in the laboratory. The \tit{dynamics} describe how the state changes in response to the actions taken by experimenters. The \tit{measurements} describe the final actions taken by experimenters, which result in a classical record of the measurement outcomes.\\

\tit{States:}\\
Every quantum system is associated with a complex Hilbert space $\scr{H}$, whose dimensionality is determined by the number of quantum degrees of freedom of the system. The state of the system is represented by a density matrix $\rho$, which is a positive semi-definite matrix with unit trace and a linear operator on the Hilbert space $\scr{H}$ of the system. The space of density matrices acting on the Hilbert space $\scr{H}$ is denoted $\scr{P}$.\\

\tit{Dynamics:}\\
The evolution of a quantum system from some initial state $\rho$ to some final state is associated with a linear completely positive and trace-preserving (CPT) map $\scr{M}:\scr{P} \mapsto \scr{P}'$ where $\scr{P}'$ is the space of density matrices acting on the final Hilbert space $\scr{H}'$ and the final state is $\rho':=\scr{M}(\rho)$. The map $\scr{M}$ has a \tit{Kraus decomposition} $\scr{M}(\rho):=\sum_i F_i \, \rho \, F^{\dagger}_i$ in terms of a set of linear operators $\{ F_i \}$ that map elements of $\scr{H}$ to elements of $\scr{H}'$. These operators satisfy $\sum_i \,F^{\dagger}_i F_i = I$ where $I$ is the identity operator on $\scr{H}$.\\

\tit{Measurements:}\\ 
A measurement on a system in the state $\rho$ with Hilbert space $\scr{H}$ and resulting in the outcome ``$k$" is associated with a CPT map $\scr{M}^{(k)}:\scr{P} \mapsto \scr{P}'$ where $\scr{H}'$ is the Hilbert space of the system after the measurement. The post-measurement state is $\frac{1}{P(k)} \scr{M}^{(k)}(\rho)$ where $P(k)$ is the probability of the outcome $k$ having occurred; $P(k)=\trc{\sum_i \, F^{(k)\dagger}_i F^{(k)}_i \, \rho }$.\\ 

The separation into these three elements is, of course, just a useful abstraction and has no fundamental significance. There is no real conceptual distinction between preparing, transforming and making measurements on a system. For example, it is quite common to incorporate measurements as part of the preparation of the system, or as part of its transformation, where later transformations may be conditioned on the outcomes of earlier measurements. Nevertheless, it is always possible to shift to an equivalent description in which all measurements are performed at the very end of the experiment. If we do this, then we do not need to know the post-measurement state and the only quantities of interest are the probabilities for different outcomes, $P(k)$. A measurement can then be characterised by specifying the operator $E_k:=\sum_i \, F^{(k)\dagger}_i F^{(k)}_i$ for every outcome $k$, with the total set $\{E_k \}$ being called a \tit{positive operator valued measure} (POVM). 

\section{Composite systems and mixed states\label{SecCSM}}
Given two systems with Hilbert spaces $\scr{H}_A$ and $\scr{H}_B$, we define the quantum state of the joint system as a density matrix $\rho_{AB} \in \scr{P}_{AB}$ acting on the joint Hilbert space $\scr{H}_A \otimes \scr{H}_B$, formed using the tensor product $\otimes$. The joint state is called \tit{separable} iff it has the form $\rho_{AB}=\sum_j p_j \sigma_A^{(j)} \otimes \sigma_B^{(j)}$ for some positive $p_i$ satisfying $\sum_j p_j=1$ and where $\{ \sigma_A^{(j)} \},\{\sigma_B^{(j)} \}$ are sets of density operators acting on $\scr{H}_A$ and $\scr{H}_B$ respectively. In the special case where $\rho_{AB}= \rho_A \otimes \rho_B$ for density matrices $\rho_A \in \scr{P}_A,\,\rho_B \in \scr{P}_B $, the joint state is called a \tit{product state}. Any joint state that is not separable is called \tit{entangled}\footnote{There is a vast literature on the different kinds of correlations that may arise in quantum mechanics, which include different types of entanglement as well as certain quantum correlations for separable states called ``discord". We will not be concerned with the particulars of these categorisations in the present thesis.}.

If one is given a composite system with joint state $\rho_{AB}$, one can similarly describe the reduced states of the subsystems. This is done by performing the partial trace over the total system. In particular, the reduced state of system $B$ is given by $\rho_B:=\ptrc{A}{\rho_{AB}}$, which is a density matrix in $\scr{P}_B$ as required. A similar procedure yields the reduced state of system $A$. 

The state of a system called \tit{pure} iff its density matrix is idempotent, $\rho^2=\rho$, otherwise it is called a \tit{mixed state}. Pure states have the form $\rho=\densop{\psi}{\psi}$, which is a projector onto a ray in Hilbert space containing the vector $\ket{\psi}$. Occasionally, for convenience, we will represent pure states by vectors in Hilbert space, but it is understood that any vector belonging to the same ray will do just as well. If we restrict our attention to pure states, then the relevant dynamical maps are \tit{unitary} CPT maps, which correspond to unitary operators on the Hilbert space of the system. For continuous time evolution generated by some Hamiltonian, one obtains $\rho(t)$ as the solution to the corresponding \tit{Schr\"{o}dinger equation}. More generally, $\rho(t)$ describes a mixed state evolving continuously in time according to the \tit{Liouville-von Neumann equation}. In this thesis, we are primarily concerned with abstract mappings from initial to final states, for which it is sufficient to talk about CPT maps. For more details about the equations of motion for continuous time parameterisation, the reader is referred to Ref. \cite{NIE}.

There are two factors that contribute to the mixedness of a state. One factor is the entanglement of the system to other systems, whose joint state is pure. In that case, the partial trace over the joint system results in mixed states for the subsystems. A mixed state resulting in this way is sometimes called an \tit{improper} mixture. Another cause of mixing is the ignorance of an agent about the actual state of the system. For example, consider an agent who ascribes probability $p$ to some system being in the pure state $\rho_{\psi}=\densop{\psi}{\psi}$ and probability $(1-p)$ for the system to be in the different state $\rho_{\phi}=\densop{\phi}{\phi}$. Strictly speaking, the agent should describe his ignorance using a probability distribution over the space of states, which might be written something like: $p\, \delta(\rho-\rho_{\psi})+(1-p)\, \delta(\rho-\rho_{\phi})$ where $\rho$ parameterises the space of density matrices $\scr{P}$ and $\delta(\rho-\rho')$ is a delta functional centred on the density matrix $\rho=\rho'$. However, due to the fact that all quantum dynamics is represented by \tit{linear} CPT maps, it turns out that an agent can simply use the new density matrix $\sigma:=p \, \rho_{\psi}+(1-p)\,\rho_{\phi}$ to describe the state incorporating her lack of knowledge about it. Such a state is sometimes called a \tit{proper} mixture. While conceptually distinct in that each type of mixture results from a different type of preparation procedure, the two kinds of mixture cannot be distinguished by any experiment in quantum mechanics. This is a direct result of the fact that only \tit{linear} CPT maps are allowed. The operational equivalence of the two kinds of mixture will be discussed in Sec \ref{SecOperational} and the breaking of this equivalence due to nonlinear dynamics is discussed in Chapter \ref{ChapNonlinBox}.

Another consequence of the linearity of CPT maps is that any such map can be regarded as the reduced dynamics of some unitary map acting on an enlarged Hilbert space. Formally, one enlarges the Hilbert space by introducing a number of hypothetical ``ancilla" systems\footnote{\tit{Trivia:} the term \tit{ancilla} originates from the Latin word meaning ``handmaiden" or ``slave-girl". In physics it is used to refer to additional quantum systems that are introduced to serve a particular purpose.}, together with a unitary map whose action on the total space reduces to the original CPT map on the relevant subspace. This is called a \tit{purification} of the dynamics\footnote{The assumption that the dynamics of any physical system can be purified in this way is sometimes called the ``Church of the Larger Hilbert Space". It is closely related to the assumption that quantum mechanics is fundamentally linear in all regimes, as in the many-worlds interpretation.}. We will see in Sec \ref{ChapNonlinBox} that for nonlinear CPT maps, it may not be possible to purify the dynamics in the normal way.   

\section{Quantum circuits\label{SecQCirc}}

In general, given some quantum state that evolves from an initial state to some final state, there are many different physical evolutions that describe what happens in between. As long as we are not interested in the state in between, we need not worry about continuous time evolution and we can focus on input-output maps as represented by the highly structured form of a \tit{quantum circuit}. Every unitary map on a finite dimensional Hilbert space -- hence every quantum mechanical evolution via its purification -- has a convenient diagrammatic representation in circuit form. A quantum circuit can be viewed as an abstraction of the dynamics, based upon the physical circuit that would need to be built in order to efficiently simulate the input-output map resulting from those dynamics. The study of what is meant by ``efficiently simulate" is called computational complexity theory. The study of computers that can efficiently simulate quantum systems (by virtue of having basic components that \tit{are} quantum systems) is called quantum computation. Despite the great simplification afforded us by representing quantum dynamics using a circuit diagram, there are in general an infinite class of circuits that produce the same input-output map on a quantum system. All circuits that are related in this way are called \tit{denotationally equivalent}. The particular choice of circuit used in any situation is then simply a matter of convention. 

\begin{figure}
 \includegraphics[width=18cm]{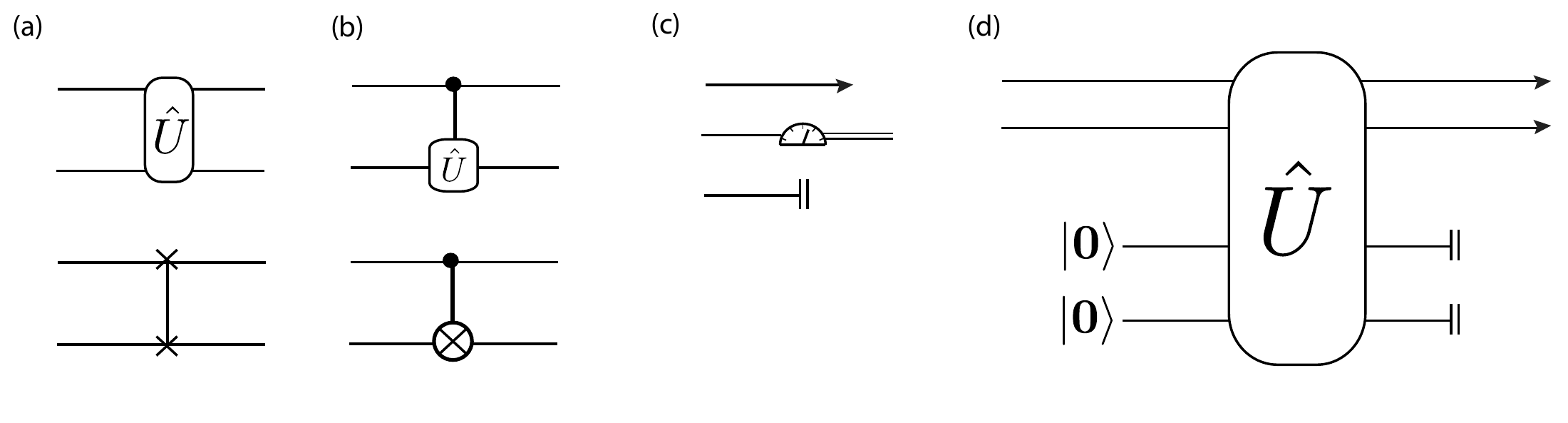}              
 \caption[Quantum circuit notation]{\label{figAllQCirc}(a) \tit{Top:} A generic unitary interaction between two rails of different Hilbert spaces. \tit{Bottom:} The symbol for the SWAP gate that interchanges two qubit rails. (b) \tit{Top:} A generic single qubit unitary acting on the Hilbert space of the lower rail, controlled by a qubit on the top rail. \tit{Bottom:} The symbol for a CNOT gate between two qubits. (c) Different rail endings. The top arrow represents a generic output that may be subject to further interactions. The middle symbol represents a measurement, resulting in classical data (represented by the double-rail). The basis of the measurement will be specified in the context. The bottom symbol indicates that the output of the rail is to be ignored, or cannot be accessed. (d) An example of a generic circuit corresponding to the purification of a CPT map on the top two rails. The lower ancilla rails are initialised in the state $\ket{\tbf{0}}$ by convention and their outputs are discarded.}
\end{figure}

In a circuit diagram, time is measured on the horizontal axis, increasing from left to right, while the vertical axis measures the degrees of freedom in Hilbert space. More precisely, the system of interest is decomposed into a tensor product of Hilbert spaces $\scr{H}_A\otimes \scr{H}_B \otimes \scr{H}_C \otimes ...$ representing the distinct subsystems of interest. Each Hilbert space is associated a horizontal line called a \tit{rail}. Nontrivial unitary interactions acting on a set of Hilbert spaces are represented by blocks called \tit{gates} that intersect the relevant rails in the diagram. We will generally allow the rails to correspond to Hilbert spaces of different dimensions, but it is usually more conventional to set all Hilbert spaces to have two dimensions. In that case, each rail corresponds to the Hilbert space of a two-dimensional quantum system, also called a \tit{qubit}. Just as bits which have binary values $0$ or $1$ are the fundamental building blocks of ordinary computer circuits, qubits are the fundamental building blocks of quantum computers and can exist in arbitrary superpositions of the orthonormal \tit{computational basis states} labelled $\ket{\tbf{0}}$ and $\ket{\tbf{1}}$, which span the qubit Hilbert space. Every quantum circuit has an equivalent representation as a circuit involving $n$ qubits, composed of gates that act on only one or two rails at a time. A set of one-qubit and two-qubit gates that is sufficient to emulate any quantum circuit is called a \tit{universal} gate set. One example of a universal gate set is a CSIGN gate (alternatively a CNOT gate) and the Pauli $\hat{X}$,$\hat{Y}$ and $\hat{Z}$ gates for single qubits. Fig. \ref{figAllQCirc} shows a list of standard symbols used to represent some common gates. In general, the operation performed by a particular gate will be discussed when it arises in the text. For common gates, such as the single-qubit Pauli gates and the CNOT or CSIGN gates, details can be found in Ref. \cite{NIE}.

\section{The operational approach\label{SecOperational}}

The philosophy behind the operational formalism is that any physical theory is essentially a recipe for telling us what is going to happen in some rationally conceivable experiment. We will adopt the operational formalism described by Spekkens \cite{SPE05}, which is notable for its generality in dealing with mixed states. The following is a brief summary of definitions used in that work, which are relevant to the present thesis. For more information, the reader is referred to Ref. \cite{SPE05}.

The basic elements of an operational theory are the preparations $P$, transformations $T$ and measurements $M$ that may be performed on a physical system, which represent lists of instructions to be carried out by the experimenter in the laboratory. An ``operational theory" is then a set of rules that determines the probabilities $p(k|P,T,M)$ for the outcomes ``$k$" given a particular preparation, transformation and measurement. This manner of describing a physical theory reminds us that whatever abstract theoretical objects we might imagine ourselves to be manipulating in an experiment -- be they wavefunctions, fields in phase space or strings -- their effects are seen purely as a mapping from the actions that we take in a laboratory to the outcome statistics that we record over many experiments. This is particularly useful in situations where there is some debate about what the ``real" objects are in the theory, as is the case in quantum mechanics. The operational formalism gives us a means of separating the ambiguous aspects of a theoretical model from the experimentally observable aspects.

Let us now make some definitions that will be useful later on. An \tit{ontological model} is a theory that aims to explain the predictions of the operational theory. In the case of quantum mechanics, Bohmian mechanics is an example of an ontological model (a.k.a. \tit{ontology}). Within an operational theory, an ontology supplies every system with an \tit{ontic state}, specified by the values of a set of variables $\lambda$. The space of possible ontic states, denoted $\Omega$, is then equal to the space of possible values taken by the variables $\lambda$. The ontic state represents the set of objective attributes assigned to a physical system, regardless of what any agent knows about that system. As we will see below, ignorance about a system's ontic state is represented by a probability distribution over the ontic states, corresponding to an agent's subjective knowledge.\\ 

A complete ontology must contain three basic elements as follows:\\
1. An ontological description of preparations. This is a set of rules that associates every preparation procedure $P$ to a corresponding probability density $p_P(\lambda)$ over the ontic states. \\
2. An ontological description of transformations. This is a set of rules that associates every transformation procedure $T$ with a transition matrix $\Gamma_T(\lambda, \lambda')$ that gives the probabilities for transitioning from the ontic state $\lambda$ to the state $\lambda'$.\\
3. An ontological description of measurements. This is a set of rules that associates every measurement procedure to an \tit{indicator function} $\zeta_{M,k}(\lambda)$ that gives the probability of obtaining the outcome $k$ given the measurement procedure $M$ is performed on the ontic state $\lambda$.\\

In general, the ontic state of a system is determined by its preparation procedure $P$, together the specification of additional \tit{hidden variables}. If there are no hidden variables, then the preparation procedure uniquely defines the ontic state. 

Given all of the above, the requirement that the ontological model reproduces the predictions of the operational theory within its regime of applicability means that we have:
\eqn{
p(k|P,T,M)=\int \mathrm{d}\lambda \, \int \mathrm{d}\lambda' \, p_P(\lambda) \, \Gamma_T(\lambda, \lambda') \, \zeta_{M,k}(\lambda') \, ,
}  
where the right hand side is the probability of obtaining the outcome $k$ in the ontological model given $P$,$T$ and $M$.

Finally, we note that within an operational theory there exist ``operational equivalence classes" of preparations, transformations and measurements, defined by the property that two elements of the same equivalence class can be interchanged in an experiment without affecting the outcomes of the experiment. For example, two preparation procedures $P$ and $P'$ are \tit{operationally equivalent} iff $p(k|P,T,M)=p(k|P',T,M)$ for all transformations and measurements. Similar definitions hold for transformations and measurements. 

In quantum mechanics, an equivalence class of preparation procedures is represented by a density matrix $\rho$, an equivalence class of transformation procedures is given by a CPT map $\scr{M}$, and an equivalence class of measurement procedures is given by a POVM $\{E_k\}$. Thus, given a preparation $P$, transformation $T$ and measurement $M$, standard operational quantum mechanics associates these with the respective elements $\rho_P$, $\scr{M}_T$ and $\{E^{(M)}_k\}$, which combine to give the probabilities: 
\eqn{ p(k|P,T,M)=\trc{\scr{M}_T(\rho_P) E^{(M)}_k\, } 
} 
Thus, ``operational quantum mechanics" can be understood to mean a very long list of preparations, transformations and measurements together with their associated equivalence classes, as determined empirically and without reference to any underlying ontological model. An ontological model is then seen as an attempt to condense this long list by providing an explanation for the observed equivalence class structure. It also provides a means of predicting what will happen in experiments that lie outside the regime of currently tested quantum mechanics. In Chapter \ref{ChapNonlinBox} we will consider a class of ontological models that are able to predict what will happen in a regime in which quantum dynamics is allowed to be nonlinear, i.e. in which the equivalence classes of transformations include the possibility of nonlinear CPT maps.

\clearpage
\pagestyle{plain}
\pagebreak
\renewcommand\bibname{{\LARGE{References}}} 
\bibliographystyle{refs/naturemagmat2012}
\addcontentsline{toc}{section}{References} 

\pagestyle{fancy}
\renewcommand{\sectionmark}[1]{\markboth{\thesection.\ #1}{}}

\chapter{The Deutsch Model}
\label{ChapDeutsch}

{\em
\parindent=10pt
\noindent ON the wide level of a mountain's head\\
(I knew not where, but 'twas some faery place),\\
Their pinions, ostrich-like, for sails outspread,\\
Two lovely children run an endless race,\\	 
\indent \indent A sister and a brother!\\	         
\indent \indent This far outstripp'd the other;\\	 
\indent Yet ever runs she with reverted face,\\	 
\indent And looks and listens for the boy behind:\\	 
\indent \indent For he, alas! is blind!\\	 
O'er rough and smooth with even step he pass'd,\\	  
And knows not whether he be first or last.\\
}

Samuel Taylor Coleridge\\

\newpage

\section*{Abstract}
This chapter serves as a review of the literature on Deutsch's toy model of nonlinear quantum mechanics in the presence of causality violation; the reader is referred to Ref. \cite{DEU91} for more details. We begin in Sec \ref{SecCTCBG} with a brief overview of wormhole metrics in the literature, focusing on the issue of their stability within a semi-classical setting. In Sec \ref{SecCausViol} and Sec \ref{SecDeutschCirc} we describe the essential assumptions of Deutsch's model. This allows us to identify which physical properties are intrinsic to the model and which properties may depend on additional assumptions that are often left implicit in the literature. Sec \ref{SecGenDeutsch} considers the limitations of the toy model and directions for possible generalisations. Finally, in Sec \ref{SecEquivCirc} we review the ``equivalent circuit" formulation of Deutsch's model, details of which can be found in Ref. \cite{RAL10}. This model will be the starting point for the relativistic extension of Deutsch's model outlined in Chapter \ref{ChapCTC1} and Chapter \ref{ChapCTC2}. The conclusion and outlook is summarised in Sec \ref{SecDeutschConcl}.

\section{A brief history of time-travel\label{SecCTCBG}}

One of the curious features of Einstein's field equations in GR is that, at least on paper, any metric tensor whatsoever is permitted: one simply plugs the desired shape of spacetime into the metric tensor and then solves the equations to find out what sort of matter distribution is required to produce it. Early on, it was pointed out by G\"{o}del \cite{GOD49} that CTCs could exist in a rotating, dust-filled universe. Indeed, there are many ways in which time-travel could arise in GR, some less practical than others; we will focus on one particular class of metrics, namely the `wormhole' metrics\footnote{For a guide to constructing wormhole spacetimes, see eg. Yodzis \cite{YOD72}.}, which are the most relevant to Deutsch's model and are arguably the most plausible candidate for a physical realisation of time-travel.  

A `wormhole' can be visualised as a geometric tunnel connecting two remote parts of spacetime, having two `mouths' that serve as the entry and exit and a `throat' which is the tunnel itself. An object entering one mouth and emerging from the other may arrive at a remote part of the universe long before it could have reached that same point by travelling externally through space-time. Given our understanding of relativity, it is perhaps not surprising that such wormholes are inextricably linked to the possibility of time-travel, allowing matter to traverse a loop in time and arrive in its own past. Geroch\cite{GER67} showed in 1967 that creation of wormholes must be accompanied by CTCs or else by the impossibility of labelling past and future in a continuous way. Morris, Thorne and Yurtsever \cite{MOR88} took this a step further in 1988, demonstrating that one could hypothetically create a CTC from a wormhole by accelerating one of the wormhole mouths on a relativistic trajectory. Thus, the possibility of creating a stable wormhole implies the possibility of time-travel in GR.

One might hope to rule out time-travel by stability arguments. The Cauchy horizon is the name given to the boundary in the wormhole metric that separates the CTC-containing region from normal spacetime. Classically, propagating waves tend to get trapped at such horizons and `pile up', producing a divergent stress-energy tensor that would cause the horizon to collapse into a black hole. In fact, all examples of Cauchy horizons prior to the case under consideration had proved to be unstable in this way \cite{THO91}. Even the analog of the wormhole metric in 2-D Misner space has a classically unstable Cauchy horizon. It is remarkable, therefore, that the metric considered by Morris et. al. avoids this source of instability. The reason is that, provided the width of the wormhole throat is sufficiently small compared to the distance between its two mouths, the metric acts to disperse and diminish the waves piling up on the Cauchy horizon, preventing them from accumulating and causing a catastrophe \cite{MOR88}.

While the wormhole might be classically stable, it is also necessary to include quantum effects. Surprisingly, quantum effects seem to be essential to the possible existence of a stable traversable wormhole. That is because the matter distribution required to produce the wormhole metric necessarily violates the Averaged Null Energy Condition (ANEC). The ANEC requires the average energy density along every null geodesic through the metric to remain strictly positive, but this is not so for wormholes. There is a consensus that classical matter fields cannot violate the ANEC, which implies that there can be no stable classical wormholes. However, it is accepted that quantum fields can violate a localised version of the ANEC, as seen in well-known examples such as the Casimir effect. Such local violations suggest that quantum matter might also violate the ANEC in more general situations. Therefore, a stable wormhole remains a possibility, provided there exists quantum matter with sufficient negative energy density to support the wormhole throat against collapsing \cite{MOR88}. 

Unfortunately for wormhole enthusiasts, the introduction of quantum effects is a double-edged sword. Contrary to the behaviour of classical waves, the size of quantum fluctuations is still expected to diverge as one approaches the Cauchy horizon, leading to a corresponding divergence in the stress-energy tensor and presumably to a fatal instability. However, once again it is possible to pull a ``hat trick" to avoid this conclusion, by taking advantage of two crucial facts. First, the predicted divergence is exceedingly weak, so that even when one is very close to the Cauchy horizon, the fluctuations are negligible. Indeed, according to Kim and Thorne \cite{KIM91} one would have to get closer than the Planck length to the horizon (as seen from the rest frame of the wormhole mouth) before the divergence would become significant. But since the semi-classical analysis is no longer valid at this scale, the authors argue that there is a quantum gravity cutoff that saves the wormhole from oblivion. In response to this work, Stephen Hawking presented a counter-analysis in which he argued that the introduction of quantum gravity effects does not occur at the point claimed by Kim and Thorne, but rather it occurs when one considers observers whose speed is such that the distance between the wormhole mouths is Lorentz-contracted to the Planck length. In this case, it may well be that the spacetime can be treated classically close enough to the Cauchy horizon such that the divergence of the stress-energy tensor renders the horizon unstable. This claim is the core of Hawking's Chronology Protection Conjecture\footnote{Hawking famously quipped that his conjecture was supported by the observed absence of tourists from the future. However, within the constraints of the wormhole model, the absence of time-travellers only leads to the unexciting conclusion that nobody has built a time machine yet.} \cite{HAW92}.

Ultimately, the question of the Cauchy horizon's stability is a moot point, since it seems to only be resolvable by a theory of quantum gravity. However, there is another way in which the wormhole might prove to be unstable: the hypothetical negative energy fields supporting the wormhole throat might be insufficient to keep the wormhole open long enough for anything to get through. Indeed, it might turn out that quantum fields do not violate the ANEC in wormhole spacetimes, thereby ruling out stable wormholes altogether. Fortunately, this question is amenable to semi-classical analysis and has been the subject of much work in the literature. So far it has been shown that the ANEC holds for quantum scalar fields in a wide variety of spacetimes, but that it can also be violated in special circumstances, such as spacetimes closed in one spatial dimension \cite{KLI91}. Thorne speculated in 1991 that we would know definitively whether the ANEC rules out stable wormholes ``within a few years". Although some progress has been made, it remains an open problem at the time of writing, some 22 years later.

To summarise, much of the literature to date has been a competition between proposals for no-go theorems on stable wormholes and proposals for evading these theorems. The debate has not been fruitless, however: we now know that there are stringent conditions that must be obeyed by any candidate for a stable traversable wormhole. Some of these requirements, such as the necessary violation of the ANEC, might be sufficiently `distasteful' to physicists to indicate the ultimate impossibility of time travel, although ultimately the matter might only be decidable within a complete theory of quantum gravity. In the absence of such a theory, recent literature has begun to ask how physics might handle time-travel, assuming that it could happen. We will turn our attention to one such proposal, in hope of shedding light on quantum gravity itself.

\section{Causality violation in the circuit picture\label{SecCausViol}}

The standard formulation of quantum mechanics, including relativistic quantum field theory, assumes a background spacetime that is globally hyperbolic. In particular, this excludes spacetimes containing closed timelike curves, despite the fact that these spacetimes are allowed by general relativity. In seeking a theory of quantum gravity, one could either reformulate GR in a manner that forbids such spacetimes before quantizing\footnote{This is the approach taken by loop quantum gravity, for example.} or else one could attempt to reformulate quantum mechanics to accommodate violations of causality. The latter approach is the basis of Deutsch's toy model.

Being a toy model, Deutsch's model does not provide an explicit account of the propagation of quantum fields in non globally hyperbolic metrics. Instead, one arrives at the model by making a number of simplifying assumptions, to hone in on the essential physics. The first assumption of Deutsch's model is that the CTC-containing spacetime possesses an `unambiguous future' and an `unambiguous past'. Roughly speaking, this means we can define a spacelike hypersurface $S_P$ in the past and a spacelike hypersurface $S_F$ in the future such that all points in $S_P$ are in the causal past of all events in $S_F$, and all CTCs exist between these two surfaces. This allows us to formulate causality violation as a scattering problem from an initial to a final state, which will be extremely useful. While this assumption excludes metrics containing cosmological CTCs, for example G\"{o}del's metric \cite{GOD49}, it does include the class of ``wormhole" metrics introduced by Morris, Thorne and Yurtsever \cite{MOR88}. 

Next, we assume that all systems of interest are pointlike. This means that any quantum mechanical degrees of freedom are confined to a localised wavepacket, such that the expectation values of the system's position and momentum define an approximately classical trajectory through spacetime. Particle wavefunctions with large variances in position or momentum are excluded from the toy model\footnote{Note, however, that no such restriction applies to the ``internal" degrees of freedom of the system, such as polarisation and energy -- the system may exist in arbitrary superpositions of these.}. The benefit of this assumption is that we can directly map the spacetime diagram onto a quantum circuit, as we now describe.

It is convenient to divide the systems of interest into two types according to the path taken through the metric. The first type constitutes those systems whose trajectories intersect both the past hypersurface $S_P$ and the future hypersurface $S_F$. These are called ``chronology-respecting" or ``CR" systems. The remaining systems are those whose trajectories form closed loops in time. These trajectories are contained entirely between the past and future hypersurfaces and do not intersect either of them. We refer to these as ``chronology-violating" or ``CTC" systems.

It is straightforward to map the trajectories of the CR systems onto the rails of a circuit diagram. One simply draws a line from left to right, with time-ordering along the horizontal axis. For quantum systems, the initial joint state of these rails is simply the joint state of all of the CR systems on the initial hypersurface $S_P$, and the final state is the state of all CR systems on the future hypersurface $S_F$. The map from the initial to final state remains to be determined by the interactions between these systems and the CTC systems. 

\begin{figure}
 \includegraphics[width=12cm]{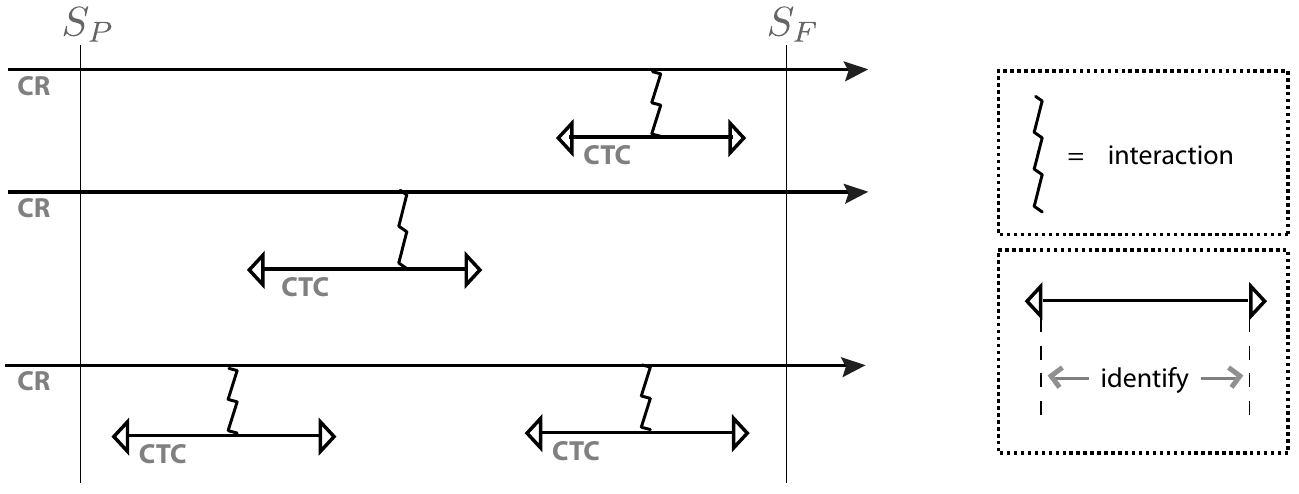}              
 \caption[A general circuit representation of systems interacting with closed timelike curves]{\label{figMetricToCircuit}A circuit diagram representing chronology respecting (CR) systems interacting with systems traversing closed timelike curves (CTCs).}
\end{figure}

The CTC systems themselves are more problematic, because they represent causal loops. A causal loop has the property that, given any two distinct events on the loop, each event has a causal influence on the other. This is in contrast to the causal ordering in a circuit diagram where, given any two events selected from a single rail, it is the event on the left that has a causal influence on the event to its right and not the converse. We therefore seek a method of embedding a causal loop into the standard framework. To do so, we imagine bisecting the trajectory at an arbitrary point on the loop (henceforth called the ``temporal origin") and designating the state at this point as the ``initial state" and the state on the other side of the cut as the ``final state". Continuity around the loop demands that the initial state be identical to the final state; we refer to this as the ``consistency condition". The loop, now separated, can be spread out in a line with the initial state placed to the left of the final state on the circuit diagram. The identification of the two endpoints by the consistency condition makes this trajectory a causal loop, as opposed to a system that merely appears out of nowhere and then vanishes at a later time. It is crucial that whatever rule is taken to enforce the consistency condition must have the property that the arbitrary placement of the temporal origin has no effect on the physics. Finally, we can describe the interacting CR and CTC systems using the circuit diagram shown in Fig. \ref{figMetricToCircuit}. Using this circuit as our guide, the next step is to interpret it as a quantum circuit and define an appropriate consistency condition.

\section{Deutsch's circuit\label{SecDeutschCirc}}

\begin{figure}
 \includegraphics[width=12cm]{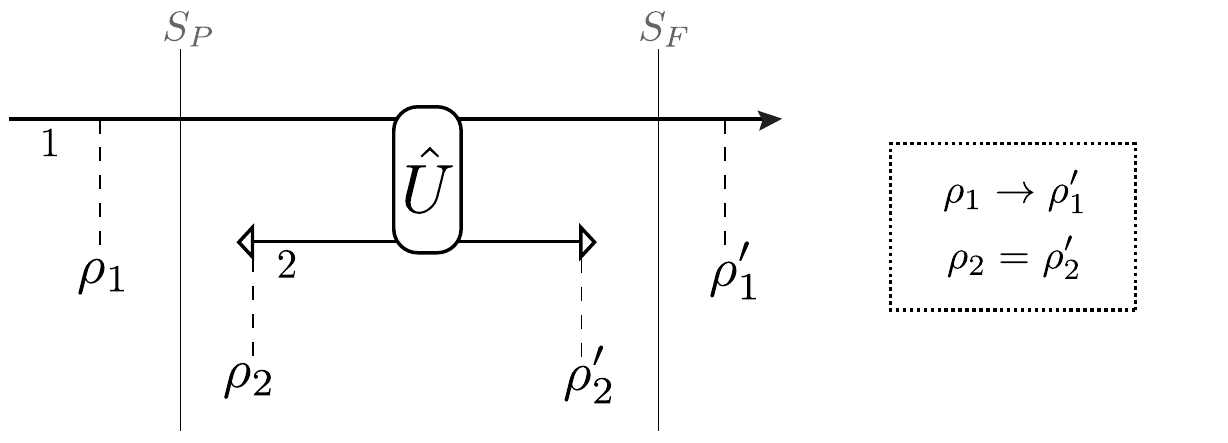}              
 \caption[A quantum circuit containing a single CTC]{\label{DeutschCirc}A quantum circuit representing a unitary interaction between a group of CR systems (rail 1) and a group of systems traversing a single CTC (rail 2). The consistency condition on the CTC rail determines the input--output relation for the CR rail.}
\end{figure}

Let us assume that both the CR and CTC systems are quantum systems, with initial and final states given by (possibly mixed) density matrices\footnote{We concentrate on systems that are known to exist within the regime of quantum mechanics, such as electrons, photons, atoms, etc. The question of whether CTCs can lend insights into fundamental issues regarding the quantum nature of macroscopic objects will be touched on later in the thesis, in Sec \ref{SecHeisCut}.}.
Furthermore, assume that only a single CTC is present, so that all CTC systems traverse the same CTC and their joint state is subject to a single consistency condition. The quantum circuit of Fig. \ref{DeutschCirc} represents the trajectories of a group of CR systems (rail 1) interacting with a group of CTC systems (rail 2). Each rail is associated with a Hilbert space, labelled $\scr{H}_1$ and $\scr{H}_2$, respectively. The total interaction is specified by a unitary operator $\hat{U}$ on the joint space $\scr{H}_1 \otimes \scr{H}_2$. The specific details of propagation through the background metric are assumed to be contained in the total unitary evolution\footnote{The detailed laws governing the propagation through the metric may ultimately impose constraints on the unitary $\hat{U}$\cite{HAW92}. Lacking such a detailed theory, we assume the interaction is unconstrained.}.
Following Deutsch, we will assume that the initial joint state of the CR and CTC systems is of the product form $\rho_1 \otimes \rho_2$ \footnote{The basis of this assumption is discussed in Sec \ref{SecInitState}.}. The experimenters in this hypothetical scenario have access to the initial and final states of the CR systems, but the CTC systems are assumed to exist in a region that is not directly accessible\footnote{This ensures that the problem can be easily cast in operational form, which will be particularly useful in Chapter \ref{ChapNonlinBox}.}. 

Given an input state for the two rails of the product form $\rho_1 \otimes \rho_2$, we can obtain the joint output state according to the usual procedure in the Schr\"{o}dinger picture:
\eqn{\rho'_{1,2}=\hat{U} (\rho_1 \otimes \rho_2) \hat{U}^\dagger \, .}
The reduced states of the outputs on each rail can be obtained by taking the partial trace. While the input state of the CR system is specified by the initial conditions (i.e. it is the state prepared in the laboratory by experimenters in the unambiguous past), the initial state on the CTC, $\rho_2$, is to be determined by Deustch's consistency condition:
\eqn{\label{consistency} \rho_2 &=& \ptrc{\neq 2} { \rho'_{1,2}} \nonumber \\ &=& \ptrc{\neq 2} {\, \hat{U} (\rho_1 \otimes \rho_2) \hat{U}^\dagger \, } \, ,}
where the trace is over all systems except those traversing the CTC. For a given choice of $\rho_1$ and $\hat{U}$, Deutsch showed that there always exists some $\rho_2$ satisfying this requirement\cite{DEU91}. It follows that the output of the circuit is the reduced state of the CR qudit after the interaction, namely:
\eqn{\label{output} \rho'_1 &=& \ptrc{2}{ \rho'_{1,2} } \, \\ &=& \ptrc{2} { \hat{U} (\rho_1 \otimes \rho_2 ) \hat{U}^\dagger } \, .}

The equations \eqref{consistency} and \eqref{output} define the map from input to output, denoted $\mathcal{N}_D: \rho_1 \rightarrow \rho'_1$ and referred to henceforth as ``Deutsch's map". As expected, the placement of the temporal origin can be shown to have no measurable effect on the output\cite{BAC04}. The fact that the map always has at least one fixed point for any given $\rho_1$ and $U$ implies that there are no logically inconsistent interactions or ``grandfather paradoxes", as will be shown in Sec \ref{SecGFParadox}. In certain cases, there are multiple fixed points, which leads to an ambiguous output. Deutsch suggested removing this ambiguity by selecting the solution with maximum entropy. Deutsch's map, together with the consistency condition \eqref{consistency} and the maximum entropy rule, constitute what we will refer to as ``Deutsch's model". Note that, apart from the necessary assumptions outlined above, we do not make any further assumptions regarding the model's interpretation\footnote{One of the aims of the thesis is to ascertain what additional assumptions are needed to remove certain ambiguities from the model; see Chapter \ref{ChapNonlinBox} and in particular Sec \ref{SecDeutschBox}.}. 

Deutsch's map has a number of interesting properties, the discussion of which will form the basis for the rest of this chapter. We review them briefly here. First, this map is a nonlinear map on the space of density operators\footnote{To see this, consider the term $\rho_1 \otimes \rho_2$ on the RHS of \eqref{output}. Since $\rho_2$ is itself dependent on $\rho_1$ through the consistency equation \eqref{consistency}, there is a nonlinear overall dependence on the input $\rho_1$.}. It is also nonunitary and takes pure states to mixed states in general. Despite these features, Deutsch showed that mixed states never evolve into pure states and that the second law of thermodynamics is respected by the total evolution\cite{DEU91}. We will see in Sec \ref{SecEquivCirc} that the model can be recast in a special form called the ``equivalent circuit". This form of the model allows us to perform calculations in the Heisenberg picture, while also leading to a derivation of Deutsch's maximum entropy rule. Finally, the motivation behind Deutsch's consistency condition is discussed in Sec \ref{SecSignal}, in connection with superluminal signalling and the ``entanglement-breaking" effect described in Sec \ref{SecEntBreak}. This will set the scene for the work described in Chapter \ref{ChapNonlinBox}.

\subsection{The Grandfather Paradox\label{SecGFParadox}}

\begin{figure}
 \includegraphics[width=12cm]{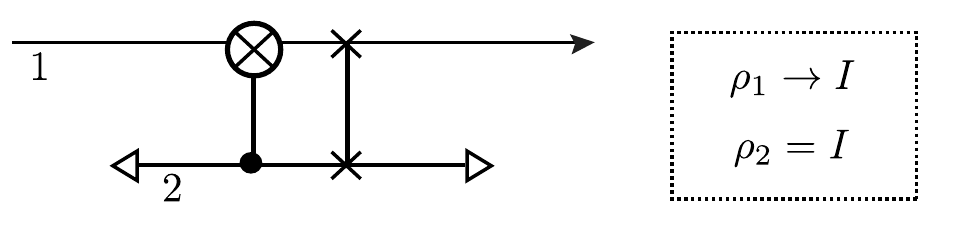}              
 \caption[A quantum circuit that performs the grandfather paradox for a pair of qubits]{\label{GPcirc}A quantum circuit that performs the ``grandfather paradox" for a pair of qubits. Deutsch's model resolves the paradox by assigning the mixed state $I$ to the control qubit.}
\end{figure}

Consider the circuit in Fig. \ref{GPcirc}. Here, the CR and CTC systems are qubits. The interaction is a quantum CNOT with the CR system as target, followed by a SWAP gate that interchanges the two systems. Consider the case where the input is prepared in the pure state $\rho_1=\densop{1}{1}$. According to Deutsch, we must solve the consistency condition \eqref{consistency} to obtain the CTC state $\rho_2$ and thereby obtain the output $\rho'_1$. What is especially interesting about this circuit is that there is no consistent \tit{pure state} solution for $\rho_2$ for this choice of input. 

To see why, note that if the consistent solution for $\rho_2$ is a pure state then it must be in the computational basis, since any other basis would lead to entangled outputs and hence a mixed state in the CTC. Therefore, let $\rho_2=\densop{1}{1}$. This implies that the CNOT was not triggered. But since the state $\ket{1}$ triggers the CNOT, we have a contradiction. The second option, $\rho_2=\densop{0}{0}$, implies that the CNOT was triggered, but the state $\ket{0}$ then becomes the control bit and ensures the converse, so we have another contradiction. This circuit therefore provides a concrete realisation of the grandfather paradox, in which the emergence of a time traveller in the past (here the control bit of the CNOT) ensures its own demise by altering its previous state.

Let us now apply Deutsch's model. Solving for \eqref{consistency}, we find that the initial (and final) state of the CTC system is the maximally mixed state:
\eqn{\rho_2 = I = \frac{1}{2}( \densop{0}{0}+\densop{1}{1})\, \label{GPsol}.}
There is an important caveat in regarding this as a valid solution to the paradox. In terms of classical probability theory, this state would represent either the classical bit ``0" or the bit ``1", with equal likelihood. In any case, the mixed state represents a state of ignorance of the real underlying state of the system, which is definitely either one bit or the other. The situation in quantum mechanics is more subtle, for it is well known that we can rewrite the maximally mixed state as a completely different ensemble of pure states:
\eqn{I = \frac{1}{2}( \densop{+}{+}+\densop{-}{-})\, .} 
Which ensemble is the correct one? Depending on the preparation procedure, it might be any one of the infinite possible pure-state decompositions. It could also be none of them: in the case where the system is half of a maximally entangled pair, its reduced density matrix will also be given by $I$, but in this case there will be no natural choice of decomposition into pure states. Whatever the situation, quantum mechanics alone does not give us any experiment by which we could determine the difference between these possibilities. However, this distinction becomes very important for the resolution of the grandfather paradox as described above.

Let us first imagine that the solution given by \eqref{GPsol} is actually a qubit in either the $\ket{0}$ or $\ket{1}$ state with equal probability. This is referred to an an \tit{epistemic} interpretation of $\rho_2$, which is then a \tit{proper mixture}. In this case there is still a paradox, because neither of the underlying state-assignments makes any sense, as we have seen; we have merely given a description of the problem that obscures the paradox. This inability to treat $\rho_2$ as a proper mixture in Deutsch's model is discussed by Wallman and Bartlett\cite{WAL10}. If we cannot treat $\rho_2$ as a state of ignorance about underlying pure states, then we are forced to adopt the remaining option: that the density matrix represents an \tit{ontic} state\footnote{As we will see in Chapter \ref{ChapNonlinBox}, any proposal for dynamics that takes the form of a nonlinear map on the space of density operators includes the \tit{a priori} assumption that the density matrix, pure or mixed, is an ontic object. Since an agent's subjective knowledge of a state cannot contribute to this density matrix, one must resort to the more general representation of states as probability distributions over the ontic states. For a relevant discussion of the distinction between the epistemic and ontic interpretations of the density matrix in quantum mechanics, the reader is referred to \cite{SPE05}.} 
 (see Sec \ref{SecOperational}).

What can this mean, physically? In the case of dead or alive grandfathers, this is a difficult question. Deutsch himself was in favour of a ``multiple universe" interpretation, whereby the state $I$ corresponds to a multitude of identical grandfather copies, half of which are dead and half of which are alive. We will not commit ourselves to any particular interpretation for the moment, but simply emphasise that the density matrix appearing in Deutsch's equations is an ontic object. This raises the important question: given some preparation procedure of a system, what is the density matrix that must be objectively ascribed to that system in Deutsch's model? We return to this question in Chapter \ref{ChapNonlinBox}.

\subsection{The Information Paradox\label{SecInfParadox}}

Let us consider a case in which Deutsch's condition \eqref{consistency} is not sufficient to uniquely determine the output of the circuit. Consider the same circuit as in Fig. \ref{GPcirc}, but this time let the input be prepared in the pure state $\rho_1=\densop{+}{+}$. Let us make a guess at the possible consistent solution. There is one trivial possibility, for which $\rho_2=\densop{+}{+}$. The CNOT does nothing to the state $\ket{+}_1 \otimes \ket{+}_2$, so the consistency condition $\rho'_2=\rho_2$ is satisfied, and the output is just the same as the input, $\ket{+} \rightarrow \ket{+}$. Alternatively, we observe that the CNOT maps the state $\ket{+}_1 \otimes \ket{-}_2$ to $\ket{-}_1 \otimes \ket{-}_2$. This again satisfies the consistency condition with the CTC state being $\ket{-}$, and the output also being $\ket{-}$. Hence there are already two possible pure-state solutions to the consistency condition, leading to completely different outputs! When we consider more mixed density matrices, there is a continuous family of different outputs that satisfy Deutsch's equations. This provides a concrete realisation of another paradox familiar from science fiction: the ``information paradox".

One version of the paradox runs as follows: a person goes back in time to meet Shakespeare, bringing along his published copy of \tit{Hamlet} in the hopes of getting it autographed. Unfortunately, he goes back too far, arriving some years before Shakespeare has written the play. Upon seeing the play, Shakespeare steals it and passes it off as his own. The question is, who wrote the original play for which Shakespeare has taken credit?

The curious feature of this paradox is that it is not as severe as the grandfather paradox: the latter entails a logical inconsistency, whereas the former describes a perfectly self-consistent chain of events. The problem in this instance is that information seems to appear out of nowhere without arising by any evolutionary process. In the absence of any widely accepted principle of ``information conservation", or some similar law that would enforce the linearity of quantum dynamics in all situations, it is difficult to argue that such behaviour should be forbidden in nature. Nevertheless, we should feel uncomfortable about it and endeavour to understand what its physical implications would be.

Deutsch suggested that in such situations one should choose the solution with maximum entropy, in this case $\rho_2=I$. He reasoned that the CTC should not have any more information than is specified by the initial data and hence the solution should be minimally informative (maximally entropic). Deutsch's approach bears some resemblance to the problem in decision theory where the maximum entropy principle allows one to make a ``best guess" about the possible state of a system about which we have incomplete knowledge. In that case, however, we are making guesses about an objective state of reality of which we are partially ignorant. In the context of the Deutsch model, we are not trying to guess the output but rather we are \tit{defining} it by appending the maximum entropy principle as an additional axiom. 

This resolution seems unsatisfactory for at least two reasons. First, the theory is weakened by the addition of an extra rule that does not follow from the existing axioms. It would be preferable if the maximum entropy rule could be derived from physical considerations rather than being imposed \tit{ad hoc}. Second, as pointed out by Bacon\cite{BAC04}, it remained to be shown that the maximum entropy rule always yields a unique solution. 

Politzer\cite{POL94} proposed that the ambiguous solutions are not physically meaningful, because they only occur for singular points in the space of possible circuits\footnote{The unitaries leading to ambiguous solutions is a set of measure zero in the space of all possible unitaries\cite{LIV07}.}. One might therefore hope that a unique solution exists in the limit approaching that singular point. One realisation of this hope is the ``equivalent circuit" approach detailed in Sec \ref{SecEquivCirc}. There, a natural embedding of Deutsch's circuit into a larger Hilbert space enables one to derive Deutsch's maximum entropy rule by considering small perturbations of the CTC interaction; moreover, the solution is unique\cite{RAL10}.

However, Bacon points out that \tit{``...[t]here is another manner in which this consistency paradox can be alleviated: one can assume that the freedom in the density matrix of the CTC systems is an initial condition freedom. One recalls that there are initial conditions which evolve into the CTC qubit; i.e., the specification of conditions such that the compact region with CTCÕs is generated. It is not inconsistent to assume that some of the freedom in the initial conditions which produce this CTC qubit are exactly the freedoms in the consistency condition. Such a resolution to the multiple-consistency problem puts the impetus of explaining the ambiguity on an as-yet [un]codified theory of quantum gravity. It is interesting to turn this around and to ask if understanding the conditions for a resolution of the multiple-consistency problem can tell us something about the form of any possible theory of quantum gravity which admits CTCÕs"}\cite{BAC04}.\\

Bacon's quote reminds us that the Deutsch model is, after all, only a toy model and we should not expect too much from it. We will do well to remember this as we encounter more ambiguities and missing pieces in the model -- these absences are not so much flaws as they are an invitation to extend and generalise the model, perhaps ultimately leading us towards the missing quantum gravity degrees of freedom that Bacon alludes to.

\subsection{Super-computation with CTCs\label{SecInfParadox}}

According to the literature, the nonlinear nature of Deutsch's map \eqref{output} enables a CTC-assisted computer to perform a number of information theoretic tasks that would be impossible with an ordinary quantum circuit. Bacon showed that it might be possible for a quantum computer with access to multiple sequential CTCs to solve NP-complete problems in polynomial time, even in the presence of noise\footnote{There is an important assumption made by Bacon, regarding the extension of the Deutsch model to multiple CTCs, which we discuss in Sec \ref{SecMultiCTC}.}\cite{BAC04}. Subsequently, Aaronson \& Watrous argued that quantum computers with access to CTCs could efficiently solve any problem in PSPACE\footnote{The authors also claimed to prove the same result for CTC-assisted classical computers. However, they implicitly assumed that probability distributions of classical bits could be treated as ontic objects for the purposes of the self-consistency condition, which may be a dubious assumption; besides, it is quite different to Deutsch's consistency condition.}\cite{AAR09}.

In addition, Brun\cite{BRU09} devised a circuit using just a single CTC that could distinguish non-orthogonal quantum states. Such a circuit could then be used to break the BB84 protocol of quantum cryptography, which is impossible using only standard quantum mechanics. Later results based on Brun's circuit also indicated that a single CTC could be used to break the no-cloning bound\cite{AHN10}.

It is an interesting philosophical question whether the laws of nature should permit such forms of ``super-computation"\footnote{This is not to be confused with the distinct concept of ``hypercomputation", which refers to a process not computable by a Turing machine.}. Some authors have suggested that these conclusions are based on erroneous assumptions about the interpretation of mixed states in Deutsch's model\cite{BEN09}. We will review this argument briefly in Sec \ref{SecBennett} and it will be taken up again in Chapter \ref{ChapNonlinBox}. It has also been argued that nonlinear theories in general lead to similar effects, apparently independently of whether or not time travel is involved\cite{ABR98}. It is hoped that the framework of nonlinear boxes discussed in Chapter \ref{ChapNonlinBox} may shed light on the latter considerations, but that is left to future work.

\subsection{Open trajectories and entanglement breaking\label{SecEntBreak}}

\begin{figure}
 \includegraphics[width=16cm]{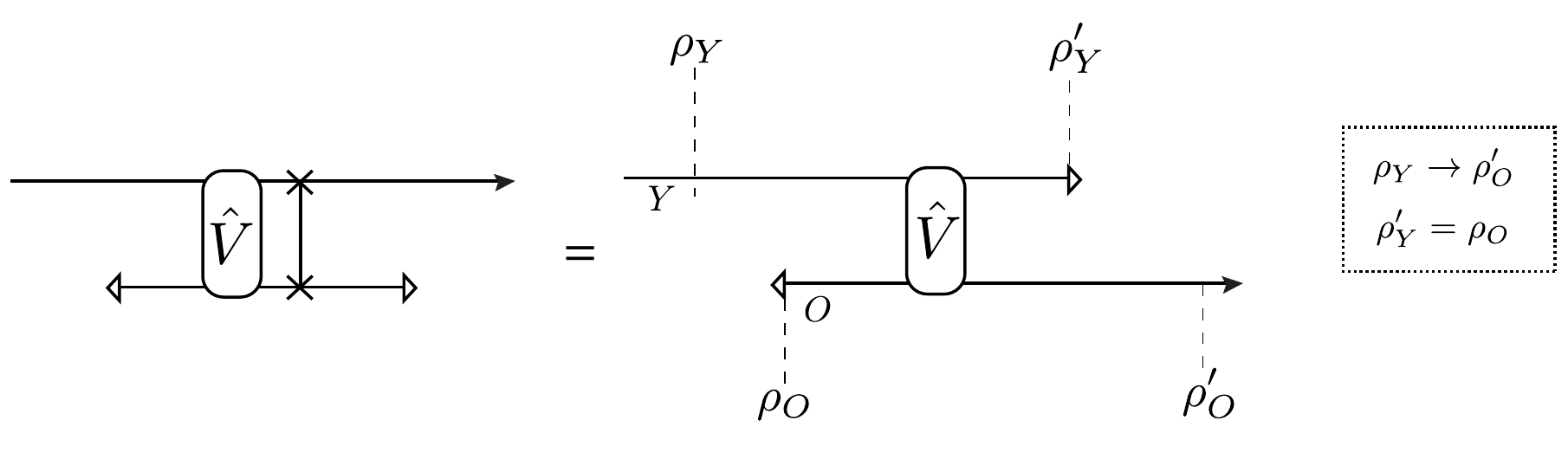}              
 \caption[A time-travelling system interacting with itself]{\label{DeutschCirc2}An equivalent arrangement of the interaction between CR and CTC systems. Now the interaction occurs  between the younger ($Y$) and older ($O$) versions of the same time-travelling system.}
\end{figure}

Let us rewrite Deutsch's circuit of Fig. \ref{DeutschCirc} in a more suggestive form by choosing the convention $U := V \otimes \trm{SWAP}$ and rearranging the rails as shown in Fig. \ref{DeutschCirc2}. Instead of an interaction between a separate CR system and CTC system, we can view this circuit as representing the interaction between ``older" and ``younger" versions of the \tit{same} system. The associated interpretation is that the incoming system ``jumps backwards in time" and interacts with itself before escaping to the unambiguous future. To adapt our terminology to this situation, we will refer to this system as a ``time traveling" (TT) system, this being equivalent to any CR system that interacts with a CTC system along its route. We adopt this alternative representation for convenience, but we emphasise that this is purely a matter of convention and it is always a straightforward matter to revert to Deutsch's original picture.
  
\begin{figure}
 \includegraphics[width=16cm]{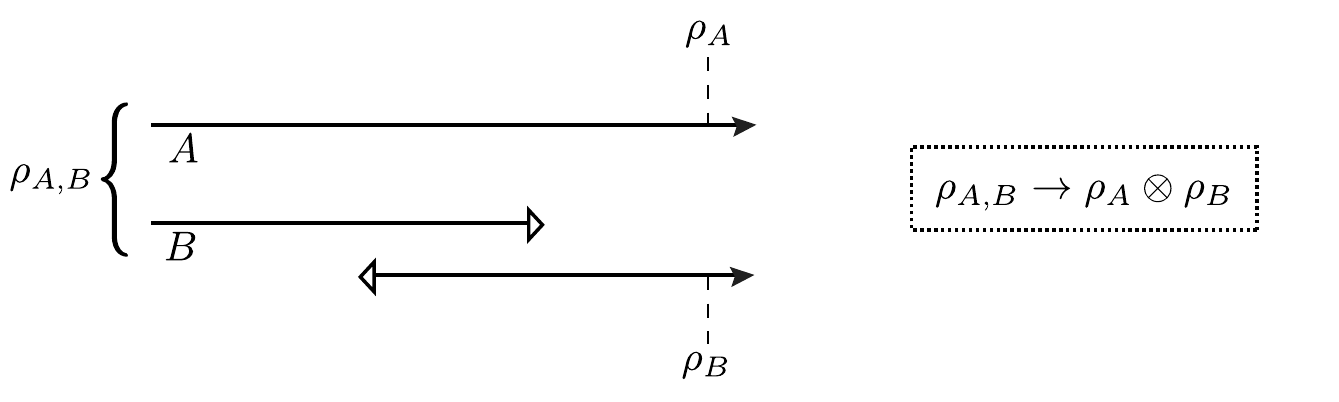}              
 \caption[A pair of initially entangled systems lose their entanglement when one of them is sent back in time]{\label{figOTC}A pair of initially entangled systems lose their entanglement when one of them is sent back in time, even though there is no interaction between its older and younger versions.}
\end{figure}
  
It will be instructive to consider a particularly simple instance of this circuit, shown in Fig. \ref{figOTC}. Here, we have a CR system labelled ``A" and a time-travelling system labelled ``B". System B jumps back in time without interacting with itself, while system A evolves trivially\footnote{Equivalently, in Deutsch's original picture, system B is a CR system that interacts with a CTC via a SWAP gate.}. The lack of any interaction between older and younger parts of the time-travelling system gives the appearance of an open loop, leading us to refer to this circuit as an ``open timelike curve" (OTC)\footnote{Note that this terminology refers to a particular choice of interaction and not to the properties of the underlying spacetime. In particular, the metric is still assumed to contain a closed timelike curve along which the system could  propagate.}. Let us assume that the systems are initially prepared in the joint state $\rho_{A,B}$. Using Deutsch's recipe, we see that the input undergoes the map:
\eqn{\label{OTC1} \rho_{A,B} \impm \rho_A \otimes \rho_B \label{EqEntbreak}}
where
\eqn{\rho_A = \ptrc{B}{ \rho_{A,B} } \, \nonumber ,\\  \rho_B = \ptrc{A}{ \rho_{A,B} } \, \nonumber. }
In the case that the systems are initially separable, i.e. $\rho_{A,B}=\rho_A \otimes \rho_B$, the map \eqref{OTC1} is equivalent to the identity. However, if the systems A and B are initially entangled, then the map \eqref{OTC1} completely destroys all entanglement between the two systems. This effect cannot be detected by local experiments on each subsystem, but it certainly affects experiments performed on the joint system at the output. Despite the resemblance to ordinary decoherence, this effect is a nonlinear map that cannot be replicated by any familiar decoherence process in standard quantum mechanics\footnote{The nonlinear dependence on $\rho_{A,B}$ can be seen by substituting the expressions for $\rho_A$ and $\rho_B$ into the RHS of \eqref{EqEntbreak}.}. This ``entanglement breaking" effect is a necessary consequence of Deutsch's consistency condition, which is restricted to the reduced state of the CTC system. To avoid it, one would have to employ an alternative consistency condition that applies to both the CTC system and to any CR systems with which it is entangled; however, this can lead to nonlocal effects as we discuss in Sec \ref{SecSignal}.

\section{Extensions of the Deutsch model\label{SecGenDeutsch}}

It would perhaps be an understatement to say that the Deutsch model leaves many questions unanswered. Nevertheless, it illustrates that one can construct a consistent toy model in the presence of causality violation, which is a remarkable result in itself. In this section we discuss what ingredients are missing from the Deutsch model and what directions have been taken in the literature to fill these gaps. One proposal, the ``equivalent circuit" model, has particular relevance to the present work and is reviewed in the following section.

\subsection{Relativistic extensions\label{SecRelExtens}}

\begin{figure}
 \includegraphics[width=18cm]{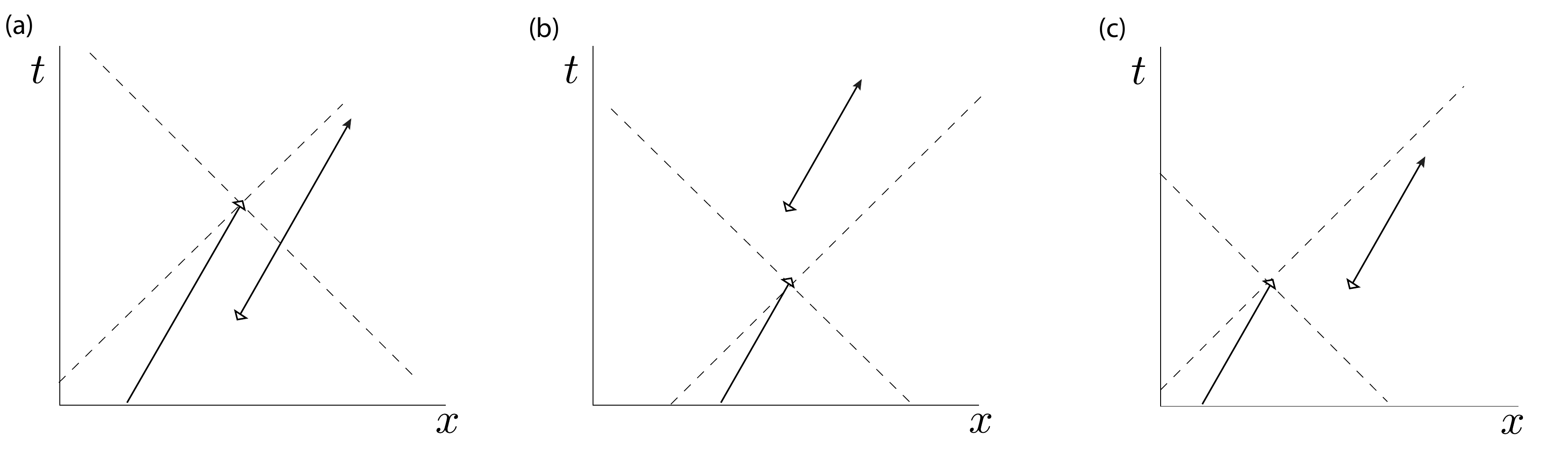}              
 \caption[Variations of the Deutsch model based on different spacetimes]{\label{figTimeVsLightST}Three different ways to arrange the endpoints of a ``CTC" trajectory. The solid line represents the world line of a time-travelling system, while the dashed lines indicate the boundaries of light-cones. Diagram (a) shows a standard CTC trajectory, where the system travels into its own past light-cone. In (b), the system jumps into the future, while in (c) it is teleported to a spacelike separated location.}
\end{figure}

Deutsch's model is intended to describe quantum dynamics in the presence of metrics containing CTCs. However, one might entertain the possibility of extending the model to describe more general effects. For example, consider the case of an OTC as described in Sec \ref{SecEntBreak}. Let us label the event of the system being sent back in time as $F$, and the event of the system appearing in its own past as $P$. For a CTC, the events $P$ and $F$ are necessarily timelike or lightlike separated, depending on whether the system is massive or massless, and $F$ lies in the future light-cone of $P$. But we might consider cases where $P$ lies in the future light-cone of $F$, corresponding to a sort of ``wormhole into the future" (Fig. \ref{figTimeVsLightST} (b)), or we might let $P$ and $F$ be spacelike separated, as would apply to a metric containing a wormhole between two distant parts of space\cite{MOR88} (Fig. \ref{figTimeVsLightST} (c)). In each of these cases we can define a consistency condition connecting the state of the system at two different events in spacetime and thereby obtain a corresponding toy model. 

Aside from considering other metrics, one would also like to make the Deutsch model compatible with special relativity. As it stands, the model assumes that the spacetime dependence of the particle wavefunctions can be neglected. A relativistic quantum field theory based on the Deutsch model and describing wavepackets with arbitrary spacetime dependence might provide an alternative to the standard paradigm of quantum field theory on a curved spacetime background. In Chapter \ref{ChapCTC2} we will make a proposal along these lines. 

\subsection{The initial state assumption\label{SecInitState}}
There is an interesting asymmetry in Deutsch's model. Although we allow the output of the interaction to be entangled, the initial states of the CR and CTC systems are assumed to be in the product form $\rho_1 \otimes \rho_2$. Is there a physical reason for making this assumption? Deutsch argues\cite{DEU91}:\\
\tit{``This assumption about the form of the initial and final states of interacting systems is a general one that underlies quantum measurement theory, namely that physically separated systems get into entangled quantum states only \tit{after} interacting with each other."}\\
Deutsch's invocation of an unambiguous meaning of the term ``after" in explaining the asymmetry seems quite reasonable given our \tit{a priori} assumption of an unambiguous past and future for the CR systems. Nevertheless, Politzer\cite{POL94} has pointed out that, were we to adopt the point of view of the system on the CTC, then the two systems certainly \tit{would} have interacted previously. We might therefore be tempted to construct a more symmetric picture which permits entanglement between the two inputs. In fact, the `equivalent circuit' version of the model discussed in Sec \ref{SecEquivCirc} could provide a natural framework for relaxing the assumption of an initial product state. Such a generalisation would also likely be connected to the problem of generalising the model to multiple CTCs. We do not pursue this line of investigation here, but mention it as a topic for future work. 

\subsection{Multiple CTCs \label{SecMultiCTC}}
The circuit we have been considering (Fig. \ref{DeutschCirc}) only applies to a system or collection of systems traversing a single CTC. In considering more than one CTC, one faces the task of extending Deutsch's consistency condition to multiple CTCs. As it turns out, there is more than one way to perform this extension, with little guidance from physical principles as to which method is more natural.

\begin{figure}
 \includegraphics[width=18cm]{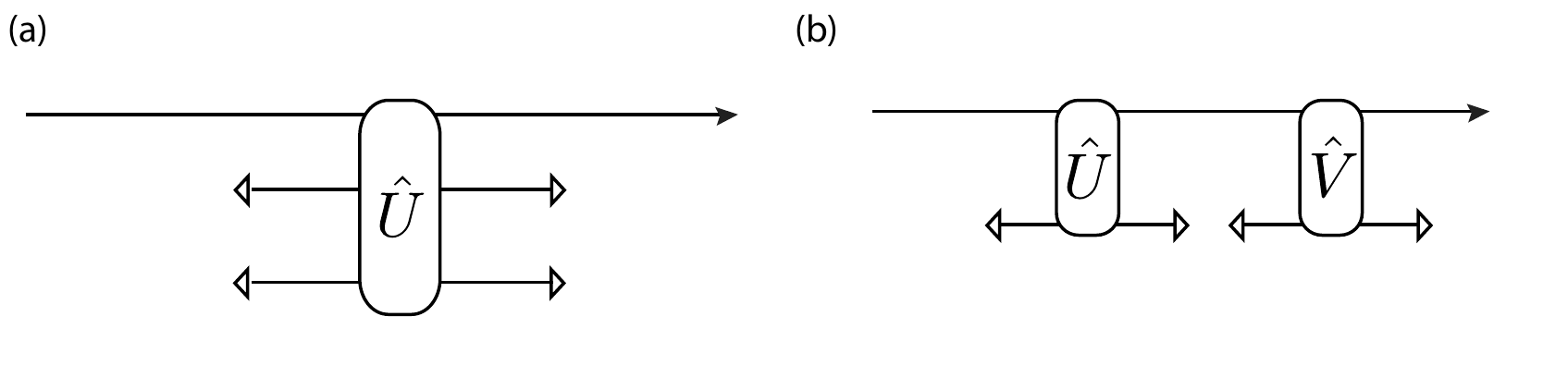}              
 \caption[Simultaneous and sequential CTCs]{\label{figMultiCTC}Two basic types of interaction between a pair of CTCs: (a) simultaneous and (b) sequential.}
\end{figure}

Consider the two basic ways in which a pair of CTCs might be related to one another in a quantum circuit. Fig \ref{figMultiCTC} (a) shows pairs of CTCs in a \tit{simultaneous} or \tit{spacelike separated} relationship, in which it may be possible for a single gate to connect both CTCs. Fig \ref{figMultiCTC} (b) shows a \tit{sequential} or \tit{timelike} relationship in which the output of one CTC forms the input of the next. We can characterise these two types of connection using the endpoints of the CTC rails in the circuit. We denote the temporal locations of the past endpoint by $P$ and the future endpoint by $F$. By definition, for a single CTC, $P$ always precedes $F$ in the temporal ordering: $P<F$. A pair of CTCs, labelled $X$ and $Y$, are called \tit{sequential} iff $F_X<P_Y$ or $F_Y<P_X$, i.e. if both endpoints of one CTC precede both endpoints of the other. If neither of these conditions is met, the CTCs are said to be \tit{simultaneous} and it is possible to connect the two CTC rails with a single gate -- provided, of course, that the trajectories of the two systems pass through a small region of spacetime localised to the interaction event. We can construct more general circuits by concatenating CTC circuits in either of these two ways. The problem that we face is to extrapolate Deutsch's consistency condition \eqref{consistency} to situations involving many CTCs with different relationships to one another in the circuit. 

For a pair of simultaneous CTCs, there are two possibilities for extending the consistency condition. \tit{Option (i)}: apply the consistency condition to the joint state $\rho_{2,3}$ of the systems on their respective CTCs, thus preserving any entanglement between them. The consistency condition is then:
\eqn{
\rho_{2,3}=\ptrc{1}{\hat{U} \,(\rho_1\otimes \rho_{2,3})\, \hat{U}^\dagger} \,.
}
\tit{Option (ii)}: apply the consistency condition separately to the reduced states $\rho_2$ and $\rho_3$ thereby breaking any entanglement between the time-travelling systems. The two conditions are:
\eqn{
\rho_{2}=\ptrc{1,3}{\hat{U} \,(\rho_1\otimes \rho_{2}\otimes \rho_{3})\, \hat{U}^\dagger} \, ,\nonumber \\
\rho_{3}=\ptrc{1,2}{\hat{U} \,(\rho_1\otimes \rho_{2}\otimes \rho_{3})\, \hat{U}^\dagger} \, .
}
(The generalisation to more than two CTCs is straightforward.) In the first case \tit{(i)} there is a total consistency condition for both CTCs, essentially treating them as two systems traversing a single CTC. In particular, this preserves entanglement between the two CTC systems, which might not seem appropriate if we are regarding these as completely independent systems. In the second case there are two independent consistency conditions, one for each CTC. While \tit{(ii)} seems more intuitive, there may not always exist a solution that satisfies the multiple consistency conditions\cite{LIV07}.

Next consider the pair of sequential CTCs shown in Fig. \ref{figMultiCTC}(b). Here, the ambiguity enters in the choice of initial condition for the joint CTC plus CR systems. For a single CTC, we assumed that the input state was of the product form, forbidding the input from being entangled to the CTC system. However, if the output of the first CTC is the input to a second CTC, ought this condition continue to hold? After all, one might entertain the possibility of ``memory effects" whereby the sequential CTCs have access to shared degrees of freedom. In that case it is not obvious that we should apply the ``no initial entanglement" rule to subsequent inputs, but the abandonment of this rule would require a generalisation of Deutsch's model along the lines discussed in Sec \ref{SecInitState}. On the other hand, it seems equally reasonable to posit that sequential CTCs should be free of such memory effects. In fact, Bacon's demonstration of computational speed-up using multiple sequential CTCs (Sec \ref{SecInfParadox}) implicitly assumes the latter. One cannot easily decide between these possibilities without a deeper understanding of what the physical CTC degrees of freedom represent.

\subsubsection{What happened to the Heisenberg picture?\label{SecNoHeis}}
Deutsch's model is formulated explicitly in the Schr\"{o}dinger picture i.e. as a dynamical map from an initial to a final state. Formally, this means that the physical quantities (expectation values) are written in the form:
\eqn{
\trc{\sigma \, \rho'}=\trc{\sigma \, \mathcal{M}(\rho)} \, ,
}
where $\rho$ and $\rho'$ are the density matrices of the initial and final states, related by the map $\mathcal{M}$, and $\sigma$ is a Hermitian matrix representing a physical observable. In standard quantum mechanics, $\mathcal{M}$ is a completely positive trace-preserving map. Without loss of generality, let us take $\mathcal{M}$ to be unitary. Then the linearity of quantum dynamics ensures that $\mathcal{M}$ can be associated with a unitary matrix $\hat{U}_M$ according to:
\eqn{\label{eqnNoHeis}
\mathcal{M}(\rho)=\hat{U}_M \, \rho \, \hat{U}^\dagger_M \, .
}
Using the cyclic property of the trace, we can rewrite the expectation values in the Heisenberg picture:
\eqn{
\trc{\sigma \, \rho'}&=&\trc{\sigma \, \hat{U}_M \, \rho \, \hat{U}^\dagger_M}\\
&=&\trc{\hat{U}^\dagger_M \, \sigma  \, \hat{U}_M \rho} \\
&:=&\trc{\mathcal{M}^{-1}(\sigma) \, \rho} \, ,
}
where $\mathcal{M}^{-1}(\sigma)$ defines the dynamic evolution of the observable, keeping the state fixed. Thus the Schr\"{o}dinger evolution of the state induces a corresponding Heisenberg evolution of the observable \tit{by linearity}. For nonlinear maps such as Deutsch's map, it is no longer clear that this can be done. The decomposition \eqref{eqnNoHeis} no longer applies in general, as the unitary $\hat{U}$ might be state-dependent. The question arises: can every instance of the $\mathcal{N}_D$ on the state space be placed in one-to-one correspondence with some map $\mathcal{N}'_D$ on the space of Hermitian operators, such that $\trc{\sigma \, \mathcal{N}(\rho)} \equiv \trc{\mathcal{N}'(\sigma) \, \rho}$ ? This might be one path towards recovering a Heisenberg picture of nonlinear maps defined in the Schr\"{o}dinger picture, but we leave the general problem to future work. For the specific case of the Deutsch model, it turns out that there is a natural way to restore the Heisenberg picture; this is discussed in Sec \ref{SecEquivCirc} and will be relevant to later chapters.

\subsection{Locality and superluminal signalling with CTCs\label{SecSignal}}

Recalling our comments in the introduction, it is worth taking a diversion to discuss the relationship between CTCs and locality. We adopt the following definition:\\

\tit{Consider two events, $e_1$ and $e_2$, such that $e_1$ exerts a causal influence on $e_2$. ``Signal locality"\footnote{In this thesis, we also frequently refer to the property of signal locality as ``no-signalling".} is satisfied iff $e_2$ lies in the future light-cone of $e_1$ for all such pairs of events in a given theory.}\\ 

(Here, a ``causal influence" from $e_1$ to $e_2$ is taken to mean that an intervention by an experimenter in the vicinity of $e_1$ changes the probabilities of observable random variables in the vicinity of $e_2$ -- for a rigorous discussion of these matters, see eg. Ref \cite{CAVW12}). In classical GR, this requirement is automatically satisfied for all pairs of events, even in globally nonhyperbolic spacetimes. Even if the two events lie on a CTC, such that there is an ambiguity about the choice of labelling of past and future light-cones, the locality requirement is satisfied for either choice of labelling.

It is important to clarify the connection between the ability to send signals faster than light and causality violation. In this thesis, the term ``superluminal signalling" refers to an intrinsically \tit{nonlocal} phenomenon, defined as being a causal influence from $e_1$ to $e_2$ such that $e_2$ lies outside the future light-cone of $e_1$. In particular, since CTCs in GR are local by our criteron, we must conclude that they cannot be used to send superluminal signals. This might seem absurd at first glance: surely we can imagine sending a message back in time using a CTC, such that the message arrives at a spacelike separated event ahead of a beam of light. Note, however, that in the presence of CTCs there might exist both a spacelike \tit{and} a timelike interval between two events, depending on the path taken. In this example, we could just as well have sent the beam of light itself through the CTC, in which case the light would still have arrived ahead of the message. Nonlocal superluminal signalling doesn't just require the transmitted message to travel faster than some particular beam of light, it has to travel faster than a beam of light \tit{that follows the optimal geodesic through the metric}.

It is known that nonlinear modifications to quantum mechanics can lead to nonlocal superluminal signalling\cite{GIS90,JOR99}, raising the question of whether such effects can occur for Deutsch's nonlinear map. If this were found to be the case, we might take it as a failure of the theory. We emphasise that it is not the causality-violating aspect of superluminal signalling that we object to, but the \tit{locality}-violating aspect.

Early path-integral approaches to quantum mechanics with CTCs relied upon a renormalisation of the final state of the system so as to preserve entanglement with other systems. This entails a consistency condition that extends to entangled systems outside the CTC region, rather than just to the reduced state of the CTC system\footnote{Recall the discussion at the end of Sec \ref{SecEntBreak}}. Hartle and Politzer both observed that, as a consequence of this consistency condition, the outcomes of experiments performed at the present time would depend upon whether a CTC will exist in the distant future\cite{HAR94,POL94}. It is not immediately clear that this violates locality, because it is not obvious how to manipulate the existence or nonexistence of a future CTC in order to send a message or exert a causal influence on the past. However, the situation is clearer in the equivalent model of postselected CTCs (P-CTCs)\cite{LLO11}. In this model, it can be demonstrated that a P-CTC in the future can be used to send signals arbitrarily far into the past, thereby violating locality\cite{RAL12}. The Deutsch model is able to avoid the above conclusion by restricting its consistency condition to the local system traversing the CTC and hence breaking all entanglement to external systems. As a consequence, it is not obvious that the Deutsch model should lead to signalling -- but nor is it obvious that the converse should be true\footnote{We are careful to distinguish our presentation of Deutsch's model from Deutsch's own formulation, which included a multiverse interpretation of the model. As noted by Deutsch, the multiverse interpretation is trivially sufficient to rid the model of ``Gisin's telegraph", the proof of which he leaves as an exercise. It is our aim to go beyond this basic observation and quantify not just what is sufficient, but what is \tit{necessary} to avoid superluminal signalling in Deutsch's model and other nonlinear theories.}. The question of signalling in general nonlinear theories will be discussed in Chapter \ref{ChapNonlinBox} and the status of Deutsch's model in this regard will be discussed in Sec \ref{SecDeutschBox}.

\subsection{Is the power of CTCs unobservable?\label{SecBennett}}

Following the results of Bacon and Brun regarding the apparent computational power of CTCs, Bennett et al. argued that the apparent power of the Deutsch model was drastically weaker than these results indicated\cite{BEN09}. Their key observation was that the protocols of earlier papers had only specified the input--output relations for a set of pure states, which is a set of measure zero in the space of possible input states. Any realistic preparation procedure always results in some small amount of error. The authors then pointed out that the standard procedure of computing the evolution of mixed states from the pure state decomposition is no longer valid for nonlinear maps such as Deutsch's map, an assumption they called the ``linearity trap". It follows that in order to evaluate the true predictions of the Deutsch model, one has to evaluate its action on arbitrary density matrices, not just a set of pure states.

At this point, it becomes apparent that the Deutsch model as stated is incomplete -- for although it specifies the dynamical map from inputs to outputs given some input density matrix, it does not provide any clear guidelines for deciding when the input should be described by a mixed density matrix or a pure state. As discussed in Sec \ref{SecInfParadox}, the density matrix cannot take into account epistemic contributions to uncertainty. Therefore the state ascribed to a system in the Deutsch model must be determined by a specification of its preparation procedure, according to some as-yet unspecified rules.

Whether or not Bennett et. al. appreciated this subtlety, they adopted the simplest point of view: that any density matrix that can be used to describe the input state in standard quantum mechanics should also suffice in Deutsch's model. This is not obviously an unreasonable assumption, but it quickly leads into absurdity, since standard quantum mechanics does not distinguish operationally between mixing due to subjective ignorance and mixing due to objective properties of the system. Ordinarily this distinction is irrelevant, but in the Deutsch model it becomes essential. Without it, agents who have different subjective knowledge about the system make inconsistent predictions about the outcomes of experiments! The inconsistency of Bennett et al's assumption was exposed by Cavalcanti and Menicucci using a model-independent ``Dutch-book" argument\cite{CAV10}. They showed that the assumption of ``verifiable" nonlinear dynamics was inconsistent with Bennett et al's assumption that the state of a system can be represented by its density matrix alone. Thus we must either rule out nonlinear dynamics by fiat, or else adopt a more sophisticated method for representing states in the theory. Taking the latter approach, the question of the information-processing power of CTCs remains open. We will return to these matters in Chapter \ref{ChapNonlinBox}.

\section{The equivalent circuit\label{SecEquivCirc}}

In this section, we review an alternate formulation of Deutsch's model using an ``equivalent circuit", due to Ralph \& Myers\cite{RAL10}. Remarkably, the equivalent circuit is both unitary and linear. However, it comes with a caveat: we require a large number of additional copies of the unknown input state. These may be regarded as just a convenient abstraction, or else we might posit that the copies represent real physical systems. The advantages of the formalism are discussed in Sec \ref{SecECDeriv} and the interpretational implications are briefly considered in Sec \ref{SecECInterp}

\subsection{Formal Derivation\label{SecECDeriv}}

\begin{figure}
 \includegraphics[width=16cm]{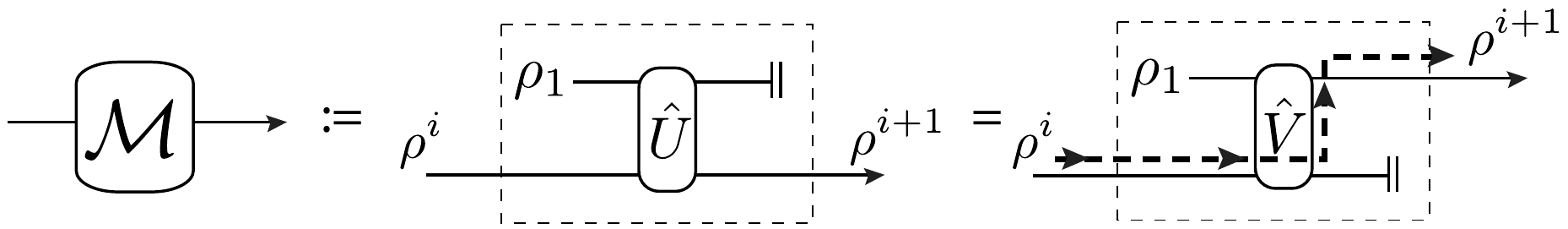}              
 \caption[Deutsch's map as a state-dependent CPT map]{\label{DMapCirc}Deutsch's consistency condition requires that $\rho_2$ is a fixed point of the CPT map $\mathcal{M}$, which can be represented as a circuit that depends on $\rho_1$ and the interaction $\hat{U}$ (or $\hat{V}:=\hat{U}\otimes \trm{SWAP}$ using the convention of Sec \ref{SecEntBreak}).}
\end{figure}

\begin{figure}
 \includegraphics[width=16cm]{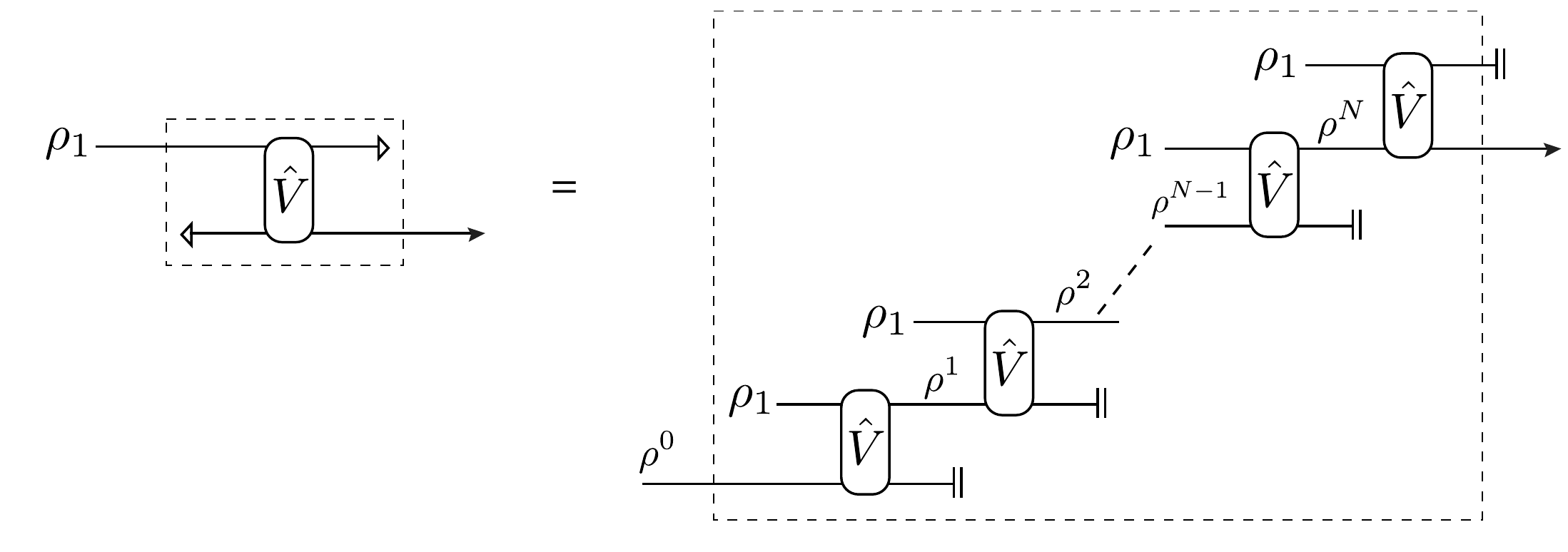}              
 \caption[The equivalent circuit to Deutsch's model]{\label{EquivCirc}The evolution of an arbitrary initial state $\rho^0$ through many iterations of the circuit corresponding to $\mathcal{M}$ produces the state required by Deutsch's consistency condition. We can employ this circuit instead of directly applying the consistency condition, leading to a unitary dilation of Deutsch's map in the form of the ``equivalent circuit" shown on the right.}
\end{figure}

Recall the consistency condition \eqref{consistency} for the state $\rho_2$ on the CTC. This equation can be written:
\eqn{\rho_2=\mathcal{M}(\rho_2)\nonumber \, ,}
where
\eqn{\label{DMap} \mathcal{M}(x):=\ptrc{1}{ \hat{U} (\rho_1 \otimes x) \hat{U}^\dagger } }
is a CPT map defined by $\hat{U}$ and $\rho_1$. It is clear that $\rho_2$ is a fixed point of $\mathcal{M}$. A standard method of finding a fixed point of a given map is to begin by guessing a solution, for example $\rho_2=\rho^{0}$. Applying the map, we obtain the first iteration, $\rho^{1}=\mathcal{M}(\rho^{0})$. If $\rho^{1}$ is appreciably different to $\rho^{0}$, then we iterate further to obtain $\rho^{2}=\mathcal{M}(\rho^{1})$. As the number $N$ of iterations increases, the state approaches a fixed point. The exact solution can be obtained analytically by taking the formal limit of $\rho^N$ as $N \impm \infty$. For a map with multiple fixed points, different choices of the initial guess $\rho^0$ will lead to different fixed points in the infinite limit. It can be proven that choosing $\rho^0=I$ will result in the maximum entropy solution, thereby reproducing Deutsch's maximum entropy principle\cite{RAL10}.
Since $\mathcal{M}$ is an ordinary CPT map, it can be represented by a quantum circuit that maps $\rho^{i}$ to $\rho^{i+1}$, as shown in Fig. \ref{DMapCirc}. Note that the circuit makes use of a single copy of the state $\rho_1$ as an ancilla. Consider now the expression for the fixed point:
\eqn{\label{Iterate} \rho^N=\mathcal{M}(\mathcal{M}(\mathcal{M}(...\mathcal{M}(\rho^0)...))) \, ,}
where $N$ is arbitrarily large. The term on the RHS is also a CPT map, obtained by repeated concatenation of the map $\mathcal{M}$. It can therefore be represented by the $N$-fold concatenation of the circuit of $\mathcal{M}$, using $N$ ancillary copies of the state $\rho_1$. We can therefore interpret $\rho_2$ as the output of this circuit with $\rho^0$ as the input. Substituting this into Deutsch's circuit, we obtain the equivalent circuit shown in Fig. \ref{EquivCirc}. Note that the output must be the same as that prescribed by Deutsch's consistency condition, because the state entering the final unitary has reached a fixed point. Hence the equivalent circuit provides an alternative means of performing calculations in the model, where the consistency condition is replaced by an evolutionary process involving an arbitrarily large number of copies of the initial state.

The simplest way to view the equivalent circuit is as a mathematical abstraction that is convenient for performing calculations. This does not require any additional assumptions beyond those required for Deutsch's original circuit; the equivalent circuit is simply an alternative formal representation of Deutsch's map. The main benefit of performing calculations using the equivalent circuit is that it has a natural formulation in the Heisenberg picture. One simply provides $N$ copies of the initial state, and then expectation values can be calculated in the initial state by applying the equivalent circuit unitary map to the observables\footnote{This will be carried out explicitly in Chapter \ref{ChapCTC1}.}. 

Another benefit of the equivalent circuit formulation of Deutsch's model is that it allows one to derive Deutsch's maximum entropy rule, using the following reasoning. Any physical process must be subject to some small amount of decoherence. Introducing this decoherence into the equivalent circuit, we find that for any choice of input state $\rho^0$ there is some large number of iterations $M < N$ after which the state has been completely decohered: $\rho^M=I$. This implies that the limiting value of the fixed point for arbitrarily large $N$ is always equal to the maximum entropy solution, regardless of the initial choice of $\rho^0$ (see the Appendix of \cite{RAL10} for more details). Hence Deutsch's maximum entropy principle is not an \tit{ad hoc} addition to the model but arises naturally when one considers the equivalent circuit.

\subsection{Interpretational implications\label{SecECInterp}}
There is one final aspect of the equivalent circuit that deserves some discussion. Rather than treating it as an abstraction, we can postulate that the circuit represents the real state of affairs underlying Deutsch's model. It follows that the copies of the input state represent real physical systems, rather than just a representational convenience. While this interpretation is somewhat radical, it does have certain interesting implications.

First, this interpretation implies that Deutsch's model is only apparently nonlinear and nonunitary; the real situation involves $N$ copies of the prepared system (whose origin remains mysterious) that evolve according to standard quantum mechanics and can interact with one another only when a CTC is present. While this view has a distinctly ``multiverse" flavour, similar to Deutsch's own interpretation of his model, the physical nature of this multiverse remains to be explicated\footnote{In particular, it remains to be specified whether the multiverse consists of actual parallel universes or can be regarded as a quantum mechanical multiverse corresponding to a single wavefunction of the universe. See also the discussion at the end of Sec \ref{SecDeutschBox}}. In so doing, one would also need to explain the origin of the copies, which would presumably involve some physical constraint on the total number of copies $N$.

This interpretation would also have consequences for the possibility of computational speed-up and the breaking of quantum cryptography. The extra copies of the input state must now be counted as real additional resources, which add to the cost of any computation. At least up to the input state, therefore, the evolution is governed by standard quantum mechanics and cannot provide any additional computational advantage. Since the apparent advantage now comes from the pre-existing resources, namely the extra copies of the input state, the origin of these resources must ultimately be accounted for in a theory that is based upon this interpretation.

Finally, whereas the loss of unitarity in the Deutsch model might be construed as a fundamental loss, the equivalent circuit retains information which could in principle play a part in more complicated interactions. Thus the equivalent circuit may serve as a platform to generalising the model to multiple CTCs with ``memory effects", as discussed in Sec \ref{SecMultiCTC}, whereby coherence lost in an interaction with one CTC may be regained in interactions with subsequent CTCs. 

Our results in later chapters will require the equivalent circuit formalism, particularly to serve as a Heisenberg picture for the Deutsch model. Nevertheless, our results are independent of whether one chooses to interpret the equivalent circuit as a strictly formal device or as a description of reality underlying Deutsch's model. We leave such questions to future work. 
  
\section{Concluding remarks\label{SecDeutschConcl}}

While Deutsch's model represents a novel framework for describing quantum dynamics in the presence of causality violation, some of its physical predictions remain unclear. The locality of the model has been asserted, but not proven, and there remains the question of how to properly address the ``linearity trap" raised by Bennett et. al. These issues seem to hinge on additional assumptions of an interpretational nature that have not been spelled out; we will frame them in more precise terms in Chapter \ref{ChapNonlinBox}. Some results, like Bacon's demonstration of computational speed-up in Sec \ref{SecInfParadox}, also depend on how the model is generalised to multiple CTCs and this choice remains to be justified from physical principles.

In Sec \ref{SecRelExtens} we noted that Deutsch's model could benefit from a relativistic generalisation. A barrier to this is the fact that the model seemed to lack a formulation in the Heisenberg picture, which is the natural picture for a relativistic field theory. Fortunately, we saw that the equivalent circuit model of Sec \ref{SecEquivCirc} can be described in the Heisenberg picture and can therefore be used as the starting point for such a generalisation. This will be taken up in Chapters \ref{ChapFields}, \ref{ChapCTC1},  and \ref{ChapCTC2}.

\clearpage
\pagestyle{plain}
\pagebreak
\renewcommand\bibname{{\LARGE{References}}} 
\bibliographystyle{refs/naturemagmat2012}
\addcontentsline{toc}{section}{References} 

\pagestyle{fancy}
\renewcommand{\sectionmark}[1]{\markboth{\thesection.\ #1}{}}

\chapter{Nonlinear Quantum Boxes}
\label{ChapNonlinBox}

\begin{quote}
\em
What I did find disappointing was that [my] nonlinear alternative to quantum mechanics turned out to have purely theoretical internal difficulties. At least for the present I have given up the problem; I simply do not know how to change quantum mechanics by a small amount without wrecking it altogether.
\end{quote}
Steven Weinberg\\


\newpage

\section*{Abstract}
We have seen that Deutsch's model presents an example of a nonlinear modification to quantum mechanics by the introduction of a nonlinear dynamical map $\mathcal{N}_D:\mathcal{P}\rightarrow \mathcal{P}$ on the space $\mathcal{P}$ of density operators acting on the time travelling system's Hilbert space. In this section, we define the general class of such maps in Sec \ref{SecDefineNL}. In Sec \ref{SecNLBox}, we restrict our considerations to nonlinear effects occurring within a bounded region of spacetime (a ``nonlinear box"), enabling us to apply the operational formalism introduced in Sec \ref{SecOperational}. This allows us in Sec \ref{SecVerifiability} to define the important concept of ``operational verifiability" and in Sec \ref{SecBoxSignal} to derive a necessary and sufficient condition for verifiable nonlinear boxes to be free from superluminal signalling. In Sec \ref{SecDeutschBox} we formulate Deutsch's map as a nonlinear box and incorporate some results from the general literature, thereby refining the model and placing it in a broader context. 

This chapter incorporates many results and observations that have been made by other authors in previous work\footnote{See particularly Refs. \cite{GHE02,CZAK98,CZA98,CZAD02,POL91,KEN05}.}; however the usage of the operational formalism to unite these features and to make explicit the role of ontology in nonlinear box theories is, as far as the author is aware, original. The original parts of this chapter are based on work undertaken with E.G. Cavalcanti and N.C. Menicucci and credit is shared equally by the authors (see Ref. \cite{CAV12}). Note that the presentation of these results in this thesis differs from the notation used in Ref. \cite{CAV12}, with more emphasis now placed on the representation of states by probability densities on $\mathcal{P}$. The reason for this change is that it allows for a clearer exposition of the necessary and sufficient condition for no-signalling; however, the author accepts sole responsibility for any errors that may have arisen due to the modified presentation.

\section{Nonlinear dynamics \label{SecDefineNL}} 

Recall from Chapter \ref{ChapQM} that the dynamics of a single system in ordinary quantum mechanics is given by a linear CPT super-operator $\scr{M}$ acting on the space of density operators on the system's Hilbert space $\scr{H}$. The map takes the initial state $\rho \in \scr{P(\scr{H})}$ to the final state $\scr{M}(\rho)$. It is then straightforward to relax the assumption of linearity to obtain a general definition of nonlinear dynamics:\\

\tbf{Definition 1a (nonlinear dynamics):} Let $\mathcal{N}:\mathcal{P}\rightarrow \mathcal{P}$ be a positive and trace-preserving map on the space of density operators $\mathcal{P}$. The dynamics represented by $\mathcal{N}$ is called \tbf{nonlinear} iff it does not satisfy the property of linearity, i.e. if there exists two distinct states $\rho_1, \rho_2$ and a third state formed by their convex combination $\rho_3:=\gamma \rho_1+ (1-\gamma) \rho_2$ for some $0<\gamma<1$, such that $\mathcal{N}(\rho_3) \neq \gamma \mathcal{N}(\rho_1)+(1-\gamma) \mathcal{N}(\rho_2)$.\\

Note that any positive map on $\scr{P}$ is either linear or nonlinear\footnote{We will generally reserve $\scr{M}$ for linear maps and $\scr{N}$ to denote general super-operators that may or may not be linear -- this will be made clear by the context.}. Note also that this defines a deterministic nonlinear map, but says nothing about how one ought to classify \tit{stochastic} nonlinear maps. This is unfortunate, since one of the most important processes in quantum mechanics -- the measurement process -- can be regarded as a stochastic nonlinear map and therefore is not included in our discussion of nonlinear transformation processes. However, given the complexity of the topic, it will serve us well to be cautious and restrict our discussion to deterministic nonlinear maps for the time being. Just as standard quantum mechanics is most easily approached using linear deterministic maps before moving to the discussion of probabilistic dynamics, we hope that the present work on deterministic nonlinear maps will lay the groundwork for future extensions of the formalism to stochastic nonlinear maps.

The astute reader will have noticed that we only specified the positivity of $\scr{N}$; this is because there is a subtlety involved in defining \tit{complete} positivity for nonlinear maps, which we discuss in Sec \ref{SecNLExtend}. Finally, the definition we have given is a sufficient condition for a general super-operator to be nonlinear (namely, that its restriction to the space of density operators is nonlinear), but we have not shown that it is a necessary condition. In fact, a super-operator whose action on general linear operators is nonlinear might nevertheless act linearly on the subspace of density operators. This seems to present a problem, as such a nonlinear super-operator would count as linear by Definition 1a. To resolve this, we note that as long as we stay in the Schr\"{o}dinger picture, we are only interested in the action of the map on density operators and hence more general extensions of the map need not concern us\footnote{However, recalling the discussion of Sec \ref{SecNoHeis}, a possible extension of our definition to nonlinear dynamics in the \tit{Heisenberg} picture presumably would encompass more general maps of this kind.}.

\subsection{The ``linearity trap"\label{SecLinearityTrap}} 
In linear quantum mechanics, it is sufficient to specify the action of a dynamical map on a complete basis of states, for we can then extend the action of the map by linearity to arbitrary states. In general, this is not true for nonlinear maps. In particular, if $\scr{N}(\rho_1+\rho_2) \neq \scr{N}(\rho_1)+\scr{N}(\rho_2)$, then knowing $\scr{N}(\rho_1)$ and $\scr{N}(\rho_2)$ does not permit us to deduce $\scr{N}(\rho')$ where $\rho'=\lambda \rho_1 + (1-\lambda) \rho_2$ is a continuous family of states. This issue is important even in simple cases: consider a map $\scr{N}$ that always maps pure states to pure states. Specifying the action of $\scr{N}$ on a basis of pure states clearly fails to specify its action on arbitrary mixed states; but even worse, it also fails in general to specify the action of $\scr{N}$ on any pure states that aren't included in the basis! This is one of many ``linearity traps'" that one might fall into when discussing nonlinear theories. Such traps arise by the na\"{i}ve assumption that certain formal devices used in linear quantum mechanics will carry over to the nonlinear case. The particular linearity trap described above is the one pointed out by Bennett \tit{et. al.} in their criticism of Deutsch's model (see Sec \ref{SecBennett}). Specifically, it was pointed out that the dynamics of an arbitrary density operator $\rho$ under the map $\scr{N}$ cannot be obtained by decomposing $\rho$ into a sum of density matrices and then applying $\scr{N}$ to these substates. Thus, for an arbitrary decomposition $\rho=\sum_{i} \alpha_{i} \rho^{(i)}$ where the $\alpha_i$ are positive coefficients that sum to unity,
\eqn{\scr{N}(\rho) &=& \scr{N}\brac{\sum_{i} \alpha_{i} \rho^{(i)}} \nonumber \\ 
&\neq& \sum_{i} \alpha_{i} \scr{N}(\rho^{(i)}) \, .
} 
It follows that the action of $\scr{N}$ must be specified over the \tit{whole} space $\scr{P}$ of density operators, or else be regarded as an incomplete specification of the dynamics.

\subsection{Extensions to multiple systems\label{SecNLExtend}}

An important part of the dynamics of any physical theory is the ability to compose multiple independent systems into a single composite system. Ideally, if the systems are physically isolated from one another and their dynamics is independent, then there should be a formal way of describing the dynamics of the composite system such that the independence of its subsystems is maintained. In standard quantum mechanics, this is achieved by the tensor product together with the linearity of the dynamics and the partial trace, as discussed in Chapter \ref{ChapQM}. In the nonlinear case, we retain the notion of tensor products and partial traces (these being part of the \tit{kinematics} of the theory rather than the dynamics), but we have yet to demonstrate that a description of the \tit{dynamics} of multiple isolated systems is possible without appealing to linearity. We begin by defining the concept of a ``general positive extension" that determines the joint dynamics of multiple systems in a nonlinear context, i.e. where the total dynamics of the systems may be nonlinear. This then allows us to define a notion of complete positivity for nonlinear maps.  

\subsubsection{General positive extensions\label{SecGenPosExten}}
Consider two independent and non-interacting physical systems, labelled A and B, associated with Hilbert spaces $\scr{H}_A$ and $\scr{H}_B$. The initial reduced states of the systems are $\rho_{A}$ and $\rho_{B}$ respectively, and the systems may be entangled in general. Suppose the systems are subjected to independent possibly nonlinear dynamical maps $\scr{N}_A$ and $\scr{N}_B$, producing the respective outputs $\rho'_{A}=\scr{N}_A(\rho_A)$ and $\rho'_{B}=\scr{N}_B(\rho_B)$. We would like to define a new map $\scr{N}_{AB}$ acting on $\scr{P}\brac{\scr{H}_A \otimes \scr{H}_B}$ that defines the (possibly nonlinear) dynamics of the joint system in a manner consistent with the given dynamics of the subsystems.

In standard quantum mechanics, the dynamics of the joint system is given by the requirement:
\eqn{ \label{LinearExtend}
(\scr{M}_A\otimes \scr{M}_B) (\rho_A \otimes \rho_B) = \scr{M}_A(\rho_A) \otimes \scr{M}_B(\rho_B) \; \; \; \; \forall \rho_{A,B} \in \scr{P}, \,
}
together with the constraint that the joint map $\scr{M}_A\otimes \scr{M}_B$ be linear. This latter requirement allows one to extend \eqref{LinearExtend} by linearity to arbitrary states. If we allow nonlinear maps, then it is no longer possible to extend by linearity and we cannot constrain the joint evolution to be linear\footnote{This leads to the peculiar possibility of a theory in which two subsystems undergo linear maps $\scr{M}_A$ and $\scr{M}_B$, but where the joint dynamics $\scr{M}_A\otimes \scr{M}_B$ is nonlinear! We have already seen an example of this scenario in the previous chapter, namely the entanglement breaking OTC circuit described in Sec \ref{SecEntBreak}.}. We therefore require a new definition of a positive extension describing the dynamics of a joint system when the dynamics is allowed to be nonlinear. As discussed in Ref. \cite{GHE02}, one is led to the following definitions:\\

\tbf{Definition 2 (positive local equivalence class):} Let $\rho$ be a density operator acting on the joint Hilbert space $\scr{H}_A \otimes \scr{H}_B$. For some $\rho_A \in \scr{P}(\scr{H}_A)$ and $\rho_B \in \scr{P}(\scr{H}_B)$, the set of all $\rho \in \scr{P}(\scr{H}_A \otimes \scr{H}_B)$ satisfying $\ptrc{A}{\rho}=\rho_B$ and $\ptrc{B}{\rho}=\rho_A$ is called the \tbf{positive local equivalence class} of $\rho_A,\rho_B$ and is denoted by $\scr{E}\{ \rho_A,\rho_B \}$.\\   

Put simply, $\scr{E}\{ \rho_A,\rho_B \}$ is the set of all states on $\scr{H}_A \otimes \scr{H}_B$ which reduce to the states $\rho_A$ and $\rho_B$ on their respective subspaces $\scr{H}_A$ and $\scr{H}_B$. We can then define:\\

\tbf{Definition 3 (general positive extension):} Given two systems subjected to local dynamics $\scr{N}_A$ and $\scr{N}_B$ as described above, the joint map $\scr{N}_{AB}$ must satisfy:\\
\indent (i) $\scr{N}_{AB} \brac{\rho_A \otimes \rho_B} = \scr{N}_A(\rho_A) \otimes \scr{N}_B(\rho_B) := \rho'_A \otimes \rho'_B$, \\
\indent (ii) $\scr{N}_{AB} \brac{\scr{E}\{ \rho_A,\rho_B \}} \subseteq \scr{E}\{ \rho'_A,\rho'_B \}$,\\
for all $\rho_A \in \scr{P}(\scr{H}_A)$ and $\rho_B \in \scr{P}(\scr{H}_B)$. Any map $\scr{N}_{AB}$ satisfying these requirements is called a \tit{general positive extension} of the local dynamics to the joint system. The generalisation to multiple systems is straightforward.\\

The justification for condition (i) follows from our assumption that the systems are mutually isolated and non-interacting. The justification for (ii) is as follows: given that $\scr{N}_A$ maps $\rho_A$ to $\rho'_A$ and $\scr{N}_B$ maps $\rho_B$ to $\rho'_B$, then in the case where $\rho_A$ and $\rho_B$ are reduced parts of an entangled state $\rho_{AB}$, we expect that the extension $\scr{N}_{AB}$ should map $\rho_{AB}$ to a new state $\rho'_{AB}$ such that $\rho'_{AB}$ reduces to the states $\rho'_A$ and $\rho'_B$ on the respective subsystems.

If $\scr{N}_{AB}$ is nonlinear, we call it a \tit{positive nonlinear extension} of $\scr{N}_A$ and $\scr{N}_B$, otherwise it is a \tit{positive linear extension}. Note that $\scr{N}_{AB}$ can only be linear if the deterministic dynamics of both subsystems is linear\footnote{However, recall from Sec \ref{SecDefineNL} that we only discuss deterministic maps in this Chapter. If we allow \tit{stochastic} nonlinear maps, then it certainly is possible to find linear purifications of such maps in general. One classic example is the measurement process regarded as a nonlinear stochastic map -- this map can generally be purified to linear unitary dynamics on a larger system that includes the measuring apparatus.}, but the converse is not true: if the subsystems evolve linearly, the joint map $\scr{N}_{AB}$ may still be nonlinear. For example, consider the entanglement-breaking map defined by $\scr{N}_{AB}(\rho_{AB}):=\rho_{A} \otimes \rho_{B}$. An observer with access to the joint system would be able to detect the highly nonlinear dynamics of the entanglement, but an observer with access to only one subsystem would not notice anything out of the ordinary -- the reduced dynamics of each subsystem is entirely trivial. Using Definition 3, we can define a notion of complete positivity for general maps:

\tbf{Definition 4 (complete positivity for general maps):} Let $\scr{N}_A$ be a general (linear or nonlinear) trace-preserving and positive map acting on $\scr{P}(\scr{H}_A)$. Consider a map $\scr{N}_B$ on the auxiliary Hilbert space $\scr{H}_B$. The map $\scr{N}_A$ is said to be completely positive if there exists a general positive extension $\scr{N}_{AB}$ to the joint Hilbert space $\scr{P}(\scr{H}_A \otimes \scr{H}_B)$ for all maps $\scr{N}_B$ and for all dimensions of $\scr{H}_B$.

Interestingly, it turns out that \tit{every} positive map, linear or not, possesses a positive nonlinear extension to an arbitrary Hilbert space\cite{GHE02}. By contrast, there exist many positive maps that do not possess a positive \tit{linear} extension to arbitrarily large Hilbert spaces. Such maps, which include the familiar transpose operation, are not completely positive in linear quantum mechanics but are completely positive when nonlinear maps are introduced. 

It is interesting to note that Definition 3 does not uniquely define the extension $\scr{N}_{AB}$ in general, hence there are an infinite class of positive nonlinear extensions $\scr{N}_{AB}$. Physically, these correspond to different behaviours of entanglement in the enlarged system: depending on the choice of extension the entanglement of the enlarged system might increase, decrease or remain the same. This is potentially troublesome, as it implies that a pair of noninteracting, initially separable systems may become entangled under nonlinear dynamics! We will address this issue in the next section. Meanwhile, we note that a nonlinear theory cannot be called complete unless it specifies a general positive extension for all scenarios involving multiple systems. We will call such a specification the ``extension rule" for nonlinear theory, with the understanding that the dynamics of composite systems is undefined without such a rule.

\subsubsection{Complete separability and superluminal signalling\label{SecESAD}}
We commented in Sec \ref{SecGenPosExten} that for general choices of positive nonlinear extension, a pair of independent systems may become entangled with each other, despite having no direct interactions. To avoid such effects, it has been suggested that the extension be constrained by physical principles governing the behaviour of entanglement in nonlinear theories, such as requiring that the entanglement be non-increasing and change smoothly in cases involving a continuous dependence on time\cite{GHE02}. One example of such an extension rule is the ``Polchinski extension", originally proposed by Polchinski in connection with Weinberg's model\cite{POL91}.

Rather than commit ourselves to some particular extension rule, let us ask why the idea of two independent systems becoming entangled without interaction seems so troublesome. After all, we must be careful to distinguish those properties of a theory that might be unphysical from those that are simply counter-intuitive. The example of ``spooky action at a distance" comes to mind; we now know that the EPR thought experiment does not imply the ability to signal faster than light, but it is nevertheless considered to illustrate the nonlocality of quantum mechanics. In the present case we are faced with what one is tempted to call \tit{extremely} spooky action at a distance (ESAD) or at least \tit{highly suspicious} action at a distance (HSAD), where, insofar as an increase in entanglement can be thought of as two systems acquiring information about one another, it appears information is being exchanged nonlocally. Given that we must be willing to accept some form of nonlocality in the laws of physics (in light of Bell's theorem), the pertinent question here is whether the predicted phenomenon would violate signal locality.

First, it is important to recall that the issue of nonlinear extensions to multiple systems in the low energy regime (see the Thesis Outline) implies the existence of a preferred reference frame, namely the rest frame of the laboratory and all the stationary systems of interest. In this frame, the map $\scr{N}_{AB}$ that defines the nonlinear extension is applied simultaneously to the joint Hilbert space $\scr{H}_A\otimes \scr{H}_B$ of the systems, which may be distributed over a large distance. All actions taken prior to this plane of simultaneity are considered to be part of the preparation procedure leading up to the application of the nonlinear map. Let us first restrict ourselves to the question of whether the joint dynamics $\scr{N}_{AB}$ leads to signalling. With a little consideration, we will see that it does not.

For input--output maps, there are several basic local operations that an agent can perform on a given system (say system $B$) to send a signal. The agent can change the initial condition $\rho_B$ by local operations, or intervene to change the local dynamics $\scr{N}_B$ of the system, or both. If any of these interventions on system $B$ leads to a causal influence on system $A$ or \tit{vice versa}, then the dynamics leads to superluminal signalling, since we have assumed that systems $A$ and $B$ are mutually isolated and may be far apart. Fortunately, it is easy to see that the completely positive extension $\scr{N}_{AB}$ does not permit a causal influence by either of these methods, since the output $\rho'_A$ is explicitly determined only by the initial condition $\rho_A$ and on the choice of map $\scr{N}_A$; similarly for system $B$. Thus, signalling is not implied by allowing the joint state $\rho_{AB}$ to become more entangled after the application of the joint map, despite the lack of an interaction between the systems. 

Extensions satisfying Definition 3 belong to a more general class of nonlinear dynamics studied by Czachor, which are called \tit{completely separable}\cite{CZA98}. This notion also extends to continuous time evolution\cite{CZA98,CZAD02}. It was shown by Czachor that all completely separable theories are free from superluminal signalling, at least for the types of manipulations just described. However, we have been careful in the above argument to consider only \tit{local} interventions at the inputs to each nonlinear map. In quantum mechanics, it is generally considered possible to prepare the input state by performing projective measurements on a distant system -- a so-called ``remote preparation". While the allowance of such preparation procedures does not lead to superluminal signalling in standard quantum mechanics, this is no longer the case in nonlinear theories, as we will see in the next section. In Sec \ref{SecBoxSignal} we will outline the necessary and sufficient constraints on remote preparations in nonlinear theories required to overcome this final hurdle.  

\subsection{Gisin's telegraph\label{SecGisin}}

\subsubsection{Historical overview}
In 1989, Weinberg\cite{WEI89} proposed a model of nonlinear quantum mechanics as a foil against which to perform precision tests of quantum mechanics. He initially had hopes that his model would provide a consistent picture of nonlinear dynamics, but these hopes were dashed when Gisin\cite{GIS90} found that his model appeared to lead to superluminal signalling. While superluminal signalling is troubling enough in models that involve causality violation as we have argued in Sec \ref{SecSignal}, it is justifiably considered to be the mark of death for theories formulated in flat spacetime such as Weinberg's model. Polchinski\cite{POL91} subsequently identified two mechanisms that were responsible; the first was the lack of a completely separable extension to multiple systems (although he did not use that terminology at the time) and the second problem was the use of the projection postulate as a means of remotely preparing the initial conditions of a distant quantum state. His remedy to the first problem was the introduction of the Polchinski extension mentioned above. His remedy to the second problem was to abandon the projection postulate altogether, opting instead for Everett's many-worlds interpretation. In this way, he was able to reformulate Weinberg's theory without it leading to superluminal signalling. 

While Polchinski's extension was later adapted to more general theories by other authors, ultimately leading to the notion of complete separability, his recourse to the many-worlds interpretation was more controversial. Polchinski himself acknowledged that it seemed to lead to very strange effects, such as the phenomenon of an \tit{Everett phone}, allowing communication between experimenters in different branches of the wavefunction. In Polchinski's model, the outcome of an experiment could depend not just upon the actions actually taken by the experimenter, but also on the actions the experimenter \tit{would} have taken in alternate histories of the universal wavefunction. 

One might be prepared to accept such effects and even embrace them as a possible means of probing the many-worlds interpretation of quantum mechanics, except that a careful consideration of the theory leads to a rather uninspiring conclusion. Polchinski writes:
\tit{``It is important to note that communication between branches of the wavefunction invalidates previous attempts to analyse the experimental consequences of nonlinearities [...]. This is because these analyses ignore previous branchings of the wave function and treat macroscopic systems as though they begin in definite macroscopic states. A complete analysis requires consideration of the entire wave function of the Universe and is therefore rather complicated. Na\"{i}vely, it would seem that nonlinear effects will be very much diluted by the enormous number of branches, since the amplitude for any given branch of the wave function is exceedingly small. This leads to the discouraging conclusion that nonlinearities could be of order 1 in a fundamental theory and yet the effective nonlinearity measurable experimentally would still be unobservably small."}\\

Polchinski's findings caused subsequent researchers to wonder whether it was really necessary to do away with the projection postulate altogether. Recent work indicates that more moderate solutions are possible\cite{CZAD02,GHE02,KEN05}. However, these considerations will force us to ask questions about what exactly constitutes the state of a quantum system after some preparation procedure. To truly exorcise the demon of superluminal signalling, we will have to extend our formalism to describe the preparations and measurements of quantum states in an operationally rigorous way. This will be done in Sec \ref{SecNLBox}. In the next section, we review Gisin's protocol for signalling in nonlinear theories using remote preparations\cite{GIS90}.

\subsubsection{Signalling by remote preparation}

Let us look at a specific example of a nonlinear map $\scr{N}$ acting on $\scr{P}(\scr{H}\otimes \scr{H})$ where $\scr{H}$ is a two-dimensional Hilbert space, i.e. $\scr{N}$ takes a pair of qubits as input and gives a pair of qubits as outputs. In particular, $\scr{N}$ implements a map on the pure states $\{\ket{0} \otimes \ket{0},\, \ket{1} \otimes \ket{0},\,\ket{+} \otimes \ket{0},\,\ket{-} \otimes \ket{0} \}$ given by:
\begin{align} \label{BrunEq}
\ket{0} \otimes \ket{0} &\mapsto \ket{0} \otimes \ket{0} \,, & \ket{+} \otimes \ket{0} &\mapsto \ket{1} \otimes \ket{0} \,, \nonumber \\ 
\ket{1} \otimes \ket{0} &\mapsto \ket{0} \otimes \ket{1} \,, & \ket{-} \otimes \ket{0} &\mapsto \ket{1} \otimes \ket{1} \,,
\end{align}
For our purposes, we only care about the behaviour of the map on the states specified above and we will assume that these states can be prepared with perfect precision in the lab. Of course, this assumption might be seen as problematic in light of the linearity trap (Sec \ref{SecLinearityTrap}). Suppose that the action of $\scr{N}$ were defined to be the identity $\scr{I}$ on all other states. Since the pure states indicated in \eqref{BrunEq} are a set of measure zero in the space of density operators, any tiny amount of imprecision in the preparation process will render the map $\scr{N}$ operationally indistinguishable from the identity map. 

To address this issue, we can relax our assumption of perfect preparations quite easily while maintaining the same effective dynamics, simply by requiring that \eqref{BrunEq} applies not just to the pure states indicated, but also for slightly different states that are very close to them within some acceptable error. We then set $\scr{N}$ equal to the identity for all density matrices outside of this error bound, while retaining the map \eqref{BrunEq} for all states sufficiently close to the indicated pure states. We may then proceed with our assumption of being able to perfectly prepare the states, confident that the introduction of small errors into the analysis will not invalidate the results so derived. 

To continue, note that the map \eqref{BrunEq} allows us to perfectly distinguish states in the computational basis $\{ \ket{0},\ket{1} \}$ from those in the diagonal basis $\{ \ket{+},\ket{-} \}$. Since the latter are constructed from even superpositions of the former, it is clear that we could not reproduce this effect using linear evolution, hence the dynamics \eqref{BrunEq} satisfies the criterion of nonlinearity. To see this more explicitly, define the states $\rho_1=\densop{00}{00}, \rho_2=\densop{+0}{+0}$ and $\rho_3=\frac{1}{2}\densop{00}{00} +\frac{1}{2}\densop{+0}{+0}$. Then we have:
\eqn{ \scr{N}\brac{\rho_3} \neq \frac{1}{2} \scr{N}\brac{\rho_1} + \frac{1}{2}\scr{N}\brac{\rho_2} \,\nonumber }
because the RHS evaluates to $\frac{1}{2}\densop{00}{00} +\frac{1}{2}\densop{10}{10}$ while the LHS is $\scr{N}(\rho_3)=\rho_3$ which is different to the RHS, satisfying Definition 1a \footnote{Alternatively, note that $\frac{1}{\sqrt{2}} \scr{N}\brac{\ket{00}}+\frac{1}{\sqrt{2}} \scr{N}\brac{\ket{10}}=\ket{0+}$ whereas $\scr{N}\brac{\frac{1}{\sqrt{2}} \ket{00}+\frac{1}{\sqrt{2}} \ket{10}}=\scr{N}\brac{\ket{+0}}=\ket{10}$, which is different. I am grateful to Paul Alsing for bringing this to my attention.}.

One thing we could do with this evolution is break quantum cryptography. Specifically, Brun showed that this evolution can be used to break the Bennett and Brassard protocol (BB84), rendering quantum cryptography insecure\cite{BRU09}. There is, however, a more worrying consequence of this dynamics: the possibility of superluminal signalling\footnote{Although Brun originally obtained the pure-state map \eqref{BrunEq} from a CTC using Deutsch's model, we are here considering the consequences of this map in a more general setting. The issue of superluminal signalling in the specific context of Deutsch's model is considered in Sec \ref{SecDeutschBox}.}. Following Gisin\cite{GIS90}, consider an entangled Bell pair $\ket{\Phi} = \frac{1}{\sqrt{2}}\brac{\ket{00}+\ket{11}}$ distributed between two distant parties Alice and Bob. Suppose that Alice performs a measurement on her half of the Bell pair in either the computational or diagonal basis. 

There are essentially two conventional ways of describing this scenario in standard quantum mechanics: we can regard Bob's actual state as being in the reduced state $I=\ptrc{A}{\densop{\Phi}{\Phi}}$ immediately after Alice's measurement (in some frame), or else we can regard it as having collapsed to some definite but unknown pure state belonging to the basis chosen by Alice. Whichever interpretation we choose, Bob's state immediately after the measurement is described by the same density matrix $\rho=I$, which is why superluminal signalling is impossible in quantum mechanics, despite its nonlocal appearance. However, the nonlinear dynamics \eqref{BrunEq} breaks this equivalence: Bob can simply take his local state and append an ancilla state $\ket{0}$ before applying the dynamics $\scr{N}$. If Bob's state is really a pure state in the computational or diagonal basis, then his qubit will be mapped to $\ket{0}$ or $\ket{1}$ respectively, allowing him to distinguish Alice's choice of measurement basis and receive signals faster than light. Alternately, if Bob's state is the maximally mixed density matrix $\rho_B=I$, then $\scr{N}$ acts as the identity and his output will be random, carrying no information from Alice.

A question whose answer was inconsequential in the regime of linear quantum mechanics (namely: what is the actual \tit{ontic} state of Bob's system after Alice's measurement?) has measurable effects in a nonlinear theory. If it is true that Alice's measurement instantaneously ``collapses" Bob's state into some pure state, as in more traditional interpretations of quantum mechanics, then Alice is able to send signals to Bob faster than light. Alternatively, if there is no collapse at all, as in a many-worlds interpretation, superluminal signalling is impossible (Everett phones notwithstanding). Even the complete separability of the map $\scr{N}$ resulting from Definition 3 is powerless to prevent signalling by Gisin's protocol, because it does not entail any constraints on the ontic states resulting from remote preparations within some specified \tit{ontological model} (see Sec \ref{SecOperational} for definitions of these concepts). 

We might be forgiven for thinking that Gisin's telegraph might be particular to the evolution \eqref{BrunEq} and that we can avoid signalling simply by restricting the allowed nonlinear maps to only those that do not give rise to such effects. Unfortunately, it turns out that \tit{any} nonlinear map $\scr{N}$ can lead to signalling using remote preparations of the state.\\
\tit{Proof:} Consider an arbitrary nonlinear map $\mathcal{N}$ acting on density operators. By definition, there must exist two sets of density operators $\{\rho_1^{(i)}:i\}$ and $\{\rho_2^{(j)}:j\}$ and probability measures $\{p_i:i\}$ and $\{q_j:j\}$ such that $\sum_i p_i \rho_1^{(i)}=\sum_j q_j \rho_2^{(j)}$ but where $\sum_i p_i \mathcal{N}(\rho_1^{(i)}) \neq \sum_j q_j \mathcal{N}(\rho_2^{(j)})$. Such a map takes what would be identical inputs in linear quantum theory to decomposition-dependent nonidentical outputs. The map \eqref{BrunEq} is one example. The quantum steering theorem, as generalized by Verstraete~\cite{VER02}, states that for any decomposition of a density matrix $\sigma_B = \sum_i p_i \rho^{(i)}$, there exists a state $\sigma_{AB}$ for a bipartite system such that $\mathrm{Tr_A}(\sigma_{AB})=\sigma_B$, and there exists a generalized measurement on $A$ that, when applied to $\sigma_{AB}$, will produce with probability $p_i$ the reduced state $\rho^{(i)}$ at $B$. Thus, for any nonlinear map $\mathcal{N}$, there exists a set of remote preparations that may be used for signalling by Gisin's protocol \cite{CAV12}. $\Box$

While Polchinski's suggestion of abandoning the projection postulate in favour of a many-worlds interpretation still stands as a possible resolution, it remains unsatisfactory for the reasons already discussed. Moreover, some notable alternative solutions have been proposed in the literature that appear to resolve the problem without resorting to the many-worlds interpretation. Czachor and Doebner\cite{CZAD02} suggested modifying the projection postulate such that Bob's state is effectively the reduced state $\rho_B=I$ regardless of Alice's choices; as we saw, this is sufficient to prevent signalling. Yet another example is given by Deutsch's model, as originally interpreted by Deutsch\cite{DEU91}, although we leave it to Sec \ref{SecDeutschBox} to spell out exactly what is necessary to avoid superluminal signalling in this model.

A decisive nail in the coffin of superluminal signalling for nonlinear theories was due to Kent\cite{KEN05}, who constructed an explicitly nonlinear model that was immune to Gisin's protocol by construction. Kent considered a hypothetical box that could perform perfect, non-destructive state tomography on any input state, i.e. a ``state readout box". Given such a box, plus additional linear resources, Kent showed that one could deterministically perform any nonlinear map, in violation of standard operational quantum mechanics. The key to Kent's construction was that the ontic state of the system appearing on the readout of the box was to be conditioned on all and only those measurement outcomes that had occurred in the past light-cone of the system at the event of its entry into the box, which sufficed to free the model from all superluminal signalling. 

These successes raise the question: what are the necessary and sufficient conditions to ensure that a general nonlinear model is free of signalling? Such a criterion could potentially unite these disparate theories under a single framework. We answer this specific question in Sec \ref{SecBoxSignal}, by restricting our attention to a class of theories that can be given a clear operational meaning: nonlinear boxes.

\section{Nonlinear boxes\label{SecNLBox}}

In Deutsch's model, we considered a bounded region of spacetime in which causality violation was proposed to occur, outside of which standard quantum mechanics was assumed to hold. We will obtain similar utility in making this assumption for nonlinear dynamics in general: we consider cases in which the nonlinear evolution is confined to a bounded spacetime region which we call a ``box", outside of which standard quantum mechanics holds at the operational level (i.e. nonlinear effects, if present, are not experimentally observable in the regime outside the box).

The box accepts an input of a quantum system in some state $\rho$ and outputs the same system in the state $\rho'$, where the input--output map is given by:
\eqn{
\rho'=\scr{N}(\rho)\, ,
}
for some nonlinear map $\scr{N}$ satisfying Definition 1a. A single nonlinear box is assumed to perform the same map on any input state and at any time. We can then define a ``nonlinear box model" in which there exist multiple nonlinear boxes labelled $i=0,1,2,...$ implementing the set of nonlinear maps $\{ \scr{N}_i :i\}$, including continuous families of nonlinear boxes $\scr{N}(\kappa)$ parameterized by variables $\kappa$. The variables $i$ and $\kappa$ in any run of an experiment are freely chosen by the experimenter. For simplicity, we assume that any given system is acted on by at most a single nonlinear box in any given run of an experiment. This is convenient because it means that the only information we need in order to predict the measurement statistics on the output of a nonlinear box is the density matrix of the output state and it is irrelevant whether this output was prepared by linear or nonlinear means. If we allow sequential boxes, however, then we are effectively introducing a whole new set of preparation procedures outside the regime of linear QM. It is then no longer sufficient to specify only the density matrix of the output, as the action of later nonlinear boxes may depend on whether the state was prepared by linear or nonlinear means. Thus, while a more complete model will need to include such considerations, the discussion of experiments in which nonlinear boxes are used sequentially on the same system is excluded from the present work.

To describe the dynamics of multiple (possibly entangled) systems, some or all of which may be acted upon by nonlinear boxes, the nonlinear box model must also specify an extension rule governing the behaviour of entanglement in the joint system (see Sec \ref{SecNLExtend}).

Note that any experiment involving a nonlinear box constitutes an experiment \tit{outside} the usual regime of standard quantum mechanics. However, the experimenter is always free not to use any nonlinear boxes, in which case the model must reduce to standard quantum mechanics. A nonlinear box model thus defines an \tit{ontological model} with the following properties:\\
(a)The ontic states of the model are represented by density matrices $\rho \in \scr{P}$.\\
(b) Every transformation procedure $T_{NL}$ involving a nonlinear box is associated with a nonlinear map $\scr{N}:\scr{P}\mapsto \scr{P}$, which transforms the ontic state of the system. All other (linear) transformation procedures are associated with their usual linear CPT maps as per the rules of standard quantum mechanics.\\ 
(c) Every measurement procedure $M$ is associated with a POVM $\{ E^{(M)}_k\}$ as per the rules of standard quantum mechanics.\\
(d) Every preparation procedure $P$ is associated with a probability density functional $p_P(\rho)$ according to some set of as-yet unspecified rules. These are called the ``ontological preparation rules".\\

The requirement (a) implies that if the density matrix is mixed, this mixing is an objective property of the state $\rho$ and not due to some particular agent's ignorance. It follows that the density matrix alone is no longer sufficient to represent the most general states of the theory\cite{CAV10}. Recall from Sec \ref{SecOperational} that within an ontological model, general preparation procedures result in probability distributions over the ontic states (epistemic states). Thus within a nonlinear box model, the general states including an agent's ignorance are the probability density functionals $p(\rho)$ over the density matrices. We refer to these functionals as ``generalised states" belonging to the space $\Xi(\scr{P})$ of probability functions on $\scr{P}$. In general, we will use the notation $p(\rho)$ where $\rho$ parameterises the state space $\scr{P}$ and $p$ is a probability density function of $\rho$. As a particularly relevant example, consider the ontic mixed state obtained by tracing out one half of a maximally entangled Bell pair. This is represented by the generalised state $\delta(\rho-I)$, meaning a delta functional centred on the density matrix $\rho=I$. Alternatively, consider the proper mixture that represents an agent's lack of knowledge about whether the state is in the pure state $\ket{0}$ or $\ket{1}$. If the agent assigns equal probability to each state, then the generalised state of the system is written: $\frac{1}{2} \delta(\rho-\densop{0}{0})+\frac{1}{2} \delta(\rho-\densop{1}{1})$, i.e. the sum of delta functionals centred on the pure states weighted by their relative subjective probabilities. Note that these two examples represent operationally equivalent states under linear quantum mechanics, but may no longer be equivalent in the presence of a nonlinear map, as we will see. A nonlinear box model will in general give rise to a \tit{new} operational theory that extends the operational model of standard quantum mechanics to the regime that includes the usage of nonlinear boxes in experiments. In particular, apart from the preparations, transformations and measurements available to us in the standard quantum regime, we now have access to a new set of transformation procedures $T_{NL}$ involving nonlinear boxes in the nonlinear regime (we continue to denote a general transformation, linear or otherwise, by $T$). It follows that we now have new sets of equivalence classes of preparations, transformations and measurements, which will in general depend on the particular set of nonlinear maps allowed by a given model, and on the choice of ontological preparation rules. Within a nonlinear box ontology, a preparation procedure $P$ results in a state $p_P(\rho)$ which transforms as: 
\eqn{\label{NTlinprob}
p'_P(\rho)&:=&\scr{N}_T(\,p_P(\rho)\,) \nonumber \\
&=&p_P\brac{\scr{N}_{T}(\rho)} \, ,} 
under a transformation $T$ with associated map $\scr{N}_T$. If this transformation is linear, then $\scr{N}_{T}$ is just the appropriate linear CPT map. The assumption \eqref{NTlinprob} does not restrict the possible maps, since one is free to take the probability distributions to be delta functions centred on ontic states, i.e. $p_P(\rho)=\delta(\rho-\rho')$. This description corresponds to an agent who has complete knowledge about the ontic state. The condition \eqref{NTlinprob} simply expresses the fact that the action of the nonlinear map does not depend on an agent's subjective knowledge about the state, but only on the ontic state itself. This is sometimes expressed by saying that the nonlinear map $\scr{N}_T$ acts linearly ``at the level of probabilities". More specifically, one can expand any probability distribution over the ontic states in terms of delta functions:
\eqn{
p(\rho)=\int \mathrm{d} \rho' \, p(\rho') \, \delta(\rho-\rho')
}
so that \eqref{NTlinprob} becomes:
\eqn{ \label{NTlinprob2}
\scr{N}_T(\,p_P(\rho)\,) &=& \scr{N}_T\brac{ \int \mathrm{d} \rho' \, p_P(\rho') \, \delta(\rho-\rho')} \nonumber \\
&=&  \int \mathrm{d} \rho' \, p_P(\rho') \, \scr{N}_T \brac{ \delta(\rho-\rho') } \, ,
}
which is linear in the sense that the choice of weighting $p_P(\rho')$ does not affect $\scr{N}_T$. Since the output of the box is not subject to any further nonlinear evolution, we can treat it using standard quantum theory. To this end, let us define the simplification map $S: \Xi \mapsto \scr{P}$ which assigns to every generalised state its corresponding density matrix under standard operational quantum mechanics. More specifically, the map $S$ is defined by its action on the delta functionals according to:
\eqn{\label{simplification}
S[\delta(\rho-\rho')]:=\rho' \, .
}
The action of this map on more general probability distributions is then defined using \eqref{NTlinprob2}, i.e.
\eqn{\label{simplification2}
S [ p(\rho) ] &=& S \sqbrac{ \int \mathrm{d} \rho' \, p(\rho') \, \delta(\rho-\rho')} \nonumber \\
&=&  \int \mathrm{d} \rho' \, p(\rho') \, S [ \delta(\rho-\rho') ] \nonumber \\
&=&  \int \mathrm{d} \rho' \, p(\rho') \, \rho' \, .
}

The measurement $M$ is associated with its usual POVM $\{ E^{(M)}_k\}$ and the probability of obtaining the outcome $k$ in a general experiment involving nonlinear boxes is given by:
\eqn{\label{operationalProbs}
p(k|P,T,M)&=&\trc{ S[\,p_P\brac{\scr{N}_{T}(\rho)}\,] \, E^{(M)}_k} \, \\
&=&\trc{ S[p'_P(\rho)] \, E^{(M)}_k} \, .
}

In what follows, we will frequently employ the shorthand $\delta(\rho-\rho') \leftrightarrow \delta(\rho')$, except where confusion might arise. Based on the above, the operational model corresponding to a nonlinear box model has the following properties.

(i) The equivalence classes of measurement procedures are the same as for standard quantum mechanics. \tit{Proof:} The new equivalence classes are defined by $p(k|P,T,M)=p(k|P,T,M')$ where the new set of transformations $T$ includes the set $T_{NL}$. However, from \eqref{operationalProbs}, we see that the set of output density matrices $S[\,p_P\brac{\scr{N}_{T}(\rho)}\,]$ for all $P$ and $T$ is necessarily identical to the set generated by $S$ when only linear transformations are considered. Specifically, the set of density matrices $S[\,p_P\brac{\scr{N}_{T}(\rho)}\,]$ generated by running over all possible $p_P(\rho)$ and all possible $\scr{N}_{T}$ is isomorphic to the total set of density matrices $\scr{P}$. But this also holds true when we only consider the subset of maps $\scr{N}_{T}$ that are linear maps. Hence, the introduction of nonlinear maps does not alter the size of the space of outputs and therefore does not alter the equivalence classes of measurement procedures.$\Box$\\
(ii) From (b), it follows that there may exist new sets of equivalence classes of transformations corresponding to the $T_{NL}$. Hence the total space of equivalence classes of transformations may be enlarged compared to standard operational quantum mechanics.\\
(iii) The set of equivalence classes of preparations is in general larger than the space of density matrices $\scr{P}$. \tit{Proof:} Since the simplification map $S$ is many-to-one, a density matrix contains insufficient information to determine the equivalence class of a preparation procedure (as opposed to containing redundant information about it). Hence the new set of equivalence classes may be larger than $\scr{P}$ in general. $\Box$\\

We now briefly comment on some of the above results. From (i) and (iii) we see a curious asymmetry introduced between preparations and measurements, whereby the introduction of nonlinear boxes leads to an enlargement of the set of equivalence classes of preparations, but not of measurements. This asymmetry can be traced back to our definition of a nonlinear map in the Schr\"{o}dinger picture, together with our assumption that nonlinear boxes are not used in sequence on the same system. We therefore conjecture that a generalisation of the formalism to the Heisenberg picture and the inclusion of sequential boxes will affect the equivalence classes of measurements as well. We also note that, since (iii) implies an enlargement of the set of operationally distinguishable states, there is in general a larger effective state space when nonlinear boxes are introduced. This raises the interesting question of whether there exist any modifications of quantum theory that have a \tit{reduced} effective state space, i.e. the information contained in the density matrix \tit{over}-specifies the operational equivalence class of preparations. A possible candidate for such a theory is ``boxworld"\cite{BAR07}, also referred to as the class of ``generalized no-signalling theories" (GNST), which include the concept of ``nonlocal boxes" (also called ``PR boxes") due to Popescu and Rorlich\cite{POP94}. It is an interesting topic for future work to understand how nonlinear boxes might fit into this framework of general probabilistic theories.

Finally, Definition 1a can be replaced by the following equivalent definition in terms of generalised states:

\tbf{Definition 1b (nonlinear dynamics):} Consider a positive map $\scr{N}:\scr{P}\mapsto \scr{P}$. By the linearity of probabilities, this induces a map $\scr{N}:\Xi\mapsto \Xi$ on the generalised states $p(\rho) \in \Xi$. The map $\scr{N}$ is called ``nonlinear" iff there exist distinct generalised states $p_3(\rho)$ and $p_4(\rho) := \gamma p_1(\rho)+(1-\gamma) p_2(\rho)$ for some real number $\gamma \in (0,1)$ such that $S[p_3(\rho)]=S[p_4(\rho)]$ but $\scr{N}(p_3(\rho)) \neq \scr{N}(p_4(\rho))$.\\

In the remainder of this chapter, we will place further constraints on nonlinear box models by requiring that they satisfy the requirement of ``potential verifiability" (see Sec \ref{SecVerifiability}) and are free from superluminal signalling at the operational level\footnote{As with standard quantum mechanics, superluminal effects are allowed in the ontological model provided they do not lead to violations of signal locality (i.e. signalling at the operational level).}. 

\subsection{Operational verifiability\label{SecVerifiability}}
Let us turn to the question of the verifiability of nonlinear dynamics. If an untrustworthy salesman gave us a black box that he claimed was able to implement a nonlinear map $\scr{N}$, we would naturally ask to test it before buying it. Specifically, we would expect to observe the nonlinear input--output map under controlled experimental conditions. In principle, our task is very simple: given a specification of the nonlinear map $\scr{N}$, by Definition 1b we can identify generalised states $p_1, p_2$ and $p_3$ (suppressing dependence on $\rho$ for clarity) with the property that $S[p_3]= \gamma \,S[p_1]+(1-\gamma) \,S[p_2]$, but where the application of the map $\scr{N}$ allows us to distinguish the generalised state $p_4:=\gamma \,p_1+(1-\gamma) \,p_2$ from $p_3$. Then we could verify the nonlinear dynamics by preparing the states $p_4$ and $p_3$ and finding out whether the box can distinguish them. Since this is by definition impossible using linear transformations, we would conclude that we have a nonlinear map.

Unfortunately, the situation is not so straightforward, as simply knowing the states $p_3, p_4$ that we need to prepare in order to verify the nonlinear map does not tell us \tit{how} to prepare those states. Recall that the ``how" is supplied by the choice of ontological preparation rules that associate a generalised state $p_P(\rho)$ to every preparation procedure $P$ (see (d) above). It will be useful to introduce the notion of ``verifiability". A nonlinear box that implements the transformation $\scr{N}: T_{NL}$ is ``verifiable" iff it can be demonstrated that the map implemented by the box is $\scr{N}$ and not some linear map. This leads us to the following necessary condition for a nonlinear box to be verifiable:\\

\tbf{Definition 5a (potential verifiability):} A nonlinear box that implements the transformation $\scr{N}: T_{NL}$ is called \tit{potentially verifiable} iff there exists a set of \tit{verifying preparations} denoted $\mathcal{P}_V:= \{P_{V,i}:i\}$ and a set of \tit{verifying measurements} denoted $\mathcal{M}_V := \{M_{V,j}:j\}$ such that the set of probabilities $p(M_{V,j}^{(k)}|P_{V,i},T_{NL},M_{V,j})$---where $k$ labels the measurement outcomes---are sufficient to identify the map $\scr{N}$. If a nonlinear box is not \tit{potentially verifiable} then it is not verifiable. 

Let us translate this into the language of generalised states. To do so, we introduce the following definition:\\

\tbf{Definition 6 (verifying set of states):} A set of generalised states $\{p_{i}:i\}$, labelled by the discrete index $i$, is called a ``verifying" set of states for the map $\scr{N}$ iff the restriction of $\scr{N}$ to the set $\{p_{i}:i\}$ is a nonlinear map.\\ 

As a trivial example, the entire state space $\Xi$ is a verifying set of states for any nonlinear map. Also, any verifying set of states for a nonlinear map $\scr{N}$ must contain at least two elements, namely the states $p_3, p_4$ appearing in Definition 1b. The necessary criterion for verifiability given by Definition 5a can then be translated into the following equivalent definition using generalised states:\\

\tbf{Definition 5b (potential verifiability):} A \tit{necessary} condition for a nonlinear box to be verifiable is that there exists a set of preparation procedures, denoted $\mathcal{P}_V:=\{P^{(p_i)}_V:i\}$, such that the preparation procedure $P^{(p_i)}_V$ results in the generalised state $p_i\in \{p_{i}:i\}$ under the given \tit{ontological preparation rules}. This ensures that it is possible to physically prepare an entire set of verifying states for the map $\scr{N}$.\\ 

We note that there is no rigorously defined \tit{sufficient} criterion for verifiability. For example, it is possible to simulate a nonlinear box using hidden linear resources\cite{KEN05}\footnote{For example, imagine an untrustworthy salesman demonstrating a device that allows the perfect cloning of unknown quantum states, in apparent violation of the linearity of quantum mechanics. It might transpire that the device first identifies the input state by spying on the preparation process using a hidden camera and then prepares copies of the \tit{known} input state, using ordinary operations. To eliminate this and more elaborate possibilities, the testing of the box must be performed under progressively more rigid experimental conditions.}. Hence the observation of nonlinearity in nature could just as well be explained by the existence of additional degrees of freedom that have not been accounted for in the description of the phenomenon (this is the same principle behind the equivalent circuit interpretation of Deutsch's model discussed in Sec \ref{SecECInterp}). Only once the possibility of such extra degrees of freedom has been ruled out can we claim the nonlinearity to be truly verified. 

Given these considerations, it is important to note that just because an ontological model is explicitly nonlinear does not mean that the nonlinearity can be observed by experiments. Recall the claim of Bennett et. al. in Sec \ref{SecBennett} that the predictions of Deutsch's model cannot be observed by a realistic experiment. The arguments used by those authors were later found to be inconsistent, because they assumed that density matrices should be sufficient to represent the equivalence classes of preparation procedures in a nonlinear model, which we have seen is not true in general (recall point (iii) above). However, it \tit{may} happen that for some particular choice of ontological model, (i.e. for some particular choice of the rules assigning generalised states to preparation procedures) the action of a nonlinear box turns out to be unverifiable at the operational level. Then the operational theory including the transformations $T_{NL}$ is equivalent to standard operational quantum mechanics and the model fails the requirement of potential verifiability.

In what follows, we will not be concerned so much with whether a nonlinear model is verifiable as whether it is at least \tit{potentially} verifiable. This criterion necessarily constrains the allowed choices of ontological preparation rules. In the next section, we will further constrain the ontological preparation rules by requiring that the resulting operational theory be free from superluminal signalling.

\subsection{Nonlinear boxes without superluminal signalling\label{SecBoxSignal}}

Let us formulate Gisin's protocol for superluminal signalling in terms of a nonlinear box model. First, we define remote preparations.\\ 

\tbf{Definition 7 (remote preparation):} A ``remote preparation" of a physical system is defined to be any preparation procedure $P$ such that the resulting generalised state $p_P(\rho)$ ascribed to the system by the ontological preparation rules is contingent on variables $X$ whose values are fixed by agents outside the system's past light-cone. The set of all possible remote preparations of a given physical system is denoted $\mathcal{P}_{R}$.\\

Note that a given preparation procedure is ``remote" only with respect to a particular set of ontological preparation rules. Hence a preparation that is remote in one nonlinear box model may not be remote in another, even if the models differ only in their choice of ontological preparation rules. We can now claim the following:\\ 

\tbf{Claim:} A potentially verifiable nonlinear box model is free from \tit{all} superluminal signalling if the set of verifying preparations $\mathcal{P}_V$ excludes all remote preparations $\mathcal{P}_{R}$ of the input.\\

The justification of this claim is straightforward, as it implies that \tit{all} experiments involving remote preparations of the input state can be explained by standard linear operational quantum mechanics and hence are free from signalling. Since Gisin's protocol relies on such preparations, it cannot be used for signalling. For non-remote (i.e. local) preparations, all other protocols for signalling are ruled out by the complete separability of the model (see Sec \ref{SecNLExtend}).

While it is sufficient, the above is not a necessary condition to avoid superluminal signalling. In fact it is possible to define ontological preparation rules such that certain states belonging to a verifying set can be remotely prepared, yet cannot be used to send a superluminal signal (see the Appendix for an explicit example). Let us therefore derive a condition that is both necessary and sufficient to prevent signalling by any means. As we have assumed, for every preparation procedure $P$, the ontology associates a unique generalised state. If the preparation is remote, then by definition there exists a set of variables taking values $x \in X$, whose values are fixed by an experimenter outside the past light-cone of the system, such that different values of $x$ result in the preparation of different generalised states. For a particular choice $x \in X$ of the variables, denote the resulting generalised state by $g_x$. The set $G_P:=\{g_x: \forall x \in X\}$ contains all possible states that can be prepared remotely using the preparation procedure $P$. Every choice of ontological preparation rules corresponds to a particular specification of $G_P$ for every $P$. A necessary and sufficient condition for no-signalling can be framed as: \tit{for every remote preparation, different choices of $x\in X$ cannot alter the measurement statistics of the prepared system}. We can formalise this statement by noting that in order for a choice of remote variables $x$ to affect the measurement statistics, they must alter the operational equivalence class of the state. That is, there must exist some pair of values $a \in X$ and $b \in X$ such that the resulting states $g_a$ and $g_b$ can be distinguished by the nonlinear box\footnote{Note that, by assumption, $g_a$ and $g_b$ cannot be distinguished by any linear map, as that would imply the possibility of superluminal signalling in standard quantum mechanics.}. We have the following result:

\tbf{Theorem 1:} Two generalised states $g_a$ and $g_b$ that are operationally indistinguishable under linear maps (i.e. having the property $S[g_a]=S[g_b]$) are operationally distinguishable under a nonlinear map $\scr{N}$ iff they form a verifying set for $\scr{N}$ under the given ontological model.  \\
\tit{Proof:} First, note that if $g_a$ and $g_b$ can be distinguished by the map $\scr{N}$, then they can be used to verify the nonlinear evolution as per Definition 6 and Definition 1b. Conversely, if $g_a$ and $g_b$ are not operationally distinguishable by the map $\scr{N}$, then they cannot be used to verify the nonlinearity since there exists a linear map that reproduces the action of $\scr{N}$ on these states (and probabilistic combinations thereof). $\Box$ \\

This leads to the necessary and sufficient condition:\\

\tbf{Theorem 2:} The operational model corresponding to a verifiable nonlinear box $\scr{N}$ is non-signalling iff for every remote preparation $P$, the set $G_P$ is not a verifying set. \\
\tit{Proof:} As already noted, a necessary and sufficient condition to avoid signalling is that for every remote preparation $P$ there can be no $g_a, g_b \in G_P$ that are operationally distinguishable by $\scr{N}$. From Theorem 1, this implies that no two elements in $G_P$ can constitute a verifying set of states for $\scr{N}$. This is equivalent to requiring that $G_P$ itself is not a verifying set. $\Box$ \\ 

In the case of a nonlinear box model that includes multiple nonlinear boxes, this criterion applies to each box. In the extreme case where the model includes a universal box, i.e. a box that can implement any nonlinear map $\scr{N}$, where every choice of box $\scr{N}$ is potentially verifiable, no-signalling requires that all remote preparations be forbidden. To see this, note that for any pair of distinct generalised states, there always exists some map $\scr{N}$ for which they are operationally distinguishable. As a consequence, the above theorem requires that for every would-be remote preparation, the set $G_P$ consists of a single state, hence $X$ contains one element and is not a remotely controllable variable. To summarise: a potentially verifiable nonlinear box model is non-signalling at the operational level iff the ontological preparation rules ensure that, for every remote preparation $P$, the set $G_P$ of states that can be prepared remotely is not a verifying set for any nonlinear box in the model.

\subsection{Ontologies with a ``Heisenberg cut"\label{SecHeisCut}}

We conclude our discussion of general nonlinear box models by framing an additional thought experiment to which every complete nonlinear box model should provide an answer. We begin by observing that, within the constraints of a nonlinear box model, a many-worlds interpretation may not be applicable for at least two reasons. One reason is that the many-worlds interpretation takes the wavefunction of the universe to be an ontic state, which may be incompatible with treating the reduced density matrix of a subsystem as being ontic. Another reason is that any given subsystem is likely to be highly entangled to the rest of the universe and therefore its reduced density matrix will be very close to a maximally mixed state. This would lead to the same problems with verifiability faced by Polchinski in applying many-worlds to Weinberg's model, see Sec \ref{SecGisin}. The logical alternative is to consider an ontological model that includes objects that can be either fundamentally quantum or classical, the separation between the two regimes being demarcated by a ``Heisenberg cut". Examples in the literature include ``quantum--classical hybrid theories"\cite{REG09} and theories invoking an objective collapse, such as the GRW spontaneous collapse model\cite{GRW09} and Penrose's proposal for gravitationally induced wavefunction collapse\cite{PEN98} (it is perhaps no coincidence that these models themselves rely on nonlinear dynamics at the fundamental level).

\begin{figure}
 \includegraphics[width=18cm]{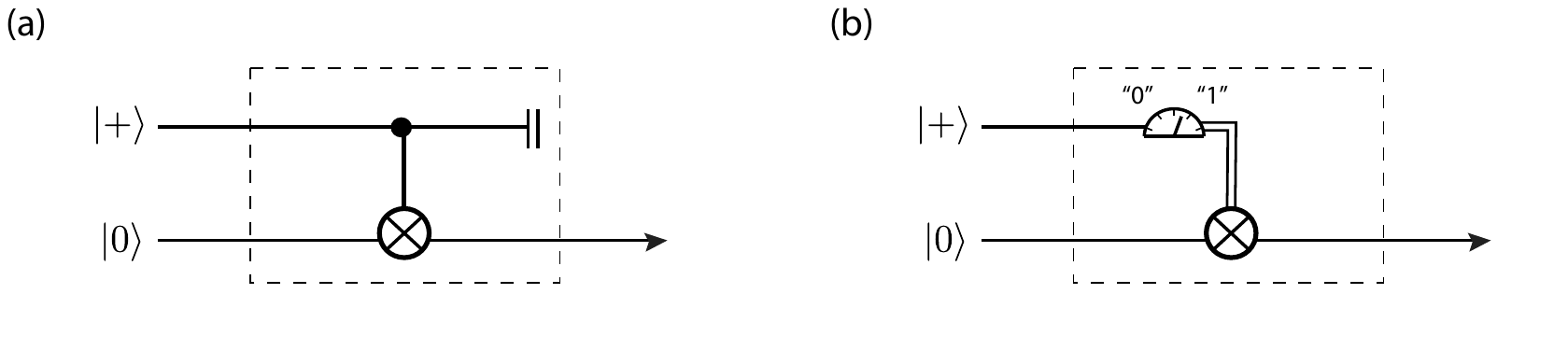}              
 \caption{\label{figHeisCut}Two preparation procedures that are operationally equivalent in standard quantum mechanics but inequivalent under nonlinear quantum mechanics. (a) The state with density matrix $\rho=I$ is prepared by a coherent CNOT interaction. (b) A state with the same density matrix is prepared by conditioning on the outcome of a classical measurement.}
\end{figure}

To illustrate the consequences of a Heisenberg cut within a nonlinear box model, consider the following two distinct preparation procedures. In the preparation procedure labelled $P^{(q)}$, a microscopic device that exists well within the quantum regime is presented with two qubits in the states $\ket{+}$ and $\ket{0}$. The device ``measures" the first qubit in the computational basis and then alters the state of the second qubit to agree with the measurement outcome. Since the device is described by quantum mechanics, this process is equivalent to a coherent entangling interaction between the device and the two qubits, as shown in Fig. \ref{figHeisCut} (a). In standard operational quantum mechanics, the output state of the second qubit is the maximally mixed state $I$. Consider now an alternative preparation procedure labelled $P^{(c)}$, in which a macroscopic device (perhaps even a human experimenter) that exists within the classical regime is presented with the same two qubits. As before, the macroscopic device measures the first qubit in the computational basis and then prepares the second qubit according to the recorded measurement outcome. Now, the process involves conditioning on a classical measurement outcome, via the incoherent process described by the circuit of Fig. \ref{figHeisCut} (b). In general, the prepared state of the second qubit is now in the state $\ket{0}$ or $\ket{1}$ with equal probability, which again can be represented by the maximally mixed state in standard quantum mechanics. It follows that the preparation procedures $P^{(q)}$ and $P^{(c)}$ are operationally equivalent with respect to the operational model of standard quantum mechanics. 

We can generalise this to a continuum of such preparations in which the intermediate device is made progressively larger or more complex, until at some stage it transitions from a quantum to a classical system. In all cases, the preparation procedure belongs to the same operational equivalence class $I$ under standard quantum mechanics. As a consequence, if a qubit is prepared by one of these procedures and presented to an agent, that agent cannot tell whether the qubit was prepared by a quantum device such as a CNOT gate or by a classical device such a human being. This is one reason why it is extremely difficult to probe the quantum-to-classical transition using the techniques available from standard quantum mechanics: the direct interaction of a quantum system with a classical system is operationally indistinguishable from an interaction with other quantum systems. 

In a nonlinear box model, this is no longer the case: the preparation $P^{(q)}$ results in the generalised state $p(\rho)=\delta(I)$ whereas $P^{(c)}$ results in $\frac{1}{2}\delta(\densop{0}{0})+\frac{1}{2}\delta(\densop{1}{1})$ (assuming that the Heisenberg cut lies between them). Thus an agent who possesses a nonlinear box capable of distinguishing these two generalised states will be able to probe the quantum-to-classical boundary directly, simply by preparing a single qubit using devices of increasing size and passing them through the nonlinear box. Hence \tit{the placement of the Heisenberg cut in a nonlinear box model has observable consequences for microscopic systems}. It follows that if nonlinear evolution were ever discovered in nature, it could be used to locate the Heisenberg cut with very little effort! These considerations reinforce our assertion that a nonlinear box model is not complete without a full specification of its ontological properties, including the precise location of the Heisenberg cut, if it is invoked as part of the model. To leave such details unspecified is to leave the physical predictions of the model ambiguous.

\section{Deutsch's model as a nonlinear box\label{SecDeutschBox}}

The Deutsch model can be straightforwardly understood as an instance of a nonlinear box, implementing Deutsch's map $\scr{N}_D$ (see Sec \ref{SecDeutschCirc}). One requirement of a nonlinear box is that it must admit an extension rule to multiple systems, as defined in Sec \ref{SecNLExtend}. Remarkably, the Deutsch model already possesses a completely separable extension to systems undergoing simultaneous linear evolution\footnote{However, the extension to multiple simultaneous CTC-boxes remains ambiguous in the Deutsch model as discussed in Sec \ref{SecMultiCTC}.}. To see this, consider the case where the input state to the CTC, $\rho_A$, is the reduced state of one system in an entangled collection of systems. Let us write the initial state of all of these systems as $\sigma \in \scr{P}(\scr{H}_A \otimes \scr{H}_R)$ such that the state of the input system is $\rho_A = \ptrc{R}{\sigma}$ and the reduced state for the rest of the systems is $\rho_R = \ptrc{A}{\sigma}$. If we label the the CTC rail as $B$, then the total dynamics of $\sigma$ is given by the extended map:
\eqn{\scr{N}_D(\sigma) = \ptrc{B}{ \hat{I}_R \, \hat{U} (\sigma \otimes \rho_B ) \hat{U}^\dagger \, \hat{I}^\dagger_R } \; \label{DeutschExtend},}
where $\hat{I}_R$ is the identity map acting on the state of the rest. It is this extension that gives the Deutsch model many of its unique properties -- in particular it ensures that when $\hat{U}$ is not an entangling gate, the entanglement is preserved in the total system $\sigma$, while in cases where $\hat{U}$ creates strong entanglement to the CTC, the entanglement between $\rho_A$ and the rest is broken.

Let us frame this in the language of general nonlinear boxes. Consider a single nonlinear box that implements the map $\scr{N}_A$ on a system with Hilbert space $\scr{H}_A$ initially in the state $\rho_A$. Consider also a set of ancillary systems with total Hilbert space $\scr{H}_B$, initially in the state $\rho_B$, which are subject to the simultaneous linear dynamics given by the map $\scr{M}_B$ on $\scr{P}(\scr{H_B})$. Finally, let $\rho_{A,B}$ be the initial state of the joint system, with $\ptrc{B}{\rho_{A,B}}=\rho_A$ and $\ptrc{A}{\rho_{A,B}}=\rho_B$. We then have the following:\\

\tbf{Definition 8 (Deutsch's extension rule):}
The \tit{Deutsch extension rule} states that the joint evolution of the composite system is given by the entanglement-breaking map $\scr{N}_{AB}$ satisfying:\\
$\scr{N}_{AB}(\rho_{A,B})=\scr{N}_A(\rho_A)\otimes \scr{M}_B(\rho_B)$\\
Compare with the entanglement breaking effect \eqref{OTC1}. Since this extension satisfies Definition 3, it is also completely separable (recall Sec \ref{SecGenPosExten}).

The next required ingredient of a nonlinear box model is an explicit set of rules that assign a generalised state to every preparation procedure. As we have defined ``Deutsch's model", these rules have been left unspecified. If we take Deutsch's model to be a nonlinear box that implements the map $\scr{N}_D$ together with Deutsch's extension to composite systems, then it seems that the box must in general permit superluminal signalling. The argument is straightforward: as shown by Brun\cite{BRU09}, there exists a choice of unitary interaction on the CTC such that the map $\scr{N}_D$ perfectly discriminates the non-orthogonal bases $\{ \ket{0}, \ket{1}\}$ and $\{ \ket{+}, \ket{-}\}$ via the pure-state evolutions \eqref{BrunEq}. As we have argued in this chapter, the assumption that this map can be verified leads to the conclusion that signalling is possible within certain ontological models. To avoid signalling in the Deutsch model, therefore, one needs to make additional assumptions about the allowed ontological models. 

Despite the simplicity of this argument, there exists a counter-argument to the effect that the Deutsch model \tit{implies} additional constraints on the choice of ontological model, such that superluminal signalling is forbidden. In what follows, we will show that in fact this argument relies on hidden assumptions beyond those stated in our minimal definition of ``Deutsch's model" (recall \ref{SecDeutschCirc}). Thus, superluminal signalling is found to be forbidden in Deutsch's model only to the extent that one considers these extra ontological assumptions reasonable. We consider the counter-argument in detail below.\\

\tit{Counter-argument}: Consider a remote preparation of the input to a CTC, where the input is one half of a distributed Bell pair and a projective measurement is performed on the other half in either the computational or diagonal bases, leading to the alternate preparation procedures $P^{(C)}$ and  $P^{(D)}$, respectively. From Sec \ref{SecEquivCirc}, the output of the CTC when the input is prepared using $P^{(C)}$ (or $P^{(D)}$) is the same as the output of the equivalent circuit with inputs consisting of $N$ independent trials of the preparation procedure $P^{(C)}$ ($N$ independent trials of the preparation procedure $P^{(D)}$). Thus if an experimenter can use a CTC to distinguish the preparation procedures $P^{(C)}$ from $P^{(D)}$, then it follows that an experimenter could use the equivalent circuit to distinguish between a large ensemble of states each prepared using $P^{(C)}$, and a large ensemble of states each prepared using $P^{(D)}$ -- but this would imply the ability to send superluminal signals using ordinary quantum mechanics, which we know to be impossible. Hence the preparation procedures $P^{(C)}$ and $P^{(D)}$ must be operationally indistinguishable by Deutsch's model, thereby preventing signalling, regardless of the choice of ontology.\\

\tit{Response:} The questionable part of the above argument is the statement that ``the output of the CTC when the input is prepared using $P^{(C)}$ (or $P^{(D)}$) is the same as the output of the equivalent circuit with inputs consisting of $N$ independent trials of the preparation procedure $P^{(C)}$ (or $N$ independent trials of the preparation procedure $P^{(D)}$)". This is supposed to be justified by inspection of the equivalent circuit. However, the discussion in Sec \ref{SecEquivCirc} regarding the equivalent circuit makes no reference to copies of `preparation procedures' but rather refers to copies of the \tit{input state} $\rho_1$. As we have just discussed, taking Deutsch's model to be a nonlinear box entails that $\rho_1$ represents the \tit{ontic state} of the input system, and it is precisely this ontic state that is copied in the corresponding equivalent circuit. Having thus clarified the content of this statement, let us see if it does indeed restrict Deutsch's model to be non-signalling.

In the present example, regardless of the choice of ontology, the preparation procedures $P^{(C)}$, $P^{(D)}$ result in respective generalised states $p_C(\rho)$, $p_D(\rho)$ with the property that $S[p_C(\rho)]=S[p_D(\rho)]=I$. Let us consider one particular choice of ontology, labelled $O_{s}$, that appears to lead to signalling.
According to the ontological preparation rules of $O_{s}$, the preparations $P^{(C)}$, $P^{(D)}$ result in the distinct generalised states given by: 
\eqn{
p_C(\rho):=\frac{1}{2} \delta(\densop{0}{0})+\frac{1}{2} \delta(\densop{1}{1}) \nonumber \\
p_D(\rho):=\frac{1}{2} \delta(\densop{+}{+})+\frac{1}{2} \delta(\densop{-}{-}) \, .
}
Since these generalised states are distinguishable by Deutsch's map $\scr{N}_D$, which is assumed to be potentially verifiable, and since these preparations count as `remote'\footnote{The choice of whether to implement $P^{(C)}$ or $P^{(D)}$ is determined by variables outside the past light-cone of the system.}, the Deutsch model appears to allow signalling within this ontology. Let us now examine how the counter-argument given above aims to rule out this ontology. According to Sec \ref{SecEquivCirc} and the above discussion, there exists an equivalent linear circuit that produces the same output statistics as Deutsch's map, given $N$ copies of the input ontic state. If we are given a generalised state that is a discrete probability distribution of ontic states, $p(\rho)=\sum_i \, p_i \rho_i$, the corresponding input to the equivalent circuit must be $p^{(EC)}(\rho)=\sum_i \, p_i \rho^{\otimes N}_i$. We interpret this as being $N$ copies of the state $\rho_i$ with probability $p_i$. By definition, these probabilities are epistemic -- they represent an agent's lack of knowledge about the actual array of copies that must be input to the equivalent circuit to obtain the same outcome statistics as predicted by the Deutsch model. In the present ontology, the two different preparation procedures correspond to the distinct inputs to the equivalent circuit:
\eqn{
p^{(EC)}_C(\rho):=\frac{1}{2} \delta(\densop{0}{0}^{\otimes N})+\frac{1}{2} \delta(\densop{1}{1}^{\otimes N}) \nonumber \\
p^{(EC)}_D(\rho):=\frac{1}{2} \delta(\densop{+}{+}^{\otimes N})+\frac{1}{2} \delta(\densop{-}{-}^{\otimes N}) \, .
}
Since the equivalent circuit is linear, we can replace these ontic states with their density matrices under the simplification map $S$, leading to:
\eqn{\label{DistinctDens}
S[p^{(EC)}_C]:=\frac{1}{2} \densop{0}{0}^{\otimes N}+\frac{1}{2} \densop{1}{1}^{\otimes N} \nonumber \\
S[p^{(EC)}_D]:=\frac{1}{2} \densop{+}{+}^{\otimes N}+\frac{1}{2} \densop{-}{-}^{\otimes N} \, .
}
As these density matrices are distinct, they lead to distinguishable output statistics in the equivalent circuit. This agrees with the predicted output statistics of the Deutsch model, but it seems to lead to an inconsistency, for the distinct density matrices in \eqref{DistinctDens} cannot be prepared \tit{from a distance} in standard quantum mechanics. That is to say, there is no entangled state of two separated systems such that a local measurement on one of the systems can prepare either the density matrix $S[p^{(EC)}_C]$ or $S[p^{(EC)}_D]$, depending on which remote measurement is performed (if it were so, then signalling would be possible with ordinary quantum mechanics). It therefore appears that, within the ontology $O_{s}$, there is no physical preparation procedure that would produce the equivalent circuit required to simulate the action of Deutsch's model on the remotely prepared input. The substance of the counter-argument now becomes clear: the ontological model $O_{s}$ is unphysical because the equivalent circuit required to emulate the predictions of the Deutsch model for a remotely prepared input cannot be physically constructed.

We are now in a position to evaluate the validity of this argument. Why should we require that it be possible to physically construct the equivalent circuit corresponding to the Deutsch model? The motivation for this requirement is not obvious \tit{a priori} because the equivalent circuit is not assumed to represent a real state of affairs. As we emphasised in Sec \ref{SecEquivCirc}, one may choose to regard the equivalent circuit as simply a convenient alternative representation of Deutsch's map or else one may choose to regard the extra copies as being, in some appropriate sense, real. Only in the latter case is it reasonable to expect that the copies must be produced by some physical process. Indeed, this is the position taken by Deutsch in his ``proof" of no-signalling, since he explicitly advocates interpreting the CTC as being essentially a portal between multiple universes. In this context, the counter-argument given above does indeed make sense. In fact, we may consider a class of ontological models, denoted $O_{ns}$, for which the generalised states corresponding to $P^{(C)}$, $P^{(D)}$ are defined to be:
\eqn{
p_C(\rho):= \delta(I) \nonumber \\
p_D(\rho):= \delta(I) \, .
}
It is clear that the Deutsch model cannot distinguish these states, nor can the corresponding equivalent circuit whose input is $I^{\otimes N}$ (in either case) and which can of course be physically prepared in principle using standard quantum mechanics. If the Deutsch model is supplemented with the ontological model $O_{ns}$, it cannot be used for signalling by the method described above -- however, we emphasise that this entails assumptions about the ontological model beyond what is stated in our minimal definition of ``Deutsch's model" when taken as a nonlinear box. 
 
The above example considered just one particular instance of Deutsch's map, given by \eqref{BrunEq}. To define the necessary requirements for an ontological model describing Deutsch's model to be non-signalling in general, assuming all instances of Deutsch's map are verifiable, it would be necessary to enumerate the full set of nonlinear boxes that can be implemented by a single CTC. This would also allow us to answer questions such as: can a single CTC function as a universal nonlinear box, or are there nonlinear maps that cannot be performed by a CTC for any choice of unitary interaction? The task of characterising the class of maps implementable by a single CTC is left to future work. If we make the conservative assumption that the Deutsch model can implement an arbitrary nonlinear map, it follows from the discussion at the end of Sec \ref{SecBoxSignal} that there can be no preparation procedures that qualify as ``remote" under the Deutsch model's ontology (whatever it may be) if we are to avoid signalling. 

The question remains as to whether or not the requirement of signal locality over-constrains the allowed ontological modes such that the model is no longer potentially verifiable. To see that it this is not the case, consider an ontological model involving an explicit Heisenberg cut, as described in Sec \ref{SecHeisCut}. By assumption, the Deutsch model only applies to microscopic quantum systems below the Heisenberg cut (recall the footnote at the beginning of \ref{SecDeutschCirc}), while we may assume that the inputs to a CTC can be prepared in arbitrary ontic states by the classical experimenters. The preparation procedures for doing this are just the usual measurement and postselection techniques, with the proviso that the systems performing the measurements exist above the Heisenberg cut. Within such an ontology, the Deutsch model is at least potentially verifiable, since the generalised states necessary for verifying a given instance of Deutsch's map can always be prepared. Of course, the precise predictions of the model will be affected by the placement of the cut, leaving some ambiguity in the model, which must be taken into account when using the model to make physical predictions. In summary, the Deutsch model may be both operationally verifiable and free of superluminal signalling within an ontological model that (i) admits an explicit Heisenberg cut and (ii) forbids remote preparations. 

\subsection{Conclusions and future work}

We have considered a class of nonlinear dynamics called nonlinear box theories. We reviewed some results from the literature on complete positivity and nonlinear extensions to multiple systems, adapting these definitions to our framework as needed. We proposed explicit criteria for nonlinear box models to be verifiable and free from superluminal signalling and discussed the extent to which a nonlinear box model may be ambiguous if its ontological properties are not explicitly defined. Among these properties, we identified rules for associating states to preparation procedures as being an important part of the specification of a nonlinear box model. While the Deutsch model was found to possess the main attributes of a nonlinear box model, its rules governing the preparations of states remain to be specified, leading to some potential for ambiguity in its physical predictions. We will assume that the ontological model associated with Deutsch's model ensures that the model is both verifiable and free from superluminal signalling. In the remainder of this thesis, wherever the predictions of the model might depend on ontological assumptions, we will make these assumptions explicit.

Our formalism of nonlinear box models is limited in a number of aspects. It does not account for sequential nonlinear operations, stochastic nonlinear maps, or continuous nonlinear evolution in time. Generalisations of the model along these lines are likely to be straightforward and would greatly improve the model's generality and its relevance to particular theories in the broader literature. Possible connections to quantum foundations and general probabilistic theories also remain to be elucidated.

One might also derive new results using the current formalism. In particular, we have not examined the information processing power of general nonlinear box models, but earlier results in the literature suggest that such considerations would be nontrivial\cite{ABR98}. It would be especially interesting to know whether the information processing powers of nonlinear boxes are constrained by the requirements of verifiability and signal locality. Finally, there remains the task of refining the set of possible ontologies for nonlinear theories using physical principles. We have employed verifiability and signal locality for this purpose, but other principles such as energy conservation, unitarity or information processing power may provide further constraints.

Regarding the Deutsch model, we have made some progress in defining the properties of the model within a more rigorous and general framework, but our conclusion is that the model lacks an explicit set of ontological rules for assigning ontic states to preparation procedures. This might be regarded as a symptom of the fact that the model is only a \tit{toy model} of causality violation and is therefore highly simplified. It is reasonable to expect that if Deutsch's model can be cast as an ideal limit of a more general and sophisticated model, we might gain a better insight into the appropriate underlying ontological model. The missing ingredients of Deutsch's model that we have spelled out in this chapter can therefore be taken as an invitation to design a more complete and less simplistic model of causality violation. We will turn to the task of constructing such a model in the next chapter.

\section{Appendix: remote preparations without signalling}

In Sec \ref{SecBoxSignal} it was argued that superluminal signalling in a nonlinear box model can be avoided by forbidding all remote preparations. To see that this is not a strictly necessary condition to avoid signalling, we need only provide an example in which the existence of remote preparations does not lead to signalling by Gisin's protocol. That such an example can exist should not be too surprising, because a trivial example is already given by standard linear quantum mechanics. As we have seen, the density matrix ascribed to a physical system is consistent with a large class of generalised states, many of which correspond to `remote preparations'. Depending on one's choice of ontology (in particular, ontologies invoking instantaneous collapse of the wavefunction), remote preparations may be allowed in linear quantum mechanics -- but this obviously does not lead to signalling in that case. We might therefore expect that an operational theory that is only a `little bit nonlinear', in the sense that the class of verifying states is small, might also permit some remote preparations without leading to signalling.

As an explicit case, consider the following example of a remote preparation of Bob's state by Alice. To improve our notation, consider the three mutually orthogonal bases for a two-dimensional Hilbert space: $\{\ket{z_+}, \ket{z_-} \}$, $\{\ket{x_+}, \ket{x_-} \}$, $\{\ket{y_+}, \ket{y_-} \}$. Suppose Bob possesses a nonlinear box that implements the following dynamics:
\eqn{\scr{N}(\densop{y_{\pm}}{y_{\pm}}) = \densop{z_{\pm}}{z_{\pm}}, \nonumber \\
\scr{N}(\rho) = \rho \; \; \trm{for all other states.}
} 
It is easy to verify that this box is nonlinear because it allows Bob to differentiate the nonorthogonal bases $\{\ket{y_+}, \ket{y_-} \}$ from $\{\ket{z_+}, \ket{z_-} \}$. One example of a verifying set of states for this map is $\{ \densop{z_+}{z_+}, \densop{y_+}{y_+}, \brac{ \frac{1}{2} \densop{z_+}{z_+} + \frac{1}{2} \densop{y_+}{y_+}} \} := \{ \rho_z, \rho_y, \rho' \}$, for which we have $\scr{N}(\rho') \neq \frac{1}{2} \scr{N}(\rho_z) + \frac{1}{2} \scr{N}(\rho_y)$. Verifiability requires that it is physically possible to prepare the verifying states $\{ \rho_z, \rho_y, \rho' \}$ using a set of preparations $\mathcal{P}_V$. Now, let us imagine an ontology that allows one of these verifying states to be prepared remotely. Consider the remote preparation of Bob's state by Alice's measurement on one half of a shared Bell pair. Let Bob's resulting generalised state be either $\frac{1}{2}\delta \brac{\densop{z_+}{z_+}}+\frac{1}{2}\delta \brac{\densop{z_-}{z_-}}$ or $\frac{1}{2}\delta \brac{\densop{y_+}{y_+}}+\frac{1}{2}\delta \brac{\densop{y_-}{y_-}}$ depending on whether Alice measures in the $z$ basis or the $y$ basis respectively, and $\rho_B=\delta(I)$ if Alice does anything else. These rules constitute a set of ontological preparation rules, corresponding to some ontological model. It is now clear that, although Alice can remotely prepare Bob's state in either the $z$ or $y$ basis, and although the states corresponding to the $z$ basis belong to a verifying set of states, Bob cannot receive any message from Alice because the nonlinear box cannot distinguish between the two remotely prepareable ensembles, nor can it distinguish them from the cases where Alice does anything else. 

The reader may object that if we can remotely prepare states in the $z$ or $y$ bases then we ought to be able to remotely prepare states in the $x$ basis as well, simply by first remotely preparing in one of the other two bases and then applying a unitary rotation at Bob's location. However, strictly speaking, the application of this extra unitary would constitute a new preparation procedure, in which case the ontology can be chosen to nullify the effect of the unitary, for example by specifying that any remote preparation procedure followed by a nontrivial unitary must result in the identity, providing no information. Although somewhat contrived, this example illustrates that ontological models can exist in which remote preparations do not lead to signalling.

\clearpage
\pagestyle{plain}
\pagebreak
\renewcommand\bibname{{\LARGE{References}}} 
\bibliographystyle{refs/naturemagmat2012}
\addcontentsline{toc}{section}{References} 

\newpage
\thispagestyle{plain}
\mbox{}

\part{High energy}
\label{part:HEN}

\pagestyle{fancy}
\renewcommand{\sectionmark}[1]{\markboth{\thesection.\ #1}{}}

\chapter{Relativistic quantum circuits}
\label{ChapFields}

{\em
\parindent=10pt
\noindent Dear Lord of all the fields\\
\indent \indent \indent \indent \indent \indent \indent \indent \indent \indent \indent \indent what am I going to {\em do}?\\
\indent \indent \indent \indent \indent \indent \indent \indent \indent \indent \indent Street-lights, blue-force and frail\\
As the homes of men, tell me how to do it \,\,\, how\\
\indent To withdraw \,\,\, how to penetrate and find the source\\
\indent \indent Of the power you always had\\
\indent \indent \indent \indent \indent \indent \indent \indent \indent \indent \indent \indent light as a moth, and rising\\
\indent \indent With the level and moonlit expansion\\
\indent Of the fields around, and the sleep of hoping men.\\
}

James Dickey\\

\newpage

\section*{Abstract}
We have seen that the quantum circuit formalism is a convenient way of representing the dynamics of quantum systems in the non-relativistic regime. One of the basic assumptions made by the traditional quantum circuit formalism is that the detailed spacetime dependence of the component particles can be ignored. If we wish to relax this assumption in Deutsch's model, therefore, we must first generalise the standard quantum circuit formalism.

In this chapter, we will generalise a quantum circuit to take account of the underlying quantum fields from which the circuit is constructed. This allows us to incorporate mismatch and explicitly Lorentz-transform the wavepackets to account for relativistic effects\footnote{In this respect, our work is an extension of the treatment of mismatch given in Ref. \cite{RMS}.}. Finally, we derive an expression for a single particle mode in terms of the vacuum modes of the field. This enables us to perform all calculations exclusively in the Heisenberg picture and to take all expectation values in the vacuum state. Apart from laying the groundwork for Chapters \ref{ChapCTC1} and \ref{ChapCTC2}, the tools developed in this chapter may be relevant to the literature on relativistic quantum information theory (RQI). In particular, the expression for a single-particle creation operator in terms of vacuum modes may facilitate the representation of qubits in non-inertial reference frames.

This chapter is organised as follows. In Sec \ref{SecQO} we give a brief overview of quantum field theory, particularly the quantisation of the electromagnetic field in the Coulomb gauge. This leads us into a review of quantum optics, particularly involving Gaussian states, for which the reader is referred to Ref. \cite{WEE12} for a more detailed exposition. Sec \ref{SecQAlgebra} addresses the task of defining a field theory consistent with a general quantum circuit. The deterministic creation of single photons from the vacuum for use as qubits is described in Sec \ref{SecSingPhot}. The resulting formalism is applied to specific examples in Sec \ref{SecQAExamples} and suggestions are made for more general applications. Conclusions and future work are discussed in the final section.  

\section{Quantum optics\label{SecQO}}

In order to incorporate special relativity into quantum mechanics, it is standard to use the more general formalism of \tit{quantum field theory}. The defining feature of a quantum field theory is that it allows particles to be \tit{created} and \tit{annihilated}. There are many physical processes in which particles appear and disappear, such as quasi-particles in condensed matter physics, for which a quantum field theory is useful. Note that a field theory need not be relativistic; for example, one can rewrite standard nonrelativistic quantum mechanics entirely as a field theory \cite{SRE07}.

Historically, however, the major motivation for quantum field theory was the attempt to reconcile the nonrelativistic formulation of Quantum Mechanics with Special Relativity. The problem was that, according to Special Relativity, energy and matter were interconvertible through Einstein's famous equation $E=mc^2$, implying that particles could spontaneously form from pure energy and disappear into pure energy under the right conditions. Unfortunately, Schr\"{o}dinger's equation was not equipped to deal with this behaviour, since it implied that the number of particles and the total energy were separately conserved.

Quantum field theory solves this problem by treating quantum particles as being auxiliary concepts to the more fundamental objects: \tit{quantum fields}. A quantum field is a quantum mechanical operator of the form $\hat{\phi}(\tbf{x},t)$ with explicit spacetime dependence. The spacetime functions can obey relativistic equations of motion, examples being the \tit{Klein--Gordon equation} describing spinless bosons and the \tit{Dirac equation} describing spin-half fermions such as the electron. In general, the states of the theory do not correspond to a definite number of particles, but may involve arbitrary superpositions over different particle numbers. Typically one uses the Heisenberg picture, in which the Hermitian functions of field modes describing observables evolve with time and their expectation values are taken in the \tit{vacuum state}, this being the collective zero-particle state of the fields.

Traditionally, one arrives at a quantum field theory by applying an appropriate ``quantisation procedure" to the equations governing a classical matter field. Quantisation procedures fall roughly into two categories, depending on whether one quantises the Hamiltonian or the Lagrangian equations of motion. The former case is called ``canonical quantisation" and will suit our purposes. While Hamiltonian quantisation can be done in a Lorentz covariant manner, it is generally easier to perform the quantisation in terms of formal quantities that are not Lorentz covariant and to demonstrate the covariance of the physical quantities \tit{a posteriori}\footnote{By contrast, Lagrangian quantisation is manifestly Lorentz covariant and is useful in more general settings such as high energy particle physics.}. This is the most convenient approach for certain applications such as quantum optics.

Quantum optics is the study of the quantum properties of light in the laboratory. In general, the full theory of quantum electrodynamics (QED) is needed to describe the interaction of light with matter. In quantum optics, however, we are only interested in a certain class of experiments with particular features in common, which allows us to greatly simplify the formalism. The essential features of a quantum optics experiment are beams of light and material components. The beams of light are emitted from sources and propagate through the network of components before being received at detectors. The material components, sources and detectors are generally assumed to be at rest in the laboratory; however this assumption can be relaxed, as we will see.

The theoretical formalism of quantum optics is arrived at by first quantising the free electromagnetic field in the Coulomb gauge (which is not manifestly Lorentz covariant). This governs the propagation of light in between the sources, detectors and material components. The interaction of the light field with a given component is then described by an effective Hamiltonian function of the field modes. The effective Hamiltonian of a material component is generally arrived at by applying perturbative QED and then making various approximations. We will simply introduce the effective Hamiltonians of the components as needed, referring the reader to the literature for detailed derivations. 

At this point, to avoid confusion, it is necessary to address the different usages of the terms ``linear" and ``nonlinear" to describe quantum dynamics in different settings. In Chapter \ref{ChapNonlinBox}, we discussed nonlinearity at the fundamental level, in which the superposition principle no longer holds in general. In standard quantum optics, of course, the dynamics is fundamentally linear in this respect. However, in cases where the components' Hamiltonians are linear functions of the field modes, such as beamsplitters and phase shifts, we refer to them as being ``linear" optical components. There is a class of components that are called ``Gaussian", which have the property that the Heisenberg evolutions involve only linear combinations of the creation and annihilation operators. Despite this, general Gaussian operations such as single-mode squeezing (see below) have effective Hamiltonians that are quadratic in the field modes and are therefore labelled as ``nonlinear". Finally, there are non-Gaussian operations such as deterministic single photon sources which one might label as ``highly nonlinear", even though these are still linear at least in the sense of conforming to standard quantum field theory and hence obeying the superposition principle. In the present thesis, we are primarily concerned with Gaussian components; the context will make it clear whether we are referring to nonlinear optical components such as Gaussian squeezing, or nonlinear fundamental dynamics as described in Chapter \ref{ChapNonlinBox}.

\subsection{The quantised electromagnetic field}

Maxwell's equations are traditionally given in terms of the electric field $\tbf{E}(\tbf{x},t)$ and magnetic field $\tbf{B}(\tbf{x},t)$, which represent measurable quantities. However, it is convenient to define the equations in terms of some vector potential $\tbf{A}$ and a scalar potential $\phi$ satisfying:
\eqn{
\tbf{E} &=& -\nabla \phi-\frac{\del}{\del t}\tbf{A} \, , \\
\tbf{B} &=& \nabla \times \tbf{A} \, .
}
Different choices of these fields corresponding to the same $\tbf{E}$ and $\tbf{B}$ are referred to as different \tit{gauge} conventions and do not affect the physics. If we define the four-vector $A^\mu=(\tbf{A},\phi)$ and the field strength $F^{\mu \nu}:=\del^\mu A^\nu-\del^\nu A^\mu$, Maxwell's equations in free space amount to:
\eqn{\label{Maxwell}
\del_\nu F^{\mu \nu}&=&0,\\
\epsilon_{\mu \nu \rho \sigma}\, \del^\rho F^{\mu \nu}&=&0 \, ,
}
where $\epsilon_{\mu \nu \rho \sigma}$ is the Levi--Civita tensor (see Notation). Quantisation proceeds in the ``Coulomb gauge", specified by the condition\footnote{Note that by singling out the spatial part of $A^\mu$, this condition is dependent on a choice of relativistic reference frame and hence is not explicitly Lorentz covariant.}:
\eqn{
\nabla \cdot \tbf{A}=0 \, .
}
In this gauge, a general solution to \eqref{Maxwell} is given by $\phi=0$ and a vector potential $\tbf{A}$ satisfying the massless Klein--Gordon equation:
\eqn{
\del^2 \, \tbf{A}=0 \, .
}
After quantisation, $\tbf{A}$ is replaced by its corresponding field operator $\hat{\tbf{A}}$ (and its conjugate $\hat{\tbf{A}}^\dagger$) having the general form:
\eqn{
  \hat{\tbf{A}}(x) &= \int \invard{k} \; [G(k,x)\hat{a}_{\tbf{k},\omega}+G^*(k,x)\hat{a}^\dagger_{\tbf{k},\omega}] \\
                       &\equiv \hat{A}_{G}+\hat{A}^\dagger_{G} \, ,
}
where $G(k,x)$ is a normalized solution of the Klein--Gordon equation and $x=(\tbf{x},t),\, k=(\tbf{k},\omega)$ are four-vectors, with $kx = g_{\mu \nu}k^{\mu}x^{\nu}$. The field operators satisfy the bosonic commutation relations: 
\eqn{\label{basiccomms}
[\hat{a}_{k},\hat{a}^\dagger_{k'}]=\delta(k-k') \,.
} 
The Lorentz invariant measure is defined as 
\eqn{ \invard{k}:=\frac{\mathrm{d}\tbf{k}}{\omega(\tbf{k})} \nonumber \, ,} 
and the dispersion relation is $\omega(\tbf{k})=|\tbf{k}|$ in units of $c=1$. The operator $\hat{a}_{\tbf{k}}$ is interpreted as the annihilation operator for a photon in mode $\hat{a}$ with momentum $\tbf{k}$ and frequency $\omega=|\tbf{k}|$ (the polarisation vector associated with $\hat{a}$ is left implicit). 
The vacuum state $\ket{0}$ is defined by: 
\eqn{
\hat{a}_{\tbf{k}} \ket{0}=0 \, , \,\, \forall \tbf{k} \, .
}
Any mode or function of modes that annihilates the vacuum state will often be referred to as a ``vacuum mode". Similarly, we interpret the operator $\hat{A}_{G}$ as the annihilation operator for a particle in the mode $G$, and $\hat{A}^\dagger_{G}$ as the corresponding creation operator. We will find it useful to consider wavepackets, which are localized superpositions of plane wave solutions:
\eqn{\label{packet}
G(k,x) = g(\textbf{k})e^{ikx} \, .
}
where $g(\textbf{k})$ is centered on some positive wave number $\tbf{k}_0$ and is required to be zero for $k < 0$ for a mode propagating in the positive $\tbf{x}$ direction. Normalization requires:
\eqn{ \label{WavepacketNorm}
 \int \invard{k} \; |G(k,x)|^2 &=& 1  \nonumber \\
\Rightarrow \int \invard{k} \; |g(\tbf{k})|^2 &=& 1
\, . }
Note that the function $G(k,x)$ transforms as a Lorentz scalar, i.e. $G\mapsto G'$ such that $G'(k',x'):=G(k,x)$ in the Lorentz transformed co-ordinates $(k',x')$. This will be relevant in Sec \ref{SecQAExamples} where we apply our formalism to a quantum circuit with relativistically moving parts. From the boson commutation relations \eqref{basiccomms} we obtain the same-time commutator for wavepackets:
\eqn{\label{wavcom}
 \begin{aligned}
  &\left[ \hat{A}_{G}(k,\textbf{x}_1,t), \hat{A}^\dagger_{H}(k',\textbf{x}_2,t) \right] \\
   &= \int \invard{k} \; G(k,\textbf{x}_1,t)H^*(k,\textbf{x}_2,t) \\
   & = \int \invard{k} \; h^*(\textbf{k})g(\textbf{k})e^{i \tbf{k} \cdot (\textbf{x}_1-\textbf{x}_2)} \, . 
  \end{aligned}
}
A state containing $n$ photons in the mode $\hat{A}_{G}$ is defined as:
\eqn{
\ket{n_A}= \frac{1}{\sqrt{n!}} (\hat{A}^\dagger_{G})^n \ket{0} \, ,
}
and is referred to as a ``number state" or ``Fock state". An arbitrary pure state for the mode $\hat{A}_{G}$ can be written as an infinite superposition of these states:
\eqn{\label{GenFieldState}
\ket{\Psi_A}=\sum^\infty_{n=0} c_n \ket{n_A} \, ,
}
where $\sum\limits^\infty_{n=0} |c_n| = 1$. 
The number states $\ket{n_{A_G}}$ are the eigenstates of the ``number operator" $\hat{n}_{A_G}:=\hat{A}^\dagger_{G}\hat{A}_{G}$. This is a Hermitian operator and represents an observable whose eigenvalue is the number of photons in a given mode. The expectation value of the number operator, $\bk{\hat{n}}$, is the quantity measured by detectors in quantum optical experiments and corresponds to the measured intensity of the light in the classical limit of $\bk{\hat{n}}\gg1$. If we are interested in measuring the fluctuations in photon number, it is useful to measure the second-order correlation function $g^{(2)}$, which is defined by the expectation value:
\eqn{
g^{(2)}:=\frac{ \bk{\hat{A}^\dagger \hat{A}^\dagger \hat{A} \hat{A}} }{ \bk{\hat{n}}^2 } \, .
}
Qualitatively, the value of $g^{(2)}$ is zero iff there are no fluctuations in photon number, i.e. if the state is a Fock state. In general, $g^{(2)}=1$ for coherent states (see below) and $g^{(2)}\geq1$ for classical light or ``bunched" quantum light. Cases where $g^{(2)}<1$ exhibit ``anti-bunching", a distinguishing characteristic of quantum correlations between particles.
  
Each mode $\hat{A}_{G}$ corresponds to a distinct Harmonic oscillator degree of freedom. The Hilbert space corresponding to a given mode is spanned by the number states (``Fock" states) $\ket{n_{A_G}}$, these being eigenstates of the number operator for the mode in question: $\hat{n}_{A_G}=\hat{A}^\dagger_{G} \hat{A}_{G}$. Since there is no limit to the number of bosons that can exist in the same mode, this Hilbert space is countably infinite. If we consider the tensor product of Hilbert spaces for the entire continuum of modes in the field, then the total Hilbert space is uncountable infinite. In general, however, we are only concerned with a finite set of orthogonal modes $\hat{A},\hat{B},\hat{C},...$ and their conjugates, so that the total Hilbert space of the system of interest is spanned by the joint Fock states $\ket{n_A, n_B, n_C, ...}$. The algebra of operators on this space is called the \tit{Fock space algebra} of the modes of interest. In general, we will use $\hat{1}$ to represent the identity operator acting on the entire Fock space of interest. 

Thus we see that there are two aspects to a field theory: the photon number statistics and the dependence on spacetime coordinates. The former aspect has to do with the possibility of particle creation and annihilation, as described by the Fock space algebra of the field modes. The second aspect of a field theory is the spacetime dependance, which appears in the continuous mode amplitudes $G(k,x)$. These amplitudes are needed whenever there are interactions between modes whose wavepackets have nontrivial overlaps, leading to commutation relations that may not be orthonormal (see \eqref{wavcom}). The spacetime dependence is also needed in order to perform Lorentz transformations of the modes. Usually in quantum optics, the experimenter goes to great lengths to make sure that there are minimal mismatch effects or relativistic effects. This makes the theorists very happy, because it means the amplitudes $G(k,x)$ become irrelevant to their considerations and the subscript $G$ can be omitted from the modes. Ironically, therefore, standard quantum optics is mostly nonrelativistic, even though photons, which are the basic objects of the theory, travel at the speed of light!

In what follows, we will review some transformations corresponding to common quantum optical components, and discuss the convenient phase space representation of states in quantum optics. Since we are reviewing the standard formalism of quantum optics, the spacetime dependence will be dropped for the time being. It will be reintroduced when we consider mismatch effects and relativity in Sec \ref{SecMismatch}.

\subsection{Basic components of quantum optics}

\tit{Displacement:} Consider the unitary ``Displacement" operator $\hat{D}(\alpha):=\trm{exp}\brac{\alpha \hat{A}^\dagger - \alpha^* \hat{A}}$. When applied to the vacuum state, this operator creates a ``coherent state" in the mode $\hat{A}$ with complex amplitude $\alpha$: 
\eqn{
 \ket{\alpha} &:=& \hat{D}(\alpha) \ket{0} \nonumber \\
 &=& \trm{exp}(-\frac{1}{2}|\alpha|^2) \sum^{\infty}_{n=0} \, \frac{(\alpha)^n}{\sqrt{n!}} \, \ket{n_A}  \, .
} 
More simply, we can displace the mode $\hat{A}$ in the Heisenberg picture:
\eqn{
\hat{D}^\dagger(\alpha)\,\hat{A} \, \hat{D}(\alpha) = \hat{A}+\alpha \, .
}

\tit{Single mode squeezing:} Consider the single mode squeezing operator 
\eqn{
\hat{S}(r):=\trm{exp}\sqbrac{r(\hat{A}^2-(\hat{A}^{\dagger})^2)/2} \, . 
}
Applying this operator to the vacuum results in a ``squeezed state" with the amount of squeezing determined by the real parameter $r>0$. In the Heisenberg picture, the mode transforms as:
\eqn{
\hat{S}^\dagger(r)\,\hat{A} \, \hat{S}(r)=\trm{Cosh}(\hat{A}) -\trm{Sinh}(\hat{A}^\dagger)\, .
}
In the lab, squeezing of a single mode is performed by optical parametric amplification of light, by shining a strong laser through a medium with a nonlinear response function. We will encounter the two-mode generalisation of this unitary in Sec \ref{SecSPDC}.\\

\tit{Phase rotations:} The free-space propagation of a mode results in a phase rotation relative to some phase reference in the laboratory. The phase is generated by the rotation unitary $\hat{R}(\theta):=e^{-i \theta \hat{A}^\dagger \hat{A}}$ which leads to the simple Heisenberg evolution:
\eqn{
\hat{R}^\dagger(\theta)\, \hat{A} \, \hat{R}(\theta)=e^{i \theta} \hat{A} \, .
}

\tit{Beamsplitters:} In contrast to the single-mode operations above, a ``beamsplitter" involves two modes, $\hat{A}$ and $\hat{B}$. In the lab, a beamsplitter is a piece of glass with a metallic or dielectric coating that partially reflects and partially transmits the incident light according to its \tit{reflectivity} $\eta \in (0,1)$. The Heisenberg evolutions of the modes through the beamsplitter unitary $\hat{T}$ are:
\eqn{ \label{beamsplit}
  \hat{T}^\dagger \, \hat{A} \, \hat{T}&=& \sqrt{\eta}\, \hat{A}+e^{i\phi}\sqrt{1-\eta}\, \hat{B} \, ,\\
  \hat{T}^\dagger \, \hat{B} \, \hat{T} &=& \sqrt{\eta}\, \hat{B}-e^{-i\phi}\sqrt{1-\eta}\, \hat{A} \, , 
}
where $\phi$ is the phase picked up by the reflected mode. It is also possible to create a ``multi-port interferometer" by concatenating many beamsplitters of different reflectivities. Such an interferometer generates arbitrary linear combinations of the modes that it acts upon.    

In quantum optics, it is often convenient to work with the quadrature operators $\hat{P}:=i \brac{ \hat{A}^\dagger-\hat{A} }$ and $\hat{Q}:= \brac{\hat{A}^\dagger+\hat{A}}$ instead of the number operator. Whereas the latter defines a discrete eigenspectrum (i.e. the number states that span Fock space) the quadrature operators have a continuous spectrum of real eigenvalues $p$ and $q$ respectively. Any density matrix describing a state in Fock space can be transformed into a function of $q$ and $p$, called ``Wigner function"\footnote{For an explicit definition of the Wigner function, the reader is referred to Ref. \cite{WEE12}.}. This is called the \tit{phase space} representation.

Of particular interest are the ``Gaussian states" whose Wigner representations are Gaussian functions. It turns out that any single-mode Gaussian state can be created using a combination of displacements, squeezings and phase rotations described above. Interestingly, the Fock states that contain a definite number of photons are not themselves Gaussian states.

Gaussian states involving two modes have the property that they can be completely characterised by the first- and second-order moments of the quadrature operators. These are given by $\bk{\hat{Q}},\bk{\hat{P}}$ together with the covariance matrix:
\eqn{
 \left[ \begin{array}{*{2}{c}}
\trm{Var}(\hat{P}) & \frac{1}{2}\bk{\hat{P} \hat{Q}}+\frac{1}{2}\bk{\hat{Q} \hat{P}}-\bk{\hat{P}}\bk{\hat{Q}} \\
\frac{1}{2}\bk{\hat{P} \hat{Q}}+\frac{1}{2}\bk{\hat{Q} \hat{P}}-\bk{\hat{P}}\bk{\hat{Q}} & \trm{Var}(\hat{Q})
   \end{array} \right]
}
where $\trm{Var}(\hat{x}):=\bk{\hat{x}^2}-\bk{\hat{x}}^2$ and all expectation values are understood to be taken in the vacuum state. We note that any multi-mode Gaussian state can be created using a multi-port interferometer and an array of single-mode squeezers\cite{WEE12}. The Fock states and the Gaussian states will form the bases of our considerations in this thesis. The task of applying our analysis to more general non-Gaussian states is left as a topic for future work. 

\section{Circuit algebra from field modes\label{SecQAlgebra}}

Following the standard conventions of quantum optics, we will consider a field of scalar bosons, working in the Heisenberg picture and taking the field ground state to be the vacuum $\ket{0}$. In a typical quantum circuit, the details of the wavepacket profiles are suppressed. This is justified so long as the wavepackets are chosen to be sufficiently narrow, approaching the limit of an ideal circuit in which the particles are pointlike. Formally, this limit corresponds to a set of $N$ modes $\hat{A}_{i,G}, \, i \in \{0,1,2...N\}$ whose wavepackets are orthonormal, i.e.
\eqn{
\int \invard{k} \; g_i^*(\textbf{k})g_j(\textbf{k}) = \delta_{ij} \, ,
}
and the same-time commutator \eqref{wavcom} reduces to:
\eqn{ [\hat{A}_{i,G}(k,\tbf{x}_i,t),\hat{A}^\dagger_{j,G}(k,\tbf{x}_j,t)] = \delta_{ij} \, \label{wavsharp}. }
We will refer to this as the limit of \tit{sharp} wavepackets, or the \tit{ideal circuit} limit. In this limit, we obtain the usual algebra of Fock states for a set of orthonormal modes. Since the spectral distribution function of the wavepacket plays no role in the dynamics, we can drop the subscript $G$ and merely consider the index $i$, hence $\hat{A}_{i,G} \rightarrow \hat{A}_{i}$. In later sections we will relax the assumption of orthonormality to include mismatch between the modes. 

To make a qubit, we first require a mode containing a single particle:
\eqn{ |1_i\rangle := \hat{A}^\dagger_{i}|0\rangle \, . }
Ideally, if we are to take all expectation values in the vacuum, we will need to account for the deterministic creation of such single-particle states from scratch. This can be achieved through the use of a number of ancilla modes, as we will demonstrate in Sec \ref{SecSingPhot}. For the moment, we will simply take the existence of single photon states for granted, as per the usual approach. 

Suppose that we begin with $n$ particles, each of which is confined to a pair of orthogonal modes, $\hat{A}^\dagger_{i},\hat{B}^\dagger_{i}$, where $i=1,2,..,n$ labels the $i_{th}$ particle. Provided each particle remains confined to its two modes, every pair of modes constitutes an effective two-level system\footnote{As we generalise this framework, the possibility of higher order photon numbers and nontrivial mode overlaps will require us to increase the number of degrees of freedom of each mode. For the moment, we ignore these additional effects and focus on the essential qubit algebra.}. Hence the collection of such particles corresponds to an $n$-qubit system. For the $i_{th}$ qubit in the two-mode number basis $\ket{\hat{n}_A \hat{n}_B}_i$, we define the computational basis states $|\textbf{0}\rangle,|\textbf{1}\rangle$:
\eqn{ |\textbf{0}\rangle_i := \hat{A}^\dagger_i \ket{00} = |10\rangle_i; \quad |\textbf{1}\rangle_i := \hat{B}^\dagger_i |00\rangle = |01\rangle_i \, . }
An arbitrary superposition state is: 
\eqn{ |\psi_i \rangle &\equiv& \alpha|\textbf{0}\rangle_i + \beta|\textbf{1}\rangle_i \nonumber \\
&=& \alpha |10\rangle_i + \beta |01\rangle_i \, , \\
 }
where $|\alpha|^2+|\beta|^2=1$. This is called the ``dual rail" encoding. It is conventional to choose a circuit ``ground state" in which all of the qubits are initialised at the same value, and assumed to be separable. In the present case, we will take this state to be: $\ket{\psi_0}:=\ket{\tbf{0}}_1\, \ket{\tbf{0}}_2 ... \ket{\tbf{0}}_n$. Notice that this state is not the \tit{field} ground state, since it consists of $n$ qubits and hence contains $n$ particles. To avoid confusion between this circuit ground state and the field vacuum state, we will refer to this state simply as the \tit{initial} state, with the understanding that this state must be physically prepared from the vacuum state prior to implementing the quantum circuit.

Given the initial state, a quantum circuit specifies a series of quantum gates acting on sets of qubits. The most general description of the circuit dynamics is as a total unitary operator on the circuit Hilbert space. To describe this operator in terms of fields, we consider the following operations on the $i_{th}$ qubit:
\[ \hat{I}_{i}=\hat{A_i}^\dagger \hat{A_i}+\hat{B_i}^\dagger \hat{B_i}  \]
\[ \hat{Z}_{i}=\hat{A_i}^\dagger \hat{A_i}-\hat{B_i}^\dagger \hat{B_i}  \]
\[ \hat{X}_{i}=\hat{A_i}^\dagger \hat{B_i}+\hat{B_i}^\dagger \hat{A_i}  \]
\begin{equation} \label{pauli}
\hat{Y}_{i}=\textbf{\textit{i}}\hat{B_i}^\dagger \hat{A_i}-\textbf{\textit{i}}\hat{A_i}^\dagger \hat{B_i} \, .
\end{equation}
These are the quantum Stokes operators \cite{KOR02} . They are unitary Hermitian operators on the state space, and it is straightforward to verify using \eqref{wavsharp} that they satisfy the usual Pauli commutation relations:
\eqn{
&& [ \hat{J}_a,\hat{J}_b] = 2 i \sum_c \, \epsilon_{abc} \hat{J}_c \, , \\
&& [ \hat{J}_a,\hat{I} ] = 0 \, ,
}
where $(a,b,c) \in (1,2,3)$ and $\hat{J}_1=\hat{X}, \,\hat{J}_2=\hat{Y}, \, \hat{J}_3=\hat{Z}, \,$ and $\epsilon_{abc}$ is the Levi-Civita symbol in three dimensions (see Notation).

The construction \eqref{pauli} represents a possible decomposition of the Pauli matrices into field operators. Due to the fact that we are confined to the one-particle sector of Fock space, spanned by the states $\ket{10}$ and $\ket{01}$, there are many other ways to decompose matrices into field operators. For example, we can construct a different representation for the identity $\hat{I}_{i}$ by noting that an equivalent operator is:
\eqn{\hat{X}_{i} \hat{X}_{i} =  \hat{A}^\dagger \hat{B}\hat{A}^\dagger \hat{B}+\hat{B}^\dagger \hat{A}\hat{A}^\dagger \hat{B}+\hat{A}^\dagger \hat{B}\hat{B}^\dagger \hat{A}+\hat{B}^\dagger \hat{A}\hat{B}^\dagger \hat{A} \nonumber \, .}
This operator must satisfy the same algebraic relations as the operator $\hat{I}$, hence it is an equivalent representation. Note that this algebraic equivalence class structure depends upon the commutation relations \eqref{wavsharp}. Let us replace the operators in (\ref{pauli}) with their equivalence classes, these being represented by the abstract Pauli matrices:
\[
 \hat{I}_{i}= \left[ \begin{array}{*{2}{c}}
    1 & 0 \\
    0 & 1
   \end{array} \right]_i 
\]
\[
 \hat{X}_{i}= \left[ \begin{array}{*{2}{c}}
    0 & 1 \\
    1 & 0
   \end{array} \right]_i 
\]
\[
 \hat{Y}_{i}= \left[ \begin{array}{*{2}{c}}
    0 & -\textbf{\textit{i}} \\
    \textbf{\textit{i}} & 0
   \end{array} \right]_i 
\]
\[
 \hat{Z}_{i}= \left[ \begin{array}{*{2}{c}}
    1 & 0 \\
    0 & -1
   \end{array} \right]_i 
 \, . \]
It is a result of linear algebra that any hermitian operator acting on the total $n$-qubit Hilbert space can be expressed as a linear superposition of tensor products of the individual qubit Pauli matrices. Hence we can use any field decomposition of the Pauli matrices, such as the one given by \eqref{pauli}, to decompose an arbitrary quantum circuit into an expression involving only field modes. 
As a simple example, consider the case of a CSIGN gate acting on a pair of qubits whose initial state is:
\eqn{\ket{\psi_{in}}= \ket{\tbf{0}_1}\ket{\tbf{0}_2} = \hat{A}^\dagger_1 \hat{A}^\dagger_2 \ket{0} \, .}
To describe the dynamics, we first decompose the circuit into Pauli gates,
\eqn{\hat{U}_{\trm{CSIGN}} =  \frac{1}{2}\brac{\hat{I}_1\hat{I}_2+\hat{I}_1\hat{Z}_2+\hat{Z}_1\hat{I}_2-\hat{Z}_1\hat{Z}_2} \, ,}
and then replace the individual Pauli operators with their field expressions \eqref{pauli} to obtain a representation of the gate in terms of field modes. As we have seen, there are many different such representations for the same gate. Finally, an arbitrary observable $\hat{J}$ on the $n$-qubit space can similarly be decomposed into fields using its Pauli representation, allowing us to write all expectation values $\bra{\psi_{in}} \hat{U}^{\dagger} \, \hat{J} \, \hat{U} \ket{\psi_{in}}$ purely in terms of field modes. A general $2^n \times 2^n$ matrix representing an operator on the qubit state space can be decomposed as: 
\eqn{ \label{GenMatDecomp}
  &\left[ \begin{array}{*{3}{c}}
    \gamma & \delta & \hdots \\
    \rho & \sigma & \hdots \\
    \vdots & \vdots & \kappa
   \end{array} \right]_i \\
  &\equiv \gamma \hat{A_1}^\dagger \hat{A_1}\hat{A_2}^\dagger \hat{A_2}...\hat{A_n}^\dagger \hat{A_n}\\ 
  &+ \delta \hat{A_1}^\dagger \hat{A_1}\hat{A_2}^\dagger \hat{A_2}...\hat{A_n}^\dagger \hat{B_n} \\
  &+...+\kappa \hat{B_1}^\dagger \hat{B_1}\hat{B_2}^\dagger \hat{B_2}...\hat{B_n}^\dagger \hat{B_n} \, .
}
Of course, in the present idealisation there is little use for such an expression, since it is far less cumbersome to work directly with matrices rather than working at the level of individual field operators. Let us therefore turn to a more general scenario in which the field operator expression becomes relevant: the case of mode mismatch.

\section{Dealing with mismatch\label{SecMismatch}}

In general, the mode amplitudes describing the sources, components and detectors may be different. Explicitly, the initial state can have the general form:
 \eqn{
\ket{\psi_{in}} &=& \ket{\tbf{0}_1 \, \tbf{0}_2 \, \tbf{0}_3 \, ... } \nonumber \\
&=& \hat{A}^\dagger_{G_1},\hat{A}^\dagger_{G_2},\hat{A}^\dagger_{G_3},...\ket{0} \, ,
 }
where the $i_{th}$ qubit is associated with the modes $\hat{A}_{G_i},\hat{B}_{G_i}$ with spacetime dependence $G_i(k,x)$. Similarly, the circuit is represented by a total unitary evolution:
\eqn{
\hat{U}_{total}:=e^{i \hat{H}^{(1)}}\,e^{i \hat{H}^{(2)}}\,e^{i \hat{H}^{(3)}}...
} 
where the $j_{th}$ Hamiltonian $\hat{H}^{(j)}$ depends on the modes $\{ \hat{A}_{K^{(j)}_1},\, \hat{B}_{K^{(j)}_1},\, \hat{A}_{K^{(j)}_2},\, \hat{B}_{K^{(j)}_2},\, \hat{A}_{K^{(j)}_3} ... \}$ with arbitrary spectral amplitudes $K^{(j)}_i(k,x)$. Finally, an observable is a Hermitian function $f(\hat{A}_{L_1},\hat{B}_{L_1},\hat{A}_{L_2},\hat{B}_{L_2},... )$ where $L_i(k,x)$ is the response function of the $i_{th}$ detector.
In general, therefore, an arbitrary expectation value can be written as:
\eqn{\bra{\psi_{in, G_i}} \hat{U}^{\dagger}_{K^{(j)}_i} \, \hat{J}_{L_i} \, \hat{U}_{K^{(j)}_i} \ket{\psi_{in, G_i}} \label{mismatchEV} \, ,}
where the subscripts $G_i, K^{(j)}_i, L_i$ refer to the modes in the field decomposition of the relevant qubit, gate or observable, respectively. We can evaluate this quantity in principle by writing out the full field decomposition and using the general commutation relations \eqref{wavcom} to solve the equations of motion in either the Schr\"{o}dinger or Heisenberg picture. This procedure is far more tedious than the simple matrix manipulations that one can use in the absence of mismatch. Unfortunately it is not obvious that we can recover a matrix representation in the presence of the more general commutation relations \eqref{wavcom}. In order to generalise the matrix representation to account for mode mismatch, we will find it convenient to use a particular choice of field decomposition, described in the following section.

\subsection{Choosing a field decomposition\label{SecFieldDecomp}}
In the ideal limit, the different ways to decompose a matrix into field operators had no observable consequences because the sharp commutation relations preserved the algebraic structure. With the more general commutation relations \eqref{wavcom}, different field decompositions may become inequivalent. In practice, the correct field decomposition is the one that corresponds to the actual physical implementation. For the present purposes, we will simply choose one possible implementation that is mathematically simple whilst still being physically realisable in principle. To this end, we assume that an arbitrary gate is implemented using combinations of CSIGN gates and single-qubit operations, the latter being composed of combinations of beamsplitters and phase rotations\footnote{The proof that every quantum gate has such a decomposition can be found in Ref. \cite{NIE}.}. The Hamiltonian for a beamsplitter acting on any particular qubit has the form:
\eqn{\label{HamBS}
\hat{H}_{\trm{bs}}:=\kappa (\hat{a}_{\tbf{k}} \hat{b}^\dagger_{\tbf{k}}-\hat{b}_{\tbf{k}} \hat{a}^\dagger_{\tbf{k}}) \, , 
}
and for a phase rotation: 
\eqn{\label{HamPhase}
 \hat{H}_{\trm{phase}}:=\theta_a  \, \hat{a}^\dagger_{\tbf{k}} \hat{a}_{\tbf{k}}+\theta_b \, \hat{b}^\dagger_{\tbf{k}} \hat{b}_{\tbf{k}} \, . 
} 
For the implementation of the CSIGN gates, a natural description applicable to optical, microwave and ionic qubits\cite{MIL89,ROO08}, is the strong nonlinear cross-Kerr effect between two modes. The external pump mediating this effect for the $i_{th}$ CSIGN gate is described by the amplitude $H_i(k,x)$. The unitary operator for the CSIGN is given by: 
\begin{equation} \label{kerr}
 \begin{aligned}
   \hat{U}_{Kerr} &= \trm{exp} [ -i \pi \hat{n}_{B_1}\hat{n}_{B_2}] \\
                  &= \hat{1} + [ e^{-i \pi} - 1]\hat{n}_{B_1}\hat{n}_{B_2}\\
                  &= \hat{1} -2\hat{n}_{B_1}\hat{n}_{B_2}\\
                  &= \hat{1} -2\hat{B}^\dag_{H,1}\hat{B}_{H,1}\hat{B}^\dag_{H,2}\hat{B}_{H,2} \, ,
 \end{aligned}
\end{equation}
where we have used the property:
\eqn{
 \left( \hat{n}_{B_1}\hat{n}_{B_2} \right)^p \equiv \hat{n}_{B_1}\hat{n}_{B_2} \, , 
 }
for any integer $p$. This follows from our restriction to the one-particle sector of Fock space for each qubit, since the modes $\hat{B}_{H,1}, \hat{B}_{H,2}$ can contain only $0$ or $1$ particle each. Finally, we assume that the observables being measured at the output of the circuit are insensitive to the spectral properties of the qubits. This corresponds to the use of ``bucket detectors" in quantum optics, which register a click when one or more photons are incident, regardless of their spectral distribution. This completes our convention for decomposing a quantum circuit into fields. 

\subsection{Recovering a circuit picture\label{SecOverlapCircuit}}

In this section we will define a matrix representation for an arbitrary circuit in the presence of mismatch, assuming the field decomposition described in the previous section. In the case of linear optical quantum circuits, mismatch has been described in the literature using the ``eigenmode decomposition"\cite{RMS}. Given some reference mode $\hat{A}_{H}$, we can construct a complete orthonormal basis of modes $\{ \hat{A}^{(i)}_{H} : i = 0,1,2,...\}$, one of which is equal to the reference mode -- we use the convention $\hat{A}^{(0)}_{H} := \hat{A}_{H}$. The functions $H^{(i)}(k,x)$ are orthonormal, leading to the commutator:
\eqn{ [\hat{A}^{(i)}_{H},\hat{A}^{(j)\dagger}_{H}] = \delta_{ij} \, . }
We can therefore decompose an arbitrary mode $\hat{A}_{G}$ into parts that are either perfectly matched or perfectly orthogonal to $\hat{A}_{H}$ according to the \tit{eigenmode decomposition}:
\eqn{\hat{A}_{G} &=& \sum_i \lambda^{(i)} \, \hat{A}^{(i)}_{H} \nonumber \\
&=& \lambda_0 \, \hat{A}_{H} + \sum_{i > 0} \lambda^{(i)} \, \hat{A}^{(i)}_{H} \nonumber \\
&:=& \lambda_0 \, \hat{A}_{H} + \sqrt{1-|\lambda^{(0)}|^{2}} \, \hat{A}_{H/G} \label{eigenmode} \, ,}
where the coefficients are defined by:
\eqn{\lambda^{(i)} := \int \invard{k} \; G(k,x)H^{(i)*}(k,x) \, ,}
and in the final step of \eqref{eigenmode} we have grouped all of the orthogonal parts of the mode into a single field mode:
\eqn{\label{EmodeOrthog}
\hat{A}_{H/G} :=\frac{1}{{\sqrt{1-|\lambda^{(0)}|^{2}}}}\sum_{i > 0} \lambda^{(i)} \, \hat{A}^{(i)}_{H} \, ,
}
which satisfies the expected commutation relation, $[\hat{A}_{H/G},\hat{A}^\dagger_{H/G}] = 1$.

Using the present framework, we can use the eigenmode decomposition to recover a matrix representation of general circuits in the presence of mismatch. The wavepacket of the $i_{th}$ qubit can be written as:
\eqn{ 
G_i(k,x):= g_i(\tbf{k})e^{ik(x-x_i)} \, , 
} 
where $x_i$ represents possible timing and spatial errors. From \eqref{wavcom}, the partial overlap between the wavepackets of the $i_{th}$ and $j_{th}$ qubits at time $t$ is given by:  
\eqn{\label{wavcom2}
  &\left[ \hat{A}_{G_i}(k,\tbf{x},t), \hat{A}^\dagger_{G_j}(k',\tbf{x},t) \right] \\
   &= \int \invard{k} \; G_i(k,\tbf{x},t)G^*_j(k,\tbf{x},t) \\
   & = \int \invard{k} \; g_i(\tbf{k})\, g^*_j(\tbf{k})\,e^{i \tbf{k} \cdot (\tbf{x}_i-\tbf{x}_j)} \, . 
}
Similarly, let the $l_{th}$ CSIGN gate have the pump amplitude:
\eqn{
H(k,x):=h(\tbf{k}) e^{i k(x-x_l)} \, ,
}
where $x_l=(\tbf{x}_l,t_l)$, i.e. the pump amplitude is peaked at location $\tbf{x}_l$ at time $t=t_l$. The pump envelope $h(\tbf{k})$ is taken to be the same for all CSIGNs. If the $l_{th}$ CSIGN mediates the interaction between the rails $i=i'$ and $i=j'$ at time $t_l$, then the functioning of the gate will depend on the overlaps:
\eqn{\label{CSIGNoverlaps}
\left[ \hat{A}_{G_{i}}(k,\tbf{x},t), \hat{A}^\dagger_{H}(k',\tbf{x},t) \right] &=& \int \invard{k} \; g_{i}(\tbf{k})h^*(\tbf{k})\, e^{i \tbf{k}\cdot (\tbf{x}_{i}-\tbf{x}_l} ) \, ; \, \, \, \, (i=i',j') \, .
}
In the absence of all timing and displacement errors (i.e. setting $\tbf{x}_{i'}(t_l)=\tbf{x}_{j'}(t_l)=\tbf{x}_l$ for each CSIGN gate) the wavepackets of the qubits are centered on the gates at the time of interaction. In this case, mismatch occurs only if the envelope functions $g_i(\tbf{k}),g_j(\tbf{k}),h(\tbf{k})$ are different shapes. In general, the qubit wavepackets may also be shifted off-center at the time of interaction, as a result of timing errors or spatial displacements of the beam\footnote{An example is the Hong--Au--Mandel experiment, well known in quantum optics, where two interfering photons at a beamsplitter can be made to miss each other by placing a delay on one of them, leading to a loss of coherence.}.

Using the eigenmode decomposition \eqref{eigenmode} we can associate four independent modes to the $i_{th}$ qubit, namely $\{ \hat{A}_{i,H},\hat{B}_{i,H},\hat{A}_{i,H/G_i},\hat{B}_{i,H/G_i} \} := \{ \hat{A}_{i},\hat{B}_{i},\hat{C}_{i},\hat{D}_{i} \}$. The modes $\hat{A}_{i},\hat{B}_{i}$ are matched to the CSIGN gates, i.e.,
\eqn{
\hat{A}_{i,H} :=  \int \invard{k} \; h(\tbf{k}) \, e^{i k(x-x_l)} \hat{a}_{i,\tbf{k}}
}
and similarly for $\hat{B}_{i}$. The extra modes $\hat{C}_{i},\hat{D}_{i}$ represent the components of $\hat{A}_{G_i},\hat{B}_{G_i}$ that are completely orthogonal to the CSIGN gates, via the eigenmode decomposition. We can construct Pauli operators from the orthogonal field modes in the usual way, for example using \eqref{pauli} we have $\hat{X}_{i,CD} := \hat{C_i}^\dagger \hat{D_i}+\hat{D_i}^\dagger \hat{C_i}$. 

The initial state of the $i_{th}$ qubit is: 
\eqn{\ket{\psi_{i,G_i}} &=& \alpha_i \ket{\tbf{0}_{G_i}} + \beta_i \ket{\tbf{1}_{G_i}} \nonumber \\
&=& \alpha_i \, \hat{A}^\dagger_{i,G_i} \ket{0} + \beta_i \, \hat{B}^\dagger_{i,G_i} \ket{0} \, ,}
where $|\alpha_i|^2+|\beta_i|^2=1$. Applying the eigenmode decomposition with respect to the reference amplitude $H(k,x)$, we can rewrite this state as:
\eqn{\ket{\psi_{i,G_i}} 
&=& \zeta_i \, \brac{ \alpha_i \hat{A}^\dagger_{i}+\beta_i \hat{B}^\dagger_{i} } \ket{0} + \sqrt{1-|\zeta_i|^2} \, \brac{ \alpha_i \hat{C}^\dagger_{i}+\beta_i \hat{D}^\dagger_{i} } \ket{0} \nonumber \, \\ 
&:=& \zeta_i \, \ket{\psi_{i,H}} + \sqrt{1-|\zeta_i|^2} \, \ket{\psi_{i,H/G_i}} \label{qubitdecomp} \, ,
}  
where 
\eqn{\label{OverlapZeta}
\zeta_i &:=& \left[ \hat{A}_{G_{i}}(k,x), \hat{A}^\dagger_{H}(k',x) \right] \nonumber \\
&=& \int \invard{k} \; G_i(k,x) \, H^*(k,x) \, \nonumber \\
&=&  \int \invard{k} \; g_i(\tbf{k}) \, h^*(\tbf{k}) \,e^{i \tbf{k}(\tbf{x}_i-\tbf{x}_l)} \, ,
}
(compare to \eqref{CSIGNoverlaps}). The mismatched qubit \eqref{qubitdecomp} can therefore be regarded as being in a superposition of being perfectly matched or perfectly orthogonal to the CSIGNS that it encounters in the circuit, where the amplitude for each case is determined by the overlap $\zeta_i$. It follows that the initial state of the mismatched qubit can be obtained from a perfectly matched initial state by applying a unitary rotation $V_i(\zeta_i)$ on the enlarged Hilbert space $\scr{H}_{AB} \otimes \scr{H}_{CD}$. Specifically, we define $\hat{V}_i$ acting on the $i_{th}$ qubit to be:
\eqn{  
 \hat{V}_{i}= \left[ \begin{array}{*{2}{c}}
    \zeta_i \, \hat{I} & \sqrt{1-|\zeta_i|^2} \, \hat{I} \\
    -\sqrt{1-|\zeta_i|^2} \, \hat{I} & \zeta_i \, \hat{I}
   \end{array} \right] \, ,
}
where the basis for the matrix representation of the $i_{th}$ qubit has been extended to $\ket{\hat{n}_A \, \hat{n}_B \, \hat{n}_C \, \hat{n}_D}_i$.
It can easily be checked that, acting on a perfectly matched initial state $\ket{\psi_{i,H}} = \alpha_i \hat{A}^\dagger_{i} \ket{0}+\beta_i \hat{B}^\dagger_{i} \ket{0}$, the gate $\hat{V}_i$ produces the mismatched state \eqref{qubitdecomp}:
\eqn{\hat{V}_i \ket{\psi_{i,H}} = \ket{\psi_{i,G_i}} \, .}
Hence we can describe a circuit involving arbitrary qubit mismatch simply by extending the Hilbert space and applying the appropriate set of mismatch unitaries $\otimes^n_i \hat{V}_i$ to the perfectly matched initial state $\ket{\psi_{in}} = \otimes^n_i \ket{\psi_{i,H}}$.

\begin{figure}
 \includegraphics[width=16cm]{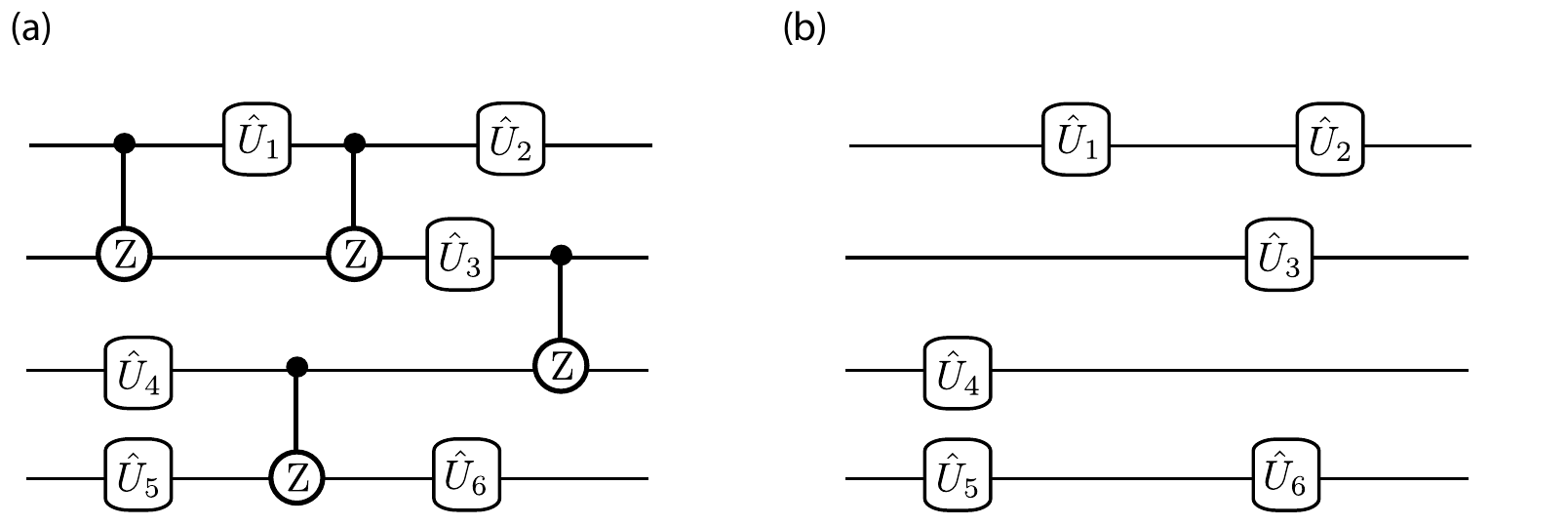}              
 \caption[Matched and orthogonal parts of an arbitrary circuit with mismatch]{\label{figOrthogCirc}(a) An arbitrary circuit decomposed into CSIGN gates and single-qubit gates. This is the effective circuit as `seen' by the component of the state that is perfectly matched to the gates. (b) The effective circuit seen by the orthogonal component. The linear single-qubit gates act as normal, but the CSIGN gates have no effect.}
\end{figure}

If the original circuit is represented by a unitary $\hat{U}$ acting on the $N$-qubit state space, then in the presence of mismatch we must replace this unitary with the extended circuit $\hat{U}_{AB} \otimes \hat{U}_{CD}$ acting on the enlarged space $\scr{H}_{AB}^{\otimes n} \; \scr{H}_{CD}^{\otimes n}$. The extended circuit must reduce to the ideal circuit on the subspace spanned by the matched components $\ket{\psi_{i,H}}$, that is, we require $\hat{U}_{AB} = \hat{U}$. To determine the circuit $\hat{U}_{CD}$ that acts on the orthogonal components $\ket{\psi_{i,H/G_i}}$, we refer to the field decomposition of the CSIGN gates, given by \eqref{kerr}. Projecting this operator onto the orthogonal subspace, the matched term $\hat{B}^\dag_{H,1}\hat{B}_{H,1}\hat{B}^\dag_{H,2}\hat{B}_{H,2}$ vanishes, leaving just the identity acting on the unmatched subspace. Hence a CSIGN gate acting on the Hilbert space $\scr{H}_{A_1,B_1}\otimes \scr{H}_{A_2,B_2}$ corresponds to $\trm{CSIGN}_{AB}\otimes \scr{I}_{CD}$ on the enlarged space $\scr{H}_{A_1,B_1}\otimes \scr{H}_{A_2,B_2}\otimes \scr{H}_{C_1,D_1}\otimes \scr{H}_{C_2,D_2}$. For the single qubit gates, it is easy to verify that the Heisenberg evolution of a qubit through the beamsplitters and phase shifts, given by the linear Hamiltonians \eqref{HamBS} and \eqref{HamPhase}, is independent of the wavepacket amplitudes $G_i(k,x)$. It follows that the representations of these gates by unitary matrices $\hat{T}(\eta,\phi)$ and $\hat{R}(\theta)$ are trivially extended to $\hat{T}_{AB}\,\hat{T}_{CD}$ and $\hat{R}_{AB}\,\hat{R}_{CD}$ on the enlarged Hilbert space. Accordingly, the unitary describing the circuit $\hat{U}_{CD}$ is obtained from the ideal circuit $\hat{U}_{AB}$ by replacing all of the CSIGN gates in the gate decomposition of $\hat{U}_{AB}$ with the identity, leaving only the single qubit gates (Fig. \ref{figOrthogCirc}).

Finally, the detection of the observable $\hat{J_i}$ on the $i_{th}$ qubit is represented in the expanded circuit by $\hat{J}_{i,AB}\, +\hat{J}_{i,CD}$. For example, in a two qubit circuit, the observable $\hat{J} = \hat{X}_1 \hat{Z}_2$ is replaced by $\hat{J}_{ABCD} := (\hat{X}_{1,AB}\, +\hat{X}_{1,CD})\,(\hat{Z}_{2,AB}\, +\hat{Z}_{2,CD})$. This follows from our assumption that the particle detectors in the physical implementation are insensitive to the mode amplitude, and hence respond to photons in both matched and orthogonal modes.

We can combine these results to obtain the generalised expectation values in the presence of mismatch. Formally, we can replace \eqref{mismatchEV} with:
\eqn{
\bra{\psi_{in}} (\otimes^n_i \hat{V}_i)^{\dagger}\, (\hat{U}_{CD} \hat{U}_{AB})^{\dagger} \, \hat{J}_{ABCD} \, (\hat{U}_{AB} \hat{U}_{CD}) \, (\otimes^n_i \hat{V}_i) \ket{\psi_{in}} \, \label{mismatchEV2},
} 
where the mismatch enters through the individual qubit overlaps $\zeta_i$ appearing in the rotations $\hat{V}_i$. This expression is much easier to work with than \eqref{mismatchEV} because it explicitly admits a matrix representation on the extended Hilbert space $\scr{H}_{AB}^{\otimes n} \; \scr{H}_{CD}^{\otimes n}$. 

\section{Producing qubits from the vacuum \label{SecSingPhot}}
In the previous section we saw that a qubit could be constructed from a single particle with two independent modes. For simplicity, we assumed that the initial state was a state of $n$ qubits, hence containing $n$ particles. This assumption is unproblematic in the nonrelativistic case, because the creation of such $n$-particle states can be treated in the Schr\"{o}dinger picture without difficulty. However, in relativistic scenarios, particularly involving observers following non-inertial trajectories, quantum field theory quickly becomes intractable in the Schr\"{o}dinger picture. As an example, consider the Bogolyubov transformation. This is a unitary transformation, well known in relativistic quantum field theory, that relates the modes of a stationary observer to the modes of another observer in uniform acceleration. Since it is unitary, the action of this transformation on the field vacuum state is trivial. However, when the transformation is applied to a multi-particle state, the resulting expression becomes extremely complicated. Working in the Heisenberg picture, it is possible to apply the transformation to the observables whilst keeping the initial state in the vacuum. This leads to tractable calculations that have found recent applications in the area of relativistic quantum information\cite{DOW12}. By contrast, such calculations would not be tractable if the initial state was assumed to contain a nonzero number of particles.

In order to make our field description of a quantum circuit truly relativistic, therefore, we need to replace the $n$ particle initial state with the field vacuum state. This amounts to finding an abstract transformation of the modes $\hat{A}_i, \hat{B}_i, ...$ appearing in the expectation values \eqref{mismatchEV2} such that we can rewrite the expectation value in the vacuum state. Formally, consider the general expectation value:
\eqn{ \bra{\psi_{in}} f\brac{\hat{A}_i, \hat{B}_i, ...} \ket{\psi_{in}} \, ,}
where $f$ is an arbitrary function representing an observable and $\ket{\psi_{in}}$ is the initial state containing $n$ particles with spectral amplitude $H(k,x)$. We seek a transformation of the modes $\hat{A}_i \rightarrow \hat{A}'_i$ such that we can rewrite this expectation value as:
\eqn{ \bra{0} f\brac{\hat{A}'_i, \hat{B}'_i, ...} \ket{0} \, ,} 
where the initial state is now the field vacuum state $\ket{0}$. In effect, this transformation takes a given vacuum mode $\hat{A}$ containing no particles:
\eqn{
  \langle 0| \hat{A}^\dag \hat{A} |0\rangle = 0 \,
}
and deterministically transforms it into a mode $\hat{A}'$ containing exactly one particle:
\eqn{ \label{SPproperties}
  & \langle 0|\hat{A}'^\dag \hat{A}'|0\rangle = 1, \\
  & \langle 0|\hat{A}'^\dag \hat{A}'^\dag \hat{A}' \hat{A}' |0\rangle=0 \, . \\
}
The process represented by this transformation should be physically realisable in principle with finite resources. Note that we can write the transformation in terms of some unitary $\hat{U}_{S}$ according to:
\eqn{ \label{SPunitary}
\hat{A}' \brac{\hat{A},\hat{v}_1, ... , \hat{v}_m} = \hat{U}^{\dagger}_{S} \hat{A} \hat{U}_{S}
}
where the modes $\{\hat{v}_i \}$ are a set of mutually orthonormal ancillary vacuum modes, i.e. 
\eqn{
[ \hat{v}_i, \hat{v}^{\dag}_j ] = \delta{ij} \, , \nonumber . 
 }
The hypothetical unitary transformation $\hat{U}_{S}$ is difficult to determine, since it is a highly nonlinear function of the field modes. A more direct method is to devise a circuit that deterministically performs the same function as $\hat{U}_{S}$, but which uses non-unitary procedures such as measurement and feeding forward of classical data. Such a circuit represents an adaptive process whereby the operations performed in later steps of the circuit are conditioned on the outcomes of measurements performed in previous steps. Circuits involving adaptive measurements and feed-forwards derive from the literature on quantum computation where they have proven useful for performing complicated operations deterministically. 

\begin{figure}
 \includegraphics[width=16cm]{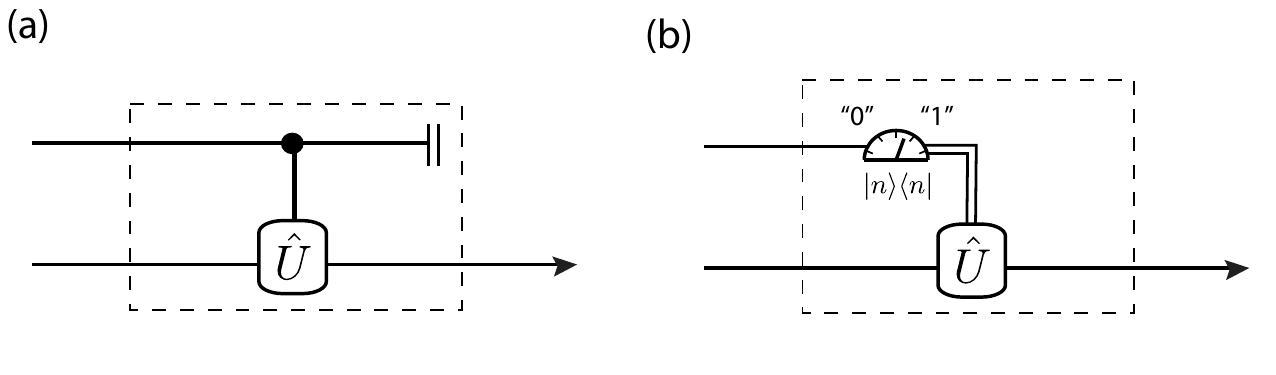}              
 \caption[An example circuit containing measurement and feed-forward]{\label{figFeedFwd}(a) A coherent control gate that implements $\hat{U}$ on the bottom rail when a particle is present on the top rail. (b) An equivalent representation using measurement and feed-forward. The number of particles measured in the top rail (zero or one) conditions the unitary applied to the bottom rail.}
\end{figure}

As an example of the technical description of a measurement and feed-forward process, consider the simple circuit shown in Fig. \ref{figFeedFwd}. This circuit implements the unitary $\hat{U}:=e^{i \hat{H}}$ depending on whether or not there is a particle present in the mode $\hat{b}$ (top rail). Formally, we can rewrite this circuit as a conditional operation by letting the operator $\hat{n}_b = \hat{b}^\dagger \hat{b}$ serve as a placeholder for its eigenvalue realised in the measurement. In the present case, we define the conditional operator: $e^{i \, \hat{n}_b \,  \hat{H}}$. Conditional on the mode $\hat{b}$ containing either zero or one photon, $\hat{n}_b$ can be replaced by the value $0$ or $1$ resulting in the application of either $\hat{U}$ or $\hat{I}$ to the target rail, respectively. Using the operator $\hat{b}^\dag \hat{b}$ is a convenient means of representing the outcome because it allows us to write down the output mode $\hat{a}_{out}$ as a quantum operator without having to specify the actual classical outcome of the measurement on which it is conditioned. In this example, we can simply write:
\eqn{\hat{a}_{out} = e^{-i (\hat{b}^\dag \hat{b}) \,  \hat{H}^\dag} \; \hat{a} \; e^{i (\hat{b}^\dag \hat{b}) \,  \hat{H}} \label{feed-forward} \, .} 
For more involved circuits, it can be easier to obtain expressions like \eqref{feed-forward} based on conditional operators rather than referring to the unitary purification of the circuit. This is the approach we adopt in deriving the single photon operator $\hat{A}'$.

\begin{SCfigure}
 \includegraphics[width=8cm]{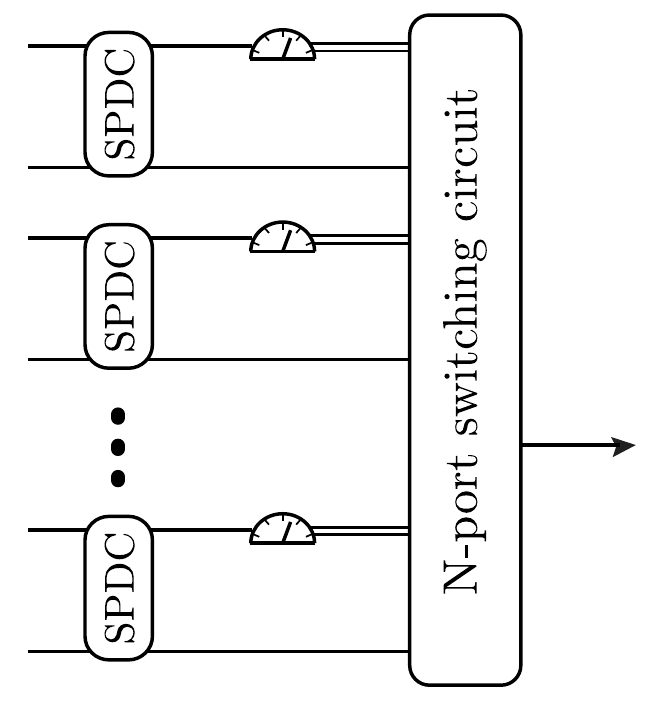}              
 \caption[A circuit for producing single photons on demand]{\label{figSPOD}A circuit for producing single photons on demand, using measurements and feed-forwards. An array of $N$ parametric down converters spontaneously produce pairs of photons. Conditioned upon the detection of one photon from a given pair, the switching circuit selects the other photon as a potential output. For large $N$, there is a high probability of at least one photon being available at any time.}
\end{SCfigure}

Consider the network of measurements and feed-forwards shown in Fig. \ref{figSPOD}. This circuit represents a single-photon on-demand source (SPOD). This particular SPOD was first proposed by Migdall, Branning \& Casteletto\cite{MIG02} for its possible experimental application. Here, we utilise their model for the theoretical purpose of obtaining an expression for the output mode as a function of the input vacuum modes. The SPOD circuit consists of two key elements: a large number $N$ of heralded spontaneous parametric down converters (SPDCs) and a switching circuit for dealing with the measurements and feed-forwards. The $i_{th}$ SPDC is represented by its unitary gate $\hat{S}_i$ which acts on a pair of modes initially in the vacuum state. The output mode corresponding to each $\hat{w}_i(\tbf{k})$ is then measured in the number basis by a bucket detector, which registers a click when the mode contains one or more particles, and nothing otherwise. The outcomes of the $N$ simultaneous detection events are fed forward into a classical switching circuit that processes the information according to an algorithm, producing an output mode containing a single photon.

It is worth emphasising that our goal is to derive an \tit{abstract} expression for $\hat{A}'$ such that it satisfies the algebraic properties of a single-photon mode, \eqref{SPproperties}. There are many other possible algebraic expressions for $\hat{A}'$ that would serve equally well and would give the same results (for example, an expression for the unitary \eqref{SPunitary}), but the SPOD provides us with a simple algorithm for deriving such an expression. Since the SPOD source is an abstraction, we are free to make various idealisations about its functioning without impacting the generality of the final result. In particular, we will assume that there are no mismatch effects within the SPOD. In the next section, we briefly review the SPDC unitary and its physical interpretation in quantum optics. In subsequent sections we describe the action of the switching circuit and use it to derive an expression for the output $\hat{A}'$. This will be shown to correspond to a single photon mode with a probability approaching unity for large $N$.

\subsection{Spontaneous Parametric Down Conversion\label{SecSPDC}}

Roughly speaking, an SPDC consists of a classical laser beam that interacts with a crystal, stimulating the spontaneous emission of photon pairs from the crystal. Formally, the process of down conversion is described by the effective unitary $\hat{S}(\chi)$, given by: 
\eqn{ \label{spdcgen}
 \hat{S}(\chi) := \trm{exp} \sqbrac{ \iint \invard{k}\invard{l} \, \chi \, P^*(k,l) \, \hat{u}^\dagger(\tbf{k})\hat{v}^\dagger(\tbf{l}) - H.c.}
}

In general, the spectral amplitude $P(k,l)$ is complex, non-separable, asymmetric and is normalised according to:
\eqn{ \iint \invard{k}\invard{l} \, |P(k,l)|^2 = 1 \, .}
For our purposes, it is convenient to make the simplifying assumption $P(k,l):=H(k,x)H(l,y)$. In terms of the wavepacket operators, i.e. $\hat{u}_H := \int \invard{k}\,H(k,x)\hat{u}(\tbf{k})$, we can rewrite this unitary as: 
\eqn{\label{spdcgen}
 \hat{S} = \trm{exp} \sqbrac{ \chi \, \hat{u}^\dagger_H \hat{v}^\dagger_H - \chi^* \, \hat{u}_H \hat{v}_H } \, ,
}
which is also called the \tit{two-mode squeezing} operator (compare to the single-mode case in Sec \ref{SecQO}). The modes $\hat{u}_H, \hat{v}_H$ commute with each other and satisfy the usual commutation relations 
\eqn{ [\hat{u}_H,\hat{u}^\dagger_H)] = [\hat{v}_H,\hat{v}^\dagger_H] = 1 \, .}
Since the spectral distributions play no role in the following discussion, we will drop the wavepacket subscripts and just write eg. $\hat{v}_H \rightarrow \hat{v}$. 
The Heisenberg evolutions of the modes $\hat{u},\hat{v}$ are given by:
\eqn{\label{EqSPDC}
  \hat{u}' = \hat{S}^\dagger \, \hat{u} \, \hat{S} = \trm{sinh}(\chi)\, \hat{u}^{\dagger}+\trm{cosh}(\chi)\, \hat{v}\\
  \hat{v}' = \hat{S}^\dagger \, \hat{v} \, \hat{S}= \trm{sinh}(\chi)\, \hat{v}^{\dagger}+\trm{cosh}(\chi)\, \hat{u} \, .
}
The probability of producing a photon pair (one in each output mode) is proportional to $|\chi|^2$, with higher order terms corresponding to higher photon numbers in the outputs.

A typical application of the SPDC is as a heralded nondeterministic single-photon source. If we set $\chi \ll 1$ in order to suppress higher photon numbers, the probability of creation of single photons is necessarily small. Fortunately, the appearance of a photon in the output (signal) mode $\hat{v}'$ is always accompanied by a photon in the heralding (idler) mode $\hat{u}'$. Therefore one can use a single SPDC to produce single photons by postselecting on the detection of a photon in the idler mode and taking the signal photon as output. However, this process is nondeterministic. To produce a single photon deterministically, we will require the more elaborate setup of the SPOD source.

\subsection{Single photons on demand}

We now return to the setup proposed by Migdall et al. for producing single photons on demand, shown in Fig. \ref{figSPOD}. The circuit consists of $N$ SPDCs with their outputs connected to a large switching circuit, which operates according to a simple algorithm, described below.
 
At any given time-step, on average, a fraction of the SPDCs proportional to $|\chi|^2$ spontaneously produce photons. In each case, the single photon in the idler mode triggers a detector, which feeds forward the information to the switching circuit. For large $N$, there is a high probability that at least one SPDC fires, heralding the presence of a single photon in its signal mode. The switching circuit simply registers which SPDCs have produced photons, based on the information fed forward from the detectors, and then selects one of the corresponding signal modes as the output. For any nonzero set of detection events, we assume that the switching circuit chooses the signal mode from the SPDC with the lowest index $i$ out of the set. If there are no detection events, then the switching circuit outputs an empty mode -- but this outcome becomes increasingly unlikely for large $N$.

Given this description of the circuit's operation, we can derive an expression for the output mode from the circuit. We will consider a single vacuum mode $\hat{a}$ which is to be populated with a single photon by the SPOD. Our goal is to write down an expression for the output mode $\hat{a}'$ as a function of the vacuum modes $\{ \hat{a}, \hat{u}_i, \hat{v}_i \}$ and their conjugates, where $i=1,2,...N$ labels the $i_{th}$ SPDC. This expression will then be shown to satisfy the requirements of a single photon mode, \eqref{SPproperties}. We proceed by considering the specific cases for low $N$ and then generalise to arbitrary $N$.

Consider the case $N=1$. Since there is only one SPDC labelled $i=1$, the switching circuit simply outputs the signal mode $\hat{v}'_1$ conditional on the detection of one (or more) photons in the idler mode, and produces an empty mode otherwise. Hence the output mode can be written as:
\eqn{ \label{out1}
 \hat{a}' &=& \brac{ \hat{u}'^\dagger_1 \hat{u}'_1} \hat{v}'_1+ \hat{c}\hat{a} \nonumber \\
 &=& \hat{n}_{u'_1} \, \hat{v}'_1 + \hat{c}\hat{a} \, ,
}
where the operator $\hat{c}$ is to be fixed by the requirement that the output satisfies the correct commutation relation $[\hat{a}',\hat{a}'^\dagger] = 1$. Note that we have assumed the possible eigenvalues of $\hat{n}_{u'_1}$ to be either $0$ or $1$ in this expression, which is only valid if we neglect terms of higher order than $|\chi|^2$. This will be problematic for our eventual calculation of $g^{(2)}$ on the output mode, where terms of order $|\chi|^4$ play a role. Let us therefore modify the conditional expression to account for the slightly larger set of possible eigenvalues $\hat{n}_{u'_j}\mapsto 0,1$ or $2$ for the $j_{th}$ SPDC. To do this, we note that the detector is a bucket detector and therefore is insensitive to photon number. The appropriate function to use instead of $\hat{n}_{u'_i}$ is therefore: 

\eqn{\hat{d}_{j} := \frac{1}{2}(3-\hat{n}_{u'_j})\hat{n}_{u'_j} \, , \label{bucketD}}

which takes the value $\hat{d}_{j} \mapsto 0$ for zero photons and $\hat{d}_{j} \mapsto 1$ if $\hat{n}_{u'_j}\mapsto1$ or $2$. Therefore the conditional output mode, correct to order $|\chi|^4$, is given by:
\eqn{ \label{out1b}
 \hat{a}' = \hat{d}_{1} \, \hat{v}'_1 + \hat{c}\hat{a} \, ,
}
Note that in this instance, $\hat{a}'$ is just equivalent to the output $\hat{v}'_1$ from the single SPDC, since the conditioning by the switching circuit does not affect the physical expectation values. In particular, the expected photon number in the output is still $\bra{0} \hat{n}_{a'} \ket{0}=|\chi|^2$. In order to increase this value to unity, we need to add more SPDCs to the circuit.

In the case $N=2$, the impact of the switching circuit is less trivial. If one or the other SPDC fires, the corresponding signal mode is output. If both of them fire, then the output is the signal mode from the first SPDC. Finally, if neither of them fire, the output is the original vacuum mode $\hat{a}$ as before. We can now express the conditional output as:
\eqn{ \label{out2}
 \hat{a}' = \hat{d}_{1} \, \hat{v}'_1 +\hat{d}_{2} \, \hat{v}'_2 \, \brac{\hat{1}-\hat{d}_{1}} + \hat{c}\hat{a} \, .
}
It can easily be verified that this operator performs as described, simply by substituting the appropriate eigenvalues for $\hat{d}_{1}$ and $\hat{d}_{2}$ appropriate to each scenario. Following similar reasoning, the conditional output for $N=3$ is given by:
\eqn{ \label{out3}
 \hat{a}' = \hat{d}_{1} \, \hat{v}'_1 +\hat{d}_{2}\,  \hat{v}'_2 \, \brac{\hat{1}-\hat{d}_{1}} +\hat{d}_{3}\,  \hat{v}'_3 \, \brac{\hat{1}-\hat{d}_{2}}\brac{\hat{1}-\hat{d}_{1}}+ \hat{c}\hat{a} \, .
}
Finally, we can extrapolate this expression to arbitrary $N$, resulting in the conditional output:
\eqn{ \label{aoutN}
  \hat{a}' = \sum_{j=1}^{N} \, \hat{d}_{j} \, \hat{v}'_j \, \prod_{i=0}^{j-1} \, \left( 1 - \hat{d}_{i} \right) + \hat{c}\hat{a} \, ,
}
where we have introduced the zeroth term $\hat{d}_{0} := 0$ for notational convenience. Setting the commutator for $\hat{a}'$ equal to $1$ results in the condition:
\eqn{\hat{c}^2 = \hat{1} - \sum_{j=1}^{N} \hat{n}^2_{u_j}\prod_{i=0}^{j-1} \left( 1 - \hat{n}_{u_i} \right)^2 \, , }
which constrains the coefficient $\hat{c}$.

We are now in a position to check that the output mode \eqref{aoutN} satisfies the requirements of a single photon mode for sufficiently large $N$ and small $\chi$. This consists of calculating the quantities $\bra{0} \hat{n}_{a'}\ket{0}$, $\bra{0} \hat{a}'^\dagger \hat{a}'^\dagger \hat{a}' \hat{a}'\ket{0}$ and demonstrating that they approach $1$ and $0$, respectively.

It is easy to justify these results heuristically. In the case of the expected photon number, we observe that, provided at least one SPDC fires, the output will contain at least one photon, by construction. Since the probability of at least one SPDC firing asymptotes to $1$ as we increase $N$, the output becomes equivalent to the output of a single SPDC conditioned on the heralding event. For small $\chi$, this corresponds to a single photon, as we have discussed. This also ensures that $g^{(2)}=0$ since higher order terms are negligible. From the explicit calculations carried out in the Appendix, we obtain the results shown in Fig. \ref{figSingphotGraphs}.

\begin{figure}
 \includegraphics[width=18cm]{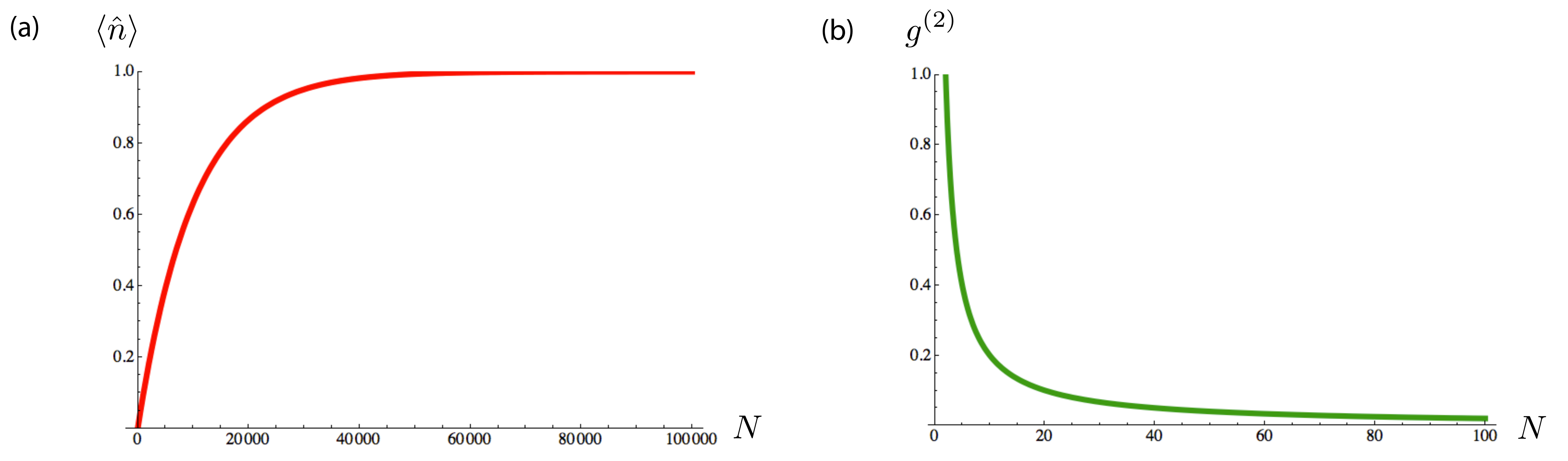}              
 \caption[Expected photon number and second-order coherence for the single photon source]{\label{figSingphotGraphs}Plots of (a) the expected photon number $\bk{\hat{n}}$ and (b) the second order correlation function $g^{(2)}$ for the single photon mode given by \eqref{aoutN}, with $\chi=0.01$. For sufficiently large $N$, we see that $\bk{\hat{n}}\approx 1$ and $g^{(2)}\approx0$. Note that $g^{(2)}$ is already negligible for $N>50$ but about a thousand times more down converters are required to bring the expected photon number up to unity, at around $N>50000$.}
\end{figure}

\section{An example: the relativistic CNOT\label{SecQAExamples}}

Relativity enters the present formalism in two essential ways -- through Lorentz transformations of the fields and through Bogolyubov transformations of the vacuum state. We will consider only the former example, which pertains to inertial observers in flat space. The second application deals with observers in non-inertial motion and is discussed briefly as a topic for future work.

The question of Lorentz invariance at the level of fields is well understood. We have been working in the Coulomb gauge, because that is the most convenient choice for quantum optics. While this gauge is not explicitly Lorentz invariant, we commented in Sec \ref{SecQO} that explicit Lorentz transformations leave the physical quantities invariant. Nevertheless, observers may disagree on the width and timing of wavepackets, since these are subject to length contraction and time dilation effects. This may lead to nontrivial effects in situations where different parts of a quantum circuit are in relative motion to one another. While such a scenario might seem far-fetched, it is not unusual for quantum information protocols using light to be carried out over large distances, possibly involving satellites in orbit around the earth. The consideration of quantum circuits with relativistically moving parts is therefore plausible and may be increasingly useful as technology improves.

Since our construction of a circuit from quantum fields is relatively simple, the effect of Lorentz transformations will enter only as an effective mismatch between parts of the circuit. In the Conclusion (\ref{SecFieldConc}) we will mention ways in which our formalism might be generalised. We will proceed with a simple example to illustrate the effects of relativistically induced mismatch within our model.

Suppose that Alice is performing a computation using a circuit consisting of a CNOT gate between two qubits. Alice prepares the control qubit with mode amplitude $H(k,x)$ chosen to match the pump amplitude driving the gate interaction. Bob, who has been instructed to prepare the target qubit, is running late and therefore prepares his qubit while in constant relative motion running towards the laboratory, using his own Lorentz-transformed co-ordinates instead of the lab co-ordinates. In addition, there is the possibility that Bob arrives too early or too late with his qubit, giving rise to timing errors. We can calculate the outcome of Alice's measurements on the circuit, taking into account the mismatch arising from Bob's inertial motion and poor timing.

\begin{figure}
 \includegraphics[width=16cm]{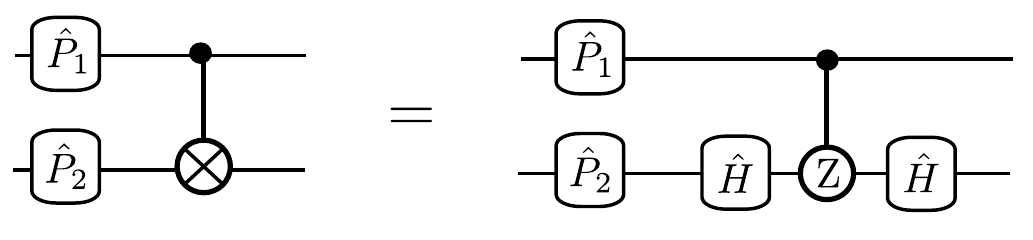}              
 \caption[The decomposition of a CNOT gate into CSIGN and Hadamard gates]{\label{figCNOTDecomp}The decomposition of the CNOT gate into a CSIGN and two Hadamard gates. Alice and Bob prepare their qubits using the single-qubit gates $\hat{P}_1$ and $\hat{P}_2$ respectively.}
\end{figure}

The CNOT is decomposed into a CSIGN gate and single-qubit gates as shown in Fig. \ref{figCNOTDecomp}, where the Hadamard gate is defined as $\hat{H}=\brac{\hat{X}+\hat{Z}}/\sqrt{2}$. The lab frame co-ordinates are denoted $(\tbf{x},t)$, or $(\tbf{k},\omega)$ in momentum space. In these co-ordinates, the pump driving the interaction is associated with the amplitude:
\eqn{ H(k,x) = h(\tbf{k}) e^{i k (x-x_l) } \,}
centred on the interaction event $x_l := (\tbf{x}_l,t_l)$. The Gaussian envelope takes the form
\eqn{ \label{GaussianWP} 
h(\tbf{k}):=N \, e^{-(\tbf{k}-\tbf{k}_0)^2 / \sigma^2} \, ,
} 
centred on $\tbf{k}_0>0$ with variance $\sigma$. The normalisation requirement \eqref{WavepacketNorm} implies $N^2=\sqrt{\frac{2}{\pi}}\frac{\omega}{\sigma}$.  

Alice prepares her single photon qubit in the mode $\hat{A}_{1,H}$, perfectly matched to the pump amplitude $H(k,x)$, while Bob produces a single photon in the mode $\hat{A}_{2,G_B}$ where $G_B(k,x)$ is the amplitude of Bob's photon in the lab frame. In Bob's frame, his wavepacket is:
\eqn{
H_B(k',x')= h(\tbf{k}') e^{i k' (x'-x'_B)} \, ,
} 
where $x'_B$ contains possible timing errors. We can convert this expression into the lab frame co-ordinates using the passive momentum-space Lorentz transformations:
\eqn{
\tbf{k}'&=&\gamma \, (\tbf{k}+\tbf{v} \omega) \, , \nonumber \\ 
\omega'&=&\gamma ( \omega + \tbf{v}\cdot \tbf{k}) \, ,
} 
where $\gamma:=1/\sqrt{1-v^2}$ and $v=|\tbf{v}|$ is the magnitude of Bob's velocity relative to the lab. Since Bob is travelling directly towards the laboratory, we set $\tbf{v}=+v \hat{\tbf{k}}$, where $\hat{\tbf{k}}$ denotes a unit vector in the $\tbf{k}$ direction \footnote{Not to be confused with a quantum mechanical operator.}. In addition, we note that the scalar product of four-vectors is Lorentz invariant, i.e. $k' x'=k x$. Transforming Bob's wavepacket $H_B(k',x')$ into the lab co-ordinates, we then obtain: 

\eqn{ G_B(k,x) &=& \brac{ \sqrt{\frac{2}{\pi}}\frac{\omega'}{\sigma}} \trm{exp}\brac{-\frac{(\tbf{k}'-\tbf{k}'_0)^2}{\sigma^2} + i k' (x'-x'_B)}  \nonumber, \\
&=& \brac{ \sqrt{\frac{2}{\pi}}\frac{\omega}{\bar{\sigma}}} \trm{exp}\brac{-\frac{(\tbf{k}-\tbf{k}_0)^2} {\bar{\sigma}^2} + i k (x-x_B) } \, .
}
This result is again a Gaussian centred at $\tbf{k}_0$, but with a modified standard deviation,
\eqn{ 
\bar{\sigma} := \frac{\sigma}{\gamma (1+v)} \nonumber 
}
Note that $\bar{\sigma} < \sigma$, so that Bob's pulse is narrower as seen from the lab frame\footnote{This effect can be interpreted as a joint consequence of both the Doppler shift and relativistic length contraction acting on Bob's wavepacket, contributing the factors $(1+v)$ and $\gamma$, respectively.}. To describe Bob's timing errors, suppose that at the time of interaction $t_l$ in the lab frame, Bob's wavepacket is centred at some position $\tbf{x}_B \neq \tbf{x}_l$, hence $x_B:=(\tbf{x}_B,t_l)$. The overlap between Bob's wavepacket and the CSIGN is given by:
\eqn{\label{BobOverlap}
\zeta &=& [ \hat{A}_{G_B}(k,x),\hat{A}^\dagger_{H}(k,x) ] \nonumber \\
&=& \int \invard{k} \; G_B(k,x)H^*(k,x) \nonumber \\
 &=& \int \invard{k} \; \sqrt{\frac{2 \omega^2}{\pi \,\sigma \, \bar{\sigma}}} \; \trm{exp} \sqbrac{-\frac{(\tbf{k}-\tbf{k}_0)^2}{\sigma^2}-\frac{(\tbf{k}-\tbf{k}_0)^2}{\bar{\sigma}^2} + i \tbf{k} \cdot (\tbf{x}_l-\tbf{x}_B) } \nonumber \\ 
&=& \sqrt{ \frac{2 \, \sigma \, \bar{\sigma} }{\sigma^2+\bar{\sigma}^2 }  } \; \trm{exp} \sqbrac{ -\frac{\sigma^2 \, \bar{\sigma}^2 \, (\Delta \tbf{x})^2}{4(\sigma^2+\bar{\sigma}^2)}  +i \tbf{k}_0 \Delta \tbf{x}} \, ,
}
where $\Delta \tbf{x}=(\tbf{x}_l-\tbf{x}_B)$. This expression is Lorentz invariant, as is apparent from the covariant form of the integral as written in the second step above. To make this invariance clear in the final expression, we can define the four-vectors $X^{\mu}:=x^{\mu}_l-x^{\mu}_B$, $K^{\mu}:=(\tbf{k}_0,\omega_0)$. One arrives at the expression of \eqref{BobOverlap} by considering the mismatch on a spatial hypersurface at time $t_l$ in the lab frame, such that $X^{\mu}=(\Delta \tbf{x},0)$, leading to the covariant quantities $X^{\mu}X_{\mu}=\Delta \tbf{x}^2$ and $\sqrt{K^{\mu}_0 \, X_{\mu}}=\tbf{k}_0 \, \Delta \tbf{x}$, which are the quantities appearing in the final step of \eqref{BobOverlap}. Finally, note that the complex phase factor $e^{i \tbf{k}_0 \Delta \tbf{x}}$ does not affect the physical observables, as these depend only on the modulus squared of the commutator, $|\zeta|^2$.

Armed with the overlap $\zeta$, we can enlarge the Hilbert space of Alice's circuit to accommodate the mismatch effects on Bob's qubit as described in Sec \ref{SecQAlgebra}. Specifically, we introduce two extra degrees of freedom for Bob's qubit, denoted $\hat{C}_2, \hat{D}_2$, which represent the part of Bob's wavepacket that is non-overlapping with $H(k,x)$. Specifically, $\hat{C}_2=\hat{A}_{2,H/G}$ and $\hat{D}_2=\hat{B}_{2,H/G}$ as defined by the eigenmode decomposition \eqref{eigenmode}. The joint input state of Alice and Bob after their respective preparations can then be written as:
\eqn{\ket{\psi_{P}} := (\hat{V}_2 \hat{I}_1) \, (\hat{P}_2 \hat{P}_1) \ket{\psi_{in}}}
where the qubit initial state is $\ket{\psi_{in}}=\ket{\tbf{0}}_1\ket{\tbf{0}}_2=\hat{A}^\dagger_{1,H} \hat{A}^\dagger_{2,H} \ket{0}$, the unitaries $\hat{P}_1, \hat{P}_2$ represent arbitrary single-qubit rotations applied by Alice and Bob, respectively, and the rotation $\hat{V}_2$ represents the mismatch on Bob's qubit due to his relativistic motion. If Alice measures the arbitrary observables $\hat{J}_1, \hat{K}_2$ on the output qubits, the expectation value is computed using the expression\footnote{Do not confuse the subscripts $AB$ with Alice and Bob's qubits; the latter are labelled $1$ and $2$ while the former refers to the Hilbert space of the component whose wavepackets are matched to the circuit.}:
\eqn{ \label{ExampleOutput}
\bra{\psi_{in}} \hat{U}_{JK} \ket{\psi_{in}}:=
\bra{\psi_{P}} \, (\hat{I}_{CD} \hat{W}_{AB,1,2})^{\dagger} \, (\hat{K}_{AB,2} \hat{J}_{AB,1} +\hat{K}_{CD,2} \hat{J}_{AB,1} ) \, (\hat{I}_{CD,2} \hat{W}_{AB,1,2}) \, \ket{\psi_{P}} \, .
} 
The operators can be given a convenient representation using $8\times 8$ matrices in the basis 
\eqn{\{ \ket{\tbf{0}}_{AB,2} \ket{\tbf{1}}_{AB,2} \ket{\tbf{0}}_{CD,2} \ket{\tbf{1}}_{CD,2} \ket{\tbf{0}}_{AB,1} \ket{\tbf{1}}_{AB,1} \} \nonumber \, .} 
We obtain:
\eqn{  
 \hat{V}_{2}= 
 \left[ \begin{array}{*{2}{c}}
    \zeta \, I_2 I_1 & \sqrt{1-|\zeta|^2} \, I_2 I_1\\
    -\sqrt{1-|\zeta|^2} \, I_2 I_1 & \zeta \, I_2 I_1
   \end{array} \right] \, ,
}
where $I_2 I_1$ is the $4\times 4$ identity matrix\footnote{In general, if $X_1$ and $Y_2$ represent $2 \times 2$ matrices on independent subspaces, then $X_2 Y_1 := X_2 \otimes Y_1$ is a $4\times 4$ matrix acting on the joint space.}. All other operators in the expression have a block diagonal form, with the blocks corresponding to the matched and orthogonal subspaces. In particular,
\eqn{  
 (J_{AB,1} K_{AB,2}+J_{AB,1} K_{CD,2})= 
 \left[ \begin{array}{*{2}{c}}
    K_2 J_1 & 0 \\
    0 & K_2 J_1
   \end{array} \right] \, ,
}
\eqn{  
 (\hat{W}_{AB,1,2} \hat{I}_{CD,2})= 
 \left[ \begin{array}{*{2}{c}}
    \hat{U}_{CNOT} & 0 \\
    0 & \hat{I}_2 \hat{I}_1
   \end{array} \right] \, ,
}
\eqn{  
 (P_2 \hat{P}_1)= 
 \left[ \begin{array}{*{2}{c}}
    P_{2} P_{1} & 0 \\
    0 & I_{2}P_{1}
   \end{array} \right] \, .
}

These matrices allow us to easily compute the single $8 \times 8$ matrix $\hat{U}_{JK}$ representing the Heisenberg evolution of the joint observable $\hat{J}_1 \hat{K}_2$ through the circuit, up to the initial state $\ket{\psi_{in}}$. Using the decomposition \eqref{GenMatDecomp}, we find that all terms annihilate the initial state and vanish except for $\bra{\psi_{in}} \hat{A}^\dagger_{1,H} \hat{A}_{1,H} \hat{A}^\dagger_{2,H} \hat{A}_{2,H} \ket{\psi_{in}}=1$, whose coefficient is the top left entry of the matrix $\hat{U}_{JK}$. Hence the evaluation of \eqref{ExampleOutput} amounts to computing $\hat{U}_{JK}$ using matrix manipulations and reading off the value from the top left corner. 

For example, suppose Bob prepares the target qubit in the state $\ket{\tbf{0}}$ as before, i.e. $P_2=I$, and suppose Alice prepares the control qubit in an arbitrary superposition by applying the rotation $\hat{P}_1$ with matrix representation:  
\eqn{
P_2 :=
\left[ \begin{array}{*{2}{c}}
    \sqrt{1-\alpha^2} & -\alpha \\
    \alpha & \sqrt{1-\alpha^2}
   \end{array} \right] \, .
}
After the evolution, Alice and Bob jointly measure the observable $\hat{I}_1 \hat{Z}_2$. Using the procedure described above, we obtain the expectation value:

\eqn{ \langle \hat{I}_1\hat{Z}_2 \rangle = 1-2\alpha^2 |\zeta|^2 \, . }

If the experiment is free of mismatch, then the CNOT implements the Heisenberg evolution of the observables: $\hat{I}_1\hat{Z}_2 \mapsto \hat{Z}_1\hat{Z}_2$ resulting in the outcome $\langle \textbf{0}_1 \textbf{0}_2 | \hat{Z}_1\hat{Z}_2 | \textbf{0}_1 \textbf{0}_2 \rangle = 1-2\alpha^2$. Thus we recover the correct result when $|\zeta|^2 \approx1$, representing negligible mismatch, i.e. no relativistic distortions or timing errors. On the other hand, when the mismatch is extreme, $|\zeta|^2 \approx0$ and the CNOT does not function at all. In that case Alice obtains the outcome corresponding to $\langle \textbf{0}_1 \textbf{0}_2 | \hat{I}_1\hat{Z}_2 | \textbf{0}_1 \textbf{0}_2 \rangle = 1$ and concludes that the CNOT interaction did not occur, due to the mismatch introduced by Bob's high velocity and large timing errors.

Of course, Bob might complain that we have not been even-handed in our description of events -- after all, from his point of view, he was stationary the entire time and it was the laboratory that came rushing towards \tit{him} at high speed. Indeed, we could just as well have taken Bob's wavepacket to be the reference amplitude, and converted the pump amplitude $H(k,x)$ into Bob's co-ordinates; the Lorentz invariance of the overlap $\zeta$ ensures that we would get the same result. Note that it is also possible to perform the calculation of \eqref{ExampleOutput} taking the initial state to be the vacuum $\ket{0}$ instead of the two-photon state $\ket{\psi_{in}}$. This is done by replacing the field operators $\{\hat{A}_1,\hat{B}_1, \hat{A}_2, \hat{B}_2, \hat{C}_2, \hat{D}_2 \}$ in the decomposition of $\hat{U}_{JK}$ with the single photon modes $\hat{A}'_1,\hat{B}'_1$, etc, as given by the expression \eqref{aoutN} in Sec \ref{SecSingPhot}. One could then use the identities \eqref{SPproperties} to obtain the same result.

A situation that would require the latter approach is the case where Bob is in an accelerated (non-inertial) frame. In that case, the vacuum state in the laboratory $\ket{0}_A$ is different to the vacuum state as seen by Bob, $\ket{0}_B$. The vacuum modes of Alice and Bob are related to one another by a Bogolyubov transformation, of which the SPDC unitary \eqref{EqSPDC} is an example. The Bogolyubov transformation mixes creation and annihilation operators, causing the vacuum state in the laboratory to be populated with a thermal distribution of particles when seen from Bob's frame\footnote{This phenomenon is well known in relativistic quantum field theory, its discovery being attributed independently to Unruh, Davies and Fulling\cite{UNR76}.}.

The difference between Alice and Bob's vacuum states poses a problem for our calculations. If we attempt to compute the expectation value at the level of fields using the state $\ket{\psi_{in}}$, the Bogolyubov transformations applied to this state quickly become intractable. Fortunately, the formalism developed in Sec \ref{SecSingPhot} only requires us to compute the Heisenberg evolution of the single photon operators under the Bogolyubov transformation, since its action on the vacuum state is trivial. The Heisenberg dynamics have a much simpler form (recall eg. \eqref{EqSPDC}) and so result in tractable calculations. The application of our formalism to non-inertial effects is left for future work.

Finally, we note that in general it is possible to describe stochastic processes in the Heisenberg picture. Hence there may be an alternative to the expression, based on the stochastic description of a single SPDC in the Heisenberg picture, that performs the same function as the expression we derived from the deterministic source. It remains to future work to pursue this possibility as a means of simplifying calculations.

\section{Conclusions\label{SecFieldConc}}

In this chapter, we developed methods for generalising a quantum circuit to take into account the quantum fields of the underlying systems. As a particular application of this formalism, we showed how to calculate the output of a circuit in which relativistic effects lead to spatial and temporal mismatch between the photon wavepackets. A feature of our formalism is that it allows us to perform most calculations at the level of matrices, using the concept of an extended quantum circuit to deal with the mismatch. This generalises the work of Ref. \cite{RMS} to arbitrary quantum circuits.

Since the quantum fields on which our formalism is based are initially in the field vacuum state, we derived an expression allowing us to convert any expectation value taken in an $n$-photon state into a function of vacuum modes acting in the vacuum state. While such an expression has no role to play in situations where there is a shared vacuum state, such as cases involving only inertial observers, we anticipate that the formalism will allow calculations involving non-inertial observers to become tractable, lending insight into the effects of non-inertial motion on quantum circuits.

Our formalism was limited to scalar bosons. However, the polarisation degrees of freedom of propagating photons, which are known to be useful for encoding information, are also subject to rotations under relativistic transformations. In addition, we considered only one particular physical implementation of a quantum circuit, chosen for its simplicity, whereas there may be more realistic implementations of quantum circuits for which the analysis of relativistic effects will necessarily require a different choice of field decomposition. These limitations provide scope for future generalisations of our formalism to a wider class of scenarios.

\section{Appendix: The single photon source}

The following calculations show that the single photon mode $\hat{a}'$ given by \eqref{aoutN} in Sec \ref{SecSingPhot} satisfies $\bk{\hat{n}}\approx 1$ and $g^{(2)}\approx 0$ when $N$ is sufficiently large. For generality and ease of notation, we will rewrite the Heisenberg SPDC evolution given in \eqref{EqSPDC} using the shorthand notation 
\eqn{ \label{EqSPDCpq}
\hat{u}'_{j}&=p \, \hat{v}_{j}+q \, \hat{u}^{\dagger}_{j}\\ \nonumber
\hat{v}'_{j}&=p \, \hat{u}_{j}+q \,  \hat{v}^{\dagger}_{j} \nonumber
}
We will also replace the detector model \eqref{bucketD} with the more general form:
\eqn{
\hat{d}_{j} := (\mu \, \hat{1}-\nu \, \hat{n}_{u'_j})\, \hat{n}_{u'_j},
} 
which is equivalent to \eqref{bucketD} when $\mu=3/2$ and $\nu=1/2$.

\subsubsection{Calculation of $\bra{0} \hat{a}'^\dagger \hat{a}'\ket{0}$.}

We write the full expression for $\langle 0|\hat{a}'^\dag\hat{a}'|0\rangle$ given $\hat{a}'$ in \eqref{aoutN}: 
\eqn{\label{Eqn:nApp}
\langle\sum_{j,k=1}^{N}\prod_{l=0}^{k-1}(1-\hat{d}_{l})\, \hat{v}'^{\dagger}_{k}\, \hat{d}_{k}\, \hat{d}_{j}\, \hat{v}'_{j}\, \prod_{i=0}^{j-1}(1-\hat{d}_{i})\rangle, 
}
where it is understood that all expectation values are taken with respect to the vacuum. The summation can be broken up into three separate terms:  
\eqn{\label{Eqn:SumApp}
\sum_{j,k=1}^{N}=\sum_{j<k=1}^{N}+\sum_{j=k=1}^{N}+\sum_{j>k=1}^{N}.
}
Since $\langle  \hat{v}'^{\dagger}_{k}\, \hat{d}_{k}\rangle=0$, only the $\sum \limits_{j=k}$ term in \eqref{Eqn:nApp} survives. Given that
\eqn{\label{Eqn:f1f2App}
\langle  \hat{v}'^{\dagger}_{j}\hat{d}_{j}\hat{d}_{j} \hat{v}'_{j}\rangle&=f_{1}(p,q,\mu,\nu)\\
\langle  (1-\hat{d}_{i})^{2} \rangle&=f_{2}(p,q,\mu,\nu),
}
where $f_{1}(p,q,\mu,\nu)$ and $f_{2}(p,q,\mu,\nu)$ are both polynomial functions of $p$, $q$, $\mu$ and $\nu$, we find: 
\eqn{\label{Eqn:ApDagAp}
\langle \hat{a}'^\dag \hat{a}' \rangle=f_{1}\sum_{j=1}^{N}\prod_{l=0}^{j-1}f_{2}=f_{1}\left(\frac{f^{N}_{2}-1}{f_{2}-1}\right).
}
When we substitute $\mu=3/2$, $\nu=1/2$, $p=1$ and $q=|\chi|$, where $|\chi|\ll 1$, we obtain, to order $|\chi|^{4}$,

\eqn{ \label{na}
  \langle \hat{n}_{a'} \rangle = \frac{(4-4|\chi|^2+9|\chi|^4)((1-|\chi|^2)^N-1)}{5|\chi|^4-4} \, .
}
It is straightforward to check that this function approaches 1 for $N \gg 1$ and $|\chi| \ll 1$, hence $\langle \hat{n}_{a'} \rangle \rightarrow 1$ as desired. 

\subsubsection{Calculation of $g^{(2)}$.}

We write the full expression for $\langle \hat{a}'^\dagger \hat{a}'^\dagger \hat{a}' \hat{a}' \rangle$  given $\hat{a}'$ in \eqref{aoutN} using the fact that $[\hat{v}'^{\dagger}_{i},\hat{d}_{j}]=[\hat{v}'_{i},\hat{d}_{j}]=0$:
\eqn{\label{Eqn:g2App}
\langle \sum_{a,c,j,k=1}^{N} \, \hat{v}'^{\dagger}_{k} \, \hat{d}_{k}\hat{v}'^{\dagger}_{a} \, \hat{d}_{a} \, \hat{d}_{j} \, \hat{v}'_{j} \, \hat{d}_{c} \, \hat{v}'_{c} \prod_{l=0}^{k-1}(1-\hat{d}_{l}) \prod_{b=0}^{a-1}(1-\hat{d}_{b})\prod_{i=0}^{j-1}(1-\hat{d}_{i})\prod_{d=0}^{c-1}(1-\hat{d}_{d})\rangle, 
}
where again it is understood that all expectation values are taken with respect to the vacuum. The summation in this case can be broken up into 75 separate terms, most of which do not contribute:
\begin{itemize}
\item The 24 permutations of summations of the form $\sum\limits_{a>c>j>k}$ do not contribute since $\langle \hat{v}'^{\dagger}_{k}\hat{d}_{k}\rangle=0$. 

\item The $12\times3=36$ permutations of summations of the form $\sum\limits_{a=c>j>k}$, $\sum\limits_{a>c=j>k}$ and $\sum\limits_{a>c>j=k}$ do not contribute since $\langle \hat{v}'^{\dagger}_{k}\, \hat{d}_{k} \, \hat{v}'^{\dagger}_{k}\, \hat{d}_{k}\rangle=0$ and $\langle \hat{v}'^{\dagger}_{k}\, \hat{d}_{k} (1-\hat{d}_{k})^{2}\rangle=0$.

\item The $4\times 2=8$ permutations of summations of the form $\sum\limits_{a=c=j>k}$ and $\sum\limits_{a>c=j=k}$ do not contribute since  $\langle \hat{v}'^{\dagger}_{k} \, \hat{d}_{k} \, \hat{v}'^{\dagger}_{k} \, \hat{d}_{k} \, \hat{d}_{k} \, \hat{v}'_{k}\rangle=0$.

\item Four of the 6 permutations of summations of the form $\sum\limits_{a=c>j=k}$ are non-zero and given by
\eqn{
&\sum\limits_{a=c>j=k}&\langle \hat{v}'^{\dagger}_{a} \, \hat{d}_{a} \, \hat{v}'^{\dagger}_{k} \, \hat{d}_{k} \, \hat{d}_{a} \, \hat{v}'_{a} \, \hat{d}_{k} \, \hat{v}'_{k} \prod_{b=0}^{a-1}(1-\hat{d}_{b})^{2} \prod_{l=0}^{k-1}(1-\hat{d}_{l})^{2}\rangle \nonumber \\
&=\sum\limits_{a>k}& f_{1} f_{4} \prod_{l=0}^{k-1}f_{3}\prod_{b=k+1}^{a-1}f_{2},
} 
where 
\begin{equation}\label{Eqn:f3f4App}
\begin{aligned}
\langle  (1-\hat{d}_{k})^{4} \rangle&=f_{3}(p,q,\mu,\nu)\\
\langle  \hat{v}'^{\dagger}_{k}\hat{d}_{k}\hat{d}_{k} \hat{v}'_{k}(1-\hat{d}_{k})\rangle&=f_{4}(p,q,\mu,\nu)
\end{aligned} 
\end{equation}
are also polynomial functions of $p$, $q$, $\mu$ and $\nu$. 

\item The summation $\sum\limits_{a=c=j=k}$ is non-zero and given by
\eqn{ 
&\sum\limits_{a=c=j=k}&\langle \hat{v}'^{\dagger}_{k}  \hat{d}_{k} \, \hat{v}'^{\dagger}_{k} \, \hat{d}_{k} \, \hat{d}_{k} \, \hat{v}'_{k} \, \hat{d}_{k} \, \hat{v}'_{k} \rangle \langle \prod_{l=0}^{k-1}(1-\hat{d}_{l})^{4} \rangle\nonumber\\
&=\sum\limits_{k}& f_{5} \prod_{l=0}^{k-1}f_{3},
}
where 
\begin{equation}\label{Eqn:f3f4App}
\langle  \hat{v}'^{\dagger}_{k}  \hat{d}_{k} \, \hat{v}'^{\dagger}_{k} \, \hat{d}_{k} \, \hat{d}_{k} \, \hat{v}'_{k} \, \hat{d}_{k} \, \hat{v}'_{k}  \rangle=f_{5}(p,q,\mu,\nu)
\end{equation}
is a polynomial functions of $p$, $q$, $\mu$ and $\nu$. 
\end{itemize}

The expression for $\langle \hat{a}'^\dagger \hat{a}'^\dagger \hat{a}' \hat{a}' \rangle$ therefore reduces to:
\begin{align} 
 & 4\sum_{a>k} f_{1} f_{4} \prod_{l=0}^{k-1}f_{3}\prod_{b=k+1}^{a-1}f_{2}+\sum_{k=1}^{N} f_{5} \prod_{l=0}^{k-1}f_{3}\label{Eqn:g2FinalApp}\\
&=4 f_{1} f_{4}\left(\frac{f_{3}-f_{2}+f_{3}^{N}(f_{2}-1)-f_{2}^{N}(f_{3}-1)}{(f_{2}-1)(f_{3}-1)(f_{3}-f_{2})}\right)+f_{5}\left(\frac{f_{3}^{N}-1}{f_{3}-1}\right).\nonumber
\end{align} 
When we substitute $\mu=3/2$, $\nu=1/2$, $p=1$ and $q=|\chi|$, where $|\chi|\ll 1$, we obtain, to order $|\chi|^{4}$,
\eqn{ \label{g2}
g^{(2)} =  -\frac{2|\chi|^2(4-5|\chi|^4)^2}{(4-4|\chi|^2+9|\chi|^4)^2((1-|\chi|^2)^N-1)} \, .
}
Taking $N \gg 1$ and $|\chi| \ll 1$, we find that $g^{(2)} \rightarrow 0$ as desired.
To complete our discussion, we note that for a given value of $\chi \ll 1$, the simultaneous requirements $\langle \hat{n}_{a'} \rangle \rightarrow 1$ and $g^{(2)} \rightarrow 0$ impose the following condition on the number of SPDCs:
\eqn{(1-|\chi|^2)^N \ll 1 \, . \label{MinimumN}}
Provided $N$ is sufficiently large such that \eqref{MinimumN} holds for a given $\chi$, the output $\hat{a}'$ approximates a single photon mode.

\clearpage
\pagestyle{plain}
\pagebreak
\renewcommand\bibname{{\LARGE{References}}} 
\bibliographystyle{refs/naturemagmat2012}
\addcontentsline{toc}{section}{References} 

\pagestyle{fancy}
\renewcommand{\sectionmark}[1]{\markboth{\thesection.\ #1}{}}

\chapter{CTCs with continuous variables}
\label{ChapCTC1}

\begin{quote}
\em
It is a full, and fairely written scrowle,\\ 
Which up into it selfe, it selfe doth rowle; \\
And, by unfolding, and, infolding, showes \\
A round, which neither end, nor entrance knowes.\\
\end{quote}
George Wither\\

\newpage

\section*{Abstract}
In this chapter, we will apply Deutsch's model of CTCs directly to continuous variable systems. Specifically, we will consider CTC circuits whose rails carry a countably infinite number of dimensions, corresponding to the Fock space of a bosonic field theory such as quantum optics. We continue to assume that the field excitations are localised to pointlike wavepackets, which allows us to accommodate `field' states such as squeezed states and coherent states without leaving the domain of applicability of the original toy model. Modifications to the toy model to include the spatio-temporal properties of wavepackets are postponed until the final chapter.

The present chapter is arranged as follows. In Sec \ref{CTC1Basics} we lay the groundwork for applying the Deutsch model to quantum optics and we state the necessary assumptions. In Sec \ref{CTC1BeamSplit} we apply the formalism to a circuit in which a photon interacts with a time-travelled copy of itself on a beamsplitter. We find that energy is not conserved for single photon inputs in any single run of the experiment, but that the energy is conserved on average. We also find that squeezed states undergo rotations of the squeezing axis and coherent states traverse the circuit without any appreciable change apart from a phase shift. The physical meaning of these results is discussed. Finally, in Sec \ref{CTC1OpenCurve} we consider the apparently trivial scenario in which the CTC contains no interaction. While this circuit is less interesting for discrete variable analyses, we find that in a continuous variable setting this `open' time-like curve allows us to violate Heisenberg's uncertainty relation for quadrature measurements on a coherent state. In particular, we describe a circuit that is able to perfectly read out the amplitude of an unknown coherent state using only linear operations and open time-like curves. The implications of this result are briefly discussed. The work in this chapter contains results previously published in Refs \cite{PIE11,PIE13}, to which the reader is referred for more information.

\section{Quantum optics with CTCs\label{CTC1Basics}}

\begin{figure}
 \includegraphics{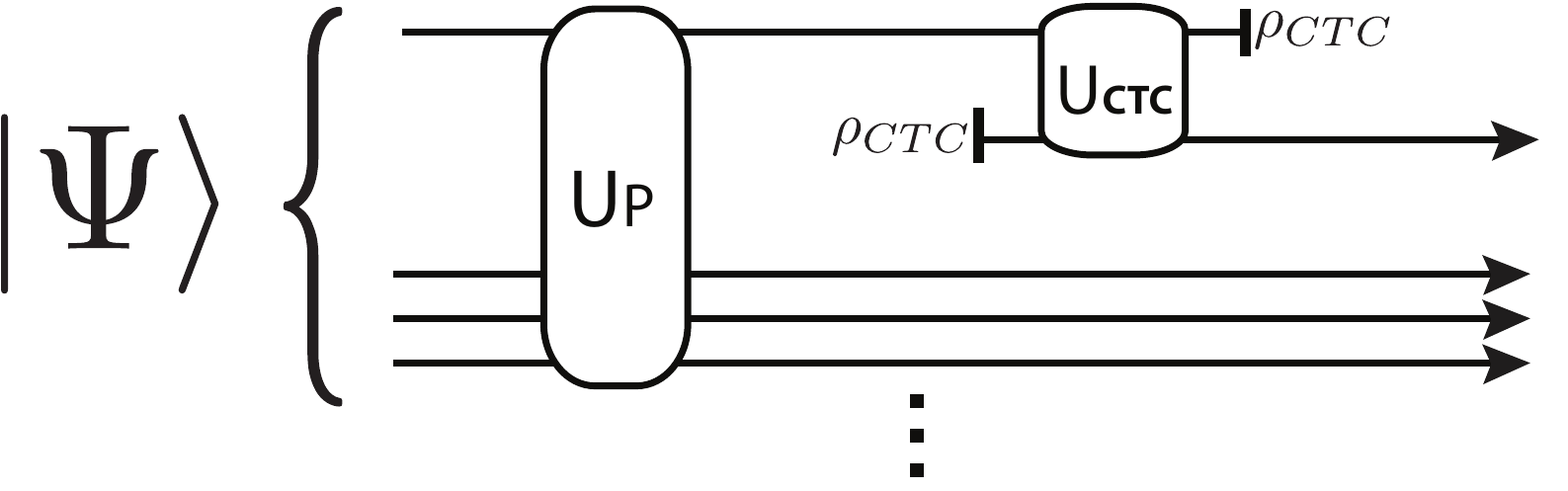}
 \caption[A general circuit containing a single CTC]{\label{genFieldCTC}A general circuit containing a single CTC.}
\end{figure}

In this chapter we will consider quantum circuits containing a single CTC of the general form shown in Fig. \ref{genFieldCTC}. Recall from Ch. \ref{ChapDeutsch} that the output of this circuit can be computed in the Schr\"{o}dinger picture with the aid of the nonlinear consistency condition \eqref{consistency} applied to the endpoints of the CTC. Previously, we have assumed that each rail is a qudit of finite dimension, with the unitary gates being defined on the joint Hilbert spaces of the qudits. Indeed, this is the approach that has been adopted almost universally in the literature on CTCs. However, there is nothing in Deutsch's formalism that prevents us from choosing the Hilbert space of each rail to be countably infinite, corresponding to the Fock space of a bosonic scalar field, for example.

In particular, we associate the rails in Fig. \ref{genFieldCTC} with non-overlapping vacuum field modes $\hat{A}_{\phi_A}, \hat{B}_{\phi_B}, \hat{C}_{\phi_C}, ... $ and their adjoints, recalling the form of a general wavepacket mode defined in the previous chapter:
\eqn{
\hat{A}_{\phi_A} = \int \invard{\tbf{k}} \, \phi_A(k,x) \, \hat{a}_{\tbf{k}} \,.
}
The Hilbert space of each rail is then spanned by the number states of its associated mode. For example, an $n$-photon state in the mode $\hat{A}_{\phi_A}$ is given by:
\eqn{
\ket{n_A}= \frac{1}{\sqrt{n!}} (\hat{A}^\dagger_{\phi_A})^n \ket{0} \, ,
}
and an arbitrary pure state for the single rail $\hat{A}_{\phi_A}$ can be written as an infinite superposition of these states:
\eqn{
\ket{\Psi_A}=\sum^\infty_{n=0} c_n \ket{n_A} \, ,
}
where $\sum\limits^\infty_{n=0} |c_n| = 1$. Since we are making the approximation of point-like wavepackets, the amplitude $\phi_A(k,x)$ will not feature in our considerations and we will omit the field subscripts from the modes from here onwards. 

The multi-rail initial state is assumed to be separable, $\ket{\Psi}=\ket{\Psi_A} \ket{\Psi_B} \ket{\Psi_C} ...$, and the `preparation' unitary $\hat{U}_P$ may generate arbitrary entanglement between the separate rails. Since we are only concerned with linear quantum optics in this chapter, the calculations are relatively simple at the level of the fields, so we do not need to add any further structure like the dual-rail qubits and gates discussed in the preceding chapter.

Within this new state space, the calculation of the outputs from the circuit Fig. \ref{genFieldCTC} can proceed as usual in the Schr\"{o}dinger picture: we obtain the input density matrix to the CTC via $\rho_{A,in} = \ptrc{B,C,...}{\rho_{in}}$ with $\rho_{in}=\hat{U}^\dagger_P (\densop{\Psi}{\Psi}) \hat{U}_P$ and use it to solve Deutsch's consistency condition for $\rho_{CTC}$. Using $\rho_{CTC}$ we can then obtain the total output:
\eqn{
\rho_{out} = \hat{U}^\dagger_{CTC} (\rho_{in} \otimes \rho_{CTC}) \hat{U}_{CTC} 
}
and in particular, 
\eqn{ \rho_{A,out}=\ptrc{B,C,...}{\rho_{out}} \, .}

\begin{figure}
 \includegraphics{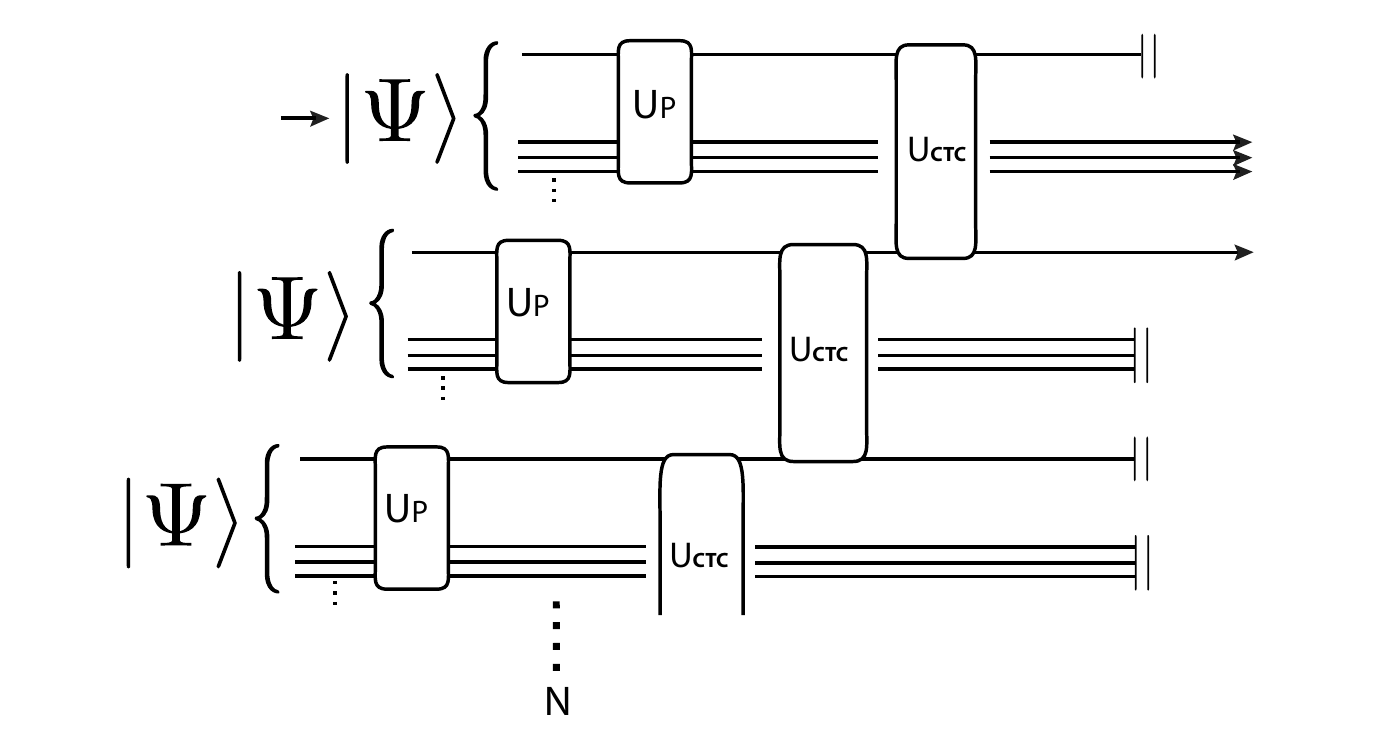}
 \caption[An equivalent circuit for a general circuit containing a single CTC]{\label{genFieldEC}An equivalent circuit to Fig. \ref{genFieldCTC}. The arrows represent the inputs and outputs that are accessible to experimenters.}
\end{figure}

While it is therefore possible to directly apply the Deutsch model using the Schr\"{o}dinger picture, calculations in field theories such as quantum optics are generally much easier to perform in the Heisenberg picture. As we saw in Chapter \ref{ChapDeutsch}, it is possible to switch to the Heisenberg picture using the equivalent circuit of the CTC circuit. For the circuit of Fig. \ref{genFieldCTC}, its equivalent circuit is Fig. \ref{genFieldEC}. Since this is just an ordinary quantum circuit that happens to begin with a large number of $N$ copies of the input state, we can take the initial state to be $\ket{\Psi_N}:= \ket{\Psi}^{\otimes N}$ and then apply the unitaries $\hat{U}_{\trm{PN}} := \brac{\hat{U}_{P}}^{\otimes N}$ and $\hat{U}_{EC}$ to the observables instead\footnote{Recall from Sec \ref{SecEquivCirc} that Deutsch's nonlinear map $\scr{N}_D$ can always be replaced by a linear, unitary map acting on the original input state plus a large set of ancilla states prepared identically to the input.}. Explicitly, if $\hat{\sigma}(\hat{A},\hat{B},...)$ is a general observable at the output, we obtain its expectation value according to:
\eqn{ 
\langle \hat{\sigma} \rangle &=& \bra{\Psi_N}\, \brac{ \hat{U}^{\dagger}_{\trm{PN}} \hat{U}^\dagger_{EC} \, \hat{\sigma} \, \hat{U}_{EC} \hat{U}_{\trm{PN}} \, } \ket{\Psi_N} \nonumber \\
&:=& \bra{\Psi_N} \, \hat{\sigma}' \, \ket{\Psi_N} 
}
In the following section, we will demonstrate the use of this formalism by performing a simple calculation in which a beam of light travels back in time and interacts with itself on a beamsplitter.

\section{The time-travelling photon\label{CTC1BeamSplit}}

\begin{figure}
 \includegraphics{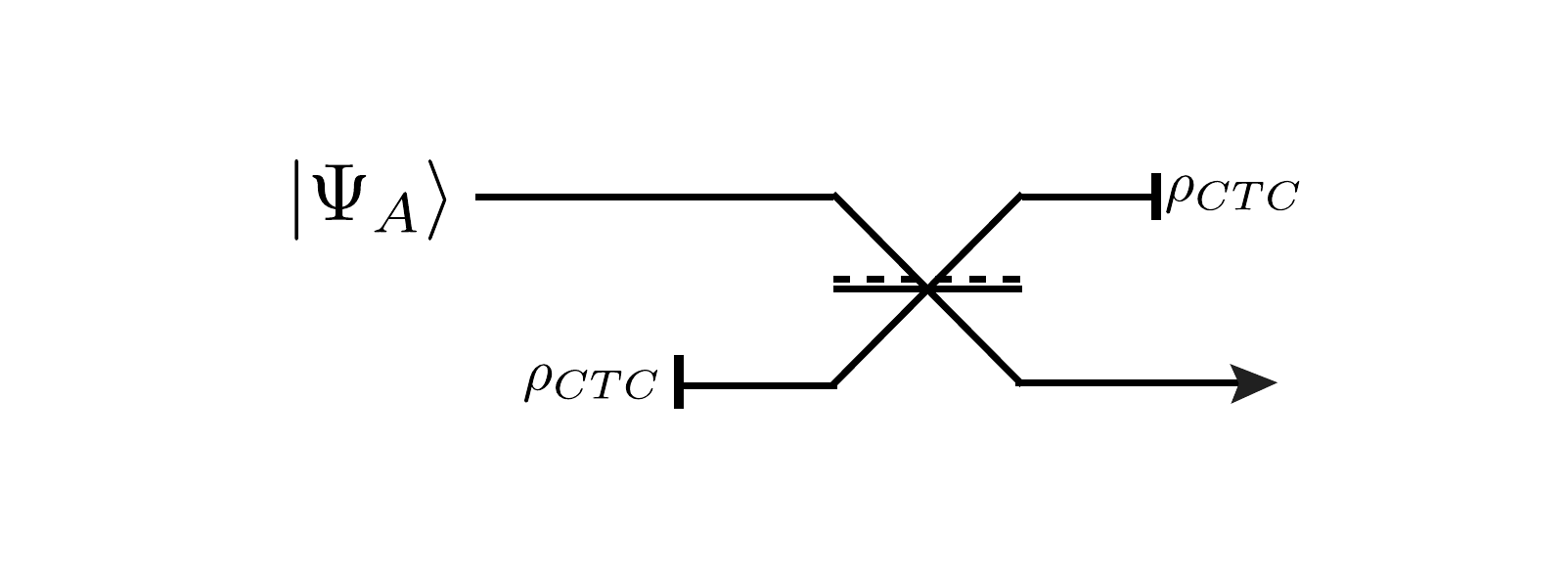}
 \caption[A time-travelling photon interacts with itself on a beamsplitter]{\label{figCTCbeamsplit}A time-travelling photon interacts with itself on a beamsplitter.}
\end{figure}

\begin{figure}
 \includegraphics{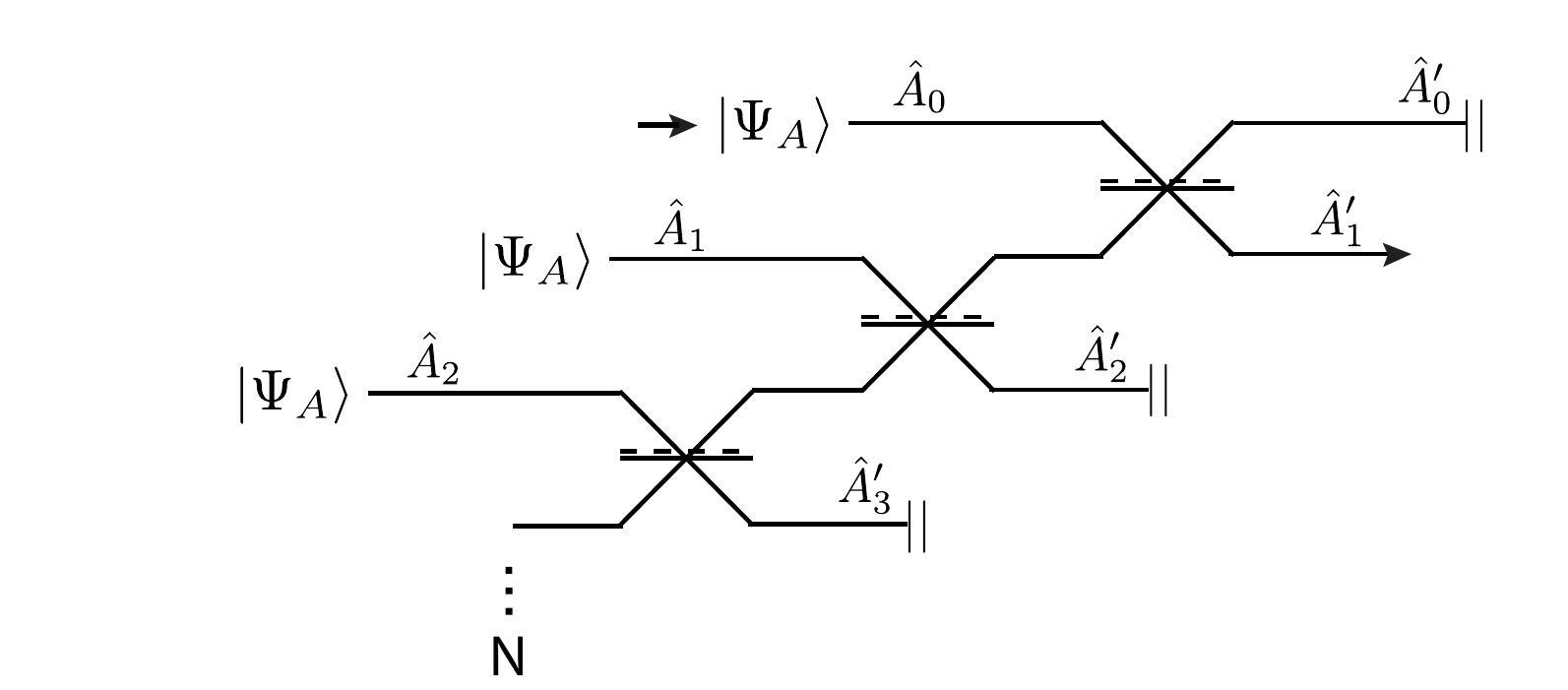}
 \caption[An equivalent circuit for the beamsplitter on a CTC]{\label{figECbeamsplit}An equivalent circuit to Fig. \ref{figCTCbeamsplit}. The modes are written above the rails are a visual guide to the calculation in the text.}
\end{figure}

Consider the scenario of Fig. \ref{figCTCbeamsplit}. The field is prepared in some arbitrary input state $\ket{\Psi_A}$, which then traverses a CTC containing a single beamsplitter of reflectivity $\eta$. The input to the lower arm of the beamsplitter is constrained to be equal to the output from the upper arm, in accordance with Deutsch's consistency condition. We proceed by writing down the equivalent circuit, shown in Fig. \ref{figECbeamsplit}, with input state $\ket{\Psi_{A,N}}=\ket{\Psi_A}^{\otimes N}$. The mode $\hat{A}_m$ corresponds to the $m_{th}$ rail in the circuit and we designate $\hat{A}_1$ as the detected mode. Our goal is then to calculate the Heisenberg evolution of the mode $\hat{A}_1$ through the circuit.

To construct the equivalent circuit unitary $\hat{U}_{EC}$, we introduce the beamsplitter unitary $\hat{T}(i,j)\, , \, (i \neq j)$ whose action on the modes $\hat{A}_i, \hat{A}_j$ is given by the Heisenberg evolutions:
\eqn{ \label{beamsplit}
  \hat{T}^\dagger(i,j) \, \hat{A}_i \, \hat{T}(i,j) &=& \sqrt{\eta}\, \hat{A}_i+e^{i\phi}\sqrt{1-\eta}\, \hat{A}_j \, ,\\
  \hat{T}^\dagger(i,j) \, \hat{A}_j \, \hat{T}(i,j) &=& \sqrt{\eta}\, \hat{A}_j-e^{-i\phi}\sqrt{1-\eta}\, \hat{A}_i \, , 
}
where $\eta \in [0,1]$ is the reflectivity and $\phi$ is an arbitrary phase. We then have:
\eqn{\label{ECbeamsplit}
\hat{U}_{EC}=\hat{T}(0,1)\, \hat{T}(1,2)\, \hat{T}(2,3)...\, \hat{T}(N-1,N) \, .
}

Using this unitary, we can now explicitly calculate the Heisenberg evolution of the mode $\hat{A}_1$ through the circuit:
\eqn{\label{HeisECBeamsplit0}
\hat{A}'_1 &:=& \hat{U}^\dagger_{EC} \, \hat{A}_1 \, \hat{U}_{EC} \nonumber \\
&=& \sqrt{\eta} \brac{e^{i \phi} \, \sqrt{1-\eta}}^{N-1}\hat{A}_N + \eta \sum^{N-1}_{m=1} \brac{e^{i \phi} \, \sqrt{1-\eta}}^{m-1} \hat{A}_m -e^{-i \phi}\sqrt{1-\eta} \hat{A}_0 \, .
}
This result holds for any finite $N$. However, recall that the equivalent circuit is only a valid approximation to Deutsch's model if $N$ is very large. We can make use of this requirement to simplify the expression. First, assume that $\eta \neq 0$, so that the term $\sqrt{1-\eta}$ is positive and strictly less than $1$. Let $X \gg 1$ be a large positive integer smaller than $N$. Then we can choose to ignore terms of order $\brac{\sqrt{1-\eta}}^X$ or smaller, provided $N$ and $X$ are sufficiently large. Then we obtain the approximation:
\eqn{
\hat{A}'_1 &=& \sqrt{\eta} \brac{e^{i \phi} \, \sqrt{1-\eta}}^{N-1}\hat{A}_N + \eta \sum^{N-1}_{m=1} \brac{e^{i \phi} \, \sqrt{1-\eta}}^{m-1} \hat{A}_m -e^{-i \phi}\sqrt{1-\eta} \hat{A}_0 \nonumber \\
&\approx& \eta \sum^{X+1}_{m=1} \brac{e^{i \phi} \, \sqrt{1-\eta}}^{m-1} \hat{A}_m -e^{-i \phi}\sqrt{1-\eta} \hat{A}_0 \nonumber \, .
}  
For the special case where $\eta=0$, all terms vanish except for the last one, so the above expression remains accurate. For most calculations in this thesis, we will take the limit $N \rightarrow \infty$, in which case one has the equality:
\eqn{\label{HeisECBeamsplit}
\hat{A}'_1 &=& \eta \sum^{\infty}_{m=1} \brac{e^{i \phi} \, \sqrt{1-\eta}}^{m-1} \hat{A}_m -e^{-i \phi}\sqrt{1-\eta} \hat{A}_0 \, .
}
Unless stated otherwise in this thesis, we will assume that $N$ is arbitrarily large and the expression \eqref{HeisECBeamsplit} holds. In principle, we can now use this expression to obtain the Heisenberg evolution of any expectation value $\sigma(\hat{A})$, since:
\eqn{ \label{expectationA1} 
\bra{\Psi_{A,N}}\, \brac{ \hat{U}^\dagger_{EC} \, \sigma(\hat{A}_1) \, \hat{U}_{EC} } \, \ket{\Psi_{A,N} } &=& \bra{\Psi_{A,N}}\, \sigma( \hat{U}^\dagger_{EC} \, \hat{A}_1 \, \hat{U}_{EC}) \, \ket{\Psi_{A,N} } \nonumber \\
&=& \bra{\Psi_{A,N}}\, \sigma(\hat{A}'_1) \, \ket{\Psi_{A,N} } \, .
}
In the following sections we will discuss the output of the CTC for two possible choices of input state: an arbitrary Gaussian pure state, and a single photon state. 

\subsection{Arbitrary Gaussian pure states\label{SecCTC1Gauss}}

Recall from Sec \ref{SecQO} that an arbitrary single mode Gaussian pure state can be written as a displaced, rotated squeezed state: $\ket{\Psi_A}= \hat{D}(\alpha)\, \hat{R}(\theta_R) \, \hat{S}(r e^{i 2 \theta_S}) \ket{0}:= \hat{U}_P \ket{0}$. Performing the Heisenberg evolution of the mode $\hat{A}_m$ through the unitary $\hat{U}_P$ we obtain:
\eqn{
\hat{V}_m &:=& \hat{U}^\dagger_P \, \hat{A}_m \, \hat{U}_P \nonumber \\
&=& \hat{S}^\dagger  \hat{R}^\dagger \hat{D}^\dagger \,  \hat{A}_m \, \hat{D}  \hat{R}   \hat{S} \nonumber \\
&=& e^{i \theta_R} \,\trm{cosh}(r)\hat{A}_m+e^{i (\theta_R-2 \theta_S)} \,\trm{sinh}(r)\hat{A}^\dagger_m + \alpha
}
so that \eqref{expectationA1} becomes
\eqn{
\bra{\Psi_{A,N}}\, \sigma(\hat{A}'_1) \, \ket{\Psi_{A,N}} = \bra{0}\, \sigma(\hat{V}'_1) \, \ket{0} 
}
where
\eqn{\label{Vprime}
\hat{V}'_1:=\eta \sum^N_{m=1} \brac{e^{i \phi} \, \sqrt{1-\eta}}^{m-1} \hat{V}_m -e^{-i \phi}\sqrt{1-\eta} \hat{V}_0 \, .
}
Since the input state is Gaussian and the equivalent circuit is just a network of beamsplitters (Gaussian operations) it follows that the output of this CTC circuit must be a Gaussian state. Hence we can characterise the state by it's first and second order moments as given by $\bk{\hat{P}}, \bk{\hat{Q}}$ and the covariance matrix:

\eqn{
 \left[ \begin{array}{*{2}{c}}
\trm{Var}(\hat{P}) & \frac{1}{2}\bk{\hat{P} \hat{Q}}+\frac{1}{2}\bk{\hat{Q} \hat{P}}-\bk{\hat{P}}\bk{\hat{Q}} \\
\frac{1}{2}\bk{\hat{P} \hat{Q}}+\frac{1}{2}\bk{\hat{Q} \hat{P}}-\bk{\hat{P}}\bk{\hat{Q}} & \trm{Var}(\hat{Q})
   \end{array} \right]
}
where $\trm{Var}(\hat{x}):=\bk{\hat{x}^2}-\bk{\hat{x}}^2$ and all expectation values are understood to be taken in the vacuum state. Since $\hat{P}:=i \brac{ \hat{V}'^\dagger_1-\hat{V}'_1 }$ and $\hat{Q}:= \hat{V}'^\dagger_1+\hat{V}'_1$ we see that all of the above quantities can be obtained from the essential moments $\bk{\hat{V}'_1}, \bk{\hat{V}'_1\, \hat{V}'_1}$ and $\bk{\hat{V}'^\dagger_1 \, \hat{V}'_1}$ using complex conjugation (eg. $\bk{\hat{V}'^\dagger_1\, \hat{V}'^\dagger_1}=\bk{\hat{V}'_1\, \hat{V}'_1}^*$) and the commutator $[\hat{V}'_1, \, \hat{V}'^\dagger_1]=1$. For example, we have:
\eqn{\nonumber
\trm{Var}(\hat{P})=1+2\bk{\hat{V}'^\dagger_1 \, \hat{V}'_1}-\bk{\hat{V}'_1\, \hat{V}'_1}-\bk{\hat{V}'_1\, \hat{V}'_1}^*+\bk{\hat{V}'_1}^2+(\bk{\hat{V}'_1}^*)^2-2\bk{\hat{V}'_1}\bk{\hat{V}'_1}^* \, .
}
It is convenient to perform calculations in the limit of $N\rightarrow \infty$, so that the terms arising from \eqref{Vprime} converge to more compact expressions. We then obtain:
\eqn{ \label{Moments}
\bk{\hat{V}'_1} &=& \frac{1-e^{-i \phi} \sqrt{1-\eta}}{1-e^{i \phi} \sqrt{1-\eta}} \, \alpha \equiv e^{i \Phi(\phi,\eta)} \, \alpha \, \\
\bk{\hat{V}'_1\, \hat{V}'_1} &=& e^{2i(\theta_R-\theta_S)}\trm{cosh}(r) \trm{sinh}(r) \brac{\frac{\eta^2}{e^{2i\phi}(1-\eta)-1}-e^{-2i\phi}(1-\eta)} \nonumber \\
&&+\, e^{2 i \Phi(\phi,\eta)} \alpha^2  \\ 
\bk{\hat{V}'^\dagger_1 \, \hat{V}'_1} &=& \trm{sinh}^2(r) + |\alpha|^2
}
The covariance matrix for a general Gaussian input state can be obtained straightforwardly from the above moments. To illustrate the properties of the output state, we will consider two special cases of the input: a coherent state and a squeezed state.

\subsubsection{Coherent state} 
If the input is a coherent state with amplitude $\alpha$, the first and second order moments are obtained by setting $r=0$ in \eqref{Moments}. The first order moments are $\bk{\hat{P}}=\trm{Im}[e^{i \Phi(\phi,\eta)} \, \alpha]$, $\bk{\hat{Q}}=\trm{Re}[e^{i \Phi(\phi,\eta)} \, \alpha]$ while the elements of the covariance matrix are trivial: 
\eqn{ \nonumber
 \left[ \begin{array}{*{2}{c}}
1&0 \\
0&1
   \end{array} \right] \, .
}
Hence the output in this case is identical to the input up to an overall phase $e^{i \Phi(\phi,\eta)}$ that depends on the beamsplitter parameters. It is curious that despite being ostensibly a circuit containing a CTC with a nontrivial interaction, the input--output map is completely trivial for coherent states. One way to explain this apparent paradox is that the chosen form of interaction, namely a beamsplitter, is non-entangling when its inputs are coherent states. In particular, if the two inputs to a beamsplitter are coherent states, then so are the two outputs. Considering Fig. \ref{figCTCbeamsplit}, this means that the input--output map can be written as:
\eqn{
\scr{M}: \densop{\alpha}{\alpha}_1 \otimes \densop{\alpha_{\trm{CTC}}}{\alpha_{\trm{CTC}}}_2 \rightarrow \densop{\alpha_{\trm{CTC}}}{\alpha_{\trm{CTC}}}_1 \otimes \densop{\alpha_{out}}{\alpha_{out}}_2 \, ,
} 
where the input on rail 1 is the coherent state $\ket{\alpha}$ and the consistent solution for the state emerging from the CTC on rail 2 is also a coherent state. By inspection, we see that this map can be implemented by a linear, unitary gate of the form:
\eqn{
\scr{M} \leftrightarrow (\trm{SWAP})\otimes(\hat{u}_1 \, \hat{I}_{2}) \, ,
}
i.e. the application of some single-mode unitary $\hat{u}$ to the input followed by swapping it with the mode emerging from the CTC. In the Heisenberg picture, this corresponds to an evolution of the operator at the output with the general form:
\eqn{
\hat{A}_{\trm{out}}=\hat{u}^\dagger_{\trm{in}}\, \hat{A}_{\trm{in}}\, \hat{u}_{\trm{in}} \, ,
}
which is completely independent of the CTC. Hence there is an effective decoupling of the system from the CTC and we do not expect to see any strange behaviour. Based on this example, one expects that, in general, the input--output map is effectively linear whenever the chosen interaction does not produce entanglement between the initial state and the consistent solution for the state on the CTC. It is a topic for future work to fully characterise the class of interactions and input states for which the action of the CTC reduces to a purely linear map.

\subsubsection{Squeezed vacuum state} 
In contrast to the previous example, squeezed states will in general become entangled on a beamsplitter. This entanglement implies a nontrivial interaction with the CTC, so we would expect to see interesting effects in this case. For simplicity, we will choose the angle of rotation to be $\theta_R \equiv \theta_S+\frac{\pi}{2}$ so that the initial squeezing is in the $\hat{P}$ quadrature. The displacement is taken to be zero: $\alpha=0$, i.e. the input state is the squeezed vacuum. Then we obtain the trivial first-order moments: $\bk{\hat{P}}=\bk{\hat{Q}}=0$ and the nontrivial covariance matrix:
\eqn{ \label{CovSqueezed}
 \left[ \begin{array}{*{2}{c}}
\trm{cosh}[2 r]-K_1(\phi,\eta) \trm{sinh}[2 r]&K_2(\phi,\eta)\, \trm{sinh}[2 r] \\
K_2(\phi,\eta)\, \trm{sinh}[2 r] &\trm{cosh}[2 r]+K_1(\phi,\eta) \trm{sinh}[2 r]
   \end{array} \right] \, ,
}
where:
\eqn{
K_1(\phi,\eta)&:=&\trm{Cos}[2 \phi ]+\frac{2 \eta \, \trm{Sin}[\phi ]^2 \brac{2 (\eta-1) \trm{Cos}[2 \phi ]+\eta } } {2+(\eta-2) \eta +2 (\eta-1) \trm{Cos}[2 \phi ]} \, ,\\
\nonumber \\
K_2(\phi,\eta)&:=&\frac{-8 (\eta-1 )^2 \trm{Cos}[\phi ] \trm{Sin}[\phi ]^3 \,}{2+(\eta-2 ) \eta +2 (\eta-1 ) \trm{Cos}[2 \phi ]} \, .
}

\begin{figure}
 \includegraphics{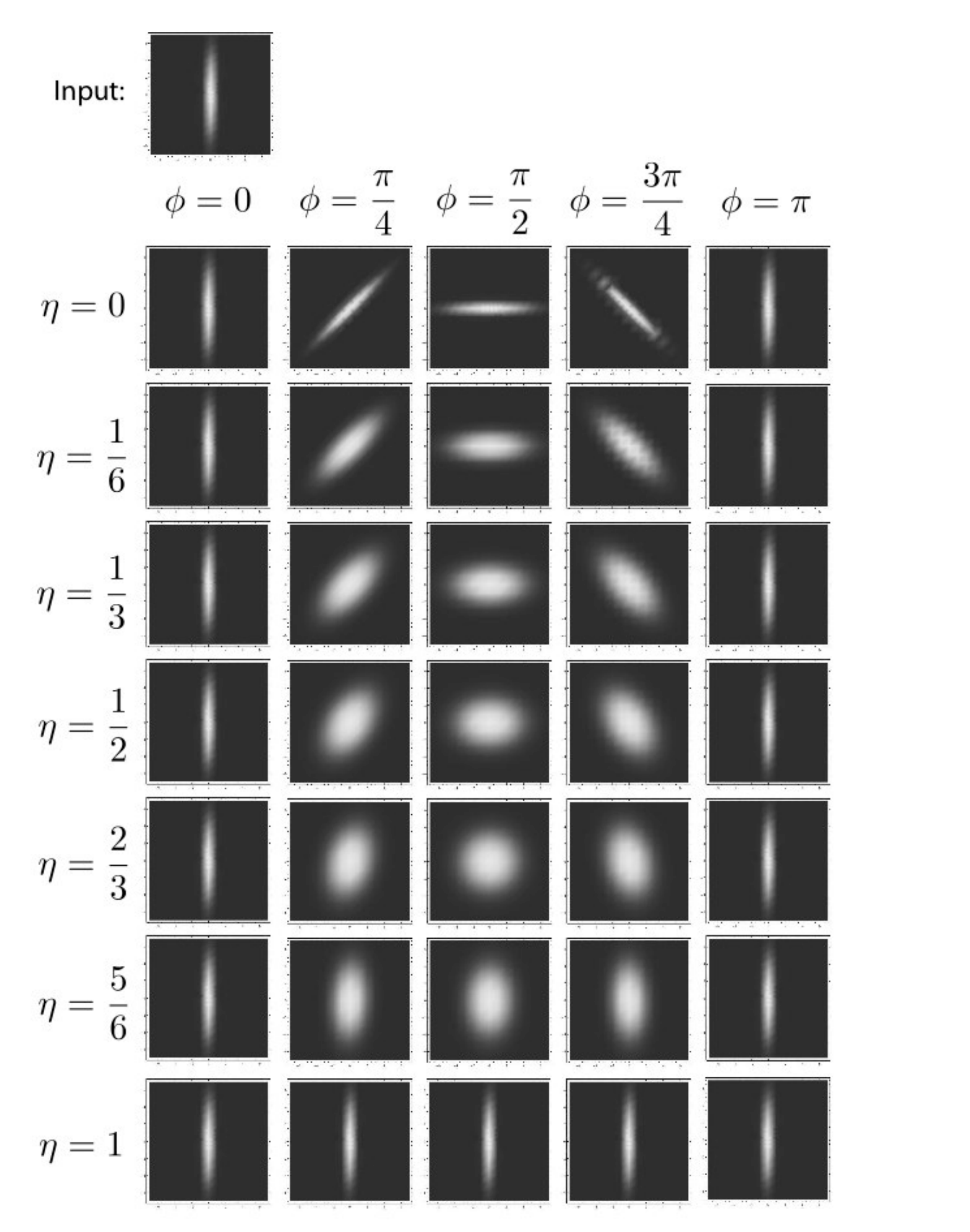}
 \caption{\label{Wigner1}The Wigner function of the output state when the input is a squeezed vacuum state (shown at top). The rows and columns correspond to different choices of the beamsplitter parameters.}
\end{figure}

This covariance matrix corresponds to a rotation of the input squeezed state with added noise. These effects can be seen by comparing the Wigner functions of the input and output states, shown in Fig. \ref{Wigner1}. We briefly comment on the behaviour for particular choices of the beamsplitter parameters. 

First, we note that when the reflectivity is either full or zero, $\eta=1,0$, there is no interaction between the modes entering the beamsplitter and the state passes through the CTC unaffected up to an overall phase. In the case of total transmittance, $\eta=0$, the additional phase corresponds to a rotation of the squeezing angle by an amount equal to $\phi$.

In more general cases where $0<\eta<1$, the evolution can still be trivial depending on the choice of $\phi$. Specifically, whenever $\phi$ is an integer multiple of $\pi$, we find that $K_1(\phi,\eta)=1$ and $K_2(\phi,\eta)=0$. This yields the same covariance matrix as the input state, indicating that the CTC has no effect. In between these limits, i.e. for $0<\eta<1$ and $n \pi<\phi<(n+1)\pi$, we find that the CTC introduces noise. The added noise is at a maximum when $\phi=\frac{\pi}{2}$ (or odd integer multiples thereof) and when $\eta=\frac{2}{3}$. In the case where both these conditions are met, the state becomes completely thermalised, i.e. the state is symmetric in phase space and all evidence of squeezing is washed out by noise.

While the introduction of noise due to the CTC is an interesting effect, it does not seem to be very unusual or explicitly nonlinear in any fundamental sense. After all, it is conceivable that one could obtain the same effect without the need for a CTC (or copies of the initial state). We might expect to see more striking effects for non-Gaussian states, whose intrinsically quantum nature is more apparent\footnote{Non-Gaussian pure states are `more quantum' than Gaussian states in the sense that their Wigner distributions have negative parts. This is related to the fact that Gaussian states and operations can be efficiently simulated on a classical computer, while non-Gaussian states and operations cannot.}. Let us therefore turn to the quintessential example of a non-Gaussian state, namely the single photon state.
 
\subsection{The single photon state}

Let us take the input state to be the single photon state $\ket{\Psi_A}=\hat{A}^\dagger \ket{0}=\ket{1_A}$. Were we to insist on taking expectation values in the vacuum state, as in the previous scenario, we could use the result \eqref{aoutN} from Sec \ref{SecSingPhot} to rewrite the single-photon expectation values as vacuum expectations of conditional operators in the Heisenberg picture. Our present calculations are not made much simpler by doing this, so we will take expectation values in the state $\ket{1_A}$. Specifically, working in the equivalent circuit, the initial state is the $N$ photon state $\ket{\Psi_{A,N}}=\ket{1}^{\otimes N}$. The following identities will prove useful:
\eqn{\label{BeamsplitSPidentities}
\bk{ \hat{A}_m } &=& \bk{ \hat{A}^\dagger_m } = 0 \\
\bk{ \hat{A}^\dagger_n \hat{A}_m } &=& 
\begin{cases} 1 & \trm{if } \, (n=m), \\
 0 & \trm{otherwise}
\end{cases} \, \\
\bk{ \hat{A}^\dagger_m \hat{A}^\dagger_n \hat{A}_p \hat{A}_q } &=& 
\begin{cases} 1 & \trm{if } \, (n=q)\neq(m=p), \\
 & \trm{or } \, (n=p)\neq(m=q), \\
 0 & \trm{otherwise}
\end{cases} \, ,
}
where now it is understood that expectation values are taken in the initial state $\ket{1}^{\otimes N}$. It is straightforward to prove these identities using the orthonormality of the Fock state basis $\braket{n}{m}=\delta_{n,m}$ and the commutator $[\hat{A}_n, \hat{A}^\dagger_m]=\delta_{n,m}$. We note that the last two identities can be expressed more succinctly as:
\eqn{ \label{BeamsplitSPidentities2}
\bk{ \hat{A}^\dagger_n \hat{A}_m } &=& \delta_{n,m} \nonumber \\
\bk{ \hat{A}^\dagger_m \hat{A}^\dagger_n \hat{A}_p \hat{A}_q } &=& \delta_{nq} \delta_{mp} + \delta_{np} \delta_{mq} - 2\, \delta_{nm}\delta_{np}\delta_{nq} \label{g2deltas} \, .
}
(Note in particular that \eqref{g2deltas} implies $\bk{ \hat{A}^\dagger_m \hat{A}^\dagger_m \hat{A}_m \hat{A}_m }=0$.)

Since we are only considering a single photon at the input, we can identify two essential properties of the output that are likely to be of interest. The first property is the average photon number at the output, given by $\bk{\hat{A}'^\dagger_1 \hat{A}'_1}$. Since only a single photon enters the CTC, we expect this number to be equal to one. Using the identities \eqref{BeamsplitSPidentities} it is straightforward to verify that, indeed,
\eqn{
\bk{\hat{A}'^\dagger_1 \hat{A}'_1} &=& \eta^2 \sum^\infty_{m=1} \brac{1-\eta}^{m-1}+(1-\eta) \nonumber \\  
&=& 1 \, .
}
We are therefore reassured that, at least on average, there is a single photon emerging from the output of the CTC. The next property of interest is the variability of the photon number, i.e. whether it is possible for more or less photons to emerge from the CTC with some non-zero probability in any given run of the experiment. A standard measure of the incidence of higher-photon terms is the second order correlation function $g^{(2)}$, which is given by:
\eqn{\label{CTC1g2}
g^{(2)} &=& \bk{\hat{A}'^\dagger_1 \hat{A}'^\dagger_1 \hat{A}'_1  \hat{A}'_1} \nonumber \\
&=& \eta^4 \sum^\infty_{n,m,p,q=1} e^{i \phi(n+m-p-q)}(\sqrt{1-\eta})^{n+m+p+q-4} \bk{\hat{A}^\dagger_n \hat{A}^\dagger_m \hat{A}_p  \hat{A}_q} \nonumber \\
&&+ \,4\, \eta^2 \sum^\infty_{n,m=1} e^{i \phi(n-m)}(\sqrt{1-\eta})^{n+m} \bk{\hat{A}^\dagger_n \hat{A}_m \hat{A}^\dagger_0 \hat{A}_0}  \nonumber \\
&=& 2 \eta^4 \brac{ \sum^\infty_{n,m=1} (1-\eta)^{n+m-2} - \sum^\infty_{m=1}(1-\eta)^{2m-2}  }+ 4\, \eta^2 \sum^\infty_{m=1} (1-\eta)^{m} \nonumber \\
&=& 2 \eta^4 \brac{\frac{1}{\eta ^2}-\frac{(1-\eta )^2}{(1-\eta)^2 (2-\eta) \eta }}+4\, \eta \brac{1-\eta} \, \nonumber \\
&=& \frac{8 \eta (1-\eta)}{(2-\eta)} \, ,
}
where we have used the identities \eqref{BeamsplitSPidentities}, \eqref{BeamsplitSPidentities2}. 
 
Note that when there is either perfect reflection or perfect transmittance, i.e. $\eta=1$ or $0$, there is exactly one photon in the output ($g^{(2)}=0$). For more general cases, however, $g^{(2)}\neq 0$ and the photon number may fluctuate. This suggests that, despite the fact that energy is conserved on average, any single run of the CTC circuit may violate energy conservation, either by causing the input photon to disappear or else by producing extra photons at the output. In what follows, we take a brief diversion to comment on the implications of this violation.

\subsubsection{A note on energy (non)conservation}

Given that our best physical theories satisfy energy conservation in one form or another, the violation of energy conservation in the present model should be taken seriously. We must ask ourselves: What is the nature energy conservation in the context of the model? Does the lack of energy conservation represent an inconsistency in the model? And can we restore energy conservation to the model, if we insist upon it?

Since the Deutsch model is intended to describe quantum effects in the presence of a CTC, our expectations for energy conservation should not be more stringent than those normally placed on CTCs in GR. The status of energy conservation in GR is already a very subtle matter, particularly for dynamic spacetimes. While the standard notion of energy conservation holds for spacetimes with a global timelike Killing vector field, in which case Noether's theorem provides an unambiguous definition of the conserved energy, even such familiar models as the expanding universe do not possess this feature. One must therefore be more specific about what one means by `energy' and `conserved' in more general situations. Given that these concepts are quite unclear in metrics containing CTCs, it is certainly not obvious that the violation of energy conservation predicted by the Deutsch model is in conflict with GR.

A more pertinent question would be whether this violation of energy conservation is somehow pathological, such that it renders the model inconsistent or unstable. The lack of energy conservation in Deutsch's model can be traced to the nonlinear entanglement-breaking effect of Deutsch's map. Suppose we place a pair of atoms in a coherent superposition such that either one or the other atom is in the excited state upon measurement in the energy eigenbasis. If we subject one of the pair to a CTC, the entanglement is broken in such a way as to preserve the local properties of each subsystem. Hence there is a finite probability afterwards of finding both systems in the ground state, or both systems in the excited state. In the former case, energy has mysteriously vanished, while in the latter case, the same amount of energy has appeared `from nowhere'.  

We note that energy seems to still be conserved on average, giving some hope that the predicted stochastic violation of energy conservation is governed by some well-defined rule. The nature of this rule becomes apparent when we consider the equivalent circuit. Here, we see that the Deutsch model can always be regarded as a unitary and energy-conserving interaction with a hypothetical environment that obeys a strict symmetry rule: the circuit is a concatenation of identical sub-circuits. This leads us to expect any violations of energy observed in one sub-circuit to be strictly constrained. Indeed, since the total energy must be conserved, the symmetry implies that it must also be equally partitioned between the sub-circuits, regardless of the choice of unitary. This heuristic reasoning might be the basis for a general proof of the conservation of energy on average in the Deutsch model, beyond the special cases that we have considered.

Despite the fact that the energy non-conservation in the model seems to be benign (or at least as benign as such an exotic effect could ever be!) there is still the option of eliminating it by taking seriously the `multiple universe' interpretation of Deutsch's model. As discussed in Sec \ref{SecECInterp}, this interpretation takes the equivalent circuit to represent the actual state of affairs, and the CTC to be a doorway between identical copies of the same sub-universe within a larger multiverse-like structure. In this scenario, one faces pressing new questions: Where do the other sub-universes reside?  Do they exist as cosmological parallel universes or are they somehow contained within the separate branches of an Everett-style multiverse? How numerous are they, and how they all came to appeal identical to one another? Whatever the precise situation, the problem of energy conservation is solved trivially, since matter lost or gained through a CTC in our universe is compensated by energy exchange within the multiverse.

Finally, perhaps the easiest solution of all is to try and incorporate the principle of energy conservation directly into the model. There are certainly cases in which energy is conserved, such as those situations in which the energy of the system can be considered a classical attribute. Since it is not subject to quantum fluctuations, the energy does not represent a degree of freedom in the system's Hilbert space to which Deutsch's map applies, and as such retains its value from the input to the output. This matches our earlier observation that energy non-conservation results from entanglement in the energy basis -- if no such entanglement is possible, then energy may be conserved. Thus, it is not inconceivable that a clever placement of a quantum-classical divide, possibly in addition to some other physically motivated rules, might result in a constrained version of the Deutsch model that would make nontrivial predictions while satisfying energy conservation. At the very least, it would be interesting to investigate the extent to which Deutsch's model might be compatible with energy conservation, similar in spirit to our analysis in Chapter \ref{ChapNonlinBox} of the model's compatibility with no-signalling. We leave this as a topic for future work.

\section{The curious case of open time-like curves\label{CTC1OpenCurve}}  

In previous sections, we saw that it is possible to apply continuous variables to circuits containing CTCs. To illustrate this fact, we calculated the outputs for a CTC with a very simple interaction, namely a beamsplitter, for various pure state inputs. Let us now turn to an even simpler situation in which the CTC contains no interaction at all: an `open time-like curve'. This situation corresponds to setting $\hat{U}_{CTC}=\hat{I}$ in Fig. \ref{genFieldCTC}.

While the dynamics of an OTC is trivial, namely $\rho_{A,out}=\rho_{A,in}$, recall the earlier discussion in Sec \ref{SecEntBreak} in which it was observed that this map does not preserve the entanglement to any external systems. In particular, this loss of entanglement is described by a non-linear map \eqref{EqEntbreak} that cannot be reproduced using ordinary quantum mechanics. It is therefore interesting to ask whether any of the more extreme non-linear effects of CTCs can be obtained entirely from OTCs. In this section, we will answer this question in the affirmative by showing that it is possible to violate Heisenberg's uncertainty relation for quantum optics using just linear optical gates and OTCs. This in turn allows us to perfectly distinguish non-orthogonal coherent states and to clone unknown coherent states.

\begin{figure}
 \includegraphics{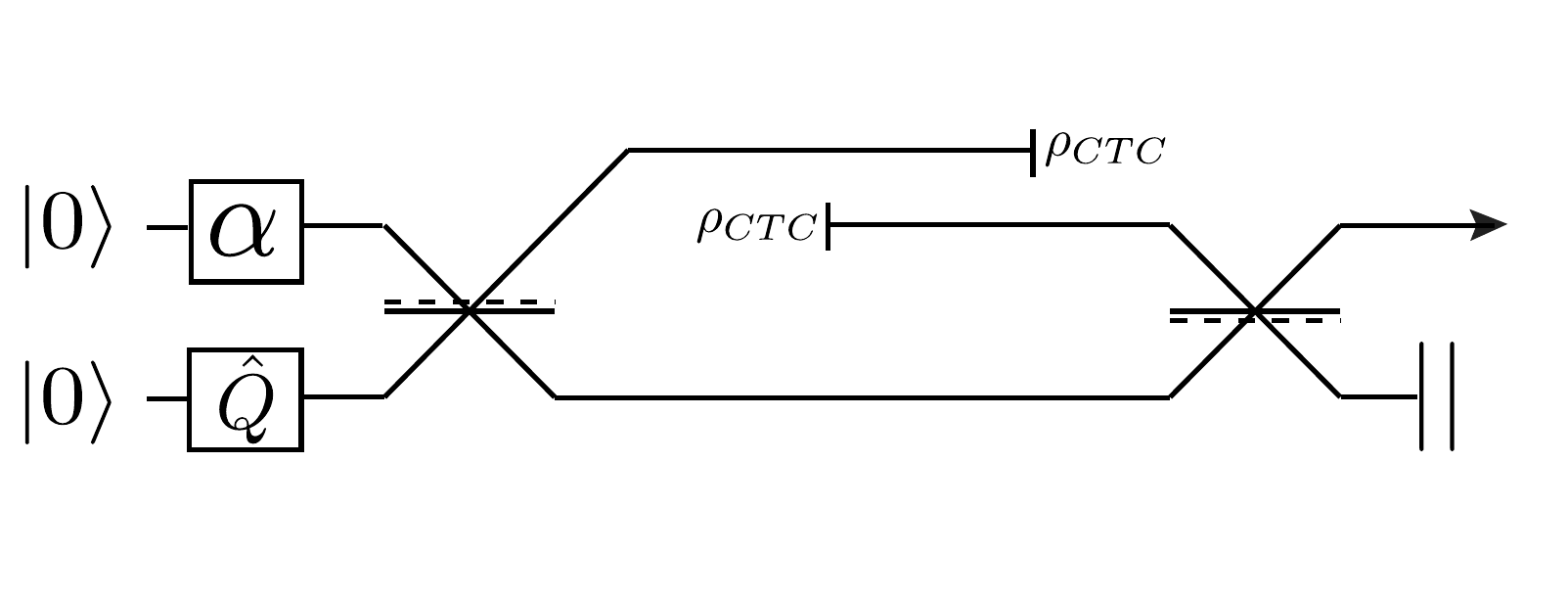}
 \caption[A quantum optical circuit containing an OTC]{\label{OTCfig1}A quantum optical circuit with an OTC.}
\end{figure}

The specific circuit that we consider is shown in Fig. \ref{OTCfig1}. An unknown coherent state in mode $\hat{A}$ is incident on a beamsplitter where it interacts with a squeezed vacuum state in mode $\hat{B}$ with known squeezing $r$ along the $\hat{Q}$ quadrature. The 50:50 beamsplitter $\hat{T}$ creates entanglement between the two modes, which is then broken by the OTC applied to the top rail. Finally, the modes recombine on a second beamsplitter, which is the inverse of the first beamsplitter, i.e. equal to $\hat{T}^\dagger$. Note that, without the OTC, the effect of the two beamsplitters would cancel out (since $\hat{T}^\dagger \hat{T} = \hat{I}$) leading to a trivial input--output map. However, the map becomes nontrivial in the presence of the OTC.

\begin{figure}
 \includegraphics{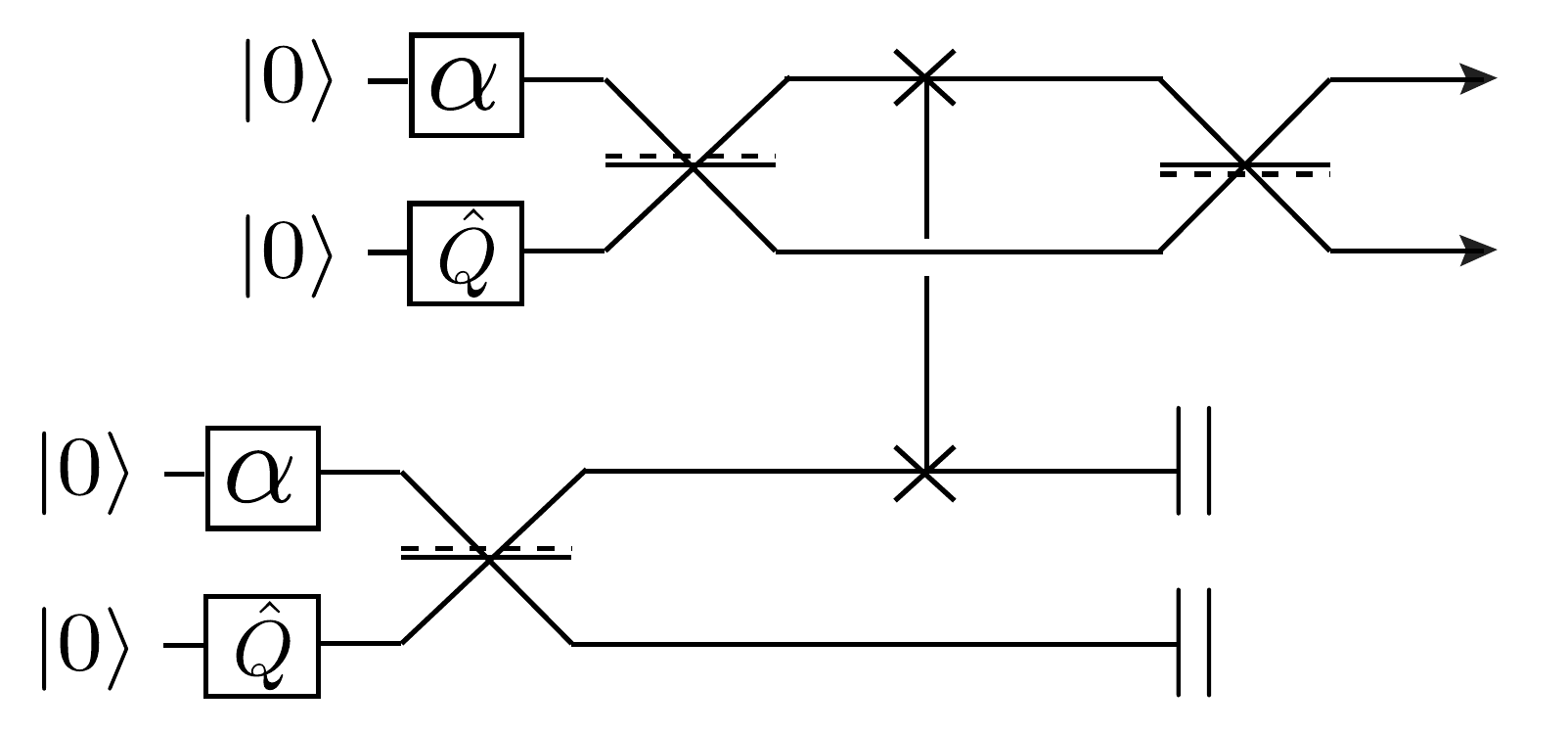}
 \caption[An equivalent circuit for a quantum optical circuit containing an OTC]{\label{OTCfig2}An equivalent circuit to Fig. \ref{OTCfig1}.}
\end{figure}

The input--output map represented by the circuit of Fig. \ref{OTCfig1} can easily be obtained using Deutsch's model, since the consistent solution $\rho_{CTC}$ is given trivially by tracing out one arm of the prepared entangled state. The key feature of the dynamics is that the two rails at the output of the circuit are no longer entangled. The map is essentially the entanglement breaking map of Sec \ref{SecEntBreak}. However, we will find it instructive to solve the problem in the Heisenberg picture, using the equivalent circuit shown in Fig. \ref{OTCfig2}. Of course, one obtains the same results in either case.

In Fig. \ref{OTCfig2}, the top two rails correspond to the modes $\hat{A}_1, \hat{B}_1$ while the lower two rails are $\hat{A}_2, \hat{B}_2$. In keeping with the correspondence to the original circuit, the only accessible outputs of the circuit are the modes $\hat{A}'_1,\hat{B}'_1$, while the remaining outputs $\hat{A}'_2,\hat{B}'_2$ must be discarded. The initial state is taken to be the vacuum $\ket{0_{A_1}}\ket{0_{B_1}}\ket{0_{A_2}}\ket{0_{B_2}}$ and the input to the circuit is generated by the preparation unitary $\hat{U}_P:=\hat{D}_{A_1}(\alpha)\hat{S}_{B_1}(r)\hat{D}_{A_2}(\alpha)\hat{S}_{B_2}(r)$ where we have taken the squeezing along the $\hat{Q}$ quadrature, $\theta_S=0$. Note that we require two copies of the unknown initial state $\ket{\alpha}$ in order to simulate the OTC with this equivalent circuit. 

The `time travel' in the equivalent circuit is implemented by the SWAP gate $\hat{U}_{S}$, which simply interchanges the modes $\hat{A}_1$ and $\hat{A}_2$. The Heisenberg evolutions for the initial beamsplitters are:
\eqn{
  \hat{T}^\dagger \, \hat{A} \, \hat{T} &=& \frac{1}{\sqrt{2}} \, \brac{ \hat{A}+\, \hat{B} }\, ,\\
  \hat{T}^\dagger \, \hat{B} \, \hat{T} &=& \frac{1}{\sqrt{2}}\, \brac{\hat{B}-\, \hat{A} } \, , 
} 
and the inverse beamsplitters are given by $\hat{T}'=\hat{T}^\dagger$. Let us consider the output $\hat{A}'_1$ of the top rail. Performing the Heisenberg evolution through the circuit, we then obtain:
\eqn{\label{OTCout1}  \hat{A}'_{1} &=& \hat{U}^\dagger_{EC} \, \hat{A}_{1} \, \hat{U}_{EC} \nonumber \\
&=& \hat{U}^\dagger_P \, \hat{T}^\dagger \, \hat{U}^\dagger_{S} \, \hat{T} \; \hat{A}_{1} \; \hat{T}^\dagger \, \hat{U}_{S}\, \hat{T} \, \hat{U}_P \nonumber \\
&=& \alpha + \frac{1}{2}(\hat{A}_{1} + \hat{A}_{2})- \frac{i}{2}\, \trm{cosh}(r)\, (\hat{B}_{1} - \hat{B}_{2}) + \frac{i}{2}\, \trm{sinh}(r)\, (\hat{B}^\dagger_{1} - \hat{B}^\dagger_{2}) \, .}
We see that $\bk{\hat{A}'_1} = \alpha$, hence there is no net displacement after evolution. The variances along the quadratures $\hat{Q}$ and $\hat{P}$ are exponentially reduced and increased, respectively: 
\eqn{
\trm{Var}(\hat{Q}) = e^{- r} \, \trm{cosh}(r) \\
\trm{Var}(\hat{P}) = e^{r} \, \trm{cosh}(r)
}
For large squeezing, $r\gg1$, the variance of the squeezed quadrature approaches $\trm{Var}(\hat{Q}) \rightarrow \frac{1}{2}$. This circuit therefore allows us to deterministically squeeze the quantum noise along $\hat{Q}$ without changing its displacement, a task that is known to be impossible using standard quantum mechanics (see eg. Ref. \cite{WAL}). We therefore conclude that this OTC circuit implements a nonlinear map.

We would now like to exploit this map to perform more dramatic information processing tasks. To do this, we would like to have access to multiple OTCs. In particular, our goal is to strengthen the squeezing effect described above by feeding the output of the OTC circuit back through an identical circuit. For this purpose, we will follow previous literature (eg. Ref. \cite{BAC04}) in assuming that there are no `memory effects' (recall the discussion in Sec \ref{SecMultiCTC}). To demonstrate the dramatic effects that are implied by OTCs alone, we now show that it is possible to modify the circuit of Fig. \ref{OTCfig1} to violate Heisenberg's uncertainty principle.

The uncertainty principle states that if $\hat{x},\hat{p}$ are two self-adjoint non-commuting operators representing observables, then the errors in a set of single measurements of these observables must satisfy $\sigma_x \sigma_p \geq \left| \frac{i}{2}[\hat{x},\hat{p}] \right|$, where $\sigma_x, \sigma_p$ are the standard deviations of the measurements (i.e. $\sigma_x = \sqrt{\trm{Var}(\hat{x})}$). The observables $\hat{x},\hat{p}$ are commonly taken to represent position and momentum, whence we obtain $\sigma_x \sigma_p \ge \frac{\hbar}{2}$; however, the uncertainty principle holds for any pair of canonical observables. In quantum optics, for example, the quadrature operators $\hat{Q},\hat{P}$ play the analogous role to the position and momentum, satisfying $\sigma_Q \sigma_P \ge 1$. 

\begin{figure}
 \includegraphics{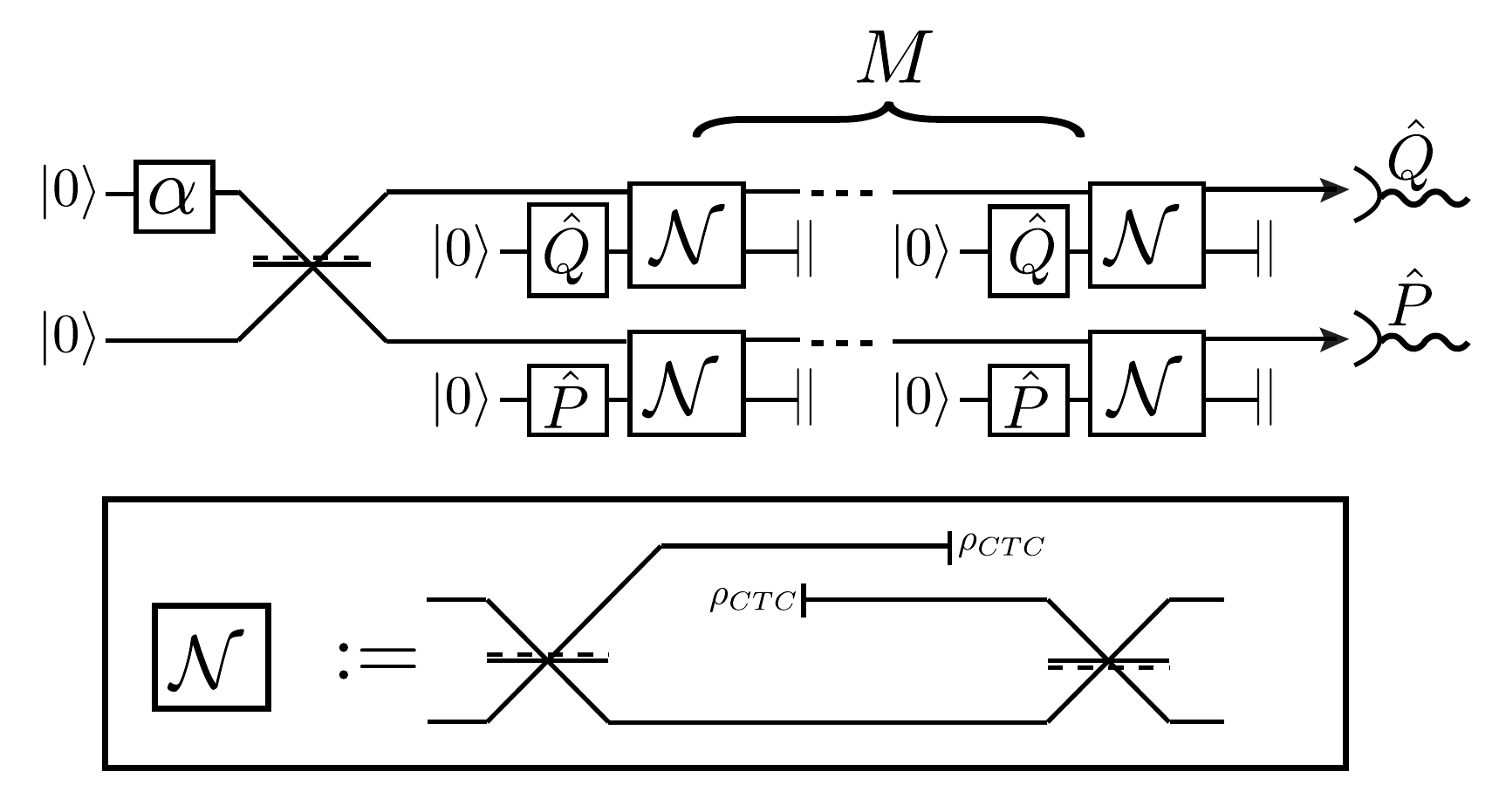}
 \caption[An OTC circuit that violates the uncertainty principle]{\label{figHeisViolate}A quantum optical circuit with multiple OTCs that violates the uncertainty principle.}
\end{figure}

Consider the circuit shown in Fig. \ref{figHeisViolate}. The input coherent state is first combined with the vacuum on a 50:50 beamsplitter, yielding two copies of the state $\ket{\alpha}\rightarrow \ket{\frac{\alpha}{\sqrt{2}}}_A\ket{\frac{\alpha}{\sqrt{2}}}_C$ in the modes $\hat{A}$ and $\hat{C}$. We send each of these states through a copy of the original circuit Fig. \ref{OTCfig1}, choosing the squeezing in each case such that $\ket{\frac{\alpha}{\sqrt{2}}}_A$ becomes squeezed in the $\hat{Q}$ direction, and $\ket{\frac{\alpha}{\sqrt{2}}}_C$ in the orthogonal $\hat{P}$ direction. Assuming we have repeated access to the OTC, we may cycle each of the outputs back through the circuit for a total of $M$ iterations, using new squeezed states on each run. For simplicity, we assume that these extra squeezed states all have the same squeezing amplitude $r$. At the end of $M$ runs, the displacement on each output will be unchanged, but the variances will be reduced in the chosen quadratures. We then make measurements of $\hat{Q}_A$ and $\hat{P}_C$ on the respective outputs, and thereby determine the components of $\frac{\alpha}{\sqrt{2}}$ along orthogonal quadrature directions.

The errors for these measurements are found to be equal to:

\eqn{\label{VarM}
\trm{Var}\bk{\hat{Q}_A}=\trm{Var} \bk{\hat{P}_C} = \frac{1+2^{M-R}-2^{-R} }{2^{M}} 
}
where $R \equiv \frac{2}{log(2)} r$ is introduced for clarity. According to Heisenberg's uncertainty principle, the errors in these measurements should satisfy $\trm{Var}\bk{\hat{Q}_A} \trm{Var} \bk{\hat{P}_C} \ge 1$. However, we see that for $r \gg M$, their product scales as $\frac{1}{2^{M}}$; thus we have measured the orthogonal components of $\alpha$ to an arbitrary degree of accuracy forbidden by the uncertainty principle. This information can then be used to identify and distinguish unknown coherent states and make clones of them as desired. We note that even for a single use of the OTC (i.e. $M=1$) the Heisenberg principle is violated at sufficiently large squeezing.

Models of CTCs have until now been considered primarily for purely theoretical reasons, but the present result shows that, were we ever to discover nonlinearities of this type in nature, we would not need to perform invasive operations within the CTC in order to exploit its information processing powers in practise -- it would be enough just to send Gaussian states through the nonlinear region and to perform operations strictly on the inputs and outputs.

It has been argued that any nonlinear theory that permits the exponential separation of nonorthogonal states with only polynomial overhead in resources must lead to advanced computational power; namely, the ability to solve both NP-complete and $\#$P problems in polynomial time\cite{ABR98}. Correspondingly, it is known that Deutsch CTCs with nontrivial gates inside the CTC can be used to efficiently solve NP-complete problems \cite{BAC04}. A natural question is whether this effect carries over to circuits containing just OTCs. Given that we have just demonstrated perfect state discrimination, it seems we are already half way to proving computational speedup -- we need only check whether our procedure can be done using only polynomial resources. In what follows, however, we show that the resources needed to distinguish nonorthogonal states using the circuit of Fig. \ref{figHeisViolate} in fact scale exponentially.

Following \cite{ABR98}, consider this task: we are given a black box that takes a string $x$ of $N$ bits as input, and gives just a single bit $f(x) \in \{0,1\}$ as output according to some unknown map $f(x)$. To efficiently solve NP-complete problems, it is sufficient to determine whether there exists an input $x'$ such that $f(x')=1$, using resources that scale polynomially in $N$. Using just a single query of the box plus ordinary quantum mechanical operations, it is possible to create the state:
\eqn{ 
\ket{\psi}&=&\frac{(2^N-s)}{A}\ket{0}+\frac{1}{A}\ket{1}  \, , \nonumber \\
A &\equiv& \frac{1}{\sqrt{(2^N-s)^2+s^2}} \, ,
}
where $s$ is the number of solutions $x'$ satisfying $f(x')=1$. The interesting cases are those for which there are only very few solutions, $s \ll 2^N$, for which we obtain the simplification:
\eqn{
\ket{\psi} \approx \ket{0}+\frac{1}{2^N}\ket{1}\, , 
}
which is equal to a coherent state $\ket{\alpha(N)}$ with amplitude $\alpha=\frac{1}{2^N} \ll 1$. The problem of efficiently determining whether $s$ is nonzero then reduces to the task of discriminating between the nonorthogonal states $\ket{0}$ and $\ket{\alpha(N)}$ with only polynomial overhead in the problem size $N$. To make the states distinguishable using our circuit Fig. \ref{figHeisViolate}, we reduce the variance of $\ket{\alpha(N)}$ until it is comparable to $\alpha$, i.e. Var$\bk{\hat{Q}_A} = \frac{K}{2^N}$, for some constant of proportionality $K>0$. Applying this condition to \eqref{VarM}, we obtain:
\eqn{\frac{1+2^{M-R}-2^{-R} }{2^{M}} = \frac{K}{2^N} \, .}
The most direct way to satisfy this equation is to set $M=N$, hence:
\eqn{ && K = 1+2^{-R}(2^{N}-1) \approx 1+2^{N-R} \, .}
For $K$ to be constant, the exponent $N-R$ should be independent of $N$. This is satisfied by choosing $R=N$, i.e. the squeezing $r$ scales linearly with $N$. To quantify the necessary resources, consider the average number of photons required to create a squeezed state. For large squeezing $r\gg 1$, the average photon number scales as $\bk{\hat{n}} \propto \frac{1}{V}$ where  $V=e^{-2r}$ is the variance of the squeezed quadrature. Therefore $\bk{\hat{n}} \propto e^{N}$ and we require an exponential number of photons to generate our squeezed state resources. We conclude that, at least in the present example, it is impossible to efficiently solve NP-complete problems, in spite of the fact that we are able to perfectly discriminate nonorthogonal states in a single shot. Nevertheless, this exponential blow-out of resources may be due to the fact that our discussion has been restricted to Gaussian states and operations. We therefore conjecture that OTCs can indeed provide increased computational power when non-Gaussian states and operations are taken into account.

\section{Conclusions}

In this chapter we directly applied the techniques of continuous variable quantum optics to circuits containing CTCs. By assuming that all systems were pointlike, we are able to apply Deutsch's consistency condition without any modifications. As a result, we showed that when a single photon interacts with its future self on a beamsplitter, violations of energy conservation may be observed, although energy appears to be conserved on average. We next applied continuous variables to the simplest scenario involving a CTC without any interaction inside, i.e. an `OTC'. While the only effect in such cases is a breaking of entanglement to external systems, we demonstrated that this effect alone could be exploited by a simple linear optical circuit to perfectly distinguish and clone unknown coherent states. This result is remarkable because all previous demonstrations of similar effects using discrete variables have required highly nontrivial gates inside the CTC. However, due to the resources required to implement this particular circuit, we noted that no computational speed-up was observed in the circuit that we considered.

While the extension of the Deutsch model to Fock space is straightforward, as we have seen in this Chapter, it has remained necessary to ignore the spacetime dependence of the fields. Arguably, this is no longer valid if the coherence time of the wavepackets is comparable to the size of the CTC. In that case, we require a model that is able to smoothly interpolate between the predictions of Deutsch's model and those of standard quantum optics, in order to describe the limit as the size of the CTC shrinks to zero. This will be taken up in the next Chapter.

\clearpage
\pagestyle{plain}
\pagebreak
\renewcommand\bibname{{\LARGE{References}}} 
\bibliographystyle{refs/naturemagmat2012}
\addcontentsline{toc}{section}{References} 

\pagestyle{fancy}
\renewcommand{\sectionmark}[1]{\markboth{\thesection.\ #1}{}}

\chapter{Beyond the Deutsch model}
\label{ChapCTC2}

\begin{quote}
\em
And if, with infirm hand, Eternity,\\ 
Mother of many acts and hours, should free\\
The serpent that would clasp her with his length; \\
These are the spells by which to reassume \\
An empire o'er the disentangled doom.\\
\end{quote}
P.B. Shelley\\

\newpage

\section*{Abstract}

In Chapter \ref{ChapCTC1} we applied Deutsch's circuit to quantum fields by extending the dimensionality of the rails to Fock space. In this final Chapter, we will modify Deutsch's model to incorporate the spacetime structure of the field modes as well. Inspired by earlier work on `event operators', in Sec \ref{SecEopsIntro} we consider an \tit{ad hoc} modification of the quantised electromagnetic field in order to accommodate the enlarged Hilbert space structure implied by the Deutsch model. By embedding the field theory in this larger Hilbert space, we show that it is possible to smoothly interpolate between Deutsch's model and standard quantum optics. The parameters that control this interpolation are the size of the CTC and the intrinsic temporal resolution of the relevant photon detector. In Sec \ref{SecEopsExamples} we apply the model to specific cases, revisiting the examples of a beamsplitter on a CTC and Gaussian quantum states interacting with OTCs. For the latter case, it is shown in Sec \ref{SecEOtoCircuit} that there exists a circuit that exactly reproduces the behaviour of the generalised formalism, at least for Gaussian states. Finally, we conjecture in Sec \ref{SecOTCNewPhys} that the modifications to quantum optics seen in the presence of an OTC might also be expected to occur in causal spacetimes with sufficient curvature. It is shown that the model reproduces the predictions of `event operators' in this case, leading to the possibility of testing the model using entangled light in Earth's gravitational field.  

\section{The modified Deutsch model\label{SecEopsIntro}}

\subsection{Tensions with field theory\label{SecFieldTensions}}

In this chapter, our goal is to extend Deutsch's model to a relativistic massless field of scalar bosons. Since field theory quickly becomes intractable in the Schr\"{o}dinger picture, we will adopt the Heisenberg picture, taking as our starting point the equivalent circuit as applied to quantum optics in Chapter \ref{ChapCTC1}. Before that, we briefly discuss the essential obstacle to applying Deutsch's model to fields with nontrivial spacetime extent.

Recall from the spacetime diagrams of Chapter \ref{ChapDeutsch} that the assumption of pointlike systems allows us to apply Deutsch's consistency condition between the reduced states at two pointlike events in spacetime, labelled $P$ and $F$, which are timelike or lightlike separated such that $F$ is in the future light cone of $P$. In more general situations, this pointlike matching condition must be replaced by a matching of the fields on a pair of consecutive spacelike slices. One consequence of this is that there is now a preferred reference frame associated with the matching condition, namely the reference frame in which the two slices represent planes of simultaneity. There is nothing strictly wrong with this, since we might well expect the presence of a CTC to introduce some asymmetry, but it is worth mentioning. In addition, there must exist a rule for mapping the spacetime points on the future slice to the spacetime points on the past slice. This rule is assumed to be given by a linear function, e.g. $x^{\mu}_F \mapsto L^{\mu}_{\nu} x^{\nu}_F+z^{\mu} = x_P^{\mu}$, mapping any point $x^{\mu}_F$ in the future time slice to a corresponding point $x^{\mu}_P$ in the past slice. In what follows, we will simply use $P$ and $F$ to refer to any pair of events $x^{\mu}_P,x^{\mu}_F$ that are identified between the two slices. Following the discussion in Chapter \ref{ChapDeutsch} we restrict ourselves to the cases where the interval between $P$ and $F$ is either timelike or lightlike. The traversal of the CTC by a single field is then represented in Deutsch's model by \tit{two} fields, differentiated by some extra degree of freedom (the index `$m$' in our considerations thus far). The ingoing field has support on the part of spacetime extending from the infinite past up to the future slice, and the outgoing field is defined on the part of spacetime beginning at the past slice and extending on to infinity. The region where the two copies of the field overlap is the region in which CTC interactions between ``older" and ``younger" parts of the field can occur.

The above discussion illustrates how Deutsch's matching condition might be dealt with in the case of arbitrary fields, if one were to proceed in the Schr\"{o}dinger picture. In the Heisenberg picture, the matching condition is replaced by a large number of $N$ copies of the field, labelled by the index $m$, which evolve independently of each other except in the region between the two slices identified by the CTC where they interact via the equivalent circuit. Only one input and one output field are accessible in the distant past and future, and the input--output map is the same as that given by Deutsch's model.

The above generalisation is only part of the story, however. Note that, since the ingoing and outgoing fields have different indices $m$, operators associated with the input field necessarily commute with operators on the output field. This is in keeping with Deutsch's model for pointlike systems, where the ingoing system on the top rail is associated with an independent Hilbert space to the system on the bottom rail, even though they are ostensibly the ``same" system, by virtue of the consistency condition. This essential feature gives rise to the entanglement-breaking effect, among other things. For pointlike systems, it is perhaps plausible that during the propagation back in time, the entire system undergoes a transformation that shifts it to a new Hilbert space and breaks its entanglement to external systems. In the case of extended fields, however, this picture is difficult to maintain. In particular, if the coherence time of the propagating field exceeds the time between the two slices representing the temporal extent of the CTC, it is possible for the leading part of the wavepacket to be sent back in time before the tail end. We can then no longer take refuge in treating the time-travelled part of the field as distinct from the non time-travelled part, since these parts are now explicitly connected through the CTC. 

To make this conflict more precise, consider the fact that in quantum field theory, a photon in the mode $A(\tbf{x},t)$ is indistinguishable from a photon in the same mode but propagated by $\Delta t$ along a null geodesic, $A(\tbf{x}+c \Delta t, t+\Delta t)$. This is evident from the non-vanishing of the propagator on the light cone, or equivalently from the fact that operators localised to one point on a null geodesic do not commute with operators at any other point on the same geodesic. In the limit that the CTC shrinks to zero size compared to the coherence time of the field, the past and future slices approach one another and the region of overlap disappears. In that case, we expect the endpoints of the time-travelling system's geodesic to smoothly connect, recovering the commutation relations of standard quantum field theory. This would require a formalism that can smoothly interpolate between the case where operators acting on the ingoing and outgoing fields commute, as in the Deutsch model, and the case where they do not, as expected in the standard theory. In what follows, we will generalise the picture provided by the equivalent circuit in order to accommodate this smooth transition, within the domain of quantum optics.   
 
\subsection{Extending the equivalent circuit}

Recall that the equivalent circuit is defined by the $N$-mode unitary $\hat{U}_{EC}(\hat{A}_0,\hat{A}_1,...,\hat{A}_N)$. The algebraic structure of this circuit depends upon the modes 
\eqn{
\hat{A}_m:= \int \invard{\tbf{k}} \, \phi_A(k,x) \, \hat{a}_{\tbf{k},m} \, , \\
\trm{where} \; \; m=0,1,2,...,N \, , \nonumber
}
and in particular the commutation relations:
\eqn{
[\hat{A}_m,\hat{A}^\dagger_n]=\delta_{mn} \, .
}   
This structure means that operators acting on distinct rails in the circuit commute with one another, in accordance with the Deutsch model. In particular, this means that operators acting on the lower rail in Deutsch's circuit commute with operators acting on the upper rail. In order to recover standard quantum optics, therefore, we need to modify these operators so that they cease to commute in an appropriate limit. To this end, we introduce a new continuous degree of freedom $\Omega$, making the replacement $\hat{a}_{\tbf{k},m} \rightarrow  \hat{a}_{\tbf{k},\Omega}$ and satisfying the extended commutator:
\eqn{
[\hat{a}_{\tbf{k},\Omega},\hat{a}^\dagger_{\tbf{k}',\Omega'}]=\delta(\tbf{k}-\tbf{k}')\delta(\Omega-\Omega') \, .
}
We then re-define the commuting modes $\hat{A}_m$ by interpreting the discrete index `$m$' as a propagation by $m \, \Delta \tau$ along the co-ordinate $\tau$ that is the Fourier complement of $\Omega$. Specifically, we replace $\hat{A}_m \rightarrow \hat{A}_{(m)}$ where:
\eqn{ \label{RepEventop}
\hat{A}_{(m)}:= \int \invard{\tbf{k}} \, \phi(k,x) \, \int \mathrm{d}\Omega \, \xi(\Omega,\tau+m \Delta \tau) \, \hat{a}_{\tbf{k},\Omega} \, .
}
The function $\xi(\Omega,\tau)$ has yet to be defined, but it is assumed to be normalised such that:
\eqn{
\int \mathrm{d}\Omega \, \, \xi(\Omega,\tau) \xi^*(\Omega,\tau) = 1 \, .
}
Since the index $m$ counts the number of times that the system has traversed the CTC in the equivalent circuit, it is natural to interpret $\tau$ as an affine parameter measuring the proper distance along the system's trajectory (see Sec \ref{SecPhysInterp} for further discussion). Importantly, the replacement \eqref{RepEventop} does not affect the algebra of the equivalent circuit so long as the Fourier transform of the function $\xi(\Omega,\tau)$ is strongly localised at some value of $\tau$. For example, taking the function to be a normalised Gaussian superposition of plane waves, 
\eqn{ 
\int \mathrm{d}\Omega \, \xi(\Omega,\tau)&=& \int \mathrm{d}\Omega \; j(\Omega)e^{i \Omega (\tau-\tau_0)} \nonumber \\
&:=& \int \mathrm{d}\Omega \; \brac{\frac{2}{\pi \sigma^2_j}}^{\frac{1}{4}} \trm{exp}\brac{\frac{-\Omega^2}{\sigma^2_j} +i \Omega (\tau-\tau_0)}
}
we find that:
\eqn{\label{GaussianEOcomm}
[\hat{A}_{(m)},\hat{A}^\dagger_{(n)}] &=& \int \invard{\tbf{k}} |\phi(k,x)|^2 \; \int \mathrm{d}\Omega \,  \xi(\Omega,\tau+m \Delta \tau) \xi^*(\Omega,\tau+n \Delta \tau) \nonumber \\
&=& \trm{exp}\brac{-\frac{\sigma^2_j}{8}(n-m)^2 \Delta \tau^2} \nonumber \\ 
&:=&  \trm{exp}\brac{-\kappa^2 (n-m)^2} \, .
}   
In the final step, we have introduced the dimensionless parameter $\kappa^2=\frac{\sigma^2_j \Delta \tau^2}{8}$. It can be verified that for $\kappa^2 \gg 1$ we recover the original commutation relations of the equivalent circuit, $[\hat{A}_{(m)},\hat{A}^{\dagger}_{(n)}]\approx \delta_{m,n}=[\hat{A}_{m},\hat{A}^{\dagger}_{n}]$. In the opposite limit, we find that it is impossible to differentiate rails with different values of $m$, since in that case we have $\kappa^2 \ll 1$ and $[\hat{A}_{(m)},\hat{A}^{\dagger}_{(n)}]\approx 1$. Note that this corresponds to the correct commutation relation in quantum optics for modes acting at different points on the same geodesic. In particular, in this limit, the observables associated with the past and future versions of the system will no longer commute with each other as they do in the Deutsch model. The procedure of replacing the modes as in \eqref{RepEventop} was first proposed in Ref \cite{RAL09} and can be thought of as an embedding of the equivalent circuit into a larger Hilbert space by extending the phase space to include the extra degree of freedom $\Omega$. However, in order to construct an explicit model based on this embedding, we need to address the inevitable operator ordering ambiguities that arise from replacing a set of commuting modes with a non-commuting set. This is taken up in the following section.

\subsection{Operator ordering}

We adopt the general scenario involving a single CTC as given by the equivalent circuit of Fig. \ref{genFieldCTC}. Without loss of generality, we take the initial state to be the vacuum\footnote{The preparation of single particle states from the vacuum can be carried out using the formalism outlined in Sec \ref{SecSingPhot}. Hence any initial state can be written as a deterministic preparation $\hat{U}_P$ acting on the vacuum.}, $\ket{\Psi}=\ket{0}:=\ket{0_A}\ket{0_B}\ket{0_C}...$, thereby subsuming the entire preparation procedure into the unitary $\hat{U}_P$. In the equivalent circuit, Fig. \ref{genFieldEC}, the set of vacuum modes is $\{ \hat{A}_{m},\hat{B}_{m},\hat{C}_{m},..., : m=0,1,...N \}$ and the vacuum state is prepared by the unitary $\hat{U}_{PN}:=\hat{U}_{P}^{\otimes N}=\bigotimes\limits^{N}_{m=0} \hat{U}_{P}(\hat{A}_{m},\hat{B}_{m},\hat{C}_{m},...)$. The gate $\hat{U}_{CTC}$ inside the CTC corresponds to the unitary $\hat{U}_{EC}(\hat{A}_{0},\hat{A}_{1},...,\hat{A}_{N})$ in the equivalent circuit. An arbitrary joint observable in the equivalent circuit is a Hermitian function $\hat{\sigma}(\hat{A}_{1},\hat{B}_{0},\hat{C}_{0},...)$ while all other modes $\hat{A}_{m\neq 1}$ and $\{ \hat{B}_{m},\hat{C}_{m},..., \; : m\neq 0 \}$ are discarded\footnote{Recall that the outputs of the modes $\{\hat{A}_{1},\hat{B}_{0},\hat{C}_{0},... \}$ in the equivalent circuit are identified with the outputs of $\{\hat{A},\hat{B},\hat{C},... \}$ in the original circuit; see Sec \ref{CTC1Basics}.}.  

Expectation values take the form:
\eqn{\label{expectEO}
\trc{\hat{U}^{\dagger}_{PN} \hat{U}^{\dagger}_{EC} \,  \hat{\sigma} \, \hat{U}_{EC} \hat{U}_{PN} \; \densop{0}{0} \, } \, .
}
In practice, the observable $\hat{\sigma}$ is obtained by applying some unitary transformation $\hat{U}_{\sigma}$ to the modes before performing measurements of the number of photons in each mode, using photon detectors. In particular, we can always write:
\eqn{
\hat{\sigma}(\hat{A}_{1},\hat{B}_{0},\hat{C}_{0},...) := \hat{U}^{\dagger}_{\sigma} \, (\hat{A}^\dagger_{1} \hat{A}_{1}) (\hat{B}^\dagger_{0} \hat{B}_{0})(\hat{C}^\dagger_{0} \hat{C}_{0}) ... \, \hat{U}_{\sigma} \, ,
}
so that the Heisenberg picture expectation value takes the form:
\eqn{\label{expectEOHeis}
\trc{  (\hat{A}'^\dagger_{1} \hat{A}'_{1}) (\hat{B}'^\dagger_{0} \hat{B}'_{0})(\hat{C}'^\dagger_{0} \hat{C}'_{0}) ... \; \densop{0}{0} \, } \, .
}
where the detected modes are defined by:
\eqn{ \label{detectmodes}
\hat{A}'_{1} &:=& \hat{U}^{\dagger}_{PN} \hat{U}^{\dagger}_{EC} \,  \hat{U}^{\dagger}_{\sigma} \; \hat{A}_{1} \; \hat{U}_{\sigma} \, \hat{U}_{EC} \hat{U}_{PN} \, , \nonumber \\
\hat{B}'_{0} &:=& \hat{U}^{\dagger}_{PN} \hat{U}^{\dagger}_{EC} \,  \hat{U}^{\dagger}_{\sigma} \; \hat{B}_{0} \; \hat{U}_{\sigma} \, \hat{U}_{EC} \hat{U}_{PN} \, , \nonumber \\
\; &&...\trm{etc.}
}
This will have relevance to our particular choice of operator ordering below. We can now frame our task as follows: given some choice of observable $\hat{\sigma}$, as a function of the modes $\hat{A}_{1,\phi_A},\hat{B}_{0,\phi_B},\hat{C}_{0,\phi_C}...$, (where we have made the spectral amplitudes explicit) and given some CTC circuit that acts as the unitary $\hat{U}_{EC} \hat{U}_{PN}$ for pointlike particles, we aim to replace the modes $\{ \hat{A}_{m,\phi_A},\hat{B}_{m,\phi_B},\hat{C}_{m,\phi_C}... : m=0,1,...,N \}$ appearing in the expression \eqref{expectEOHeis} with the generalised modes $\{ \hat{A}_{(m),\phi_A,\xi_A},\hat{B}_{(m),\phi_B,\xi_B},\hat{C}_{(m),\phi_C,\xi_C}... : m=0,1,...,N \}$ according to \eqref{RepEventop}. The resulting expectation value is then interpreted as a generalisation of \eqref{expectEOHeis} that applies to arbitrary field modes, with the original expectation value being recovered in the limit of pointlike wavepackets. In order to proceed, we must address an ambiguity in the ordering of the operators. For example, in the case of an OTC, we have $\hat{U}_{EC}=\hat{I}$ and $\hat{U}_{PN}=\hat{U}_{P,0} \hat{U}_{P,1}$ where $\hat{U}_{P,0}$ is a function of the modes $\hat{A}_{0}, \hat{B}_{0}, \hat{C}_{0}, ...$ and $\hat{U}_{P,1}$ is a function of $\hat{A}_{1}, \hat{B}_{1}, \hat{C}_{1}, ...$. From \eqref{detectmodes} we have the output of mode $A$ given by:
\eqn{\label{OpOrdExample}
\hat{A}'_{1} &:=& \hat{U}^{\dagger}_{P,0}\hat{U}^{\dagger}_{P,1} \, \hat{A}_{1} \, \hat{U}_{P,1} \hat{U}_{P,0} \, .
}
Since $\hat{U}_{P,1}$ and $\hat{U}_{P,0}$ commute, we could rewrite this expression in a number of equivalent ways. For example, in the above expression we could change the relative ordering of $\hat{U}_{P,1}$ and $\hat{U}_{P,0}$, or we could commute $\hat{U}^\dagger_{P,0}$ over to the right-hand side and use $\hat{U}^\dagger_{P,0}\hat{U}_{P,0}=\hat{I}$ to eliminate it altogether. Additionally, we could completely solve the Heisenberg equation of motion for $\hat{A}_{1}$, leading to an expression that is some function of the modes $\hat{A}_{1}, \hat{B}_{1}, \hat{C}_{1}, ...$. However, all of these expressions are equivalent only so long as the commutators $[\hat{A}_{0},\hat{A}^\dagger_{1}],\; [\hat{B}_{0},\hat{B}^\dagger_{1}],\; [\hat{C}_{0},\hat{C}^\dagger_{1}] ,\;...$, etc, vanish. Under the replacement \eqref{RepEventop}, these commutators become nontrivial and we will obtain physically distinct results depending on how we arrange the expression \eqref{OpOrdExample} just before making the replacement. It follows that we need to impose an operator ordering convention to remove this ambiguity. Formally, prior to making the replacement \eqref{RepEventop} in the expression \eqref{expectEOHeis}, we must place the modes in ordered form, eg:
\eqn{\label{Ordering1}
 \, \hat{A}'_{1} &\rightarrow& \mathcal{O}[\hat{A}'_{1}] \, \nonumber \\
 &=& \mathcal{O}[ \, \hat{U}^{\dagger}_{PN} \hat{U}^{\dagger}_{EC} \,  \hat{U}^{\dagger}_{\sigma} \; \hat{A}_{1} \; \hat{U}_{\sigma} \, \hat{U}_{EC} \hat{U}_{PN}  \, ] \, .
}
where $\mathcal{O}[...]$ indicates that the expression in the square brackets is ordered according to the rule $\mathcal{O}$. In this thesis, as in previous work on this topic, we will restrict our attention to the case where the detected modes $\hat{A}'_{1}, \hat{B}'_{0}, \hat{C}'_{0},...$ can be expressed as a linear combination of functions of the input modes and their conjugates, such that each function in the series only contains terms with a single value of $m$. Explicitly, we will assume that each mode can be written in the form\footnote{This condition is trivially met by the two cases that we will be concerned with in this thesis, namely the case of a beamsplitter on the CTC and the case of no interaction on the CTC (i.e. an OTC).}:
\eqn{\label{Ordering2}
\hat{A}'_{1}= \sum^{N}_{m=0} f_m(\hat{A}_{m},\hat{B}_{m},\hat{C}_{m},...) \, ,
}
for some set of functions $f_m$. We then define our operator ordering rule, denoted $\mathcal{O}_L$, as the requirement that the modes be placed in this form before applying the replacement \eqref{RepEventop}. We note that whether or not the output modes take this form depends only on the choice of the CTC interaction, because anything that happens prior to the CTC is contained within the preparation unitary $\hat{U}_P$, which is a tensor product $\otimes^N_{m=0} \, \hat{U}_{P,m}$ and so cannot mix terms of different $m$. In particular, we expect all cases of Gaussian interactions on the CTC to be expressible in the form \eqref{Ordering2}. For more general non-Gaussian interactions, such as optical implementations of controlled gates, the outputs will not be expressible in the form \eqref{Ordering2} and the operator ordering rule will need to be generalised.

Given the ordering rule \eqref{Ordering2}, we can now associate a unique expectation value to every observable $\hat{\sigma}$ as a function of the usual equivalent circuit plus the new functions $\xi_A, \xi_B, \xi_C,...$ and the quantity $\Delta \tau$. For a particular choice of these quantities, we have shown that the commutators corresponding to the usual Deutsch model can be recovered. This limit occurs when the `smearing' of the functions $\xi$ along $\tau$ is very narrow compared to $\Delta \tau$, which suggests a possible connection to the spacetime degrees of freedom of the field modes. However, it still remains to define these quantities in terms of the relevant physical parameters and to show that standard quantum optics can be obtained as another limit. We turn to this task in the next section.

\subsubsection{Physical interpretation and recovery of quantum optics\label{SecPhysInterp}}

So far, we have only demonstrated that it is possible to replace the equivalent circuit of Deutch's model with a more general construction. But what does this highly abstract formalism correspond to physically? We can characterise the departure from the standard equivalent circuit in terms of the altered commutation relations arising from the replacement \eqref{RepEventop}. Before making this replacement, the only nontrivial commutators take the form:
\eqn{ \label{precomms} 
[\hat{A}_{m,\phi_A}, \hat{A}^{\dagger}_{n,\phi_B}] &=& \int \invard{k} \; \phi_A(k,x) \phi^*_B(k,x) \, \delta_{m,n} \, , \nonumber \\
\, [\hat{A}_{m,\phi_A}, \hat{A}^{\dagger}_{n,\phi_C}] &=& \int \invard{k} \; \phi_A(k,x) \phi^*_C(k,x) \, \delta_{m,n} \, \nonumber \\
... \, etc,
}
with similar expressions for the other modes $\hat{B}_{m,\phi_B}, \hat{C}_{m,\phi_C}, ...$. These are just the usual effects of mode mismatch between the spectral amplitudes of the wavepackets. Note in particular that modes with different values of $m$ are necessarily orthogonal. After making the replacement \eqref{RepEventop} as discussed in the previous section, the commutators between different modes $\hat{A}, \hat{B}, \hat{C}...$ remain trivial, but the commutators \eqref{precomms} are replaced by:
\eqn{ \label{postcomms}
[\hat{A}_{m,\phi_A}, \hat{A}^{\dagger}_{n,\phi_B}] &=& \int \invard{k} \; \phi_A(k,x) \phi_B^*(k,x) \, \int \mathrm{d}\Omega \; \xi_A(\Omega,\tau+m\Delta t) \xi^*_B(\Omega,\tau+n\Delta t), \nonumber \\
\, [\hat{A}_{m,\phi_A}, \hat{A}^{\dagger}_{n,\phi_C}] &=& \int \invard{k} \; \phi_A(k,x) \phi_C^*(k,x) \, \int \mathrm{d}\Omega \; \xi_A(\Omega,\tau+m\Delta t) \xi^*_C(\Omega,\tau+n\Delta t),  \nonumber \, \\
... \, etc,
}  
where there is now a nontrivial overlap between the modes with different $m$, depending on our choice of the functions $\xi_A(\Omega,\tau), \xi_B(\Omega,\tau), ...$. An obvious choice is to define these functions directly in terms of the wavepackets of their corresponding modes. For example, $\xi_A$ is uniquely determined by $\phi_A$, $\xi_B$ is determined by $\phi_B$ and so on. However, this method of defining the functions $\xi_A, \xi_B,...$ is problematic, as we now show.

We expect our formalism to be equivalent to standard quantum theory in the case where $\hat{U}_{\trm{CTC}}=\trm{SWAP}$ and the CTC completely decouples from the circuit. Commutators between modes with different values of $m$ now do not occur and the observables depend only on the modes $\hat{A}_{0}, \hat{B}_{0}, ...$. We therefore require that the replacement \eqref{RepEventop} does not alter the commutation relations between these modes. However, setting $m=n=0$ in \eqref{precomms} and \eqref{postcomms}, we see that in general the two will not be equivalent. 

One way to fix this problem is to renormalise the modes, so that \eqref{postcomms} becomes:
\eqn{ \label{postcommsRenorm}
[\hat{A}_{m,\phi_A}, \hat{A}^{\dagger}_{n,\phi_B}] &=& \int \invard{k} \; \phi_A(k,x) \phi_B^*(k,x) \, \frac{ \int \mathrm{d}\Omega \; \xi_A(\Omega,\tau+m\Delta t) \xi^*_B(\Omega,\tau+n\Delta t) } { \int \mathrm{d}\Omega \; \xi_A(\Omega,\tau) \xi^*_B(\Omega,\tau)}, \nonumber \\
... \, etc.
}  

We see that whenever $m=n$, the dependence on the functions $\xi_A, \xi_B, ...$ then becomes trivial and we have agreement with \eqref{precomms}, as required. However, this prescription may lead to problems with the physical interpretation of the theory. In particular, the renormalisation procedure is only well-defined provided the function
\eqn{\label{RenormQuantity}
 \frac{ \int \mathrm{d}\Omega \; \xi_A(\Omega,\tau+m\Delta t) \xi^*_B(\Omega,\tau+n\Delta t) } { \int \mathrm{d}\Omega \; \xi_A(\Omega,\tau) \xi^*_B(\Omega,\tau)}
} 
takes values in between $0$ and $1$, these being the physical limits of the commutator corresponding to Deutsch's model and standard quantum theory, respectively. In cases where the wavepackets $\phi_A$, $\phi_B$, ..., and hence the amplitudes $\xi_A$, $\xi_B$, ..., are Gaussians, this condition is met, since the overlap between two Gaussians that are initially centred on the same point must be strictly decreasing as they are shifted apart. However, in general, there exist choices of the different $\xi$ functions in which \eqref{RenormQuantity}, and hence the commutator, becomes greater than one. To avoid this, while still allowing the $\xi_A$, $\xi_B$, ..., to differ from each other would require an additional fine-tuning of the commutator to ensure that it stays within the required range of values for all possible functions. Fortunately, we need not go down this path if we can find a reason to replace the different functions $\xi_A, \xi_B, ...$ by a single function $\xi(\Omega,\tau)$ that is uniquely defined in a given physical scenario. In that case, the renormalisation constants are trivially equal to 1. Of course, this resolution is contingent on the identification of a natural candidate for the function $\xi$.

Since any expectation value represents the expected value of an outcome registered by a specific detector, it seems natural to single out the detector's response function as a candidate for determining $\xi$. In a general scenario, however, we are faced with as many detectors as there are independent modes, hence the many different response functions $\phi_A, \phi_B, \phi_C, ...$. We therefore require a physical basis for singling out one of these detectors from the rest. Fortunately, the location of the CTC in Fig. \ref{genFieldCTC} can be used to break this symmetry. Note that we have singled out by convention the mode $\hat{A}$ as being the mode that traverses the CTC. In the equivalent circuit, this means that the CTC unitary is applied only to the `internal' modes $\hat{A}_m$, having nothing to do with the remaining modes $\hat{B}_m,\hat{C}_m, ...$ that represent the `chronology respecting' parts of the field. We therefore conclude that, at least in the case of a single CTC that we have been considering, it is most natural to postulate that the function $\xi$ be determined by the response function $\phi_A$ of the detector that is tuned to the mode emerging from the CTC. 

We are now in a position to define the quantities $\Delta \tau$ and $\xi(\Omega,\tau)$. For the remainder of this section, we will consider only the spatial component along the direction of propagation and restrict to 1+1 dimensions. Thus we return to using $x^{\mu}$ to designate a four-vector, reserving `$x$' for the positive spatial component in the direction from $P$ to $F$. Correspondingly, we consider only the components $(x,t)$ of the full set of co-ordinates $x^{\mu}=(x,y,z,t)$, and Lorentz transformations will be performed in this plane only. The temporal size of the CTC as seen by the detector is measured by the time it takes for a beam of light to travel from $P$ to $F$ in the detector's reference frame. In particular, let $(x',t')$ denote the rest frame of the detector and $(x'_p,t'_p)$ and $(x'_f,t'_f)$ the co-ordinates of $P$ and $F$, respectively. We then define $\Delta \tau$ to be the proper time between the two events $x^{\mu}_A=(0,t'_p)$ and $x^{\mu}_B=(0,t'_f)$.

Next, consider the function that characterises the detector's response in spacetime:
\eqn{ \Phi(x,t):= \int \iinvard{k} \; \phi(k,x,t) \, . \nonumber }
We have dropped the subscript `$A$', which is implicit in all that follows. Our goal is to define the function $\xi$ such that it captures the temporal variance of $\Phi$ in a way that can be compared to $\Delta \tau$. To do so, we consider the restriction of $\Phi$ to the time-axis of a set of co-ordinates which are at rest in the detector frame and whose origin is centred at the detection event $x^{\mu}_d=(x_d,t_d)$. In the detector's co-ordinates, the function confined to this slice is given by:
\eqn{
\Phi'|_{\trm{slice}}(t') = \int \iinvard{k} \; \phi(k,x'_d,t'-t'_d) \, .
}
or in explicitly covariant form:
\eqn{
\Phi|_{\trm{slice}}(\tau) = \int \iinvard{k} \; \phi(k,x^{\mu}_0(\tau)-x^{\mu}_d) \, .
}
where $x^{\mu}_0(\tau)$ is an invariant parameterisation of the slice such that $\tau=t'$ in the detector co-ordinates. It is then natural to compare the variance of $\Phi|_{\trm{slice}}(\tau)$ with the temporal displacement $\Delta \tau$. Let us interpret this function as the Fourier complement of $\xi(\Omega,\tau)$, hence:
\eqn{\label{eventopFunctionDef1}
\Phi|_{\trm{slice}}(\tau) &:=& \int \mathrm{d} \Omega \; \xi(\Omega,\tau) \, ,
}

or, after some manipulations:
\eqn{\label{eventopFunctionDef2}
\xi(\Omega,\tau) &:=& \int \mathrm{d} s \; \Phi|_{\trm{slice}}(s) \; e^{i \Omega (s-\tau)} \, \nonumber \\
&=& \int \mathrm{d} s \; \int \iinvard{k} \; \phi(k,x^{\mu}_0(s)-x^{\mu}_d) \; e^{i \Omega (s-\tau)} \, .
}

This completes our definition of $\xi(\Omega,\tau)$. It is important to note that, in taking $\tau$ as complementary to $\Omega$, we are treating $\tau$ as an additional degree of freedom independent of the other co-ordinates $(x',t')$. This is in accordance with our introduction of $\Omega$ as an independent spectral degree of freedom. However, by construction, this extra degree of freedom $\tau$ has no observable role to play in the laws of physics except when a CTC is present. We will therefore take \eqref{eventopFunctionDef2} as a postulate for the time being, with the physical interpretation of $\tau$ being left to the more speculative discussion at the end of this chapter (see Sec \ref{SecOTCNewPhys}).

\subsubsection{Comparison to a zero-delay feedback loop\label{SecFBLoop}}

So far, we have been describing a CTC as a unitary interaction between two systems, where the outputs of the top rail is sent back in time to become the input on the bottom rail. We can therefore think of a CTC as being a feedback loop with \tit{negative} delay, i.e. the particle is allowed to arrive at the input on the bottom rail \tit{before} it has left the output of the top rail\cite{DEU91}. Deutsch's consistency condition then ensures that the input on the bottom rail is identical to what the output from the top rail \tit{will be}. The generalised version of Deutsch's model described above allows us to shrink the size of this delay until the interaction becomes effectively instantaneous. In that limit, we simply require that the input on the bottom rail be the same as the output on the top rail at the instant of interaction. Of course, this limit can also be arrived at from the \tit{positive} direction. Imagine that the same unitary interaction from the CTC is implemented as an ordinary feedback loop, whereby the output of the top rail is fed into the input at the bottom rail after some positive time delay. As this delay shrinks to zero (which also requires the spatial locations of the respective output and input coincide) we should approach the same limit as that obtained for a CTC of vanishing size\cite{YAN}. In the examples that follow, we will see that this requirement is met.

\section{Examples\label{SecEopsExamples}}
In this section we will apply our formalism to the examples of previous sections, namely the case of a beamsplitter on a CTC from Sec \ref{CTC1BeamSplit} and the HUP-violating OTC circuit of Sec \ref{CTC1OpenCurve}. In each case, we show that the generalised formalism developed in previous sections smoothly tunes between the results obtained using Deutsch's model and the limit of a zero-delay feedback loop in standard quantum optics. As we have seen, the latter limit corresponds to a CTC whose temporal size is much smaller than the temporal resolution of the detector.

\subsection{The beamsplitter on a CTC revisited}
Recall from Sec \ref{CTC1BeamSplit} the case of a single mode traversing a CTC that contains a beamsplitter. The equivalent circuit is show in Fig. \ref{figECbeamsplit} of Sec \ref{CTC1BeamSplit}, with $\ket{\Psi_A}=\hat{U}_P \ket{0}$. From that section, the output mode can be written as:
\eqn{\label{EOBSoutUP}
\hat{A}'_1 &=&  \eta \sum^N_{m=1} \brac{e^{i \phi} \, \sqrt{1-\eta}}^{m-1} \hat{V}_m -e^{-i \phi}\sqrt{1-\eta} \hat{V}_0 \nonumber \\
&:=& \sum^N_{m=0} \, j_m \, \hat{V}_m \, ,
}
where $\eta$, $\phi$ are beamsplitter parameters and in the last step we have introduced the coefficients 
\eqn{\label{jmCoeff}
j_m &:=& \eta \brac{e^{i \phi} \, \sqrt{1-\eta}}^{m-1} \; \; \; \; \; \trm{for} \; m \neq 0 \; , \nonumber \\
j_0 &:=& -e^{-i \phi}\sqrt{1-\eta} \,  .
}
The modes resulting from the preparation procedure are:
\eqn{
\hat{V}_m &:=& \hat{U}^\dagger_P \, \hat{A}_m \, \hat{U}_P \nonumber \\
&=&  \hat{U}^\dagger_{P,m} \, \hat{A}_m \, \hat{U}_{P,m} \, ,
}
for some choice of preparation unitary $\hat{U}_P$. Note that, regardless of the choice of preparation, the expression \eqref{EOBSoutUP} is ordered according to the prescription $\mathcal{O}_L$, since $\hat{U}_{P,m}$ depends only on modes with a single value of $m$. 
 
\begin{figure}
 \includegraphics[width=14cm]{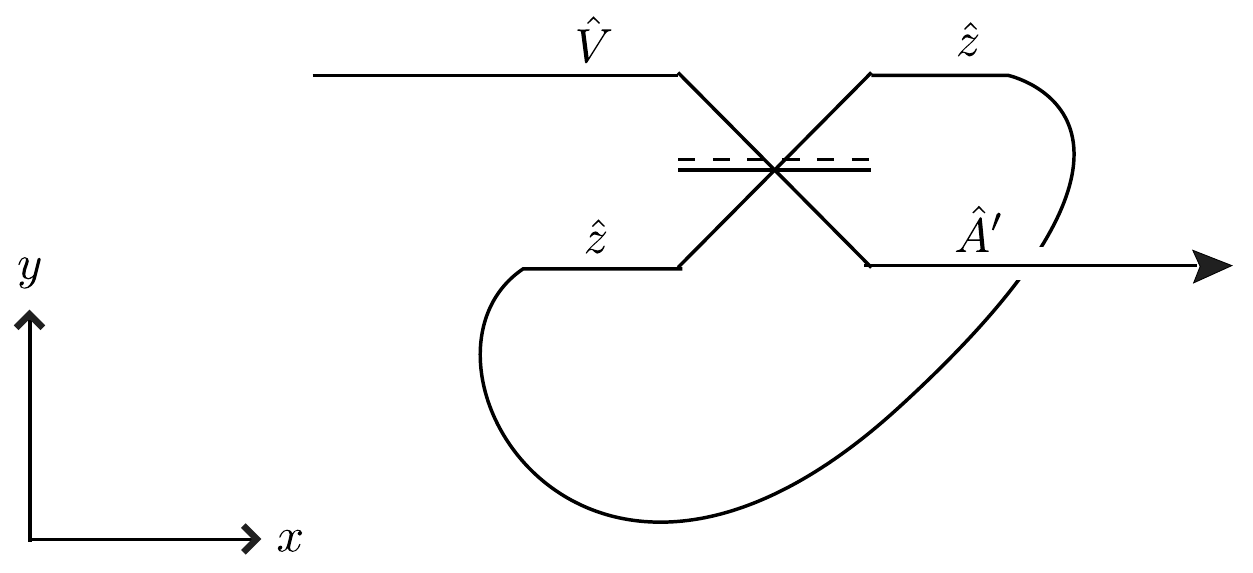}
 \caption[A beamsplitter with a feedback loop of positive delay]{\label{figFBloop}A beamsplitter with a feedback loop of positive delay (compare to Fig. \ref{figCTCbeamsplit}). Note that both axes on this diagram represent spatial dimensions, with the temporal properties left as implicit. The output of the top port of the beamsplitter takes a positive amount of time to reach the input of the lower port. When this delay approaches zero (by directly connecting the two ports) we obtain the same limiting behaviour as for the case when the two ports are connected through a CTC of zero size.}
\end{figure}

As in Chapter \ref{ChapFields} we will consider two classes of preparations: Gaussian pure states and a single photon states. In each case, we proceed by writing $\hat{A}'_1$ in the form \eqref{EOBSoutUP} and then replacing the vacuum modes according to \eqref{RepEventop} to obtain $\hat{A}'_{(1)}$. Arbitrary expectation values $\hat{\sigma}(\hat{A}'_1)$ are replaced by their generalised counterparts $\hat{\sigma}(\hat{A}'_{(1)})$ which can then be used to calculate the expectation values for modes with general spatial properties and CTCs of arbitrary size. Before considering specific cases, let us consider what happens in the limit of $\kappa \rightarrow 0$, i.e. for CTCs of vanishing size compared to the temporal uncertainty of the mode. In this limit, we can drop the index $m$ and replace $\hat{A}_{(m)}\mapsto \hat{A}$, leading to the output:
\eqn{\label{FBfromCTC}
\hat{A}'&=& \brac{ \sum^N_{m=0} \, j_m } \, \hat{V} \nonumber \\
&=& \frac{1-e^{-i\phi}\sqrt{1-\eta} }{1-e^{i\phi}\sqrt{1-\eta}}\, \hat{V} \nonumber \\
&:=& e^{-i\Phi(\eta,\phi)} \, \hat{V} \, .
}
Hence, in this limit, the CTC has no effect apart from the residual phase $\Phi$. We argued in Sec \ref{SecFBLoop} that this phase should be identical to that obtained in the limit of a positive-delay feedback loop as the delay shrinks to zero; let us check that this is true. Consider the beamsplitter with a feedback loop shown in Fig. \ref{figFBloop}. In the limit of zero delay, the mode exiting the top port is the same as the mode entering the lower port of the beamsplitter. This leads to the self-consistent evolution:
\eqn{
\hat{A'}=\sqrt{\eta} \, \hat{z} -e^{-i\phi}\sqrt{1-\eta} \hat{V} \, ,
}
where $\hat{z}$ satisfies:
\eqn{
\hat{z}=\sqrt{\eta} \hat{V}+e^{i\phi}\sqrt{1-\eta} \, \hat{z} \, .
}
We can solve iteratively for $\hat{z}$ leading to the output:
\eqn{
\hat{A'}=\brac{ \eta \, \sum^{\infty}_{m=0} (e^{i \phi} \sqrt{1-\eta})^m-e^{-i\phi}\sqrt{1-\eta} } \hat{V} \, ,
}
It can be verified by inspection that this is equivalent to \eqref{FBfromCTC}. The result \eqref{FBfromCTC} can also be obtained using the ``fractional linear transformation" that is more usual in the literature on coherent feedback control theory, see eg. Ref \cite{GOU08}. We conclude that in the case of a beamsplitter on a CTC, the generalised formalism gives the same predictions as a zero-delay feedback loop in the limit $\kappa \rightarrow 0$ of standard quantum theory.

\subsubsection{Gaussian pure states\label{SecBSGaussian2}}
Let $\hat{U}_P$ prepare the mode $\hat{A}_{\psi}= \int \invard{\tbf{k}} \, \psi(k,x) \, \hat{a}_{\tbf{k}} $ in an arbitrary Gaussian pure state. If we assume that this mode is matched to the detector response function, that is if $\psi(k,x)\equiv \phi(k,x)$, then we can ignore the effects of mismatch and simply write:
\eqn{\label{EOGauss}
\hat{V}_m &=& e^{i \theta_R} \,\trm{cosh}(r)\hat{A}_m+e^{i (\theta_R-2 \theta_S)} \,\trm{sinh}(r)\hat{A}^\dagger_m + \alpha \nonumber \\
&:=& p \, \hat{A}_m+q \, \hat{A}^\dagger_m +\alpha \, ,
} 
where the coefficients $p$ and $q$ have been introduced for convenience, and the implicit spacetime dependence is given by:
\eqn{
\hat{A}_{m,\phi}:= \int \invard{\tbf{k}} \, \phi(k,x) \, \hat{a}_{\tbf{k},m} \, .
}
The preparation $\hat{U}_P$  therefore creates a Gaussian pure state in the same field mode $\phi(k,x)$ seen by the detector. More generally, we might like to include the effects of mismatch, for $\psi(k,x)\neq \phi(k,x)$. This makes the problem much more complicated. However, we can simplify the mathematics if we assume that the squeezing is very small, $r\ll1$, such that the evolution of the mode $\hat{A}_{\phi}$ through the source unitary $\hat{U}_P$ can be truncated to first order in $r$. Then \eqref{EOGauss} can be replaced by:
\eqn{\label{EOGaussMismatch}
\hat{V}_m=\hat{A}_{m,\phi}+\bar{q} \, \hat{A}^\dagger_{m,\psi} +\bar{\alpha} \, ,
} 
where the mismatch is carried by
\eqn{
\bar{q}&:=&e^{i (\theta_R-2 \theta_S)} \,\trm{sinh}\brac{r \, \int \invard{\tbf{k}} \, \phi(k,x) \psi^*(k,x) } \nonumber \\
\bar{\alpha}&:=&\int \invard{\tbf{k}} \, \phi(k,x) \psi^*(k,x) \, \alpha \, .
} 
Previously, we have had to assume that the spacetime dependence of the modes are irrelevant to the physics, i.e. that the Fourier transforms of functions such as $\phi(k,x), \psi(k,x)$ in the spacetime domain are all strongly localised, hence `pointlike'. In relaxing this assumption to allow arbitrary field modes on spacetime, we will employ the generalised formalism that we have developed earlier in this chapter. Thus, it is the detector response $\phi(k,x)$ that will determine $\xi(\Omega, \tau)$ and hence the degree of observed coupling to the CTC in more general cases. 

Proceeding to the generalised model, we perform the replacement \eqref{RepEventop} to obtain:
\eqn{\label{EventopBS}
{A}'_{(1)}:=\sum^N_{m=0} \, j_m \, \brac{ p \, \hat{A}_{(m)}+\bar{q} \, \hat{A}^\dagger_{(m)} +\bar{\alpha} } \, ,
} 
where the spacetime dependence now includes the additional component $\xi(\Omega, \tau)$ and the CTC size $\Delta \tau$ as defined in the previous section.
Calculating expectation values in the vacuum state, the first and second-order moments are:
\eqn{\label{EventopMoments}
\bk{\hat{A}'_{(1)}} &=& \frac{1-e^{-i \phi} \sqrt{1-\eta}}{1-e^{i \phi} \sqrt{1-\eta}} \, \alpha \equiv e^{i \Phi(\phi,\eta)} \, \bar{\alpha} \, \\
\bk{\hat{A}'_{(1)}\, \hat{A}'_{(1)}} &=& \sum^N_{m=0}\sum^N_{n=0} \, j_m  j_n \, p \bar{q} \, C_{mn} +\, e^{2 i \Phi(\phi,\eta)} \bar{\alpha}^2 \\
\bk{\hat{A}'^\dagger_{(1)} \, \hat{A}'_{(1)}} &=& \sum^N_{m=0}\sum^N_{n=0} \, j^*_m  j_n \, |\bar{q}|^2 \, C_{mn}  + |\bar{\alpha}|^2 \, ,
}
where we have introduced $C_{mn}:=[\hat{A}_{(m)},\hat{A}^\dagger_{(n)}]$ and we have evaluated the summations analytically in the limit $N\rightarrow \infty$ where possibile. Comparing these to the moments obtained for the unmodified equivalent circuit in Sec \ref{CTC1BeamSplit}, we see that they are equivalent when $C_{mn}=\delta_{mn}$, as expected. Note that the dynamics is completely unchanged in the special case of a coherent state (setting $p=q=0$). As already discussed in the case of the equivalent circuit for point particles, coherent states do not interact with the CTC through the beamsplitter, so it is natural that this result also holds in the more general case.

The interesting cases are those for which the beamsplitter generates entanglement, such as when the input to the CTC is the squeezed vacuum. In this case, the CTC introduces extra noise, as we saw in Sec \ref{CTC1BeamSplit}. In our modified formalism, therefore, we expect this noise to disappear as the size of the CTC approaches zero compared to the variance of the modes. For simplicity, we will ignore mismatch effects and take $\psi(k,x)=\phi(k,x)$ in this calculation. We will furthermore concentrate our attention on the case where the effect of the CTC is greatest, choosing the beamsplitter phase to be $\phi=\frac{\pi}{2}$. In general, it is no longer a simple matter to evaluate the summations appearing in the second-order moments \eqref{EventopMoments}, due to the appearance of the nontrivial function $C_{mn}$. Even when this function is taken to be a Gaussian, as in \eqref{GaussianEOcomm}, it is a highly nontrivial matter to evaluate the summations analytically in the limit as $N$ tends to infinity\footnote{Believe me, I tried.}. Fortunately, by taking it to be a Gaussian, we observe that the term $C_{mn}$ must become very small whenever the exponential factor $\frac{1}{8}\sigma^2 \Delta{\tau}^2(m-n)^2$ is very large. This means that we can truncate the summation by ignoring all terms beyond a finite cutoff $|m-n|=X$, where $X$ is sufficiently large to render contributions from terms with larger values of $|m-n|$ negligible. We will still need to compute the remaining terms, but as we will see, they can be computed analytically for $N \rightarrow \infty$.

\begin{figure}
 \includegraphics[width=16cm]{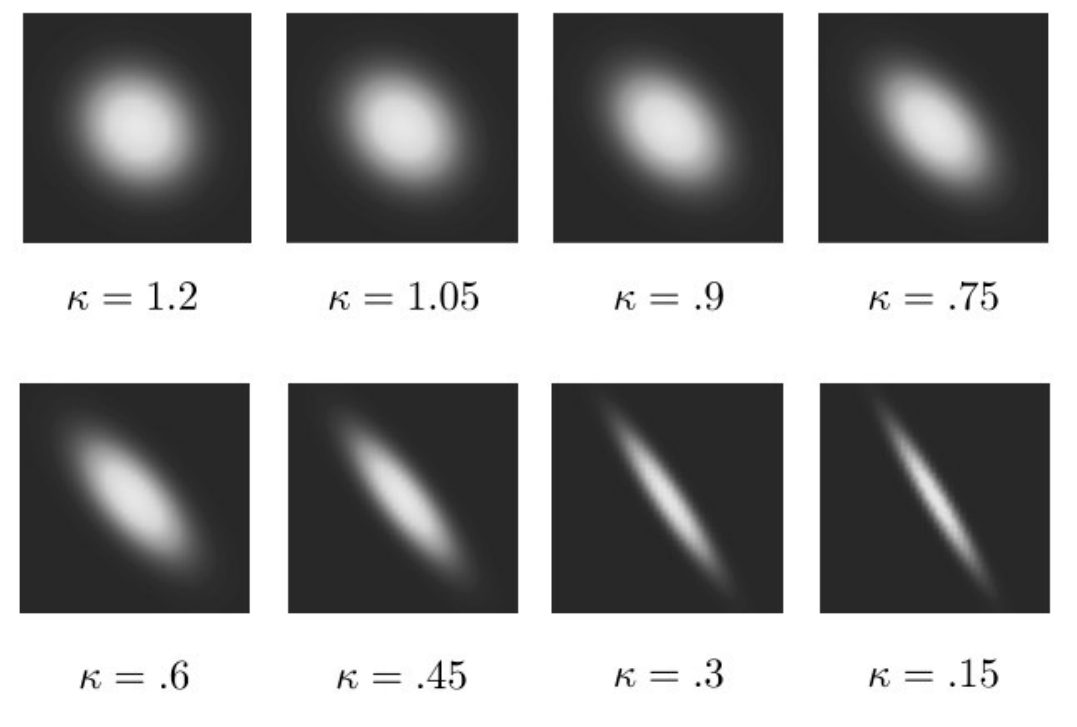}
 \caption[The Wigner function of the output from a CTC when the interaction is a beamsplitter and the input is the squeezed vacuum]{\label{figEOInterpGauss}The Wigner function of the output from a CTC when the interaction is a beamsplitter and the input is the squeezed vacuum. The beamsplitter parameters are held fixed at $\eta=\frac{2}{3}$, $\phi=\frac{\pi}{2}$ while the coupling to the CTC as measured by $\kappa$ is varied (see text). For $\kappa \gg 1$ we obtain the same results as predicted by the Deutsch model (compare to Fig. \ref{Wigner1} in Chapter \ref{ChapCTC1}). As $\kappa \rightarrow 0$ we smoothly recover the original squeezed state, up to a phase rotation, indicating that the CTC has become decoupled.}
\end{figure}

First, we will find it convenient to expand the terms:
\eqn{
\bk{\hat{A}'_{(1)}\, \hat{A}'_{(1)}} &=& pq\brac{  \sum^N_{m=1}\sum^N_{n=1} \, j_m  j_n \,e^{-\kappa^2 (m-n)^2}
+2 \sum^N_{m=1} \, j_m  j_0 \, e^{-\kappa^2 m^2} +j_0^2} \; , \label{EOexpandAA} \\     
\bk{\hat{A}'^{\dagger}_{(1)}\, \hat{A}'_{(1)}} &=& |q|^2 \, \brac{ \sum^N_{m=1}\sum^N_{n=1} \, j^*_m  j_n \, e^{-\kappa^2 (m-n)^2}
+ \sum^N_{m=1} \, j^*_m  j_0 \, e^{-\kappa^2 m^2}+\sum^N_{m=1} \, j^*_0  j_m \, e^{-\kappa^2 m^2} +|j_0|^2} \, , \nonumber \\.
&&  \label{EOexpandAdA}
}
where $\kappa^2:=\frac{1}{8}\sigma^2\Delta \tau^2$. The terms containing only a single summation can be straightforwardly truncated, eg:
\eqn{
\sum^N_{m=1} \, j_m  j_0 \, e^{-\kappa^2 m^2} \approx \sum^X_{m=1} \, j_m  j_0 \, e^{-\kappa^2 m^2}, 
\; \; \; \; X \gg \frac{1}{\kappa} \, .
}
Consider the double summation appearing in $\bk{\hat{A}'^{\dagger}_{(1)}\, \hat{A}'_{(1)}}$ in \eqref{EOexpandAdA}. 
To evaluate this term, we change indices to $v:=(n-m)$, $u:=(n+m)$; then using \eqref{jmCoeff} and setting $\phi=\frac{\pi}{2}$ we obtain:
\eqn{\label{DoublesumAdA}
&&\sum^N_{m=1}\sum^N_{n=1} \, \eta^2 (-1)^{m-1}(i \sqrt{1-\eta} )^{m+n-2} \,e^{-\kappa^2 (m-n)^2}  \nonumber \\
&=&\frac{\eta^2}{1-\eta} \, \sum^{1-N}_{v=N-1}\, e^{-\kappa^2 v^2} \, \sum^{2N}_{u=2} \, (-1)^{\frac{1}{2}(u-v)}(i \sqrt{1-\eta} )^{u} \nonumber \\
&\approx& \frac{\eta^2}{1-\eta} \sum^{X}_{v=-X}\, (-1)^{-\frac{1}{2}v}e^{-\kappa^2 v^2} \, \sum^{2N}_{u=2} \,(-1)^{\frac{1}{2}u}(i \sqrt{1-\eta} )^{u}  \, ,
}
where the index $u$ takes values from $2+|v|$ to $2N-|v|$ in increments of 2. For example, at $v=-5$, the index $u$ takes the values:
\eqn{
u=7,\,9,\,11,\, ...\,,\,2N-7,\,2N-5 \;.
} 
In the final step, we truncated the summation in $v$, leaving the remaining summation over $u$ to be evaluated analytically. To simplify this expression, we introduce a new index $w:=1+|v|,\,2+|v|,\,...\,,N-1,N$ that starts from $1+|v|$ and increases to $N$ in increments of 1, such that $u:=2w-|v|$. We then obtain:
\eqn{
&&\sum^{2N}_{u=2} \, (-1)^{\frac{1}{2}u}(i \sqrt{1-\eta} )^{u}  \nonumber \\
&=& \sum^{N}_{w=|v|+1} \, (-1)^{\frac{1}{2}(2w-|v|)}(i \sqrt{1-\eta} )^{2w-|v|} \, \nonumber \\
&=& (-1)^{-\frac{1}{2}|v|}(i \sqrt{1-\eta} )^{-|v|} \, \sum^{N}_{w=|v|+1} \,(-1)^{w}(i \sqrt{1-\eta} )^{2w} \, \nonumber \\
&=& (-1)^{-\frac{1}{2}|v|}(i \sqrt{1-\eta} )^{-|v|} \, \brac{ \frac{1}{\eta}(1-\eta)^{|v|+1} } \, ,
}
where we have evaluated the summation in the limit $N\rightarrow \infty$ in the final step. Substituting this result into \eqref{DoublesumAdA} yields:
\eqn{\label{TruncAA}
&& \eta^2 \, \sum^N_{m=1}\sum^N_{n=1} \, (-1)^{m-1} (i \sqrt{1-\eta} )^{m+n-2} \,e^{-\kappa^2 (m-n)^2}  \nonumber \\
&\approx& \eta \,\sum^{X}_{v=-X}\, (i)^{-v-|v|}\,(i \sqrt{1-\eta} )^{-|v|}(1-\eta)^{|v|}\, e^{-\kappa^2 v^2} \, .
} 
Following a similar procedure, we can simplify the double summation appearing in $\bk{\hat{A}'_{(1)}\, \hat{A}'_{(1)}}$ to the truncated form:
\eqn{\label{TruncAdA}
&& \eta^2 \, \sum^N_{m=1}\sum^N_{n=1} \, (i \sqrt{1-\eta} )^{m+n-2} \,e^{-\kappa^2 (m-n)^2}  \nonumber \\
&\approx& \frac{\eta^2}{2-\eta} \, \sum^{X}_{v=-X} \, (i \sqrt{1-\eta} )^{-|v|}(1-\eta)^{|v|}\, e^{-\kappa^2 v^2} \, .
} 
Finally, we substitute the truncated summations \eqref{TruncAA} and \eqref{TruncAdA} into the full expressions \eqref{EOexpandAA} and \eqref{EOexpandAdA} to obtain:

\eqn{
\bk{\hat{A}'_{(1)}\, \hat{A}'_{(1)}} &=& pq\, \frac{\eta^2}{2-\eta} \, \sum^{X}_{v=-X} \, (i \sqrt{1-\eta} )^{-|v|}(1-\eta)^{|v|}\, e^{-\kappa^2 v^2}  \nonumber \\
&+& 2pq\, \eta \sum^X_{m=1} \, (i \sqrt{1-\eta})^m \, e^{-\kappa^2 m^2} -pq\,(1-\eta) \; , \label{EOtruncfullAA} \\     
\bk{\hat{A}'^{\dagger}_{(1)}\, \hat{A}'_{(1)}} &=& |q|^2 \,  \eta \,\sum^{X}_{v=-X}\, (i)^{-v-|v|}\,(i \sqrt{1-\eta} )^{-|v|}(1-\eta)^{|v|}\, e^{-\kappa^2 v^2} \nonumber \\
 &-& |q|^2 \, \eta \,\sum^X_{m=1} \, (-i \sqrt{1-\eta})^m \, e^{-\kappa^2 m^2}-|q|^2\, \eta \,\sum^X_{m=1} \, (i \sqrt{1-\eta})^m \, e^{-\kappa^2 m^2} +|q|^2\,(1-\eta) \, . \nonumber \\
&& \label{EOtruncfullAdA}
}
With the convention $\theta_R=\theta_S+\frac{\pi}{2}$, we have $pq=-\trm{sinh}(r)\trm{cosh}(r)$ and $|q|^2=\trm{sinh}(r)^2$. The moments \eqref{EOtruncfullAA}, \eqref{EOtruncfullAdA} can be used to compute the covariance matrix of the output state and hence the Wigner function, for various choices of the parameters. In Fig. \ref{figEOInterpGauss} we have plotted the Wigner function for $\eta=\frac{2}{3}$ and varying choices of $\kappa^2=\frac{1}{8}\sigma^2\Delta \tau^2$, truncating after an appropriate number of terms in each case. Note that the spatial extent of the mode traversing the CTC is characterised by $1/\sigma^2$, i.e. by the inverse of the variance of $\xi(\Omega,\tau)$. Hence when the spatial variance of the mode is very large compared to the size $\Delta \tau$ of the CTC, we have $\kappa^2 \rightarrow 0$ and the CTC has no effect apart from a residual phase rotation. Conversely, when the mode is very narrow (pointlike) in space compared to the CTC, we have $\kappa^2 \rightarrow \infty$ and the extra noise predicted by Deutsch's model. From Fig. \ref{figEOInterpGauss}, we can see that our model is able to smoothly interpolate between these two limits as a function of $\kappa^2$.

\subsubsection{Single photon states\label{SecSingPhotCTC2}}
\begin{figure}
 \includegraphics[width=18cm]{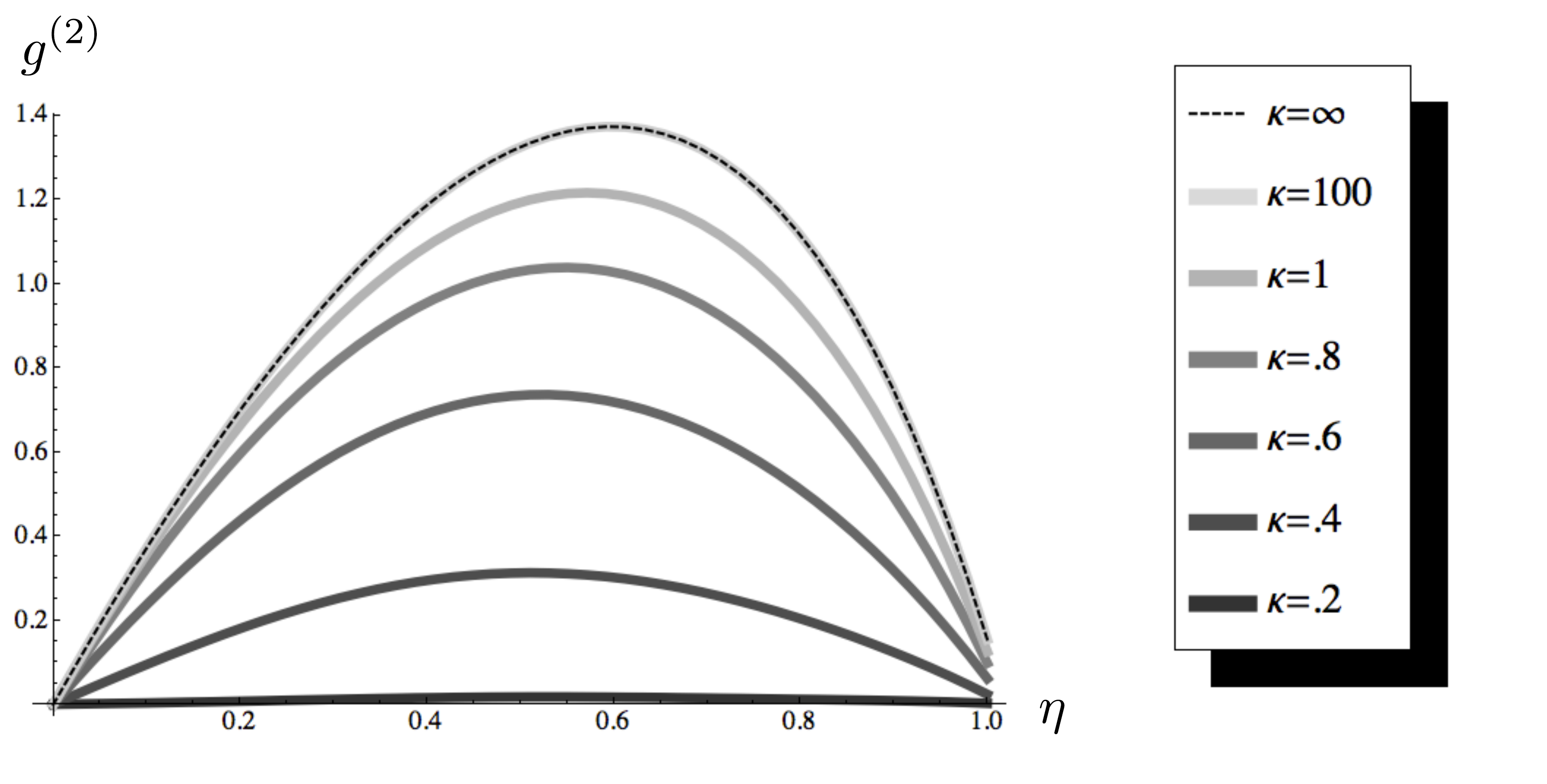}
 \caption[The fluctuations in photon number at the output of a CTC when the input is a single photon]{\label{figEOG2}The fluctuations in photon number at the output of a CTC, as measured by the second order correlation function $g^{(2)}$, when the input is a single photon. The value of $g^{(2)}$ is plotted here as a function of the beamsplitter reflectivity $\eta$ for different values of $\kappa$. We see that large values of $\kappa$ recover the functional dependence predicted by Deutsch's model (see \eqref{CTC1g2}), whereas for small $\kappa$ the noise disappears as the system decouples from the CTC.}
\end{figure}

Next, we would like to perform a similar analysis for the case where the input to the CTC is a single photon state, prepared from the vacuum. In this case, the output mode for the equivalent circuit will have the form:
\eqn{
\hat{A}'_1= \sum^N_{m=0} \, j_m \, \hat{V}_m \, \nonumber,
}
as before, except that now we take the mode $\hat{V}_m$ to contain a single photon. We can use our analysis of single photons in the Heisenberg picture from Chapter \ref{ChapFields} to write this mode in terms of a large set of ancillary vacuum modes:
\eqn{\label{EOsingphotfield}
\hat{V}_m = \sum_{j=1}^{M} \, \hat{d}_{j,m} \, \hat{v}'_{j,m} \, \prod_{i=0}^{j-1} \, \left( 1 - \hat{d}_{i,m} \right) + \hat{c}\hat{A}_m
}  
using the definitions of Sec \ref{SecSingPhot}. We apply the transformation \eqref{RepEventop} to obtain:
\eqn{\label{singphotEOsmall}
\hat{A}'_{(1)}= \sum^N_{m=0} \, j_{(m)} \, \hat{V}_{(m)} \, \nonumber,
}
for which we are interested in the physical quantities:
\eqn{\label{EOSPquantities}
\bk{\hat{n}_{A'_{(1)}}}&=& \bk{ \hat{A}'^\dagger_{(1)}\hat{A}'_{(1)} } \, , \\
g^{(2)}&=& \bk{\hat{A}'^\dagger_{(1)}\hat{A}'^\dagger_{(1)}\hat{A}'_{(1)}\hat{A}'_{(1)}} \, ,
}
where expectation values are in the vacuum state. Fortunately we do not need to use the full expression \eqref{EOsingphotfield} to calculate these quantities. It will suffice to use the following identities:
\eqn{\label{EOSPidentities}
&&\bk{ \hat{A}_{(m)} }=0 \, ,\\
&&\bk{ \hat{A}^\dagger_{(m)}\hat{A}_{(m)} }=1 \, ,\\
&&\bk{\hat{A}^\dagger_{(m)}\hat{A}^\dagger_{(m)}\hat{A}_{(m)}\hat{A}_{(m)}}=0 \, ,
}
which are characteristic of the single photon state (see Sec \ref{SecSingPhot}). From these, we now derive two further identities that will be useful. Recall the eigenmode decomposition from Chapter \ref{ChapFields}, whereby any mode $\hat{A}_{(m)}$ induces an orthonormal basis of modes $\{ \hat{A}^{(i)}_{(m)} \}$ for which we take $\hat{A}^{(0)}_{(m)}=\hat{A}_{(m)}$. Then an arbitrary mode $\hat{A}_{(n)}$ can be decomposed in terms of this basis as:
\eqn{
\hat{A}_{(n)}&=&\sum^{\infty}_{i=0} \, \lambda^{(i)}_{n/m}\, \hat{A}^{(i)}_{(m)} \nonumber \\
&=& C_{nm} \, \hat{A}_{(m)}+\sum^{\infty}_{i=1} \, \lambda^{(i)}_{n/m}\, \hat{A}^{(i)}_{(m)} \, ,
}
where 
\eqn{
\lambda^{(i)}_{n/m}:=[\hat{A}_{(n)},\hat{A}^{(i)\dagger}_{(m)}] 
}
and in particular, $\lambda^{(0)}_{n/m}=C_{nm}$. Recalling \eqref{EmodeOrthog}, we can then define the mode:
\eqn{
\hat{A}_{n/m} &:=& \frac{1}{\sqrt{1-|C_{nm}|^2}}\,\sum^{\infty}_{i=1} \, \lambda^{(i)}_{n/m} \, \hat{A}^{(i)}_{(m)} \nonumber \\
&=&\frac{1}{\sqrt{1-|C_{nm}|^2}}\, \brac{ \hat{A}_{(n)}-C_{nm}\, \hat{A}_{(m)} } \, ,
}
as the part of $\hat{A}_{(n)}$ that commutes with $\hat{A}_{(m)}$. For convenience, let us define $C_{n/m}:=\sqrt{1-|C_{nm}|^2}$ as the coefficient of the orthogonal component.Together with the identities \eqref{EOSPidentities} we then have:
\eqn{\label{AdACNM}
\bk{\hat{A}^\dagger_{(m)}\hat{A}_{(n)} }=\bk{\hat{A}^\dagger_{(m)} \brac{C_{nm} \hat{A}_{(m)}+C_{n/m}\, \hat{A}_{n/m} } } 
=C_{nm} \, ,
}
and
\eqn{\label{AdAdAACNM}
&&\bk{ \hat{A}^\dagger_{(m)}\hat{A}^\dagger_{(n)}\hat{A}_{(r)}\hat{A}_{(s)} } \nonumber \\
&=&\bk{ \hat{A}^\dagger_{(m)} \brac{ C_{mn}\hat{A}^\dagger_{(m)}+C_{n/m}\hat{A}^\dagger_{n/m} }   \brac{ C_{mr}\hat{A}_{(m)}+C_{r/m}\hat{A}_{r/m} }  \brac{ C_{ms}\hat{A}_{(m)}+C_{s/m}\hat{A}_{s/m} } } \nonumber \\
&=& C_{mr}\,C_{n/m}\,C_{s/m} \,\bk{\hat{A}^\dagger_{n/m} \hat{A}_{s/m} }+C_{ms} \,C_{n/m}\,C_{r/m} \,\bk{\hat{A}^\dagger_{n/m} \hat{A}_{r/m} } \nonumber \\
&=& C_{mr} \bk{ \brac{ \hat{A}^\dagger_{(n)}-C_{nm}\, \hat{A}^\dagger_{(m)} } \brac{ \hat{A}_{(s)}-C_{sm}\, \hat{A}_{(m)} }  } + C_{sm} \bk{ \brac{ \hat{A}^\dagger_{(n)}-C_{nm}\, \hat{A}^\dagger_{(m)} } \brac{ \hat{A}_{(r)}-C_{rm}\, \hat{A}_{(m)} }  } \nonumber \\
&=& C_{mr} \brac{C_{ns}-C_{mn}C_{ms} } + C_{ms} \brac{C_{nr}-C_{mn}C_{mr} } \nonumber \\
&=& C_{mr} C_{ns}+C_{ms}C_{nr}-2 C_{mn}C_{mr}C_{ms} \, .
}
These results are simply a generalisation of the identities \eqref{BeamsplitSPidentities2} derived in Sec \ref{SecSingPhot}. (Note that our decision to take $\{ \hat{A}^{(i)}_{(m)} \}$ as our basis is a matter of convention; the same results just derived can be obtained using an arbitrary basis $\{ \hat{A}^{(i)}_{(x)} \}$, although the calculations become longer.) We can now evaluate the expressions:
\eqn{
\bk{ \hat{A}'^\dagger_{(1)}\hat{A}'_{(1)} } &=& \sum^N_{m=0} \sum^N_{n=0} \, j^*_m j_n \, C_{mn} \label{AdAexpr} \\
\bk{\hat{A}'^\dagger_{(1)}\hat{A}'^\dagger_{(1)}\hat{A}'_{(1)}\hat{A}'_{(1)}}&=& \sum^N_{m=0} \sum^N_{n=0}\sum^N_{r=0} \sum^N_{s=0} \, j^*_m j^*_n j_r j_s \, \bk{\hat{A}^\dagger_{(m)}\hat{A}^\dagger_{(n)}\hat{A}_{(r)}\hat{A}_{(s)}} \label{AdAdAAexpr}\, ,
}
using the results \eqref{AdACNM}, \eqref{AdAdAACNM} and truncating the summations where needed, as in the previous section. From \eqref{EOtruncfullAdA}, we immediately have:
\eqn{
\bk{\hat{A}'^{\dagger}_{(1)}\, \hat{A}'_{(1)}} &=&   \frac{\eta^2}{2-\eta} \, \sum^{X}_{v=-X} \, (i \sqrt{1-\eta} )^{-|v|}(1-\eta)^{|v|}\, e^{-\kappa^2 v^2} \\
 &-& \, \eta \,\sum^X_{m=1} \, (-i \sqrt{1-\eta})^m \, e^{-\kappa^2 m^2}- \eta \,\sum^X_{m=1} \, (i \sqrt{1-\eta})^m \, e^{-\kappa^2 m^2} +(1-\eta) \, . \nonumber \\
 && \label{EOPhotnum}
}
The expression for $\bk{\hat{A}'^\dagger_{(1)}\hat{A}'^\dagger_{(1)}\hat{A}'_{(1)}\hat{A}'_{(1)}}$ is large and tedious to calculate. The full calculation and resulting expression can be found in the Appendix. From \eqref{EOPhotnum}, one finds numerically that $\bk{\hat{A}'^{\dagger}_{(1)}\, \hat{A}'_{(1)}}\approx 1$, independently of the choice of $\kappa$ or $\eta$. In fact, it is possible to show that, at least for the case of the beamsplitter interaction, the average photon number is always conserved, for \tit{any} choice of preparation $\hat{U}_P$. To prove this, note that
\eqn{\bk{\hat{A}'^\dagger \hat{A}'} &=& X\bk{\hat{V}^\dagger_{(m)} \hat{V}_{(m)}}+Y\bk{\hat{V}^\dagger_{(m)}} \bk{\hat{V}_{(m)}}, \nonumber \\
X &\equiv& \eta^2 \sum \limits^{\infty}_{m,n=0} e^{i(n-m)\phi}\sqrt{1-\eta}^{(n+m)}C_{nm}+(1-\eta), \nonumber \\
Y &\equiv& \eta^2 \sum \limits^{\infty}_{m,n=0} e^{i(n-m)\phi}\sqrt{1-\eta}^{(n+m)} \sqrt{1-|C_{nm}|^2} \nonumber \\
&-&\eta \sum \limits^{\infty}_{n=0}e^{i(n+1)\phi}\sqrt{1-\eta}^{(n+1)} C_{n,-1}-H.c.
}
Evaluating the summations numerically, one finds that $X\approx 1$ and $Y\approx 0$, leading to the result:
\eqn{\label{conserv} \bk{\hat{A}'^\dagger \hat{A}'}=\bk{\hat{V}^\dagger_{(m)} \hat{V}_{(m)}},}
which holds for any choice of input $\hat{V}_{(m)}=\hat{U}^\dagger_P \, \hat{A}_{(m)} \, \hat{U}_P$ and independently of the overlap $C_{nm}$. We have already seen that the presence of a CTC does not alter the expected photon number in Deutsch's model. The above result indicates that this remains true in between the limits of Deutsch's model and standard field theory. Regarding the behaviour of $g^{(2)}$, we recall from Sec \ref{SecSingPhot} that in the limit of Deutsch's model, the CTC acts as an absorber and emitter of photons, leading to a nonzero $g^{(2)}$. However, we expect these fluctuations in photon number to disappear as the size of the CTC shrinks to zero, or equivalently, as the modes become much longer than the CTC. In Fig. \ref{figEOG2} we have plotted $g^{(2)}$ as a function of $\eta$ for $\phi=\pi/2$ and different values of $\kappa$. We see that as $\kappa \rightarrow 0$, the fluctuations disappear and $g^{(2)}$ smoothly goes to zero as expected. 

\subsection{The OTC revisited}

In Sec \ref{CTC1OpenCurve} we showed that even when the mode traversing the CTC does not interact with itself -- a situation we referred to as an ``OTC" -- it is nevertheless possible to exploit the entanglement-breaking effect of Deutsch's model to violate Heisenberg's uncertainty principle. In this section we show that within our generalised formalism, a sufficiently small CTC can no longer be used to violate the uncertainty principle by this method.

Recall the circuit shown in Fig. \ref{OTCfig1}, consisting of an OTC, a single ancillary squeezed state, and two beamsplitters. Its equivalent circuit is shown in Fig. \ref{OTCfig2}. If the input is prepared in a coherent state of some unknown displacement $\alpha$, we obtain the output mode \eqref{OTCout1}:
\eqn{\label{EOsharpOTC}
\hat{A}'_1&=& \alpha + \frac{1}{2}\brac{ \hat{A}_{0} - i\, \trm{cosh}(r)\, \hat{B}_{0} + i \, \trm{sinh}(r)\, \hat{B}^\dagger_{0} } \nonumber \\
&+& \frac{1}{2} \brac{ \hat{A}_{1} + i \, \trm{cosh}(r)\, \hat{B}_{1} - i \, \trm{sinh}(r)\,\hat{B}^\dagger_{1} } \, ,
}
which has the form required by the ordering rule $\mathcal{O}_L$. Recall that this evolution results in the deterministic squeezing of the state along the chosen quadrature ($\hat{Q}$ by convention) without changing the average value, which is impossible to perform using standard quantum optics with only a single copy of the unknown input state. The violation of Heisenberg's uncertainty principle follows from a direct amplification of this effect. To generalise this evolution to arbitrary wavepackets, we perform the replacement \eqref{RepEventop}:
\eqn{\label{EOfuzzyOTC}
\hat{A}'_{(1)}&=& \alpha + \frac{1}{2}\brac{ \hat{A}_{(0)} - i\, \trm{cosh}(r)\, \hat{B}_{(0)} + i \, \trm{sinh}(r)\, \hat{B}^\dagger_{(0)} } \nonumber \\
&+& \frac{1}{2} \brac{ \hat{A}_{(1)} + i \, \trm{cosh}(r)\, \hat{B}_{(1)} - i \, \trm{sinh}(r)\,\hat{B}^\dagger_{(1)} } \, ,
}
where for simplicity we continue to ignore the effects of mismatch by setting the source and detector amplitudes to be equal, $\psi(k,x)=\phi(k,x)$. It is easy to see that the evolution represented by \eqref{EOfuzzyOTC} can be replicated by standard quantum optics in the limit of $\kappa \rightarrow 0$, i.e. when the CTC is small compared to the spatial extent of the mode. In this limit, the commutators become $[\hat{A}_{(m)},\hat{A}^\dagger_{(n)}]\approx1, [\hat{B}_{(m)},\hat{B}^\dagger_{(n)}]\approx1$ regardless of the indices $m,n$. Hence we can drop the indices and simply replace $\hat{A}_{(m)}\rightarrow \hat{A}, \, \hat{B}_{(m)}\rightarrow \hat{B}$ in this limit without affecting the physics. Applying this replacement to \eqref{EOfuzzyOTC} we obtain the trivial evolution: $\hat{A}'= \alpha +\hat{A}$, which is just the usual dynamics under the displacement operator $\hat{D}(\alpha)$ without any anomalous squeezing. Finally, to see that this limit is obtained smoothly in the limit of small CTCs, one can make use of the eigenmode expansion. Expanding with respect to the reference modes $\hat{A}_{(0)}, \hat{B}_{(0)}$ (for example) one finds that the matched part of the expression \eqref{EOfuzzyOTC} corresponds to the limit of standard theory given above, while the orthogonal part of the modes represents the anomalous squeezing due to the CTC. Since the coefficient of the orthogonal term is $\sqrt{1-|C_{1,0}|^2}$, this component gradually vanishes as $C_{1,0}\rightarrow 1$. Hence it becomes progressively harder to use an OTC to violate Heisenberg's uncertainty principle, approaching impossibility as the size of the CTC shrinks to zero.

\section{The generalised formalism as a quantum circuit\label{SecEOtoCircuit}}

So far, despite the apparent consistency of our `generalised formalism' for the cases considered, the procedure underpinning this formalism has remained somewhat mysterious. Formally, the replacement of the modes by \eqref{RepEventop} has the appearance of an embedding of the equivalent circuit into a larger Hilbert space. This embedding is somewhat analogous to the usual treatment of mismatch in quantum optics, wherein two modes that commute by virtue of non-overlapping spatial wavepackets will fail to commute if we allow more general wavepackets with nontrivial overlaps. In Chapter \ref{ChapFields} we saw how to incorporate these mismatch effects into the circuit by extending the Hilbert space of the circuit and treating the mismatch as a unitary rotation into the additional degrees of freedom. It is therefore natural to expect that the results obtained by means of our generalised formalism can be reproduced by an extended equivalent circuit, which treats the mismatch between the CTC rails as rotations parameterised by the overlaps $C_{nm}$. For the same reason that the equivalent circuit is useful for proving the internal consistency of Deutsch's model, an extended version of the equivalent circuit that matches our generalised formalism would confirm its internal consistency. However, it is not obvious that such an extended equivalent circuit exists in general. Rather than attempting to prove the general case, we will content ourselves with demonstrating that it can be done for the OTC circuit discussed in the preceding section and in Chapter \ref{ChapCTC1}. A more general assessment is left to future work.

\begin{figure}
 \includegraphics[width=16cm]{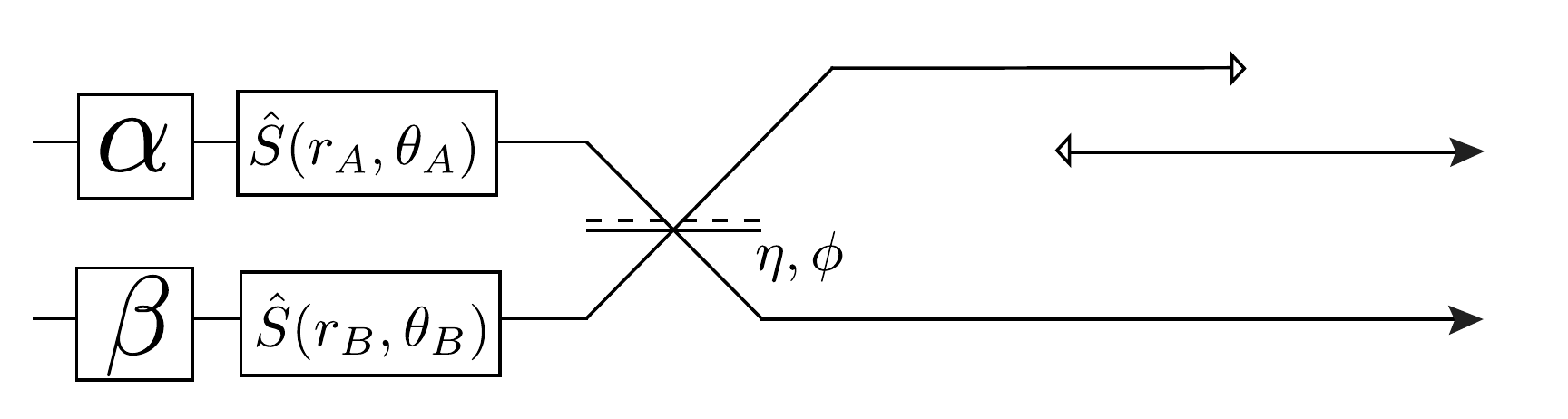}
 \caption[One arm of a general two-mode Gaussian state is sent through an OTC.]{\label{EOGaussOTC}One arm of a general two-mode Gaussian state is sent through an OTC.}
\end{figure}

\begin{figure}
 \includegraphics[width=16cm]{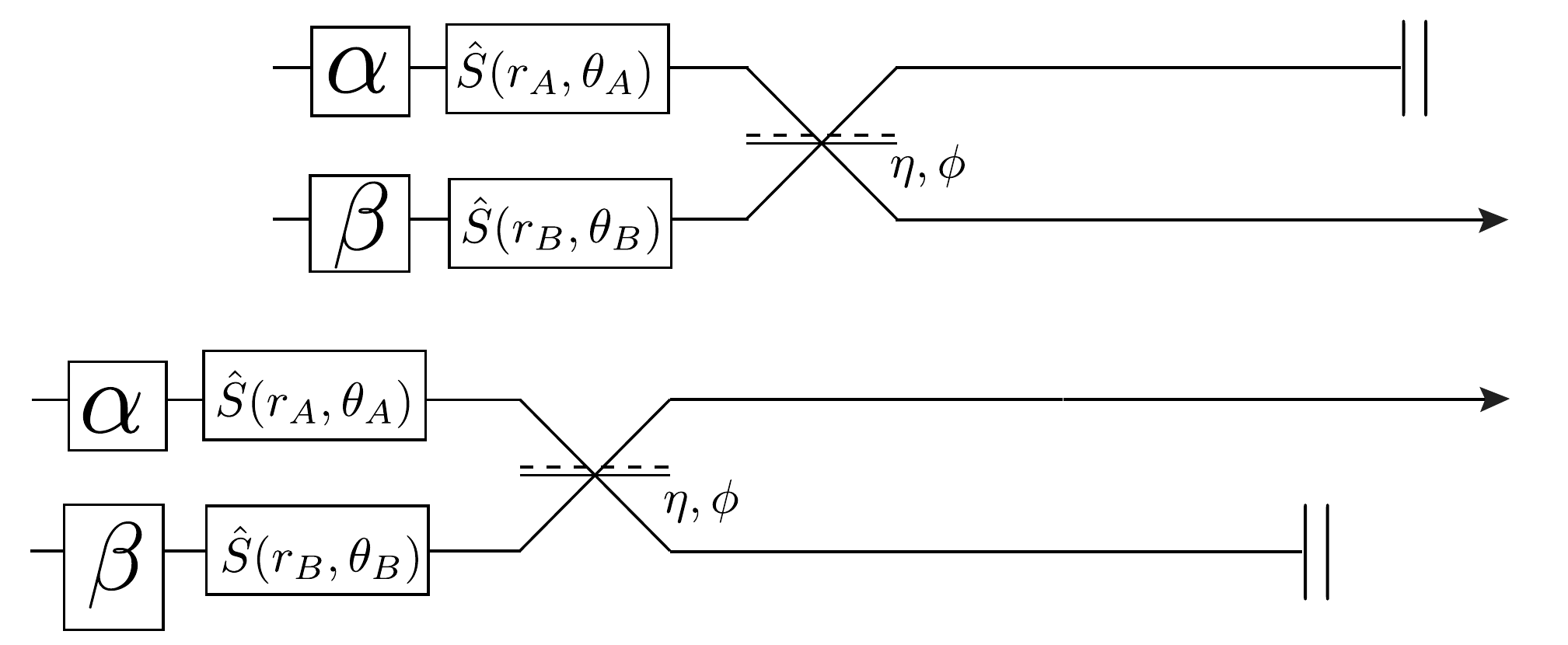}
 \caption[An equivalent circuit to the previous figure]{\label{ECOTCbase}An equivalent circuit to Fig. \ref{EOGaussOTC}.}
\end{figure}

\begin{figure}
 \includegraphics[width=16cm]{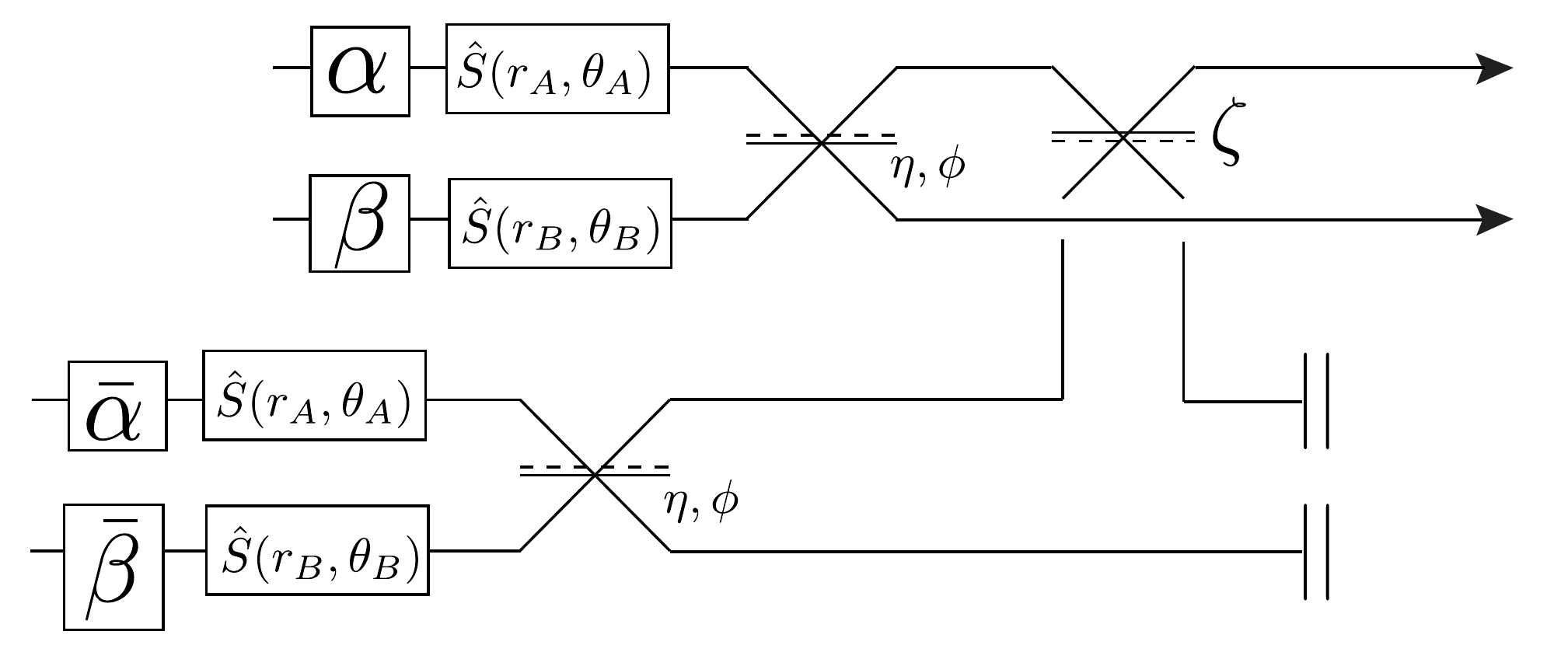}
 \caption[A circuit that reproduces the predictions of the generalised formalism]{\label{EOgeneralised} A circuit that reproduces the predictions of the generalised formalism. The reflectivity is defined as $\zeta:=|C_{1,0}|^2$ where $C_{1,0}$ is the commutator $[\hat{A}_{(0)},\hat{A}^\dagger_{(1)}]$.}
\end{figure}

Formally, we pose the problem as follows. Consider the circuit of Fig. \ref{EOGaussOTC}, which is the same as Fig. \ref{OTCfig1} except we have allowed for more general Gaussian preparations, given by the displacements $\alpha$, $\beta$ and the squeezings $\hat{S}(r_A,\theta_A)$, $\hat{S}(r_B,\theta_B)$ (arbitrary phase rotations of each mode are absorbed into the choice of phase convention for the beamsplitters). The equivalent circuit is shown in Fig. \ref{ECOTCbase}. Using our generalised formalism, this circuit yields the outputs:
\eqn{
\hat{A}'_{(1)}&=& \sqrt{\eta} \brac{\trm{cosh}(r_A) \hat{A}_{(1)}+e^{-i2\theta_A} \trm{sinh}(r_A)\hat{A}^\dagger_{(1)} +\alpha}\nonumber \\
&&+ e^{i \phi} \sqrt{1-\eta} \brac{\trm{cosh}(r_B) \hat{B}_{(1)}+e^{-i2\theta_B} \trm{sinh}(r_B)\hat{B}^\dagger_{(1)} +\beta} \nonumber \\
&:=& h_0+h_1\hat{A}_{(1)}+h_2 \hat{A}^\dagger_{(1)}+h_3\hat{B}_{(1)}+h_4 \hat{B}^\dagger_{(1)} \nonumber \\
\hat{B}'_{(0)} &=& \sqrt{\eta} \brac{\trm{cosh}(r_B) \hat{B}_{(0)}+e^{-i2\theta_B} \trm{sinh}(r_B)\hat{B}^\dagger_{(0)} +\beta} \nonumber \\
&&- e^{-i \phi} \sqrt{1-\eta} \brac{\trm{cosh}(r_A) \hat{A}_{(0)}+e^{-i2\theta_A} \trm{sinh}(r_A)\hat{A}^\dagger_{(0)} +\alpha} \nonumber \\
&:=& k_0+k_1\hat{A}_{(0)}+k_2 \hat{A}^\dagger_{(0)}+k_3\hat{B}_{(0)}+k_4 \hat{B}^\dagger_{(0)} \, ,\nonumber \\
\, \label{EOrecipe}
}
where we have introduced the coefficients $\{h_l\},\{k_l\},\, l\in\{0,1,2,3,4\}$ for clarity. If we rewrite \eqref{EOrecipe} using the eigenmode decomposition in the basis $\{ \hat{A}^{(i)}_{(0)}\},\{ \hat{B}^{(i)}_{(0)}\}$, the expression for $\hat{B}'_{(0)}$ remains unchanged while $\hat{A}'_{(1)}$ separates into:
\eqn{ \label{ArbGaussOuts}
\hat{A}'_{(1)}&=&  C_{1,0} \brac{h_0+h_1\hat{A}_{(0)}+h_2 \hat{A}^\dagger_{(0)}+h_3\hat{B}_{(0)}+h_4 \hat{B}^\dagger_{(0)} } \nonumber \\
&&+ \sqrt{1-|C_{1,0}|^2} \brac{ h_0+h_1\hat{D}+h_2 \hat{D}^\dagger+h_3\hat{E}+h_4 \hat{E}^\dagger }\, ,
}
where we have defined the additional `mismatch' modes:
\eqn{
\hat{D}:=\frac{1}{\sqrt{1-|C_{1,0}|^2}}\brac{ \hat{A}_{(1)}-C_{1,0} \hat{A}_{(0)} } \nonumber \\
\hat{E}:=\frac{1}{\sqrt{1-|C_{1,0}|^2}} \brac{\hat{B}_{(1)}-C_{1,0} \hat{B}_{(0)} } \, . 
}
We would like to interpret these outputs as arising from some equivalent unitary circuit $\hat{U}_{Eq}$ acting on the space spanned by $\{\hat{A}_{(0)}, \hat{B}_{(0)}, \hat{D}, \hat{E} \}$. Specifically, we seek a unitary $\hat{U}_{Eq}$ such that:
\eqn{\label{GeneralisedOuts}
\hat{A}'_{(1)} = \hat{U}^\dagger_{Eq}\, \hat{D} \, \hat{U}_{Eq} \, , \nonumber \\
\hat{B}'_{(0)} = \hat{U}^\dagger_{Eq}\, \hat{B}_{(0)} \, \hat{U}_{Eq} \, .
}
It turns out that the unitary represented by the circuit shown in Fig. \ref{EOgeneralised} satisfies this requirement. This circuit is similar to the original equivalent circuit of Fig. \ref{ECOTCbase}, except that the coupling to the extra rails is mediated by a beamsplitter with reflectivity $\zeta=|C_{1,0}|^2$ and the extra copies of the displacements are modified according to:
\eqn{
\bar{\alpha}:=\frac{1-\sqrt{\zeta}}{\sqrt{1-\zeta}} \alpha \, , \nonumber \\
\bar{\beta}:=\frac{1-\sqrt{\zeta}}{\sqrt{1-\zeta}} \beta \, .
}
It can easily be checked that when $C_{1,0}=0$ (hence $\zeta=0$), this corresponds to the original equivalent circuit, and when $C_{1,0}=1$ ($\zeta=1$) the circuit decouples from the CTC, leaving only a residual phase. In between these limits, the evolution of the modes agrees with \eqref{ArbGaussOuts}, which can be straightforwardly (but tediously) verified by computing the outputs \eqref{GeneralisedOuts} for this circuit and comparing them to \eqref{ArbGaussOuts}.

\section{Can OTCs lead to new physics?\label{SecOTCNewPhys}}

It is a common joke amongst physicists to say, when asked whether time travel is possible, ``Of course it is! We're travelling into the future as we speak." Indeed, according to GR, a certain degree of `time travel' is already permitted to us without requiring anything more strange than a simple gravitational well, as that produced by Earth. While generic gravitational time dilation does not permit one to travel into one's own past, it raises the important question of exactly what physical phenomenon in GR the Deutsch model purports to describe. It is generally assumed that the Deutsch model should only apply in situations where an actual CTC exists in the metric, and not otherwise. Alternatively, one might suppose that any form of temporal distortion between quantum systems brought about by the curvature of spacetime should be treated on equal footing. A CTC then loses its special status in the model, it being just another example of a metric leading to a temporal dislocation between different trajectories.

If we adopt this point of view, then it is natural to ask how Deutsch's model might be adapted to describe quantum mechanics in the presence of more general types of `time travel' within GR, particularly the phenomenon of gravitational time dilation. It is obvious that while a gravitational field produces a time difference between two clocks, it does not reproduce all of the effects of a CTC, which would also allow interactions between older and younger versions of the same system. However, the sub-class of situations in which there is no interaction on the CTC -- cases that we have deemed `OTC's -- do produce the same result as gravitational time-dilation, at least qualitatively, and from the point of view of an external observer. We might therefore attempt to design a theory of quantum mechanics in curved spacetimes that treats the effects of curvature on quantum systems in exactly as if the time difference were being generated by an OTC, rather than using the conventional methods of quantum field theory in curved spacetime. For such an alternative model to be feasible, it must agree with the predictions of standard quantum theory in the regimes for which we currently have data, namely the limit of weak curvature and for classical fields. However, for entangled systems across large gravitational gradients, we would expect the model's predictions to depart from those of standard quantum field theory; in particular we would expect to see decoherence effects due to the entanglement-breaking properties of OTCs. This opens up the possibility of performing an experimental test of such a model, which is reason enough to pursue this possibility further. In what follows, we will characterise a class of situations in which the time-dilation produced by a gravitational field on a beam of light can be reproduced by an OTC. This class of situations will then form the basis for applying the generalised formalism of this chapter to gravitational time-dilation. The possibility of extending this correspondence to more general situations, including massive fields, will also be discussed at the end.

\subsection{Correspondence between OTCs and time-dilation}

\begin{figure}
 \includegraphics[width=8cm]{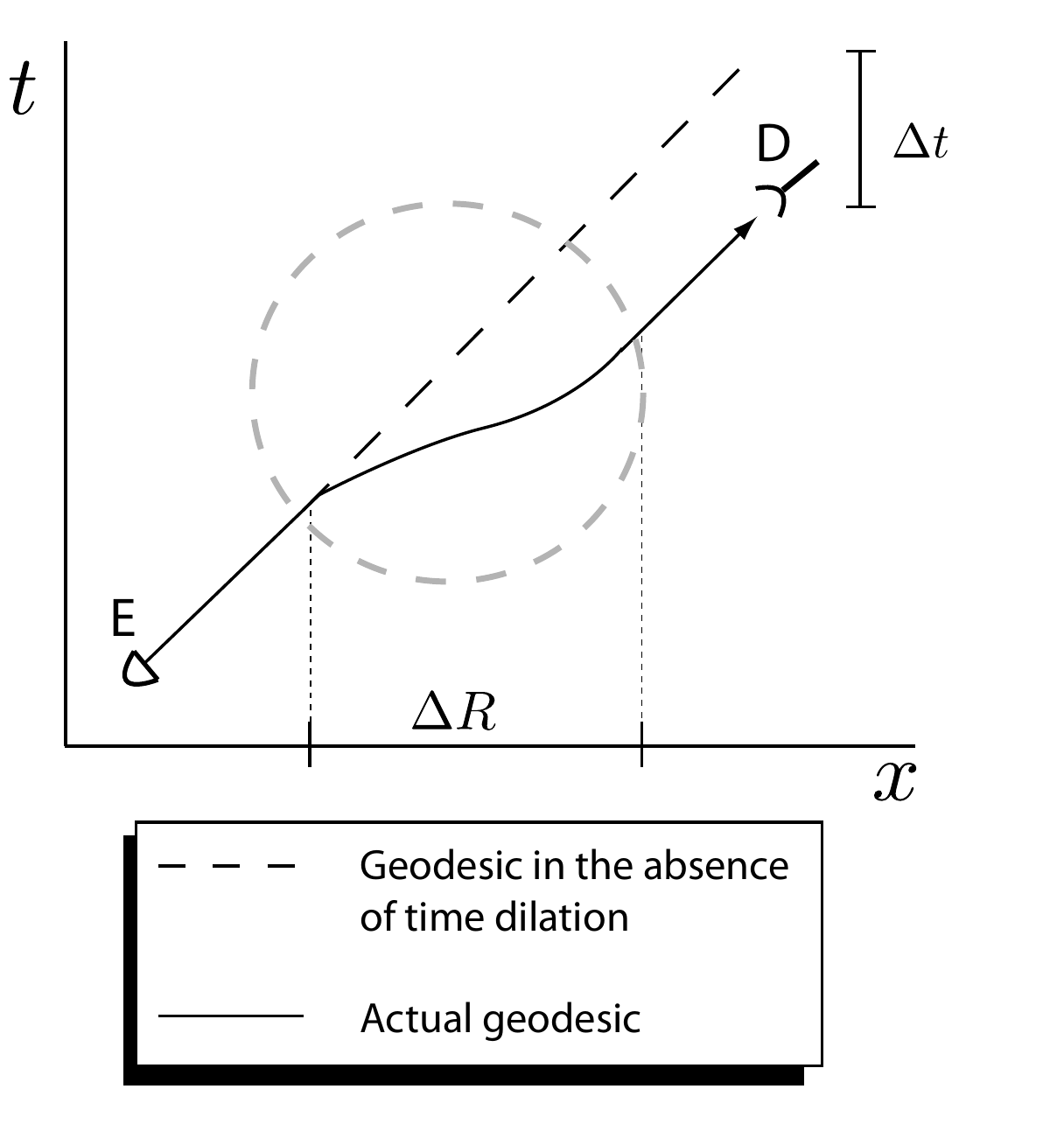}              
 \caption[Comparison of approximate null geodesics for light rays in curved spacetime]{\label{figST1}Comparison of approximate null geodesics for light rays in curved spacetime, as measured by observers situated at $E$ and $D$. The region $\Delta R$ is assumed to have significantly smaller curvature than the regions around $E$ and $D$. The effects of reduced curvature in the region $\Delta R$ causes the pulse to arrive early by an amount of time $\Delta t$ as measured by clocks at $E$ and $D$. }
\end{figure}

\begin{figure}
 \includegraphics[width=8cm]{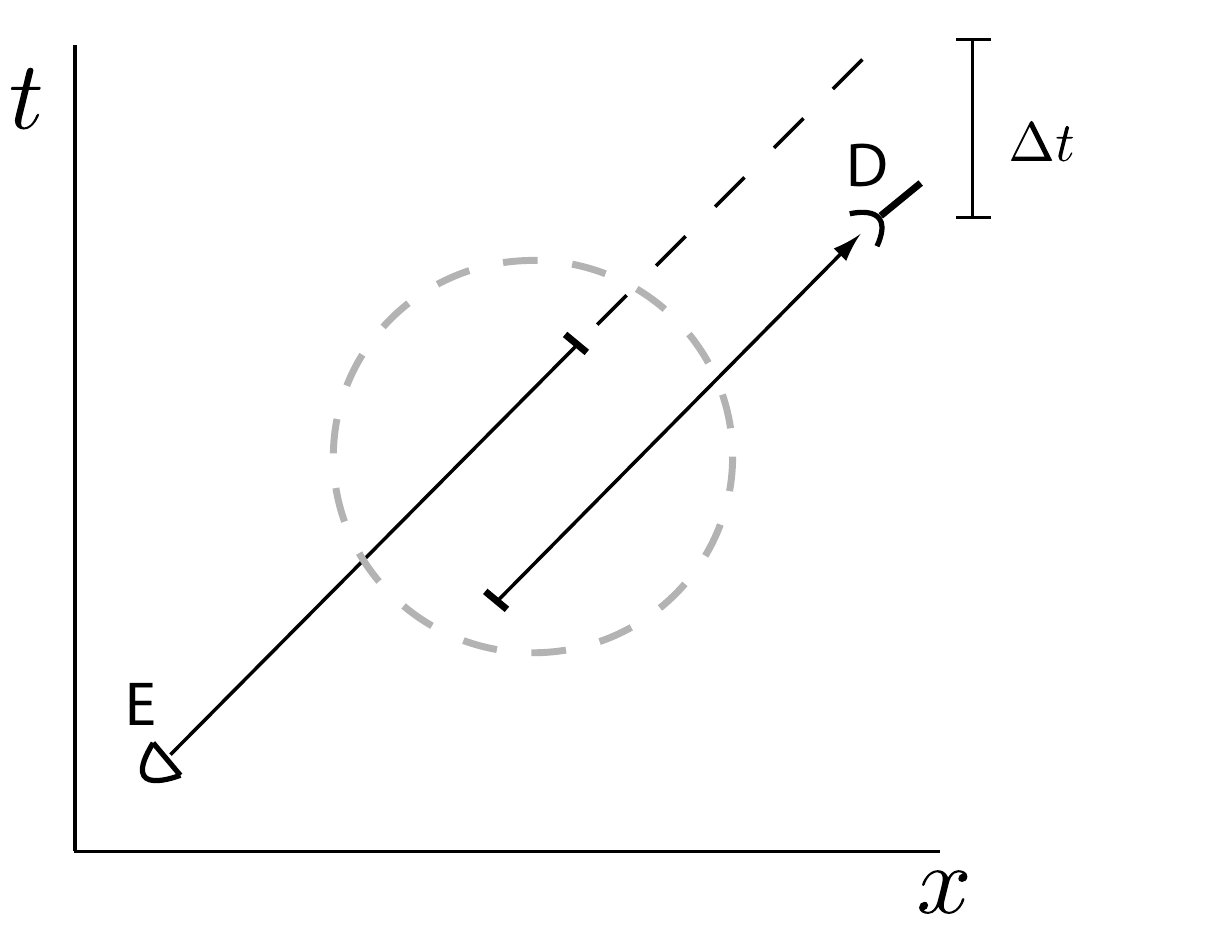}              
 \caption[An OTC that can be used to model the quantum effects of curvature]{\label{figST2}An OTC sends the pulse of light back in time by an amount $\Delta t$. Since there are no interactions inside the OTC, the only externally observable effect is that the pulse arrives early compared to flat spacetime. }
\end{figure}

Consider a pulse of light in a coherent state, emitted from an event $E$ and detected at $D$ as shown in Fig. \ref{figST1}. The metric is assumed to be static and the events $E$ and $D$ are assumed to be situated at the same gravitational potential. In between these two events lies a region $\Delta R$ in which the gravitational potential may vary, leading to distortions of the null geodesics and hence of the arrival time of the pulse. The average curvature of $\Delta R$ can then be deduced from the difference $\Delta t$ between the actual arrival time of the pulse and the expected arrival time in space with constant curvature, as measured in the reference frame of the detector. In general, there are two possibilities: either the average curvature in $\Delta R$ will cause the pulse to travel at less than the speed of light as seen from the lab co-ordinates, hence arriving late, or else the curvature will cause the pulse to apparently travel faster than the speed of light and arrive earlier than the flat space trajectory. In what follows, we will concentrate on the second possibility. In that case, the observers in this scenario cannot tell whether the time shift $\Delta t$ is due to curvature, or due to the pulse having been sent back in time by an OTC of size $\Delta t$ as shown in Fig. \ref{figST2}. We postulate that in all such scenarios where the time dilation can be simulated by an OTC, the effect on quantum systems should be modelled as though there \tit{were} an OTC present.

\begin{figure}
 \includegraphics[width=16cm]{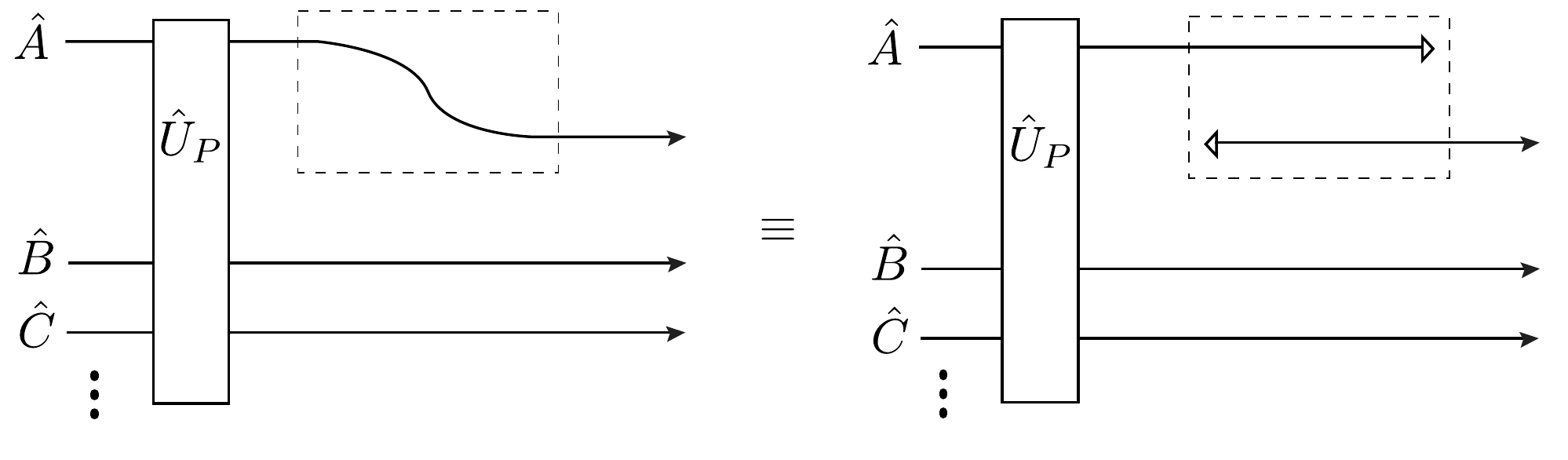}              
 \caption[Illustration of the correspondence between an OTC and time dilation]{\label{figTimeDilationCirc}The OTC--time-dilation correspondence principle. If mode $\hat{A}$ traverses a region of curved spacetime such that it gains an amount of time $\Delta t$ compared to a set of reference modes $\hat{B},\hat{C},...$, then the resulting physical effect is identical to the traversal of an OTC of size $\Delta t$ by the mode $\hat{A}$. For very large curvatures, this will lead to a breaking of entanglement between the modes as predicted by Deutsch's model. For small curvatures, the difference from standard quantum theory is undetectable (see main text for quantitative details).}
\end{figure}

In order to make this correspondence more precise, consider the circuit of Fig. \ref{figTimeDilationCirc}. Here, the modes $\hat{A},\hat{B},\hat{C},...$ are emitted at the same spacetime event $E$. The mode $\hat{A}$ follows a trajectory through the region of varying curvature $\Delta R$, while the remaining modes $\hat{B},\hat{C},...$ remain at the same gravitational potential. This can be achieved, for example, by sending $\hat{A}$ into the region $\Delta R$ and having it reflected back by a mirror. Without loss of generality, we assume that the detection events are timelike separated from one another. We then restrict ourselves to the case where the traversal of the region $\Delta R$ causes the mode $\hat{A}$ to be detected earlier than the other modes, as seen from any frame. Having thus singled out the mode $\hat{A}$, we choose the detection events to occur at the same position $x_d$ in the rest frame of the detector that receives mode $\hat{A}$. In these co-ordinates, the mode $\hat{A}$ arrives early by an amount $\Delta t$ due to the curvature. According to our postulate, this situation is equivalent to mode $\hat{A}$ traversing an OTC of size $\Delta t$.

As a specific application, suppose that a pair of entangled photons is emitted at event $E$ on Earth's surface. One can produce such a pair using weak spontaneous parametric down conversion (SPDC). The photons occupy the modes $\hat{A}$ and $\hat{B}$ respectively. Mode $\hat{B}$ is delayed at Earth's surface, while mode $\hat{A}$ is allowed to propagate to some height above Earth's surface before being reflected back by a mirror on an orbiting satellite. Due to time-dilation in the region $\Delta R$, the mode $\hat{A}$ is detected at position $x_d$ earlier than mode $\hat{B}$ by an amount $\Delta t$, in the detector frame. According to the proposed correspondence, this is equivalent to the OTC circuit of Fig. \ref{EOGaussOTC}, as described by our generalised formalism. Comparing this circuit to the predictions of standard quantum optics (simply replace the OTC with a phase shift from propagation), we see that the OTC model predicts a decorrelation of entanglement between the modes $\hat{A}$ and $\hat{B}$ not predicted by the standard theory. This effect will be greatest when the wavepacket envelope of mode $\hat{A}$ is sufficiently narrow compared to the time-shift $\Delta t$. Integrating the time along the null trajectory in the Schwarzchild metric one obtains $\Delta t = \frac{2 G M}{c^3} \trm{log}[\frac{r_e+h}{r_e}]$ where $h$ is the height above Earth's surface from which the mode $\hat{A}$ is reflected back, $M$ is the mass of the Earth and $r_e$ is its radius\cite{RAL09}. We have included the gravitational constant $G$ and the speed of light $c$ to ensure that $\Delta \tau$ has units of time. Since $\frac{h}{r_e}\ll 1$, we can approximate this quantity to first order to obtain $\Delta t \approx \frac{2 G M h}{c^3 r_e}=2(1.49\times10^{-11})\frac{h}{r_e} \, \trm{s}$. 
Based on current experimental data, we take the intrinsic temporal resolution of the detector to be $\sigma_j=2\times10^{-13}$ s. From \eqref{GaussianEOcomm} we can then calculate the overlap, $C_{0,1}=\trm{exp}\brac{-\frac{\sigma^2_j}{8}\Delta \tau^2}$, which differs significantly from $1$ when $\Delta \tau$ is comparable to $\sigma_j$, which occurs for heights of the order of $h=100$km. Therefore, significant decorrelation of entanglement will be seen if the beam is reflected back from this height or greater. We note that this result has been obtained previously in a similar calculation using `event operators' \cite{RAL09}. The connection of the present work to event operators will be discussed in the next section.

At this juncture, we should be careful to note that the similarity between the classical effects of OTCs and some instances of gravitational time dilation does not necessarily imply that these two phenomena \textit{must} be equivalent, but only that such an equivalence would not be inconsistent with present data. As such, we can use this observation as the starting point for two distinct lines of investigation. On one hand, we could have argued that the observation implies that OTC's should be modelled using semi-classical techniques\footnote{The author thanks Craig Savage for pointing this out.}. This could then be regarded as an extension of QFT to globally non-hyperbolic spacetimes. However, such a model would be of limited interest, since it would be confined to describing non-interacting fields on CTCs (i.e. OTCs) for which there already exist adequate treatments using path integrals \cite{GOL94}. Alternatively, one could take our observation as the basis for extending Deutsch's model into the accessible regimes, as we have done. We do not claim that the model should be correct in this regime, only that it is consistent with the most basic behaviour that we expect from classical GR, and therefore might be correct. Since it leads to a feasible and falsifiable alternative to the standard paradigm, the resulting model may provide a useful foil against which to test the standard theory in a new regime.

\subsection{Connection to event operators}
   
In earlier work, Ralph, Milburn \& Downes\cite{RAL09} proposed a modification of quantum optics in which the usual mode operators of the theory, eg.
\eqn{
\int \invard{k} \, \phi(k,x) \, \hat{a}_{\tbf{k}} 
}
are replaced by `event operators', namely:
\eqn{
\int \invard{k} \, \phi(k,x) \int \mathrm{d}\Omega \, \xi(\Omega,\tau) \, \hat{a}_{\tbf{k},\Omega} \, . 
}
As in the present work, the authors' motivation was to modify quantum optics so as to make it consistent with Deutsch's model in the limit of extreme curvature. While we begin from the same starting point, our discussion of the model includes many aspects left implicit in the original work, such as the explicit application of the model to CTCs and the discussion of operator ordering ambiguities. In addition, we have explicitly stated the conditions under which the model might relate to gravitational time dilation, using the OTC as physical motivation. The remaining key difference between the two formalisms is that the event operator formalism of Ref. \cite{RAL09} obtains $\Delta \tau$ as the difference in path-length between two trajectories through curved spacetime, which applies to more general cases than those considered so far in the present work. It turns out that we can obtain a model that exactly reproduces the predictions of event operators, by extending the OTC correspondence to more general situations, as we now demonstrate. 

First, instead of choosing the detection events to be timelike separated, we can also choose the convention such that they are spacelike separated and occur simultaneously in some reference frame. In this frame, we can define the invariant \tit{length} of a trajectory by the total time of travel as measured along the trajectory by a series of incremental observers, all at rest in the chosen frame. The effect of the curvature in region $\Delta R$ is then to increase the length of $\hat{A}$'s trajectory, for the cases that we are considering. In particular, if we choose the detection events to be simultaneous in the rest frame of the detector that receives $\hat{A}$, and measure the path length in this frame, then the curvature increases by an amount equal to $\Delta t$ compared to flat space. This increase in length along the geodesic can be accounted for by an OTC of the corresponding temporal size, thus recovering the original correspondence. 

Let us now propose an extension of this correspondence, based on the effect of curvature on the path length of a geodesic. Suppose, as before, that the modes $\hat{A},\hat{B},\hat{C},..$ are emitted at the same event $E$, which may be placed anywhere in the metric. The mode $\hat{A}$ traverses one trajectory through curved spacetime, while the remaining modes follow a different trajectory. As before, we take the detection events to be spacelike separated, but they may occur at arbitrary gravitational potentials, not necessarily the same as each other or the source event $E$. Let $\Delta t$ be the difference in length between the two trajectories (of which $\hat{A}$'s trajectory is assumed to be the longer) using the measure of path length defined above, in the rest frame of $\hat{A}$'s detector. The OTC correspondence principle then implies that this scenario is equivalent to $\hat{A}$ traversing an OTC of size $\Delta t$.

The advantage of this proposal is that it refers only to the difference in path length between a given pair of trajectories through spacetime, without placing any special conditions on the placement of the source and detectors in the gravitational potential. This is sufficiently general to reproduce the theory of event operators as described in Ref. \cite{RAL09}. We end this section with a few comments on the advantages and limitations of this model.

First, in reformulating the theory of event operators as an extension of Deutsch's model to time-dilation, we have improved the theory in several ways. First, we have given a precise account of the correspondence between time dilation and OTCs. This also provides an explicit connection between the treatment of optical fields on CTCs (see eg. Ref. \cite{PIE11} and Ch. \ref{ChapCTC1}) and the work on event operators, which are now seen to be two applications of the same formalism. Second, our model explicitly addresses the issue of operator ordering, which was not discussed in Ref. \cite{RAL09}, and is posed in a way that makes the relativistic invariance of the theory more explicit. Finally, by precisely characterising the class of situations in which the formalism applies, we have clarified the essential limitations of the theory and identified areas for future work.

One limitation of the model is our requirement that the longer of the two geodesics be traversed by only a single mode. In order to describe three or more entangled modes following distinct trajectories through spacetime, we would need to generalise the formalism of this chapter to accommodate multiple OTCs. This will encounter the same obstacles as the generalisation of Deutsch's model to multiple CTCs (see Sec \ref{SecMultiCTC}), as well as the potential for additional ambiguities regarding the definition of $\Delta t$ and $\xi(\Omega,\tau)$ when there are multiple OTCs present. Finally, one would hope to extend this model to massive fields and particles. For a slowly moving massive particle, the relevant indicator of curvature is the proper time measured by a physical clock carried with the particle's internal degrees of freedom. Specifically, referring to the scenario in Fig. \ref{figST1}, the time elapsed along a non-inertial trajectory of constant velocity from $E$ to $D$ through the region $\Delta R$ will be either greater or less than the time of travel measured by clocks stationed at $E$ and $D$, depending on the average curvature in $\Delta R$. In addition, one would need to decide whether the effects of special relativistic time dilation on a massive particle should also be modelled as an OTC effect, or whether the time-dilation of interest is strictly that which is due to spacetime curvature. The answers to these questions are left to future work.

\section{Appendix: Photon number fluctuations.}

In Fig. \ref{figEOG2} of Sec \ref{SecSingPhotCTC2} we showed a graph of $g^{(2)}$ based on a numerical calculation of $\bk{\hat{A}'^\dagger_{(1)}\hat{A}'^\dagger_{(1)}\hat{A}'_{(1)}\hat{A}'_{(1)}}$. Below, we include the details of this calculation. We evaluate the summations in the limit $N\rightarrow \infty$ where possible, but in most cases it is necessary to truncate the summations after an appropriate cutoff. Taking the $C_{nm}$ to have the Gaussian form \eqref{GaussianEOcomm}, the contribution from the truncated terms can be neglected to a good approximation provided the cutoff $X$ is inversely proportional to the parameter $\kappa:=\sqrt{\kappa^2}$.    

Starting from \eqref{AdAdAAexpr} and separating the coefficients $j_0$ from $\{ j_m : m\neq 0 \}$, we obtain an expression consisting of $16$ terms. By inspection of \eqref{AdAdAACNM} using \eqref{GaussianEOcomm}, we see that all terms with $(m=n=0)$ or $(m=r=s=0)$ will vanish. Of the remaining terms, those which are related by an exchange of the indices $r$ and $s$ are equivalent. We are left with eight distinct terms, which are subdivided into three terms each according to \eqref{AdAdAACNM}. The final expression can therefore be written:
\eqn{\label{APg2Expand}
g^{(2)}=\{1\}+\{2\}+\{3\}+\{4\}+\{5\}+\{6\}+\{7\}+\{8\} \, , 
}
where the individual terms will be defined below. In order to truncate the terms, we make use of the following identity:\\

\tit{Identity 1}\\
Let $v=(m-n)$ and $u=(m+n)$. Then:
\eqn{
\sum^{N}_{m=1}\sum^{N}_{n=1}\,f(m,n)=\sum^{N-1}_{v=1-N}\sum^{2N-|v|}_{u=2+|v|}\,f\brac{\frac{(u+v)}{2},\frac{(u-v)}{2}} \, ,
}
where the index $u$ increases in increments of $2$, i.e. $u=2+|v|, 4+|v|, 6+|v|, ..., 2N-|v|$. The proof of this identity follows from \eqref{DoublesumAdA} in Sec \ref{SecBSGaussian2}. Together with the standard manipulations of summations, we can use this identity to truncate the indices that appear in the exponentials, while evaluating any remaining summations in the limit of $N\rightarrow \infty$. For example,
\eqn{
\sum^{N}_{r=1} \sum^{N-1}_{v=1-N} \sum^{2N-|v|}_{u=2+|v|}\, f(v,u,r) e^{-\kappa^2 (v^2+r^2)} \nonumber \\
\approx \sum^{X}_{r=1} \sum^{X}_{v=-X} e^{-\kappa^2 (v^2+r^2)} \brac{ \sum^{\infty}_{u=2+|v|}\, f(v,u,r) } \, .
}
There is, however, one term that does not yield to this method. The term $\{ 1\}$ can be broken up into two sub-terms:
\eqn{
\{ 1\} &=& \sum^{N}_{m=1}\sum^{N}_{n=1}\sum^{N}_{r=1}\sum^{N}_{s=1}\,j^*_mj^*_nj_rj_s \, (C_{mr}C_{ns}+C_{ms}C_{nr}-2 C_{mn}C_{mr}C_{ms}) \nonumber \\
&=&  2\brac{\sum^{N}_{m=1}\sum^{N}_{n=1}\,j^*_m j_r \, e^{-\kappa^2 (m-r)^2} }^2 \nonumber \\
&&-2\sum^{N}_{m=1}\sum^{N}_{n=1}\sum^{N}_{r=1}\sum^{N}_{s=1}\,j^*_mj^*_nj_rj_s \, e^{-\kappa^2 \brac{(m-n)^2+ (m-r)^2+(m-s)^2}} \, ,
}
where we have obtained the first term by exploiting the symmetry of the functions under the exchange $r\leftrightarrow s$. The first sub-term can be truncated by first manipulating it according to \tit{identity 1} above; all terms appearing in $\{ 2\} - \{ 8 \}$ can be similarly dealt with. However, the second sub-term of $\{ 1\}$ above cannot be separated so easily into smaller parts. In fact, the author is sad to report that it was ultimately necessary to enumerate every possible relationship between the four indices and truncate each sub-sub-term individually. The method for carrying out this tedious procedure is outlined below.\\

First, consider the case $(m=n=r=s)$ for $m=1,2,...,N$. In this case, all the exponents become trivial and the sum can be evaluated analytically in the limit $N\rightarrow \infty$; setting $\phi=\frac{\pi}{2}$ and using the expressions \eqref{jmCoeff} for $j_m$ we obtain:
\eqn{
\sum^{N}_{m=1}\, |j_m|^4 = \frac{\eta^3}{2-\eta} .
}
That was the easy one. Next, consider all cases where three indices are equal and less than the value of the fourth index, eg. $(m=n=r)>s$, $(m=r=s)>n$, etc. In all, there are $\binom{4}{1}=4$ combinations. Two of these are equivalent due to the exchange symmetry of $r$ and $s$, leaving just 3 distinct terms, one of which has a factor of $2$ in front. As an example, the first term $(m=n=r)>s$ can be written in terms of the smallest index $s$ and a second index $u=(m-s)$ which also takes values $u=1,2,...,N$. Then we have:
\eqn{
\sum^{N}_{u=1} \sum^{N}_{s=1} e^{-\kappa^2 (u)^2}  \, |j_{(u+s)} |^2 \, j^*_{(u+s)} j_s \approx \sum^{X}_{u=1} e^{-\kappa^2 (u)^2} \sum^{\infty}_{s=1} \, |j_{(u+s)}|^2 \, j^*_{(u+s)} j_s \, ,
}
where we have truncated $u$ while taking the remaining summation over $s$ to infinity. Similarly, there are $\binom{4}{3}-1=3$ distinct terms in the class $m>(n=r=s)$ which can be evaluated in a similar manner. Moving along, we find $\binom{4}{2}-2=4$ distinct terms in the class $(m=n)>(r=s)$; we find $\binom{4}{2}\cdot \binom{2}{1}-5=7$ distinct terms in the class $(m=n)>r>s$, and the same for $m>(n=r)>s$ and $m>n>(r=s)$; and last but not least we find $\binom{4}{3}\cdot \binom{3}{2}\cdot \binom{2}{1}-12$ terms in the class $(m>n>r>s)$. Each of these terms is straightforward to truncate, following the strategy of summing to infinity over the smallest index, while truncating the remaining indices at $X$. For example, the term $s>m>n>r$ is given by:
\eqn{
&&\sum^{N}_{u=1}\sum^{N}_{v=1}\sum^{N}_{w=1}\sum^{N}_{r=1}\,j^*_{(r+u+v)}j^*_{(r+u)}j_{r}j_{(r+u+v+w)} \, e^{-\kappa^2 \brac{v^2+ w^2+(u+v)^2} } \nonumber \\
&&\approx \sum^{X}_{u=1} \sum^{X}_{v=1} \sum^{X}_{w=1} \, e^{-\kappa^2 \brac{v^2+ w^2+(u+v)^2}} \, \sum^{\infty}_{r=1} \, j^*_{(r+u+v)}j^*_{(r+u)}j_{r}j_{(r+u+v+w)} \, ,
}
where $n=r+u$, $m=r+u+v$ and $s=r+u+v+w$. Furthermore, since it is equal to $r>m>n>s$, we include a factor of two in front. After writing out all 44 distinct terms, with factors of two where necessary, we can perform the infinite summations with $\phi=\frac{\pi}{2}$ and the expressions for $j_m$ from \eqref{jmCoeff}, and add them together with all the remaining truncated terms to obtain an expression that yields an approximation of $g^{(2)}$ as a function of $X$, $\kappa$ and $\eta$. Evaluating this for $X=30$ and for different values of $\kappa$, we obtain the graph shown in Fig. \ref{figEOG2}.

\clearpage
\pagestyle{plain}
\pagebreak
\renewcommand\bibname{{\LARGE{References}}} 
\bibliographystyle{refs/naturemagmat2012}
\addcontentsline{toc}{section}{References} 

\chapter*{Conclusion and Outlook}
\addcontentsline{toc}{part}{Conclusion and Outlook} 

\lettrine[lines=3]{I}{n} this thesis, we have argued that Deutsch's model of causality violation for quantum systems can be given a consistent operational interpretation as a nonlinear box. In Chapter \ref{ChapNonlinBox} we derived necessary and sufficient conditions for any nonlinear box to be free from superluminal signalling, assuming that it is potentially verifiable (i.e. that it is not trivially equivalent to standard operational quantum mechanics). Using the formalism developed in Chapter \ref{ChapFields}, in Chapter \ref{ChapCTC1} we applied Deutsch's model directly to bosonic scalar fields to show that a CTC containing no interaction (an OTC) can be used to violate Heisenberg's uncertainty principle for quantum optics. Finally in Chapter \ref{ChapCTC2} we extended the model to wavepackets with arbitrary spacetime extent compared to the size of the CTC. Within this generalised framework, we showed that as the CTC shrinks to zero size compared to the coherence time of the wavepackets, we smoothly obtain the limit of a zero-delay feedback loop as described by standard quantum optics. Based on an observed correspondence between OTCs and gravitational time-dilation, we proposed that the framework be extended to causality-respecting curved metrics under certain conditions. The resulting predictions of our model were shown to be consistent to the event operator model proposed in Ref.\cite{RAL09}. Our model generalises event operators to CTCs and provides a clearer physical motivation for the formalism. We noted that the predictions of the model are testable using present technology.

If it is possible to retain signal locality in the presence of causality violation as this thesis suggests, we might entertain the possibility of causality violation at the level of Planck scale fluctuations in quantum gravity. However, this might entail violations of the unitarity and linearity as we have seen. While Deutsch's model retains a sensible probability interpretation, it does violate the uncertainty principle, allowing perfect measurements of canonically conjugate observables. While an increase in uncertainty due to fluctuations of the metric is expected in quantum gravity, the possibility that these same fluctuations might lead to more \tit{accurate} measurements is a novel possibility. 

The situation regarding unitarity of the model is similar to that in ordinary quantum mechanics, where there is some debate about whether the purification of the measurement process represents the real state of affairs or not, i.e. whether macroscopic objects are to be treated quantum mechanically. In the case of Deutsch's model, we have seen that there exists a unitary purification of the dynamics, but it requires the existence of multiple copies of the input state in order to account for the model's nonlinearity. For this reason, one is tempted to regard the purification as a mathematical convenience rather than attributing any physical significance to the extra copies, but we need not adopt this view if we wish to take unitarity as a fundamental property of nature. Related to the issue of unitarity is the status of energy conservation in the model. Although the model obeys energy conservation on average, as we have seen, energy conservation may fail in any single run of an experiment involving a CTC. This apparent non-conservation of energy does not pose a problem if one is willing to adopt the unitary picture described above, but otherwise it seems that one must relax the principle of energy conservation as it pertains to CTCs in quantum gravity. 

If we are prepared to accept the unusual features described above, then the model presented in this thesis provides a consistent framework for predicting how a probe field would interact with a microscopic CTC. It is a topic for future work to see whether this can be used to model the effects of space-time foam at the Planck scale. From an abstract point of view, the model provides an interesting conceptual perspective on the nature of time. Our model appears at least partly consistent with proposals that give time a tensor product structure, but further research is needed to elucidate this connection. Finally, in the absence of a classical background, we conjecture that our model might provide a framework for classifying the different types of causal connections allowed by quantum mechanics, including superpositions of causal structures and ``indefinite causality"\cite{ORE12}.

\clearpage
\pagestyle{plain}
\pagebreak
\renewcommand\bibname{{\LARGE{References}}} 
\bibliographystyle{refs/naturemagmat2012}
\addcontentsline{toc}{section}{References} 

\end{document}